\documentclass{msuphddissertation}
\usepackage{amsmath,amssymb,amsthm,paralist,graphicx}
\usepackage{epstopdf}
\author{Hui Wang}  
\title{Study of Particle Ratio Fluctuations and Charge Balance Functions at RHIC} 
\field{Physics} 

\begin{document}

\maketitlepage 
\begin{abstract}

The study of correlations between opposite-sign charge pairs and particle-ratio fluctuations can provide a powerful tool to probe the properties of the quark-gluon plasma (QGP). It has been suggested that the existence of a QCD phase transition would cause an increase and divergence of fluctuations. Thus the event-by-event particle-ratio fluctuations could be used to study strangeness and baryon number fluctuations near the critical point in the QCD phase diagram. On the other hand, the balance function, which measures the correlations between opposite-sign charge pairs, is sensitive to the mechanisms of charge formation and the subsequent relative diffusion of the balancing charges. The study of the balance function can provide information about charge creation time as well as the subsequent collective behavior of particles. 

For fluctuations we present dynamical $K/\pi$, $p/\pi$, and $K/p$ ratio fluctuations from Au+Au collisions at $\sqrt{s_{\rm NN}}$ = 7.7 to 200 GeV at the Relativistic Heavy Ion Collider using the STAR detector. Charge dependent results as well as multiplicity scaling properties of these fluctuation results are discussed. The STAR data are compared to different theoretical model predictions and previous experimental measurements. 

For balance functions we present results for charged particle pairs, identified charged pion pairs, and identified charged kaon pairs in Au+Au, $d$+Au, and $p+p$ collisions at $\sqrt{s_{\rm NN}}$ = 200 GeV.  These balance functions are presented in terms of relative pseudorapidity, $\Delta \eta$, relative rapidity, $\Delta y$, relative azimuthal angle, $\Delta \phi$, and invariant relative momentum, $q_{\rm inv}$.  In addition, balance functions are shown in terms of the three components of $q_{\rm inv}$: $q_{\rm long}$, $q_{\rm out}$, and $q_{\rm side}$. 

Beam energy and reaction-plane-dependent balance functions are also discussed in this paper. We will present charge balance function results for $\Delta \eta$ at $\sqrt{s_{\rm NN}}$ = 7.7 to 200 GeV. The normalized balance function width ($W$ parameter) is employed to compare different experimental measurements. The reaction-plane-dependent balance functions for Au+Au collisions at $\sqrt{s_{\rm NN}}$ = 200 GeV will be studied using the STAR detector. The reaction-plane-dependent balance function analysis is consistent with the three particle correlator analysis as expected mathematically. The model of Schlicting and Pratt incorporating local charge conservation and elliptic flow can reproduce most of the three-particle azimuthal correlation results at 200 GeV. 

\end{abstract}





\begin{acknowledgment}

First, I would like to thank my advisor, Gary Westfall, for taking me as his student and for all his support. Gary is always nice and patient, even when I blow up his computer. I really appreciate all his guidance over the past four years.

\vspace{9pt}

I would also like to thank Scott Pratt, for his great ideas and insightful discussions. He always shows me the way when I am struggling with formulas and equations. 

\vspace{9pt}

A lot of "Thank yous" to Terry Tarnowsky, for answering my questions and all the trivial games we have been to together.

\vspace{9pt}

Thanks to Soeren Schlichting for discussions.

\vspace{9pt}

Thanks to John Novak for proof reading my thesis.

\vspace{9pt}

Many thanks to the STAR Bulk Correlations Physics Working Group, especially Dan Cebra, Hiroshi Masui, Paul Sorensen and Shusu Shi for all the great suggestions. 

\end{acknowledgment}


\TOC

\LOT

\LOF


\newpage
\pagenumbering{arabic}
\begin{doublespace}

\chapter{Introduction}

\section{Quantum ChromoDynamics}

Throughout history, there is a fundamental question that has been asked: what is the smallest building block of matter? How do these building blocks interact with each other? Our best answer to this question to date is the Standard Model, which includes 12 fermions as elementary particles and gauge bosons as force carriers.  One missing piece from the Standard Model is the Higgs boson. The Standard Model also assumes two major interactions between particles - quantum electroweak and Quantum ChromoDynamics (QCD)~\cite{QCD_book}.

QCD is a theory that describes the strong interactions of the quarks and gluons. It is a non-abelian gauge field theory based on the $SU$(3) Yang - Mills theory of color-charged quarks~\cite{QCD_gauge}. In contrast, quantum electrodynamics (QED) assumes that charged particles interact by exchanging charge neutral particles. The non-abelian character of QCD does allow the force carrier, in this case gluons, to carry gauge charges and to couple to themselves.  Thus two peculiar properties of QCD are generated: color confinement~\cite{Confinement} and asymptotic freedom~\cite{Asymptotic_freedom}.

\begin{figure}
\centering
\includegraphics[width=32pc]{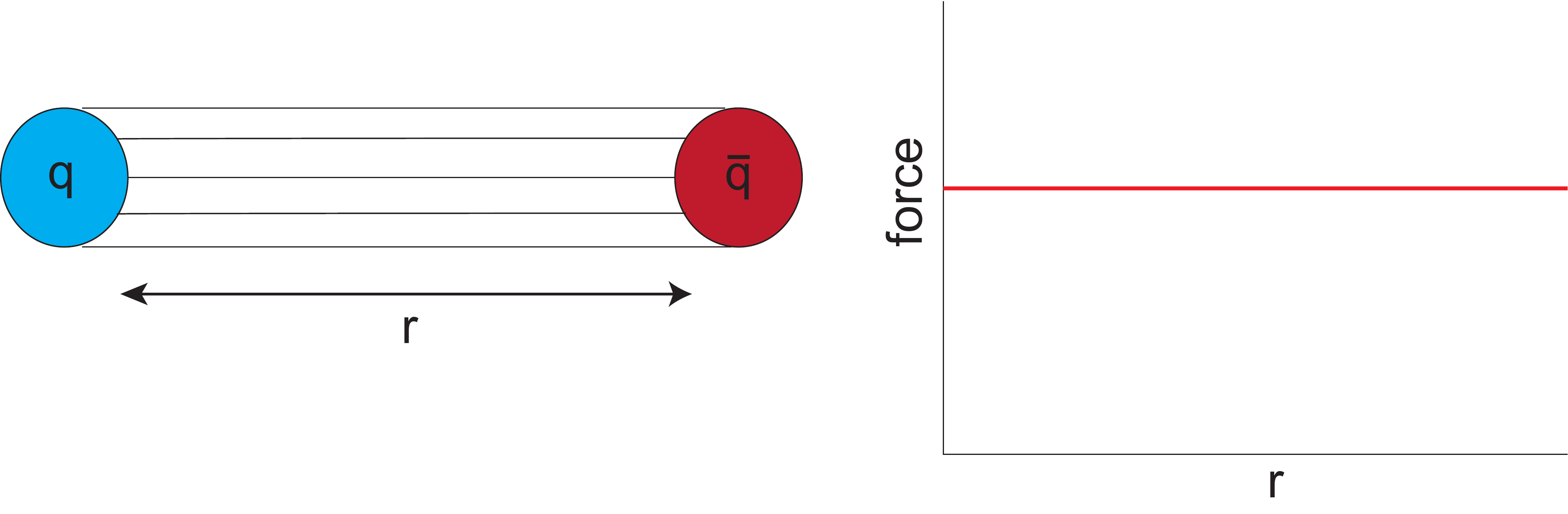}
\caption{\label{fig:fig1_1_01} A confining flux tube forms between distant quarks, which results in a constant force as a function of distance. For interpretation of the references to color in this and all other figures, the reader is referred to the electronic version of this dissertation.}

\end{figure}

The spatial potential between strongly interacting quarks can be described as

\begin{eqnarray}
\label{eq:eq_1_1}
V_{\rm s} (r) =  - \frac{4}{3}\frac{{\alpha _{\rm s} }}{r} + kr
\end{eqnarray}

\noindent
where $r$ is the quark spacing, $\alpha _{\rm s} $ is the strong coupling constant and $k$ is a constant that describes the long range interactions. If quarks interact at short distances, the first term dominates, which can be understood as the exchange of a single gluon.  This results in a 1/$r$ potential. However, as two quarks separate, due to the fact that gluons also carry color charge and can interact with themselves, the gluon fields form narrow tubes of color charge and cause a constant force between quarks as shown in Figure~\ref{fig:fig1_1_01}. This means that it would require infinite energy to remove quarks from hadrons. In particle collisions,  new quark - antiquark pairs will spontaneously appear to break the tube and form new hadrons. Thus quarks are always confined in hadrons.

\begin{figure}
\centering
\includegraphics[width=30pc]{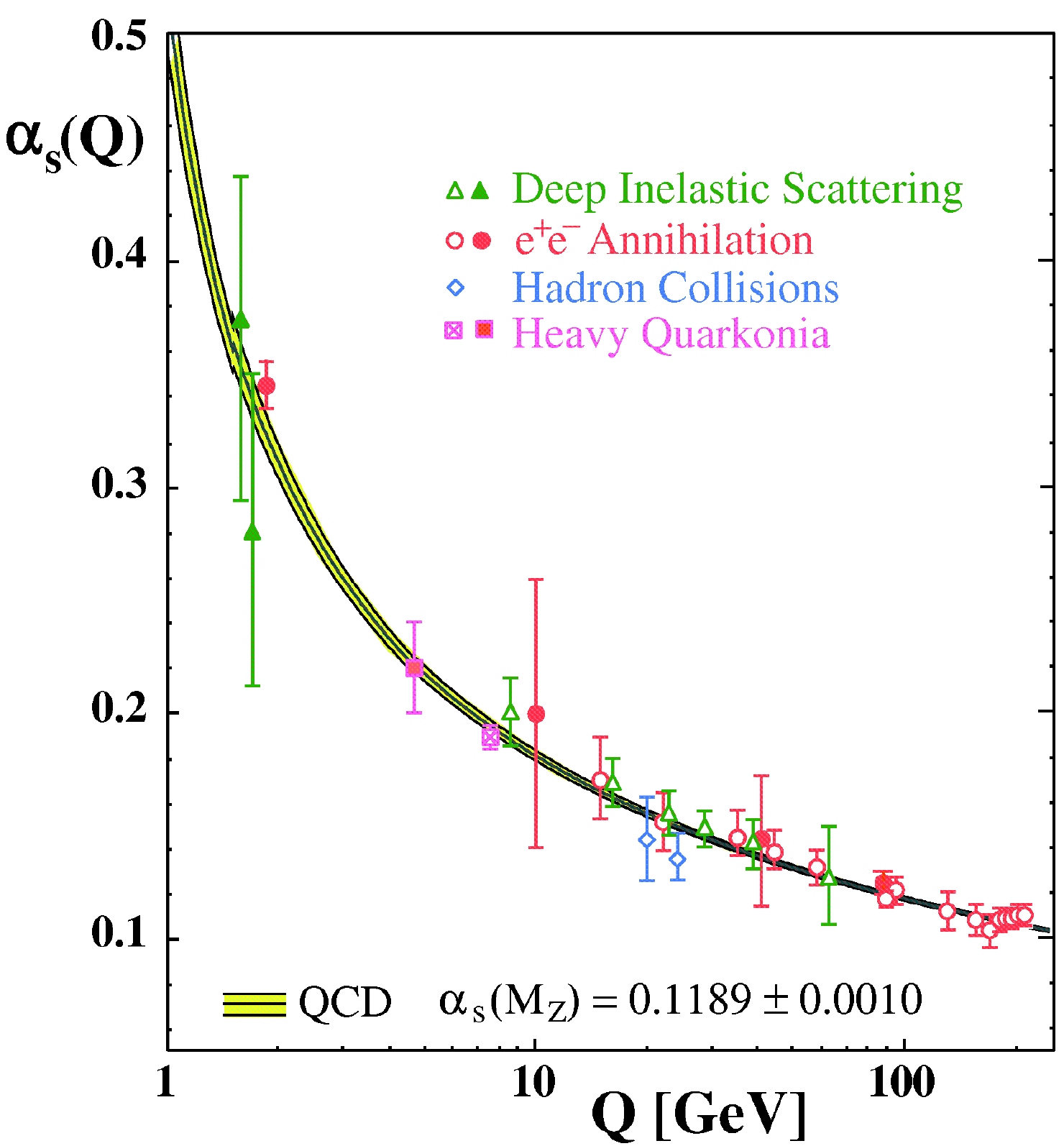}
\caption{\label{fig:fig1_1_02} Summary of measurements of $\alpha_{\rm }s$ as a function of the respective energy scale Q~\cite{alpha_paper}. }

\end{figure}

Similar to the running coupling constant in QED, the renormalized QCD effective coupling constant $\alpha _{\rm s} (\mu )$ depends on the renormalization scale. However, due to gluon self-interaction, the dependence is different.  Coupling constant $\alpha _{\rm s} (\mu )$ can be written as $\alpha _s (\mu ) \equiv \frac{{g_s^2 (\mu )}}{{4\pi }} \approx \frac{4\pi}{{\beta _0 \ln (\mu^2 /\Lambda ^2 )}}$, where $\Lambda$ is the QCD scale and $\mu$ is the momentum transfer scale. When $\beta_0 > 0$, the coupling decreases logarithmically with increasing energy, which is known as asymptotic freedom. Asymptotic freedom also suggests that the QCD can only be calculated perturbatively for high momentum transfer interactions or over short distances. Figure~\ref{fig:fig1_1_02} shows the measurements of $\alpha_{\rm s}$ as a function of the respective energy scale $Q$~\cite{alpha_paper}.

\section{Quark Gluon Plasma and the QCD Phase Diagram}

\begin{figure}
\centering
\includegraphics[width=36pc]{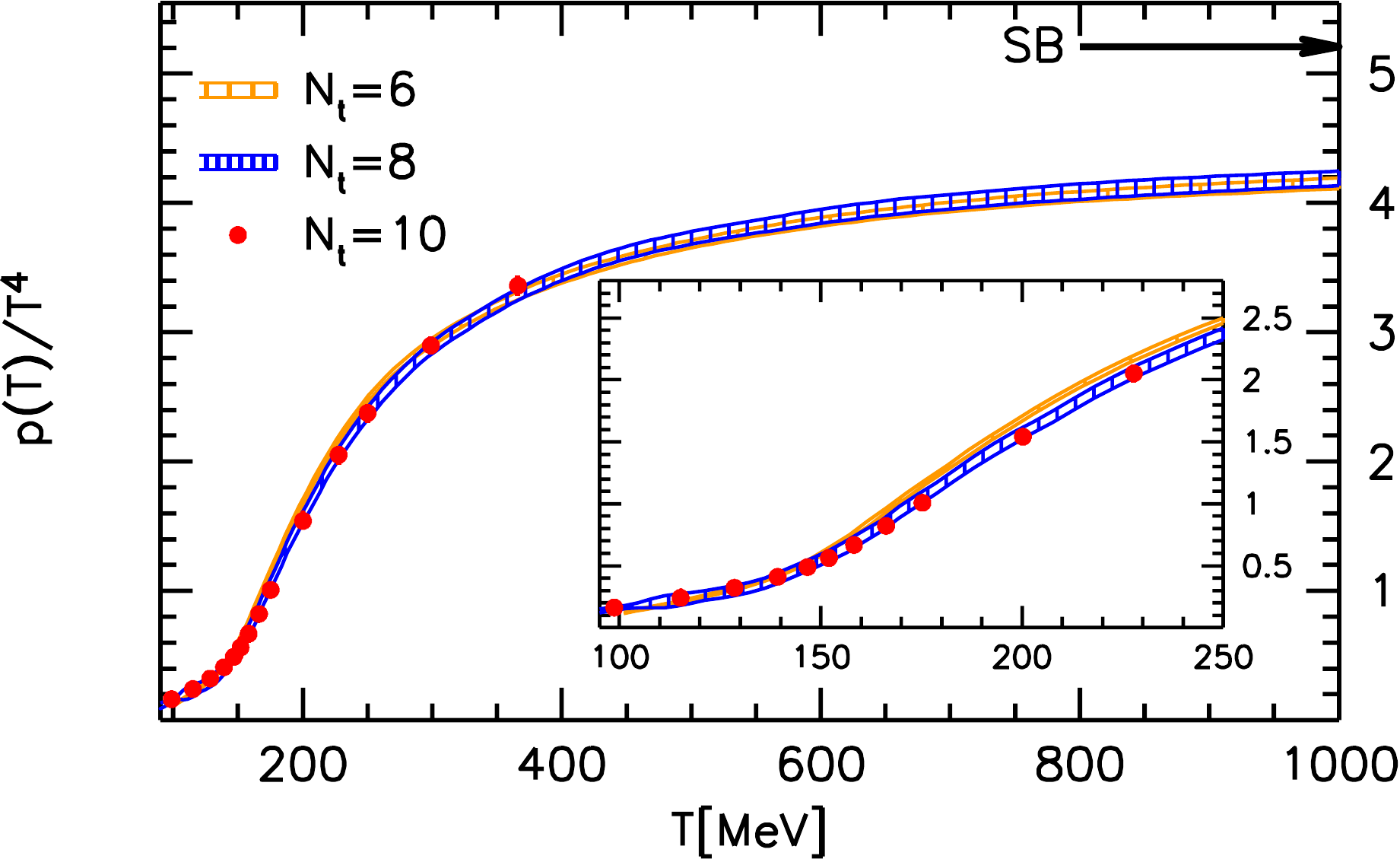}
\caption{\label{fig:fig1_2_01} The pressure normalized by $T^4$ as a function of the temperature on $N_t$ = 6, 8 and 10 lattices~\cite{QCD_eos_paper}. }

\end{figure}

Due to the color confinement, quarks can only move in the volume of a hadron at normal temperature and density. However, in extreme high density and temperature conditions, matter is compressed beyond a density where hadron boundaries overlap and merge. In this case, color charges can be screened in a manner analogous to Debye screening for electric charges.  In this way, the distance over which the long-range interaction occurs is shortened in a dense medium of charges. This screening effect results in the deconfinement of strongly interacting particles that are now free to propagate through the dense matter. Therefore the hadronic matter is transformed to a new state of matter called a Quark Gluon Plasma (QGP). The transition from a hadronic state to a QGP is accompanied by an increase in the (color) degrees of freedom, implying an increase in the entropy density and pressure as the temperature increases. Figure~\ref{fig:fig1_2_01} shows the equation of state from lattice QCD calculations~\cite{QCD_eos_paper}, obtained on temporal extent $N_\tau = 6-10$ lattices. The ratio of the pressure over $T^4$ rises rapidly when the temperature goes above the critical temperature $ T_{\rm c} \approx 160$ MeV, which indicate a phase transition from a hadronic matter phase to a QGP phase. Current experiments at the Relativistic Heavy Ion Collider (RHIC) populate the region of $T < 2T_{\rm c}$, far from the Stefan-Boltzmann limits where the system has massless, weakly interacting quarks and gluons.  We should still expect strong interactions in the QGP created experimentally.

\begin{figure}
\centering
\includegraphics[width=40pc]{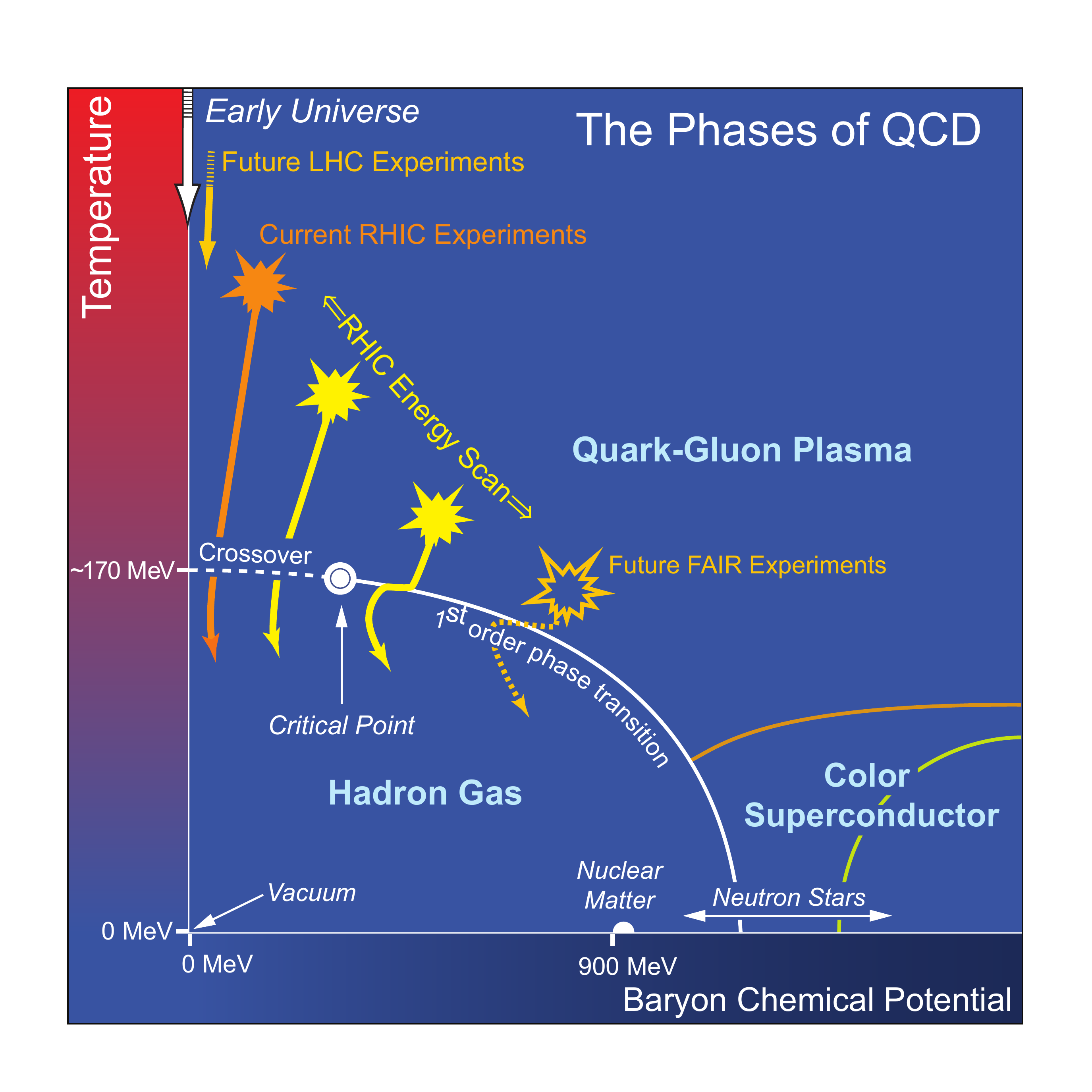}
\caption{\label{fig:fig1_2_02} The phase diagram of QCD matter~\cite{NSAC}. }

\end{figure}

Figure~\ref{fig:fig1_2_02} shows the QCD phase diagram of nuclear matter as a function of temperature and baryon chemical potential ($\mu_{\rm B}$). At low temperatures and for $\mu_{\rm B}$ around 1 GeV, matter exists in the regular hadronic state. If the temperature remains low but the $\mu_{\rm B}$ increases, ordered quark matter phases are predicted. When $\mu_{\rm B}$ is very large, the quarks start to form Cooper pairs, resulting in a weakly interacting Fermi liquid of quarks called ``color superconductivity"~\cite{Color_super}. This phase of matter is predicted to exist in the core of a neutron star.

On the other hand, if the temperature is high and $\mu_{\rm B}$ is relatively small, a deconfined quark gluon plasma phase is expected.  Both lattice QCD and experimental data indicate this transition from hadronic matter to Quark Gluon Plasma is a analytical transition (cross-over)~\cite{cross_over}, while some theoretical calculations predict that the transition at lower temperatures and high $\mu_{\rm B}$ is a 1st order phase transition~\cite{1st_order}. If there exists a phase transition at higher $\mu_B$, with a cross-over at $\mu_B=0$, the phase transition would end in a critical point at finite $\mu_B$. However, due to the difficulty of lattice QCD calculations at finite $\mu_{\rm B}$, accurate predictions of the critical point location are still lacking~\cite{critical_theory}. Therefore it falls to experiment to search for traces of the existence of the critical point of QCD~\cite{BES_write_up}.

\section{Relativistic Heavy Ion Collisions}

\begin{figure}
\centering
\includegraphics[width=34pc]{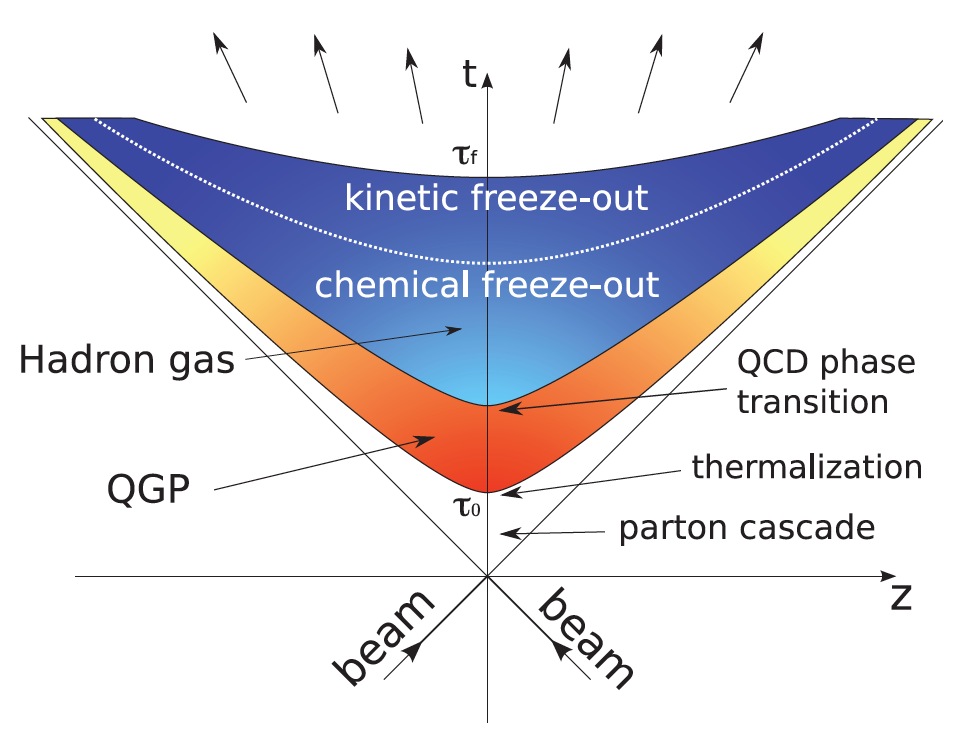}
\caption{\label{fig:fig1_2_03}Space-time evolution of a heavy ion collision. The figure is taken from Ref.~\cite{sss_thesis}. }

\end{figure}

Currently, the only way to experimentally study the deconfined QGP is through heavy ion collisions. Figure~\ref{fig:fig1_2_03} shows the space-time evolution of a typical heavy ion collision. The collision starts with two incoming nuclei traveling at nearly the speed of light. During the first few fm/$c$ of the collision, the system is dominated by hard processes like quark pair production, jet production and fragmentation. With the evolution of interactions among partons, the system reaches (local) equilibrium and forms a strongly interacting QGP. Due to the strong internal pressure, the QGP expands and cools. Once its temperature drops to around $T_{\rm c}$, the phase transition from QGP to hadronic matter occurs (hadronization). Since the phase transition is a cross-over, a mixed phase exists around $T_{\rm c}$. When the system cools further, the inelastic scattering stops and the relative ratios of different hadron species are fixed (chemical freeze-out), and finally elastic interactions between particles stop and the system comes to kinetic freeze-out. The final state particles then stream out and are detected experimentally.

\begin{figure}
\centering
\includegraphics[width=34pc]{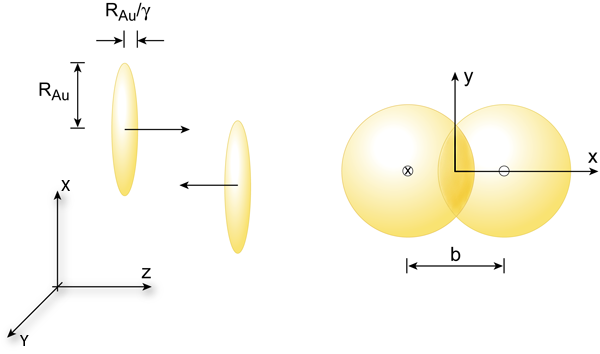}
\caption{\label{fig:fig1_2_04}Geometry of a high-energy heavy-ion collision~\cite{collision_geometry}. }

\end{figure}

One unique feature of heavy ion collisions is the collision geometry. The left panel of Figure~\ref{fig:fig1_2_04} shows the collision of two Lorentz-contracted gold nuclei~\cite{collision_geometry}. The Lorentz factor is about 100 for $\sqrt{s_{\rm NN}}$ = 200 GeV (100 AGeV in each direction), which makes the nucleus appear as a flat disk in the laboratory frame of reference. The right panel shows the same collision viewed along the beam pipe. The impact parameter, $b$, is defined as the distance between the center of two nuclei in the plane transverse to their direction. Another variable related to geometry is the number of participant nucleons, $N_{\rm part}$, which is defined as the number of nucleons that undergo at least one inelastic nucleon-nucleon collision. If a collision occurs with the two nuclei almost overlapping with each other, one would expect a small impact parameter and a large $N_{\rm part}$. This is called a central collision. On the other hand, if the overlapping volume is small, a peripheral collision with large impact parameter and small $N_{\rm part}$ will occur. Experimentally, most QGP signals only exist in central collisions where the system size is the largest.

Unfortunately, neither impact parameter nor $N_{\rm part}$ can be measured directly. One way to determine the centrality is to combine Monte-Carlo simulations and experimental data. The multiplicity of observed charged particles can then be correlated with centrality.  One finds that a larger multiplicity of charged particles corresponds to a more central collision.  Using the multiplicity, one can then determine the impact parameter and $N_{\rm part}$ using Glauber model calculations~\cite{glauber, glauber_review}. At STAR, the centrality is defined using ``reference multiplicity" and is grouped in increments of 0-5\%, 5-10\%, 10-20\%, 20-30\%, 30-40\%, 40-50\%, 50-60\%, 60-70\% and 70-80\% of the total reaction cross section.

\begin{figure}
\centering
\includegraphics[width=34pc]{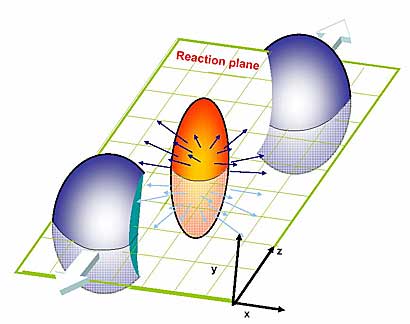}
\caption{\label{fig:fig1_2_05} Illustration of the reaction plane definition. }

\end{figure}

In non-central collisions, the overlap area of two nuclei is not symmetric. Instead, it  has a short axis in the transverse plane , which is parallel to the impact parameter, and a long axis perpendicular to it. The reaction plane is defined by the impact parameter vector and the momentum vector of the projectile as shown in Figure~\ref{fig:fig1_2_05}. Again, the true reaction plane can't be measured directly from data. But as discussed later, the event plane, which is a good approximation of reaction plane, can be measured event by event via azimuthal distributions of particle multiplicity~\cite{flow_sergei,sergei_flow_old}.

\chapter{Particle Ratio Fluctuations and the Charge Balance Function}
\section{Particle Ratio Fluctuations}
\label{fluctuation}

\begin{figure}
\centering
\includegraphics[width=28pc]{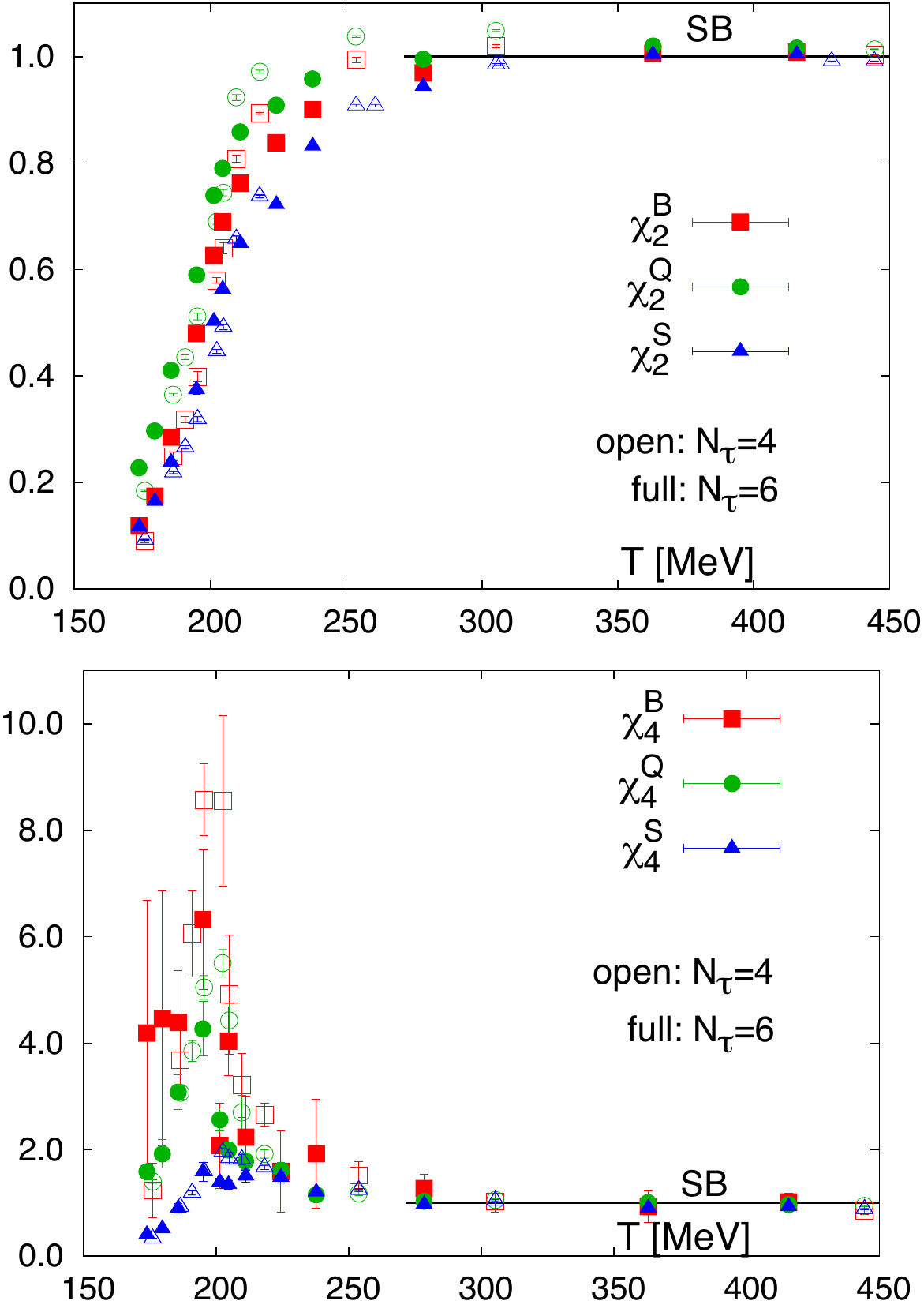}

\caption{\label{fig:fig2_1_01} Quadratic and quartic fluctuations of baryon number, electric charge, and strangeness. All quantities have been normalized to the corresponding free quark gas values \cite{LQCD_critical}.}
\end{figure}

As discussed in the previous chapter, one major challenge in relativistic heavy ion physics is to understand the structure of the QCD phase diagram, i.e., locate the end point of the first order phase transition line from the deconfined QGP phase to a hadronic gas phase. Similar to the critical opalescence phenomenon in other second-order phase transitions, a divergent susceptibility and a power-law decay of correlations should be observed near the QCD critical point. Indeed, lattice QCD calculations have indicated large fluctuations near the QCD critical temperature~\cite{LQCD_critical}\cite{LQCD_pro}. Figure~\ref{fig:fig2_1_01} shows quadratic ($\chi _2 $) and quartic ($\chi _4 $) derivatives of the QCD partition function with respect to baryon number, electric charge, and strangeness chemical potentials in terms of the system temperature.

Experimentally, the divergence of susceptibility and increased fluctuations can be related to event-by-event fluctuations of a given observable. For example, $K/\pi$, $p/\pi$, and $p/K$ fluctuations could be related to strangeness fluctuations, baryon number fluctuations and baryon-strangeness correlations~\cite{p_k_core} at mid-rapidity. The advantage of a ratio fluctuation is that the volume fluctuation cancels out.  One such observable, $\sigma_{\rm dyn}$~\cite{sigma_dyn}, is defined as the difference of measured fluctuations between real data and mixed events.

\begin{eqnarray}
\label{eq:eq_2_1_sigma_dyn}
\sigma _{\rm dyn}  = {\mathop{\rm sgn}} (\sigma _{\rm data}^2  - \sigma _{\rm mixed}^2 )\sqrt {|\sigma _{\rm data}^2  - \sigma _{\rm mixed}^2 |} 
\end{eqnarray}

\noindent
Here $\sigma_{\rm data}$ is the relative width (standard deviation divided by the mean) of the event-by-event particle-ratio distribution ($K/\pi$, $p/\pi$ or $p/K$) calculated from the data. $\sigma_{\rm mixed}$ is the same relative width calculated from mixed events. The mixed events are created by taking no more than one track from each real event so that there are no correlations between tracks in mixed events. The fluctuations of mixed events represents the statistical background within the experimental acceptance. The subtraction of $\sigma_{\rm mixed}$ from $\sigma_{\rm data}$ removes statistical background and leave us with nonstatistical fluctuations only. Due to the definition of our observable, it is possible to have $\sigma^2_{\rm data} < \sigma^2_{\rm mixed}$ (for example, decay of $\Delta  \to p + \pi$ will introduce correlations between pions and protons and reduce the width of $p/\pi$ ratio distribution). We use ${\mathop{\rm sgn}}(\sigma _{\rm data}^2  - \sigma _{\rm mixed}^2 )$ to extract sign outside of the square root in equation~\ref{eq:eq_2_1_sigma_dyn}. 

Another observable, $\nu_{\rm dyn}$, is also proposed~\cite{nu_dyn} to study the deviation from Poisson behavior. The observable $\nu_{\rm dyn}$ for kaons and pions can be written as

\begin{eqnarray}
\label{eq:eq_2_1_nu_dyn}
\nu _{{\rm dyn},K\pi }  = \frac{{ < K(K - 1) > }}{{ < K > ^2 }} + \frac{{ < \pi (\pi  - 1) > }}{{ < \pi  > ^2 }} - \frac{{2 < K\pi  > }}{{ < K >  < \pi  > }}
\end{eqnarray}

\noindent
where $K$ and $\pi$ are the number of charged kaons and charged pions in each event and the brackets represent event averages. If pions and kaons distribution are Poisson and independent of each other, one would expect $ < N_K (N_K  - 1) >  =  < N_K  > ^2 $, $ < N_\pi (N_\pi  - 1) >  =  < N_\pi  > ^2 $ and $< N_K N_\pi   >  =  < N_K  >  < N_\pi   > $, which will provide a zero $\nu _{{\rm dyn}} $.  Thus, $\nu_{\rm dyn}$ can be used to study the deviation of fluctuations from Poisson behavior.

Although $\nu_{\rm dyn}$ is calculated in a different way than $\sigma_{\rm dyn}$, it represents that same dynamical fluctuations.  The advantage of $\nu_{\rm dyn}$ is that it does not require the creation of mixed events. A Poisson simulation also shows that $\nu_{\rm dyn}$ provides more stable results compare to $\sigma_{\rm dyn}$ if the statistics is limited. With enough statistics and and a sufficiently large denominator,  it can be shown that $\nu _{\rm dyn}  \approx \sigma ^2_{\rm dyn} $. A detailed study can be found in Ref.\cite{kpi_scaling}.

\section{Balance Function}

In a typical central collision of two gold nuclei at RHIC, thousands of particles are produced.  Most of these charged particles are created during the dynamical evolution of the hot and dense media. Due to local charge conservation, charges are pair produced close to each other in space and time. The charged pairs are then pulled apart by diffusion and interactions while the system is expanding. Hence the study of correlations between opposite-sign charge pairs provides a unique tool to probe the properties of the hot dense matter created at RHIC. One such observable, the balance function, is sensitive to the correlation of balancing charges.

 In general, the balance function can be written as~\cite{balance_PRL}. 

\begin{eqnarray}
B(p_2 |p_1 ) \equiv \frac{1}{2}\{ \rho (b,p_2 |a,p_1 ) - \rho (b,p_2 |b,p_1 ) + \rho (a,p_2 |b,p_1 ) - \rho (a,p_2 |a,p_1 )\} 
\end{eqnarray}

\noindent
where $\rho (b,p_2 |a,p_1 )$ is the conditional probability of seeing particle $b$ in condition $p_2$ given condition that particle $a$ is in condition $p_1$. A like sign subtraction is applied to study only the balancing charges. Therefore the balance function can be used to study the conditional probability of a particle in condition $p_1$ being accompanied by an opposite-sign charge in condition $p_2$.

Experimentally, the conditional probability $\rho (b,p_2 |a,p_1 )$ can be calculated via

\begin{eqnarray}
\rho (b,p_2 |a,p_1 ) = \frac{{N(b,p_2 |a,p_1 )}}{{N(a,p_1 )}}
\end{eqnarray}

\noindent
where $N(b,p_2 |a,p_1 )$ is the number of pairs that satisfy both conditions and $N(a,p_1 )$ is the number of particles in $p_1$. A simplified version of balance function refers to $p_1$ as detecting a particle anywhere inside the detector and $p_2$ as having a relative pseudorapidity $\Delta \eta$ (or $\Delta \phi$, $\Delta y$, etc.) with respect to the first particle. For example, the balance function for $\Delta \eta$ can be written as~\cite{balance_PRC}.

\begin{eqnarray}
B(\Delta \eta ) = \frac{1}{2}\left\{ {\frac{{N_{ +  - } (\Delta \eta ) - N_{ +  + } (\Delta \eta )}}{{N_ +  }} + \frac{{N_{ -  + } (\Delta \eta ) - N_{ -  - } (\Delta \eta )}}{{N_ -  }}} \right\}
\end{eqnarray}

\noindent
Specifically, $N_{ +  - } (\Delta \eta )$ denotes the number of charged particle pairs in a given pseudorapidity range $\Delta \eta  = |\eta ( + ) - \eta ( - )| $, which is calculated by taking in turn each positive particle in an event and incrementing a histogram of $\Delta \eta$  with respect to all the negative particles in that event.  The distribution $N_{ +  - } (\Delta \eta )$ is then summed over all events.  A similar procedure is followed for $N_{ +  + } (\Delta \eta )$, $N_{ -  - } (\Delta \eta )$, and $N_{ -  + } (\Delta \eta )$. Also $N_{+(-)}$ is the number of positive(negative) particles integrated over all events.

Balance functions are sensitive to the mechanisms of charge formation and the subsequent relative diffusion of the balancing charges \cite{balance_theory}.   The idea is that in heavy ion collisions, most of the final state charges are created during the dynamical evolution of the system, due to local charge conservation, particles and their antiparticles are pair produced, so they are correlated initially in coordinate space. If hadronization occurs early, the created charge pairs would be expected to separate in rapidity due to expansion and rescattering in the strongly interacting medium. However, if a deconfined system of quarks and gluon is created during the collision, the observed balancing charges are then created when the deconfined system hadronizes, which reduces the effects of expansion and diffusion on the correlation of the balancing charges. The same arguments were used in discussing charge fluctuations \cite{charge_fluct}.   The narrowing of the balance function in central collisions implies high degree of correlation in coordinate space. This  has been postulated as a signal for delayed hadronization \cite{balance_theory}, which would not allow charges the opportunity to separate in coordinate space. 

Balance functions are also affected by the freeze-out temperature and by radial flow \cite{balance_distortions_jeon}. Remarkably, balance functions for central collisions have been shown to be consistent with blast-wave models where the balancing charges are required to come from regions with identical collective flow \cite{balance_blastwave}.  It has been previously presented that balance functions from Au+Au collisions at $\sqrt{s_{\rm NN}}$ = 130 GeV for all charged particles and for identified charged pions \cite{star_balance_130} narrow in central Au+Au collisions. In the following chapter we discuss further studies at additional incident energies.

The width of the balance function can be used to gain insight into the correlation between balancing charge pairs. The authors of Ref.~\cite{balance_blastwave} made the point that the observed width of the balance function in terms of relative rapidity, $\sigma_{y}$, is a combination of the rapidity spread induced by thermal effects, $\sigma_{\rm therm}$, and the separation of the balancing partners of the charge/anti-charge pair in coordinate space.  The authors of Ref.~\cite{balance_theory} stated this relationship as $\sigma_{y}^{2} = \sigma_{\rm therm}^{2}+4\beta \ln{(\tau/\tau_{0})}$, where $\beta$ is a diffusion constant, $\tau$ is the proper time after the initial collision when the charge/anti-charge pair is created, and $\tau_{0}$ is a characteristic time on the order of 1 fm/$c$.  After the initial collision, the width of the balance function decreases because the thermal width narrows as a result of cooling, while diffusion tends to increase the width of the balance function.  If production of the charge/anti-charge pairs occurs at early times, then scattering and expansion affects the partners of the charge/anti-charge pair during the entire lifetime of the system. The diffusion term is then large and significantly broadens the observed balance function.  If the production of charge/anti-charge pairs occurs late, the time during which the partners of the charge/anti-charge pair are exposed to scattering and expansion is small, which makes the effect of diffusion negligible.  Thus, in the case of late production of the charge/anti-charge pairs, the width of the balance function is determined by the thermal width.

\section{Models}

Due to the difficulty of lattice QCD calculations, semi-classical models have been generally applied as tools to investigate the collision dynamics of relativistic heavy-ion collisions. Two major types of models, hydrodynamical/statistical models and microscopic transport models, are used in this paper.

Hydrodynamical models start from the very basic physical principles of energy and momentum conservation~\cite{hydro_model}. A typical hydrodynamical model assumes local thermal equilibrium and neglects viscosity. Initial conditions are provided by outside sources. Once the initial conditions and the equation of state (EoS) are known, the expansion of the fluid is determined by hydrodynamic equations. As the system expands, at some point, the energy density become so low that local thermal equilibrium can no longer be maintained. That is the point where the hydrodynamic evolution is stopped. The freeze-out prescription are then calculated through other methods like transport models.  

At the end of expansion, hydrodynamical quantities are then linked to other methods to generate the final state observables.


Unlike hydrodynamical models that require input about initial and final state interactions, transport models deal with the entire time evolution of the heavy-ion collisions by modeling its microscopic constituents and their interactions. Most transport models are based on quantum molecular dynamics or solutions of the Boltzmann transport equation (BTE). The system's degrees of freedom are chosen to be baryons and mesons in the case of hadronic transport models or quarks and gluons in the case of parton cascades. Because microscopic transport models don't assume local thermal equilibrium, they are useful for the study of fluctuations and equilibration mechanisms.

\subsection{UrQMD}

The Ultra-relativistic Quantum-Molecular-Dynamics (UrQMD) model is a microscopic model used to simulate $p+p$, $p$+A and A+A interactions at relativistic energies ranging from Bevalac and SIS energies (1 AGeV) up to the Alternating Gradient Synchrotron (AGS) (10 AGeV), the Super Proton Synchroton (SPS) (160 AGeV) and RHIC~\cite{UrQMD_1,UrQMD_2}. The main goals of the model are to gain understanding about the following physical phenomena within a single transport model~\cite{UrQMD_web}:

\begin{itemize}
\item{ Creation of dense hadronic matter at high temperatures}
\item{ Properties of nuclear matter, $\Delta$ \& Resonance matter}
\item{ Creation of mesonic matter and of anti-matter}
\item{ Creation and transport of rare particles in hadronic matter}
\item{ Creation, modification and destruction of strangeness in matter}
\item{ Emission of electromagnetic probes}
\end{itemize} 
 
The UrQMD model is based on the covariant propagation of all hadrons on classical trajectories in combination with stochastic binary scatterings, the excitation and fragmentation of color strings, and the formation and decay of hadronic resonances. At higher energies more sub-hadronic degrees of freedom are introduced by the introduction of a formation time for hadrons produced in the fragmentation of strings and by hard (pQCD) scattering via the PYTHIA model~\cite{Pythia}. The UrQMD model is a hadronic transport model that does not incorporate a phase transition from hadronic matter to deconfined quark-gluon matter. 

Throughout this paper,  the UrQMD model is used as a hadronic reference for comparison with experimental data. We use version 3.3 and all parameters are set to default settings.  

\subsection{HIJING}

HIJING~\cite{HIJING_2}, heavy-ion jet interaction generator, is a Monte Carlo event generator used to study jet and multi-particle production in high energy $p+p$, $p$+A, and A+A collisions. It combines the pQCD approach of PYTHIA with low $p_t$ multi-string phenomenology.  The main features included in HIJING are:~\cite{HIJING}

\begin{itemize}
\item{ Multiple minijet production with initial and final state radiation}
\item{ Nuclear shadowing of parton distribution functions}
\item{ A schematic mechanism of jet quenching in hot dense matter}
\end{itemize}

\subsection{Thermal Blast-wave Model}

Unlike transport models that deal with the complete dynamic evolution of the system created in heavy-ion collisions, blast-wave models are used to study the kinetic freeze-out properties of the system. A conventional single-particle blast wave model assumes that particles are locally thermalized at a kinetic freeze-out temperature and are moving with a common collective transverse radial flow velocity field. The system can be described as~\cite{parity_soeren}

\begin{eqnarray}
\frac{{dN}}{{d^3 p  d^4 x}} \propto \exp ( - \frac{{p^\mu  u_\mu  (x)}}{{T_{\rm kin} }})\chi _{\rm S} (x)
\end{eqnarray}

\noindent
where the parameters are the freeze-out surface characteristic function $\chi _{\rm S} (x)$, the collective velocity $u_\mu  (x) $ and the kinetic freeze-out temperature $T_{\rm kin}$. These parameters can be determined by fitting to transverse momentum spectra and elliptic flow data.

In this paper, a modified single particle blast wave model is applied to compare with STAR balance function data.  In this model, an ensemble of particles with exactly conserved charges are generated. The particles are then assigned flow velocities according to the single particle blast-wave model parameterization with the additional requirement that they are emitted close to each other in space and time. This modification enforces local charge conservation inside the model, which is important for the study of correlations between charge pairs. A detailed description of the model can be found at Ref.~\cite{parity_soeren}. 

\chapter{Experimental Setup}

\section{The Relativistic Heavy Ion Collider (RHIC)}

\begin{figure}
\centering
\includegraphics[width=32pc]{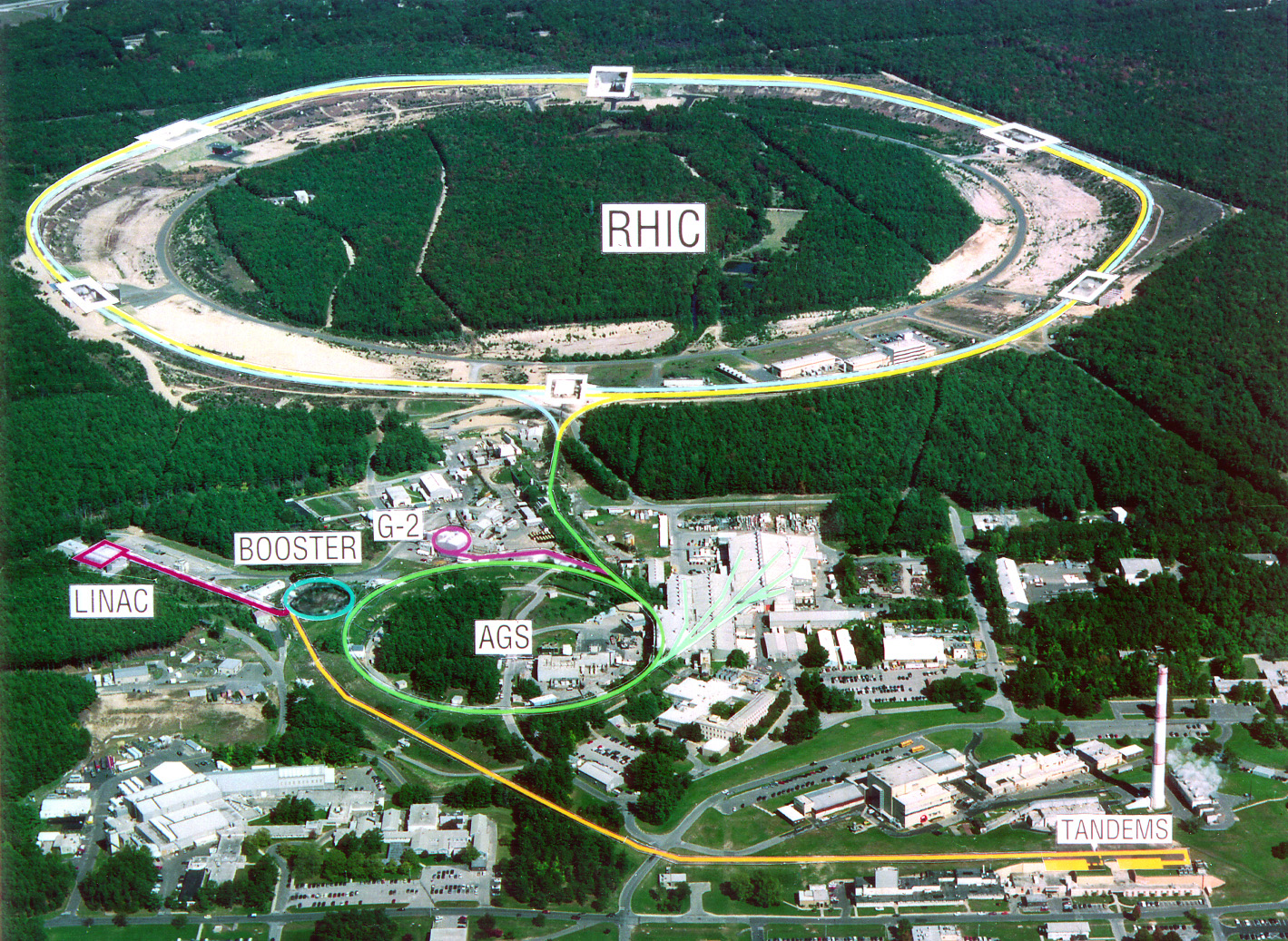}
\caption{\label{fig:fig3_1_01} Aerial view of the Relativistic Heavy Ion Collider (RHIC). }

\end{figure}

The Relativistic Heavy Ion Collider (RHIC)~\cite{RHIC}, located at Brookhaven National Laboratory (BNL), Upton, NY, is an experimental facility capable of colliding both heavy ions and polarized protons. An aerial view of RHIC is shown in Figure~\ref{fig:fig3_1_01}. The accelerator chain includes the Tandem Van de Graaff, the Booster, the AGS, and the RHIC main rings. A proton linac is used as the source of polarized protons. RHIC is a super-conducting collider comprised of two concentric storage rings in a tunnel 3.8-km in circumference with the blue ring designed for clockwise beam and the yellow ring designed for counter-clockwise beam. RHIC has six interaction regions where the beams can be collided head-on.

At RHIC, the heavy ion acceleration process starts at the Tandem.  A cesium sputter ion source injects negatively charged gold ions into the Tandem Van de Graaff. They are partially stripped of their electrons to a positive charge state using a thin carbon foil at the terminal, and then accelerated to an energy of 1 AMeV by the second stage of the Tandem. At the exit of the Tandem, gold ions are further stripped and then selected by bending magnets, resulting in a beam of charge state +32 gold ions, which are accelerated to 95 AMeV inside the Booster Synchrotron. When exiting the booster, ions are stripped again to charge state of +77 and injected into the AGS. The AGS accelerates the gold ions to an energy of 10.8 AGeV and fully strips the gold ions to a charge state of +79. After that, beams are injected into the RHIC rings through the AGS-to-RHIC Beam Transfer Line. 

Thanks to the two concentric but completely independent rings and two sources of ions, many different kinds of collisions are possible at RHIC. Table~~\ref{tab:tab3_1_01} shows the different collision energies and systems RHIC has run to date.  Collisions of both equal species of ions (Au+Au , Cu+Cu, and $p+p$) and unequal species of ions ($d$+Au) have been performed at RHIC. Although each ring at RHIC is designed to operate at a top energy of 100 AGeV ($\sqrt{s_{\rm NN}}$ = 200 GeV) for heavy ions and 250 GeV ($\sqrt{s}$ = 500 GeV) for protons, it can provide collisions at a variety of lower energies. RHIC has also been run by injecting gold ions at energies lower than the normal injection energy of $\sqrt{s_{\rm NN}}$ = 19.6 GeV without ramping the RHIC rings.  Energies down to $\sqrt{s_{\rm NN}}$ = 7.7 GeV have been produced at RHIC.

\begin{table}
\begin{center}
  \begin{tabular}{ |c | c | c | c |c |  }
    \hline
Run & Year & Species &  $\sqrt{s_{\rm NN}}$ (GeV) & Delivered luminosity \\ 
\hline
1 & 2000 & Au+Au & 56 & $< $ 0.001 ${\rm \mu b}^{-1}$\\ 
 &  & Au+Au & 130 & 20 ${\rm \mu b}^{-1}$\\ 
 \hline
2 & 2001/2002 & Au+Au & 200 & 258 ${\rm \mu b}^{-1}$\\ 
 &  & Au+Au & 19.6 & 0.4 ${\rm \mu b}^{-1}$\\ 
 &  & polarized $p+p$ & 200 & 1.4 ${\rm pb}^{-1}$\\  
 \hline
3 & 2003 & $d$+Au & 200 & 73 ${\rm nb}^{-1}$\\ 
 &  & polarized $p+p$ & 200 & 5.5 ${\rm pb}^{-1}$\\  
 \hline
4 & 2004 & Au+Au & 200 & 3.53 ${\rm nb}^{-1}$\\ 
 &  & Au+Au & 62.4 & 67 ${\rm \mu b}^{-1}$\\ 
 &  & polarized $p+p$ & 200 & 7.1 ${\rm pb}^{-1}$\\ 
 \hline
5 & 2005 & Cu+Cu & 200 & 42.1 ${\rm nb}^{-1}$\\ 
 &  & Cu+Cu & 62.4 & 1.5 ${\rm nb}^{-1}$\\ 
 &  & Cu+Cu & 22.4 & 0.02 ${\rm nb}^{-1}$\\  
 &  & polarized $p+p$ & 200 & 29.5 ${\rm pb}^{-1}$\\ 
 &  & polarized $p+p$ & 409.8 & 0.1${\rm pb}^{-1}$\\ 
 \hline
6 & 2006 & polarized $p+p$ & 200 & 88.6 ${\rm pb}^{-1}$\\ 
 &  & polarized $p+p$ & 62.4 & 1.050 ${\rm pb}^{-1}$\\  
 \hline
7 & 2007 & Au+Au & 200 & 7.25 ${\rm nb}^{-1}$\\ 
 &  & Au+Au & 9.2 & small\\ 
 \hline
8 & 2008 & $d$+Au & 200 & 437 ${\rm nb}^{-1}$\\ 
 &  & polarized $p+p$ & 200 & 38.4 ${\rm pb}^{-1}$\\ 
 &  & Au+Au & 9.2 & small\\ 
 \hline
9 & 2009 & polarized $p+p$ & 500 & 110 ${\rm pb}^{-1}$\\ 
 &  & polarized $p+p$ & 200 & 114 ${\rm pb}^{-1}$\\  
 &  & polarized pp2pp & 200 & 0.6 ${\rm nb}^{-1}$\\ 
 \hline
10 & 2010 & Au+Au & 200 & 10.3 ${\rm nb}^{-1}$\\ 
 &  & Au+Au & 62.4 & 544 ${\rm \mu b}^{-1}$\\ 
 &  & Au+Au & 39 & 206 ${\rm \mu b}^{-1}$\\ 
 &  & Au+Au & 7.7 & 4.23 ${\rm \mu b}^{-1}$\\ 
 &  & Au+Au & 11.5 & 7.8 ${\rm \mu b}^{-1}$\\ 
 \hline
11 & 2011 & polarized $p+p$ & 500 & 166 ${\rm pb}^{-1}$\\ 
 &  & Au+Au & 19.6 & 33.2 ${\rm \mu b}^{-1}$\\ 
 &  & Au+Au & 200 & 9.79 ${\rm nb}^{-1}$\\ 
 &  & Au+Au & 27 & 63.1 ${\rm \mu b}^{-1}$\\ 
\hline

  \end{tabular}
\caption{\label{tab:tab3_1_01}Summary of RHIC operating modes and total integrated luminosity delivered to 6 experiments.~\cite{run_summary}}
\end{center}
\end{table}

In this paper, we discuss particle-ratio fluctuations and charge balance function results from different RHIC runs. Specifically, Chapter 5 and Section 6.2 analyzes Run 10 and 11 Au+Au data from $\sqrt{s_{\rm NN}}$ = 7.7, 11.5, 19.6, 27, 39, 62.4 and 200 GeV. Section 6.1 analyzes Run 7 Au+Au results at 200 GeV, Run 2 $p+p$ results at 200 GeV, Run 3 $d$+Au results at 200 GeV. Section 6.3 analyzes Run 10 Au+Au 7.7, 11.5, 39 and Run 4 Au+Au 62, 200 GeV data. Section 6.4 analyzes Run 10 Au+Au 200 GeV data.

\section{The STAR Experiment}

Among RHIC's six interaction regions, there are four experiments~\cite{RHIC_5years}, BRAHMS (Broad RAnge Hadron Magnetic Spectrometers), PHENIX (Pioneering High Energy Nuclear Interactions eXperiment), PHOBOS, and STAR (Solenoidal Tracker At RHIC) located at positions around the RHIC ring of 2 o�clock, 8 o�clock, 10 o�clock and 6 o�clock, respectively.  BRAHMS and PHOBOS have completed their experimental programs and are no longer operational.

\begin{figure}
\centering
\includegraphics[width=32pc]{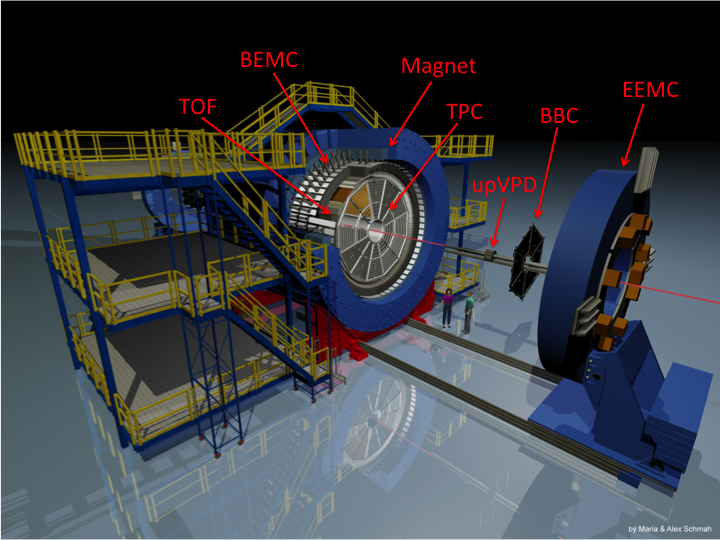}
\caption{\label{fig:fig3_1_02} A perspective view of the STAR detector.\cite{TPC_alex}}

\end{figure}

\begin{figure}
\centering
\includegraphics[width=36pc]{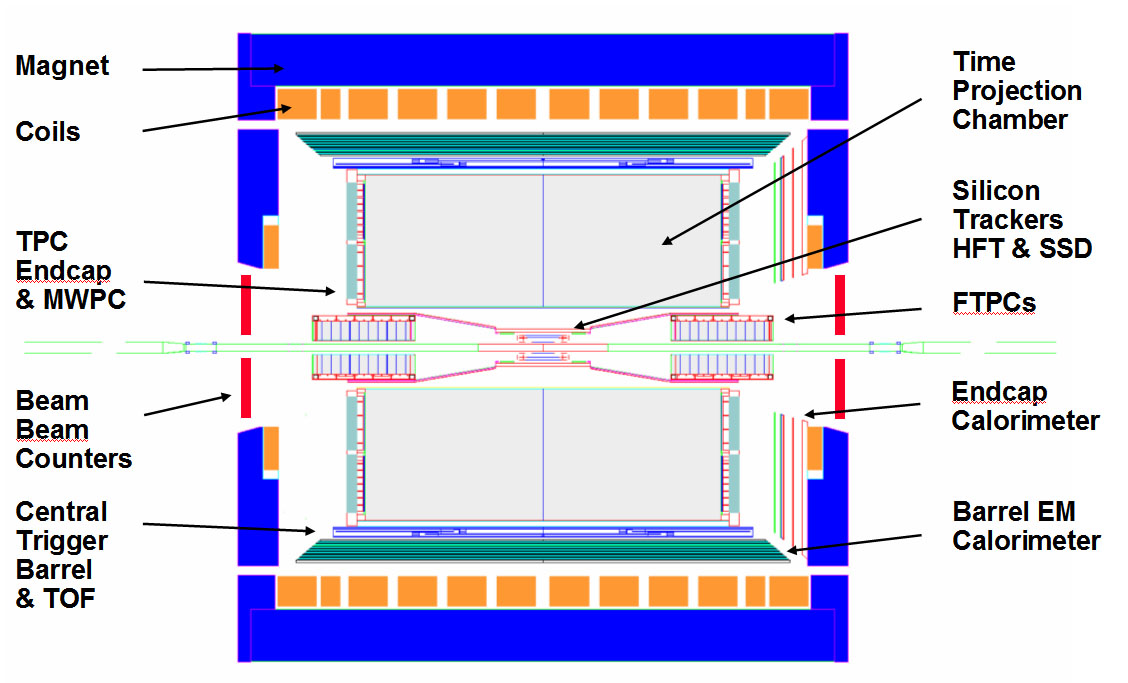}
\caption{\label{fig:fig3_1_03} a cross-section view of the STAR detector.}

\end{figure}

The STAR experiment is a multiple detector system constructed to study the properties of quark-gluon plasma created at RHIC~\cite{STAR_nim}. Figure~\ref{fig:fig3_1_02} shows a perspective view of STAR's current configuration during Runs 10 and 11 (forward time projection chamber [FTPC] not included) while Figure~\ref{fig:fig3_1_03}  shows a cross-sectional view of STAR from previous runs. Located in the center of STAR is a 5-cm radius beam pipe. It is made of beryllium in the $|z| <50$ cm region and changes to aluminum for $|z|>50$ cm. Immediately surrounding the beam pipe is the inner tracking system comprised of the Silicon Vertex Tracker (SVT)~\cite{SVT_nim} and Silicon Strip Detectors (SSD)~\cite{SSD_nim}. The SVT has three cylindrical layers located 7, 11 and 15 cm from the beam axis with a total of 216 silicon drift detectors, while the 4th layer is the SSD. Outside the SSD is the Time Projection Chamber (TPC), which acts as the main tracking and particle identification detector in STAR. The TPC is a barrel gas detector 420 cm in length, 50 cm (inner) and 200 cm (outer) in radius with full azimuthal coverage and a pseudo-rapidity range of $|\eta| < 1.8 $ (uniform acceptance within $|\eta| < 1.0 $ ). To enhance STAR's particle identification capability, a barrel Time of Flight detector (TOF) is also installed outside the TPC. The TOF consists of a total of 120 trays that cover the full azimuth and have a pseudo-rapidity range $|\eta| < 0.9 $. Outside the TOF is the barrel electromagnetic calorimeter (BEMC)~\cite{BEMC_nim}, a lead-scintillator sampling calorimeter that also covers the full azimuth and has a pseudo-rapidity range of $|\eta| < 1.0 $. All the detectors mentioned above are installed inside a room temperature solenoidal magnet, which generates a uniform magnetic field of 0.25 T to 0.50 T.

In the forward region, STAR also has multiple detectors. A pair of Forward Time Projection Chamber (FTPCs)~\cite{FTPC_nim} with full azimuthal coverage and a pseudo-rapidity range $2.5 < |\eta| < 4 $ are installed on both sides of STAR serving as the forward tracking detectors. An endcap electromagnetic calorimeter (EEMC)~\cite{EEMC_nim} covers $1.09 < \eta < 2 $. The Beam-Beam Counters (BBCs), which are 3.75 meters away from the center of STAR, are two sets of scintillator rings installed around the RHIC beam pipe. The BBCs are used as trigger detector for STAR low energy runs. Another trigger detector is the Zero-Degree Calorimeter (ZDC). The ZDCs are installed at the first bending magnets in the collider line and use  layers of lead and scintillator to detect mainly spectator neutrons. The ZDCs are used for minimum bias trigger, which is the major trigger used for this paper.

During the time period of 2007 to 2011, there were some significant changes in STAR's detector configuration, among them two changes affecting the analysis in this paper. First, the SVT and the SSD along with their utility lines were completely removed from STAR after the completion of Run 7 in 2008. The reduction of material inside the TPC reduced the low $p_{t}$ background particles. Second, the TOF detector, which improved STAR's particle identification capability at high momenta, was fully installed after Run 9 in 2010. The lower amount of material combined with the full TOF installation makes Runs 10 and 11 ideal for identified particle analysis. 

\subsection{The STAR Time Projection Chamber}

\begin{figure}
\centering
\includegraphics[width=36pc]{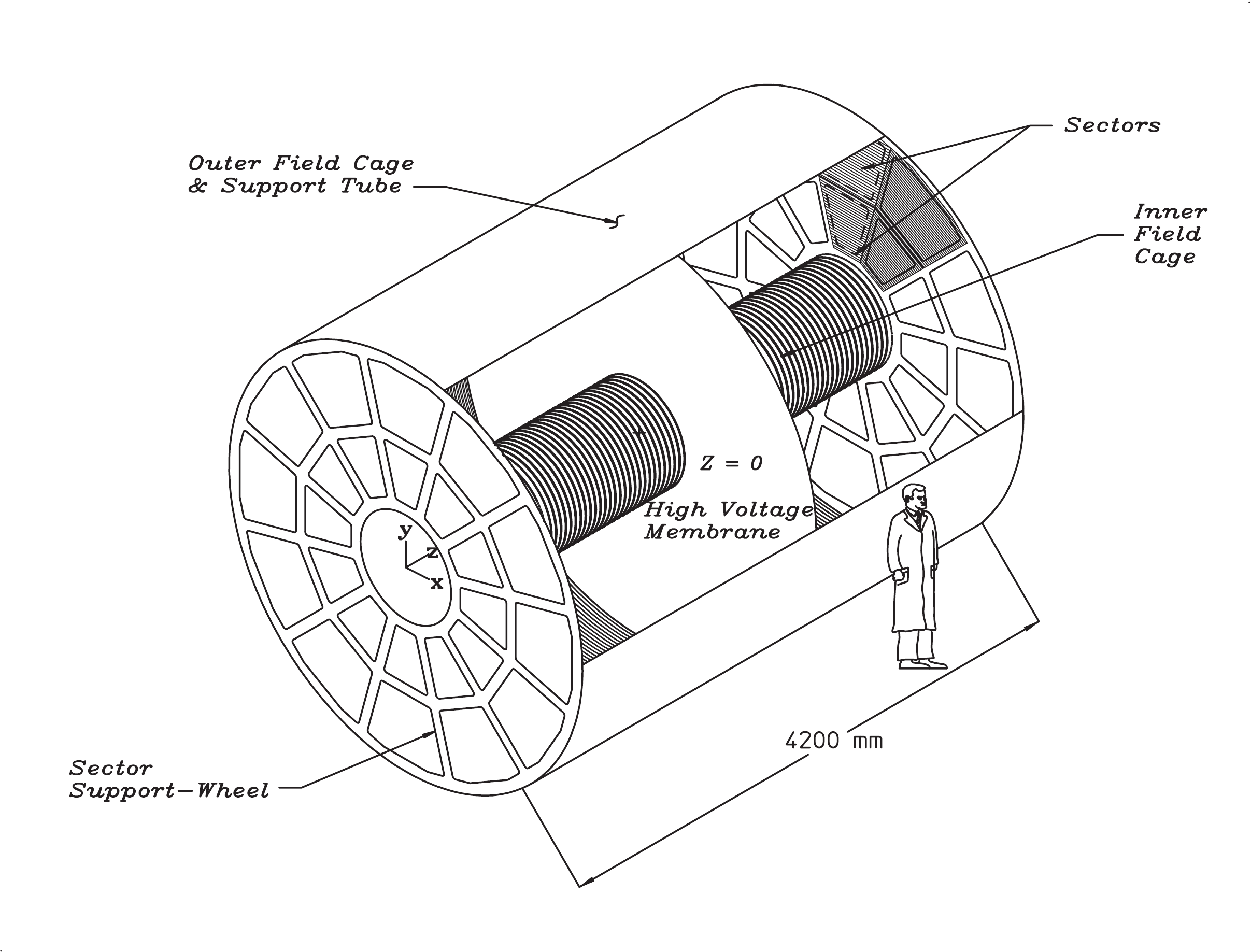}
\caption{\label{fig:fig3_1_04}STAR TPC~\cite{TPC_nim}.}

\end{figure}

The TPC, which acts as the main tracking detector at STAR, is a gas detector capable of recording tracks of particles, measuring particle momentum, and performing particle identification using the particles' ionization energy loss ($dE/dx$)~\cite{TPC_nim} combined with the measurement of the magnetic rigidity. Figure~\ref{fig:fig3_1_04} shows the schematic of STAR TPC. The detector volume is filled with P10 gas (10\% methane, 90\% argon) held at 2 mbar above atmospheric pressure. This slight over-pressure is designed to ensure that air does not contaminate the P10 in the detection volume.

When a particle travels through the TPC gas volume, it ionizes gas molecules along its path. The ionized electrons then drift to the readout end caps under an uniform electric field of $\approx$ 135 V/ cm with direction parallel to the beam pipe. The electric field is generated by the central membrane, concentric field-cage cylinders and the readout end caps. At both ends of the TPC, multi-wire proportional chambers (MWPC) with readout pads are used as the readout system. The high fields close to MWPC's anode wires cause drifting electrons to avalanche, which provides an amplification of 1000 - 3000 . The read-out pads are arranged in 12 sectors, with each sector contain a inner sub-sector and outer sub-sector. A total of 136,608 pads are used for read-out. Induced charge on the pads from the avalanche is then measured by FEE cards. Each FEE card contain 32 channels. The FEE cards are then supported by 144 larger readout boards (RDOs)~\cite{TPC_readout_nim}, which provide power and control signals, read out the data. Due to a low voltage power supply issue during RHIC Run 10, a significant amount of RDOs were masked out at various times during the run. Therefore a dead RDO number cut will be discussed later to minimize impact of the missing RDO boards.
\begin{figure}
\centering
\includegraphics[width=32pc]{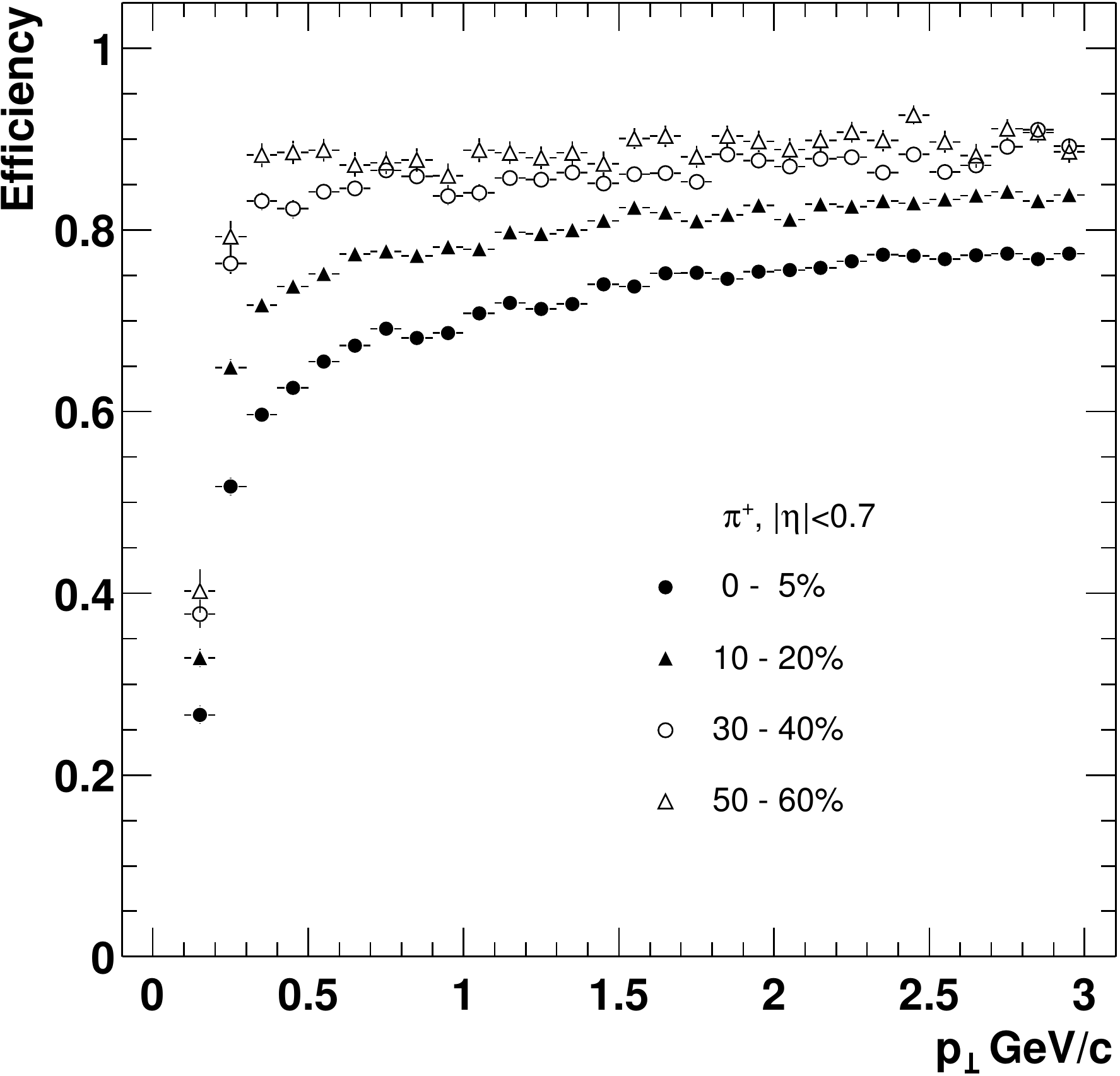}
\caption{\label{fig:fig3_1_05}The pion tracking efficiency from Au+Au events. Only tracks with $|y| < 0.7 $ are used and the magnetic field is 0.25 T~\cite{TPC_nim}.}
\end{figure}

The TPC tracking starts by finding ionization clusters along the track. The clusters drift towards two ends of TPC and are measured by read-out pads. Each cluster's $x$ and $y$ coordinates are measured by the position of read-out pads that have induced charge while the $z$ coordinate is measured by the drift time. The measured space points are then associated together to form a track. Once a track is formed, a model is used to fit the points and extract momentum information. The major factors affecting the total tracking efficiency are the imperfect acceptance due to sector boundaries, hardware failure from run to run (masked out RDOs),  and two-track separation resolution. Figure~\ref{fig:fig3_1_05} shows the the pion reconstruction efficiency as a function of transverse momentum.

\subsection{The STAR Time-of-Flight Detector}

Although the TPC has a large uniform acceptance and good momentum measurement capability, its particle identification capability is limited by the $dE/dx$ method. It is difficult to identify particles with momentum above 0.7 GeV/$c$ because the energy loss is less mass-dependent for high $p_t$ particles~\cite{TPC_nim}. About 30\% of the total charged hadrons in an event cannot be identified. Therefore, the TOF detector is applied to enhance STAR's particle identification capability~\cite{TOF_Kohei}.

The basic idea of the TOF detector system is to precisely measure the flight time of a charged particle when traveling between two space points. The start time is determined by two upgraded pseudo-vertex position detectors (upVPD) and the stop time is determined by TOF barrel itself. Each upVPD has 19 detector channels of photomultiplier tubes (PMT) with scintillators and is mounted close to the beam pipe (Figure~\ref{fig:fig3_1_02}). The upVPDs are sensitive to spectators from collisions and can be used to determine the start time as well as the vertex position (time difference between the two upVPDs). The TOF barrel is a cylindrical shell consisting of 120 trays that cover the full azimuth and a pseudo-rapidity range of approximately $|\eta|< 0.9$. A total of 3840 Multi-gap Resistive Plates (MRPC) are used to construct the TOF barrel~\cite{TOF_nim}. 

The TOF particle identification is done by combining information from both TPC and TOF. All TOF hits are matched to TPC reconstructed tracks. With the flight time $\Delta t$ from TOF and path length $\Delta s$ from TPC, the particle velocity can be calculated via

$$
\frac{1}{\beta } = c\frac{{\Delta t}}{{\Delta s}}
$$

Furthermore, with the momentum determined by TPC, we can calculate the mass of the charged particle~\cite{TOF_Kohei}

$$
m = \frac{p}{{\gamma \beta c}} = \frac{{p\sqrt {1 - \beta ^2 } }}{{\beta c}} = \frac{p}{c}\sqrt {(\frac{1}{\beta })^2  - 1} 
$$

\begin{figure}
\centering
\includegraphics[width=36pc]{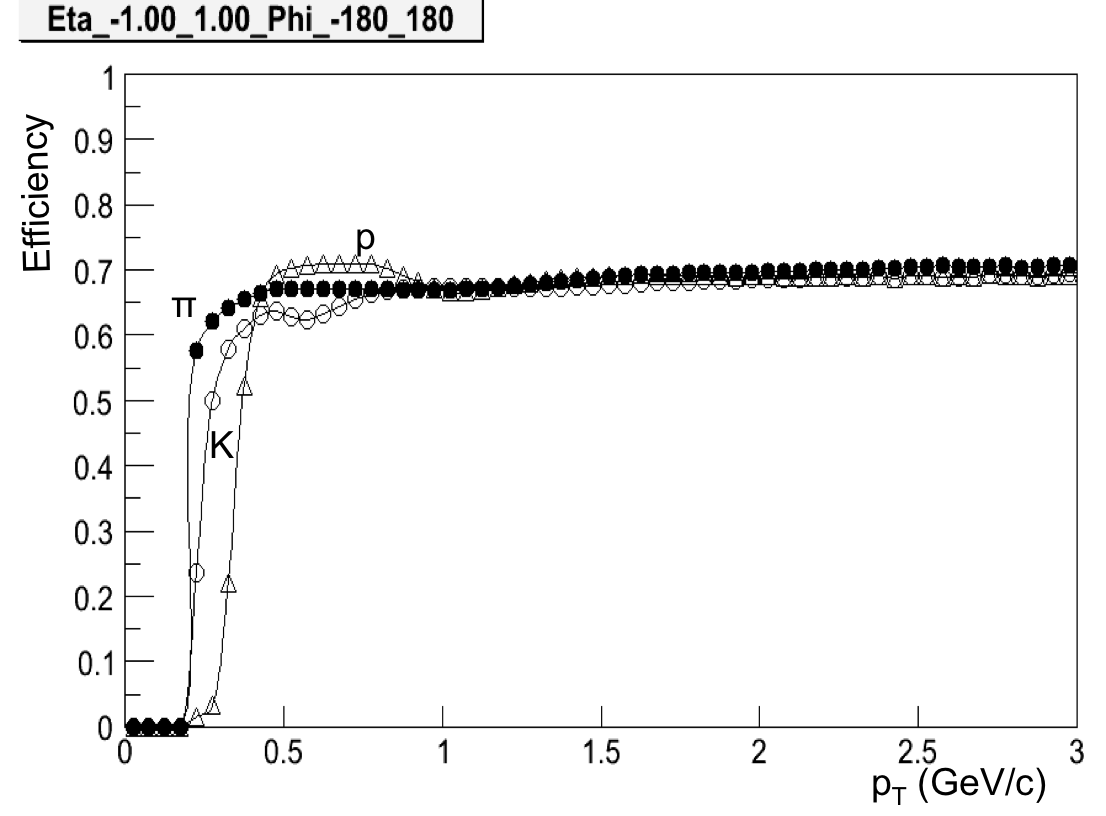}
\caption{\label{fig:fig3_1_06}The TOF matching efficiency ~\cite{TOF_eff_xing}.}

\end{figure}

Since the TOF particle identification (PID) requires TPC tracks, the matching efficiency of TPC tracks to TOF hits is important for TOF particle identification. Figure~\ref{fig:fig3_1_06} shows the TOF matching efficiency for identified pions, kaons, and protons in terms of transverse momentum $p_t$.  The efficiency is about 70\% in the high $p_t$ region and drops dramatically for low $p_t$ particles.    

For RHIC low energy runs during Runs 10 and 11, due to the relatively low efficiency of upVPDs, the TOF detector is analyzed in start-less mode: A self-calibration algorithm is applied to the TOF calibration so the upVPD information is not required for flight time determination. In this paper, the 7.7, 11.5, and 39 GeV Au+Au TOF data and 19.6 GeV Au+Au TOF data from Run 11 are calibrated using this start-less mode.

\chapter{Analysis Method}

\section{Event and track selection}

Throughout this thesis, minimum-bias data sets were used from RHIC Runs 4, 7, 10 and 11. The centrality of each collision was determined according to the measured charged hadron multiplicity within the pseudorapidity range $|\eta| < 0.5$ (reference multiplicity).The centrality bins were calculated as a fraction of the reference multiplicity distribution starting with the highest multiplicities (most central) to the lowest multiplicities (most peripheral). Nine centrality bins were used: 0-5\%, 5-10\%,10-20\%, 20-30\%, 30-40\%, 40-50\%, 50-60\%, 60-70\%, and 70-80\%.

To ensure nearly uniform detector acceptance and avoid multiplicity biases near the edges of the TPC, cuts were made on the $z$ position of the reconstructed primary vertex. These cuts are listed in Table~\ref{tab:tab4_1_01} for each experimental run. In addition, the radial position of the primary vertex was required to be less than 2 cm for Runs 10 and 11. This radial cut is more important in RHIC's low energies runs where the beam spot was relatively large and the beam pipe events were a large source of background.

\begin{table}
\begin{center}
  \begin{tabular}{|c |c | c | c | c |c |  }
    \hline
   Run  & Year  & $\sqrt{s_{\rm NN}} ~({\rm GeV})$ & Kind&  $|z|~({\rm cm})$ & No. of Events (millions)\\ \hline
  4 & 2004 & 62.4   &Au+Au & 30 & 8 \\ \hline
  4 &   2004 & 200   &Au+Au & 30 & 14 \\ \hline 
  7 &  2007 & 200   &Au+Au & 10 & 28 \\ \hline
10 &  2010 & 200   &Au+Au & 30 & 33 (220 for Chapter 6.4) \\ \hline
10 &  2010 & 62.4     &Au+Au & 30 & 17 \\ \hline
10 &  2010 & 39     &Au+Au & 30 & 10 \\ \hline
10 &  2010 & 11.5  &Au+Au & 50 & 16 (4 for ratio fluctuation) \\ \hline
10 &  2010 & 7.7    &Au+Au & 70 & 4 (3 for ratio fluctuation)   \\ \hline
11 &  2011 & 19.6 &Au+Au & 30 & 15 \\  \hline
11 &  2011 & 27 &Au+Au & 30 & 29 \\  \hline

  \end{tabular}
\caption{\label{tab:tab4_1_01} Summary of data sets, primary vertex cuts, and the number of good events used in the analysis.}
\end{center}
\end{table}

Standard STAR track quality cuts were used.  Only tracks having more than 15 hits out of a maximum of 45 measurable space points along the trajectory were considered as good. The ratio of the numbers of reconstructed space points to possible space points along the track was required to be greater than 0.52 to avoid the effects of track splitting. Tracks in the TPC were characterized by the distance of closest approach (DCA), which is the distance between the projection of the track at its closest point to the measured event vertex. Particles originating from weak decays can have larger DCAs than the direct primary particles.  All tracks were required to have a DCA of less than 3 cm. 

\section{Particle identification}

Particle identification was accomplished with both STAR's TPC and TOF detectors. A charged particle's trajectory is deflected by the external magnetic field while traveling inside the TPC gas volume so the magnetic rigidity can be used to determine the particle's momentum. Also, the charged particles interact with the gas and lose energy by ionizing electrons of the gas atoms. This specific ionization energy loss, $dE/dx$, is a function of the particle momentum and species. The TOF measures the particle's flight time precisely, which is combined with the momentum measurement from the TPC to provide particle identification.

\begin{figure}
\centering
\includegraphics[width=38pc]{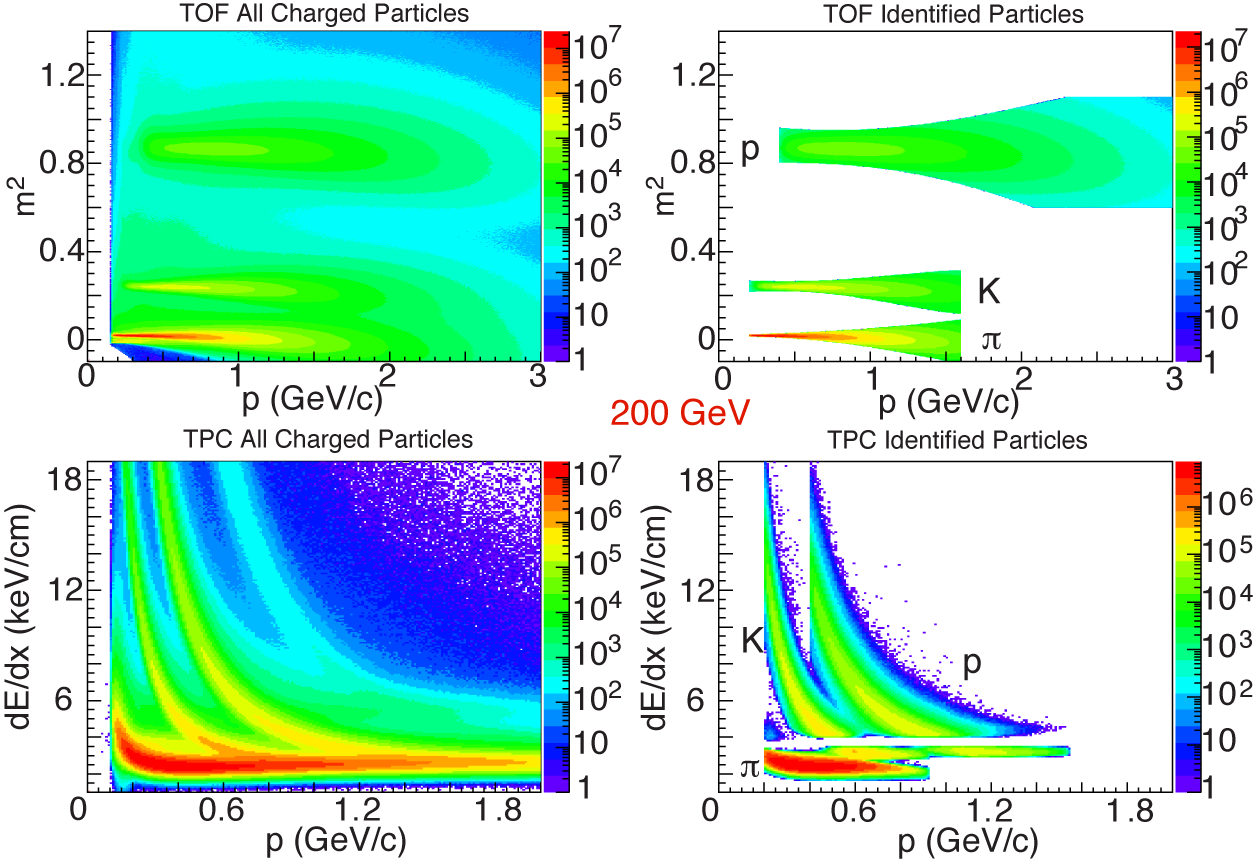}
\caption{\label{fig:fig4_2_01} Upper: $m^2$ vs. $p$ for all charged particles and identified kaons, pions, and protons using TOF PID. Lower: $dE/dx$ vs. $p$ for all charged particles and identified kaons, pions and protons using TPC PID. }

\end{figure}

Due to the fact that the TOF has excellent particle identification capability but relatively low efficiency, a combined identification method is applied to achieve the maximum efficiency and accuracy. If a particle has a TOF match, we use the particle's velocity $\beta$ extracted from the time-of-flight, otherwise we switch to the TPC and use its $dE/dx$ to make the identification. Figure~\ref{fig:fig4_2_01} shows $m^2$ vs. $p$ for all charged particles and identified kaons, pions and protons using TOF PID, as well as $dE/dx$ vs. $p$ for all charged particles and identified kaons, pions and protons using TPC PID. For TOF PID, the $m^2$ method is used, which can be written as

\begin{eqnarray}
m^2  = \frac{{p^2 }}{{\gamma ^2 \beta ^2 }} = (\frac{1}{{\beta ^2 }} - 1)p^2 
\end{eqnarray}

\noindent
where the $\beta$ is extracted from the TOF and the momentum information is taken from the TPC. The benefit of this $m^2$ method is that it does not depend on momentum so the mean value of each mass distribution should be constant as a function of momentum. The disadvantage is that the method uses the momentum from TPC so the identification uncertainty depends on momentum. In this analysis, we used Gaussian fits to all three mass peaks at each momentum bin, extracted the width from the fits, and only kept tracks that are less than two standard deviations away from the expected $m^2$ value.  A hard cut of $0.6 < m^2 < 1.1 $ is also applied to the proton mass peak due to its relatively wide distribution. For identified pions and kaons, we use a momentum cut of $p_{t} > 0.2$ GeV/$c$ and $ p < 1.6$ GeV/$c$, while for protons, we increase the lower cut to $p_{t} < 0.4$ GeV/$c$ to reduce the number of background protons knocked out from the beam pipe and the detector materials. A $p < 3$ GeV/$c$ upper cut is also used to improve the separation of kaons and protons (compared to pions and kaons) in the high momentum region.

Similarly, for TPC PID, we required that the specific energy loss be less than two standard deviations away from the energy loss predicted for the desired particle species, and would also be more than two standard deviations away from the energy loss predicted for the other particles. In addition, electrons were excluded from the analysis for all cases by requiring that the specific energy loss for each track was more than one standard deviation away from the energy-loss predictions for electrons. Due to the relatively low resolution of TPC, the $p_{t}$ cut for TPC PID is $ 0.2 < p_{t} < 0.6$ GeV/$c$ for pions and kaons, while for protons it is  $ 0.4 < p_{t} < 1.0$ GeV/$c$.

It should be noted that TOF was only fully installed after year 2009 at RHIC, so throughout this analysis, only Run 10 and 11 data use TPC+TOF PID methods, Run 4 and 7 use TPC as the only particle identification detector.

\section{Event plane method}

In practice the reaction plane angle for a given collision is not known. However, the event plane, which can be calculated from the particle azimuthal distributions, can be used as an estimation of reaction plane. As discussed in Ref.\cite{flow_sergei}, event plane is calculated by
\begin{eqnarray}
\label{eq:eq_4_3_cos}
Q_n \cos (n\psi _n ) = X_n  = \sum\limits_i {w_i \cos (n\phi _i )} 
\end{eqnarray}

\begin{eqnarray}
\label{eq:eq_4_3_sin}
Q_n \sin (n\psi _n ) = X_n  = \sum\limits_i {w_i \sin (n\phi _i )} 
\end{eqnarray}

\begin{eqnarray}
\label{eq:eq_4_3_psi}
\psi _n  = (\tan ^{ - 1} \frac{{\sum\limits_i {w_i \sin (n\phi _i )} }}{{\sum\limits_i {w_i \cos (n\phi _i )} }})/n
\end{eqnarray}

\noindent
where  $Q_{n}$ is the event flow vector and $\psi_{n}$  is the event plane angle from the $n^{th}$ harmonic of the distribution. The $\phi_i$ is the azimuthal angle of the $i$th particle and  $w_i$ are weights optimized to make the reaction plane resolution as high as possible. In this particular analysis, we use the second order event plane to take advantage of the large elliptic flow measured at RHIC. A transverse momentum weighting is also applied to maximize event plane resolution, where we use the particle�s $p_t$ up to 2.0 GeV/$c$ as the weight $w_i$.

\begin{figure}
\centering
\includegraphics[width=34pc]{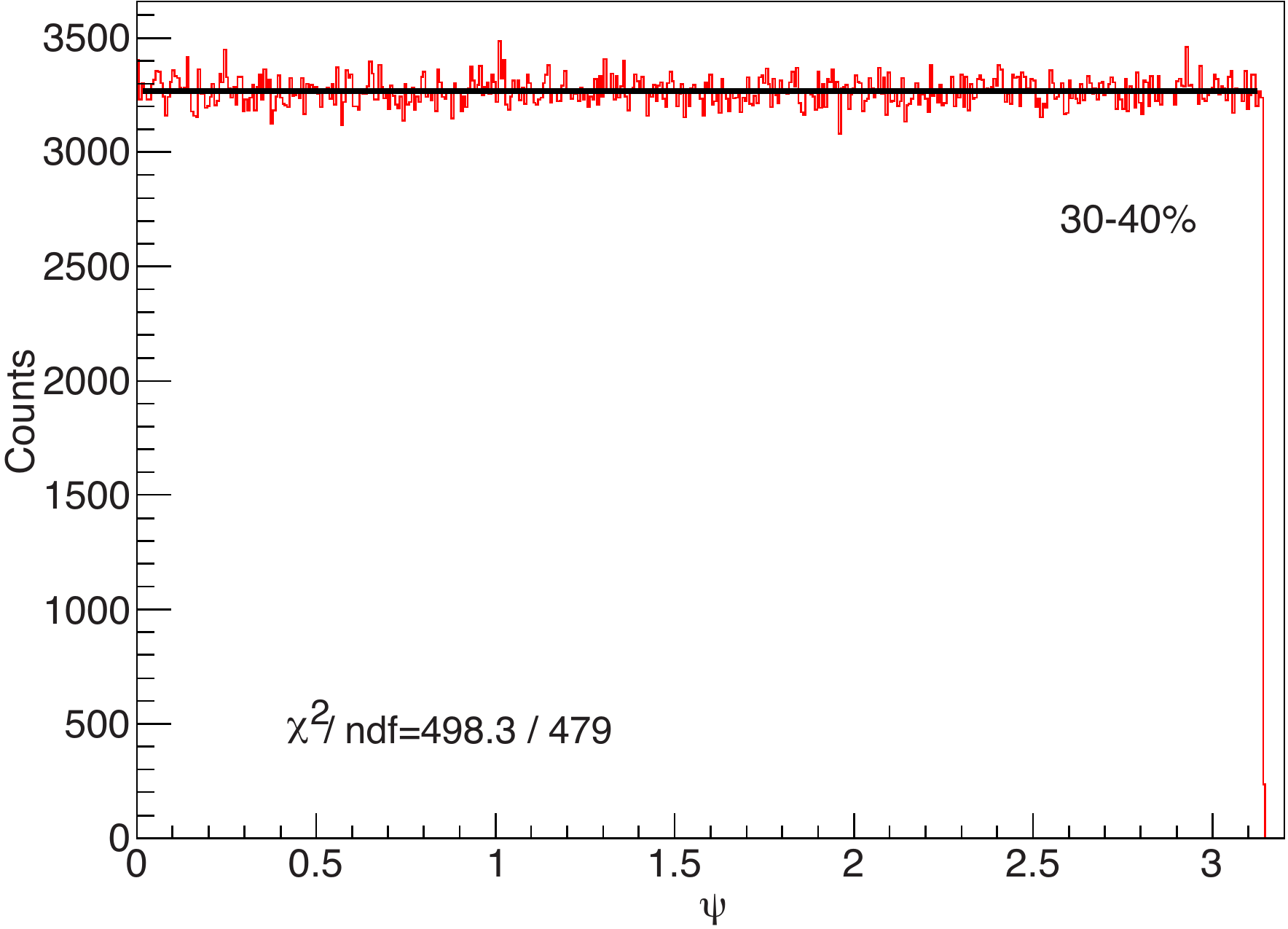}
\caption{\label{fig:fig4_3_01} Reconstructed TPC event plane distribution after $p_{t}$ and $\phi$ weighting. The data are from 30-40\% centrality, Run 4 Au+Au collision at$\sqrt{s_{\rm NN}}$ = 200 GeV.}

\end{figure}

In a case of a perfect detector where the azimuthal acceptance and efficiency is uniform, the azimuthal distribution of the event plane should be identical in all directions. However, things like TPC sector boundaries and malfunctioning electronics can cause a finite acceptance and non-uniform detection efficiency, which can result in a non-flat distribution of the event plane angle and bias the final analysis results. To flatten the event plane distribution, a $\phi$ weight is also folded into the $w_i$ of Equation \ref{eq:eq_4_3_cos}, \ref{eq:eq_4_3_sin} and~\ref{eq:eq_4_3_psi}. The $\phi$ weighting is done by inverting the $\phi$ distributions of all selected tracks for a given number of events. In this analysis, we use a run by run and centrality by centrality $\phi$ weighting to deal with different detector acceptance and efficiency in long runs and different centralities. When filling the $\phi$ distribution histogram, we weight with each track's $p_t$ to maximize the event plane resolution. Figure~\ref{fig:fig4_3_01} shows the event plane azimuthal distribution after $\phi$ weighting is applied. The data are from Run 4's Au+Au collisions at $\sqrt{s_{\rm NN}}$ = 200 GeV, only 30-40\% centrality is shown. A constant fit of this azimuthal distribution gives $\chi ^2 /$ndf$ = 498.3/479 \approx 1 $, indicating a flat event plane azimuthal distribution.

\begin{figure}
\centering
\includegraphics[width=34pc]{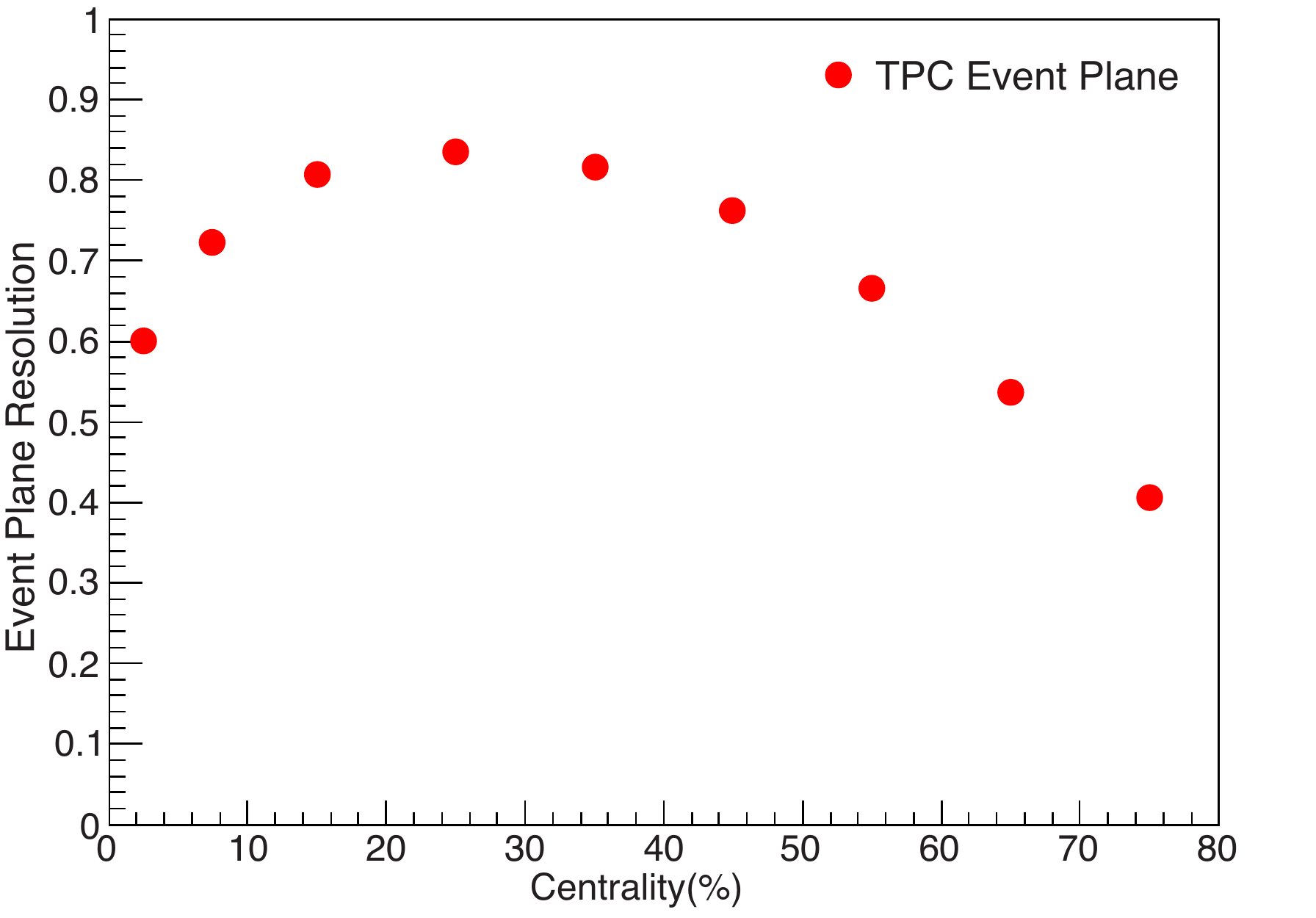}
\caption{\label{fig:fig4_3_02} TPC event plane resolution for $\sqrt{s_{\rm NN}}$ = 200 GeV Au+Au MinBias collisions. }

\end{figure}

Due to finite particle multiplicity, there are always differences between the measured event plane and the real reaction plane. This finite event plane resolution causes differences between the observed event plane dependent balance function and the real one. To correct for this resolution, an event plane resolution factor is used as \cite{flow_sergei} 

\begin{eqnarray}
\label{eq:eq_4_3_reso}
 < \cos [2(\psi _2  - \psi _r )] >  = \frac{{\sqrt \pi  }}{{2\sqrt 2 }}\chi _2 \exp ( - \chi _2^2 /4) \times [I_0 (\chi _2^2 /4) + I_1 (\chi _2^2 /4)]
\end{eqnarray}

\noindent
where $\psi_2$ is the second harmonic event plane calculated from the azimuthal distribution of the TPC tracks,  $\chi _2  \equiv v_2 /\sigma $ and $I_1$ is the first order modified Bessel function. In practice, a sub-event method is applied to calculate the event plane resolution by dividing the full event into two random sub-events and calculating the event plane angles in both events. If there are no other correlations, or if the correlations are negligible compared to the flow signal, the resolution of each of them can be calculated via

\begin{eqnarray}
 < \cos [2(\psi _2^a  - \psi _r )] >  = \sqrt { < \cos [2(\psi _2^a  - \psi _2^b) ] > } 
\end{eqnarray}

Once we have each sub-event,s resolution, we can use Equation \ref{eq:eq_4_3_reso} to extract their $\chi _2^a $. Because $\chi _2  \equiv v_2 /\sigma $ is proportional to the square root of the event multiplicity $\sqrt N $, we can calculate the full event's  $\chi _2  = \sqrt 2 \chi _2^a $ and use Equation \ref{eq:eq_4_3_reso} again to calculate the full event resolution.

Figure ~\ref{fig:fig4_3_02} shows the centrality dependence of the TPC event plane resolution for STAR Run 4 Au+Au collisions at $\sqrt{s_{\rm NN}}$ = 200 GeV. The resolution is relatively low in peripheral collisions due to low multiplicity and also low in the most central bin due to less anisotropic flow. 

\section{Mixed and Shuffled Events}

As discussed in Section~\ref{fluctuation}, the dynamical fluctuation $\sigma_{\rm dyn}$ is given by 

$$ \sigma _{\rm dyn}  = {\mathop{\rm sgn}} (\sigma _{\rm data}^2  - \sigma _{\rm mixed}^2 )\sqrt {|\sigma _{\rm data}^2  - \sigma _{\rm mixed}^2 |} $$

Note that mixed events are required to extract $\sigma _{\rm dyn}$. For balance function analysis, TPC sector boundaries introduce different acceptance for positive and negative charge particles. A mixed events subtraction is necessary to correct for this different acceptance. 

Mixed events are created by taking one track from each event, selected according to the bin in centrality and the bin in event vertex position. One single mixed event includes no more than one track from any observed event. This new mixed-event data set has the same number of events and the same multiplicity distribution as the original data set but all correlations are removed.

For balance function calculations, shuffled events are also created.  These shuffled events are produced by randomly shuffling the charges of the particles in each event. The shuffled events thus have all the momentum correlations and the same total charge observed in the original event, but the charge momentum correlations are removed. Because shuffling uniformly distributes a particle�s balancing partner across the measured phase space, balance functions calculated using shuffled events can be used to gauge the widest balance functions that one can measure using the STAR acceptance for the system under consideration.

\chapter{Results for Fluctuations}
\section{$K/\pi$, $p/\pi$ and $p/K$ fluctuations}

\begin{figure}
\includegraphics[width=38pc]{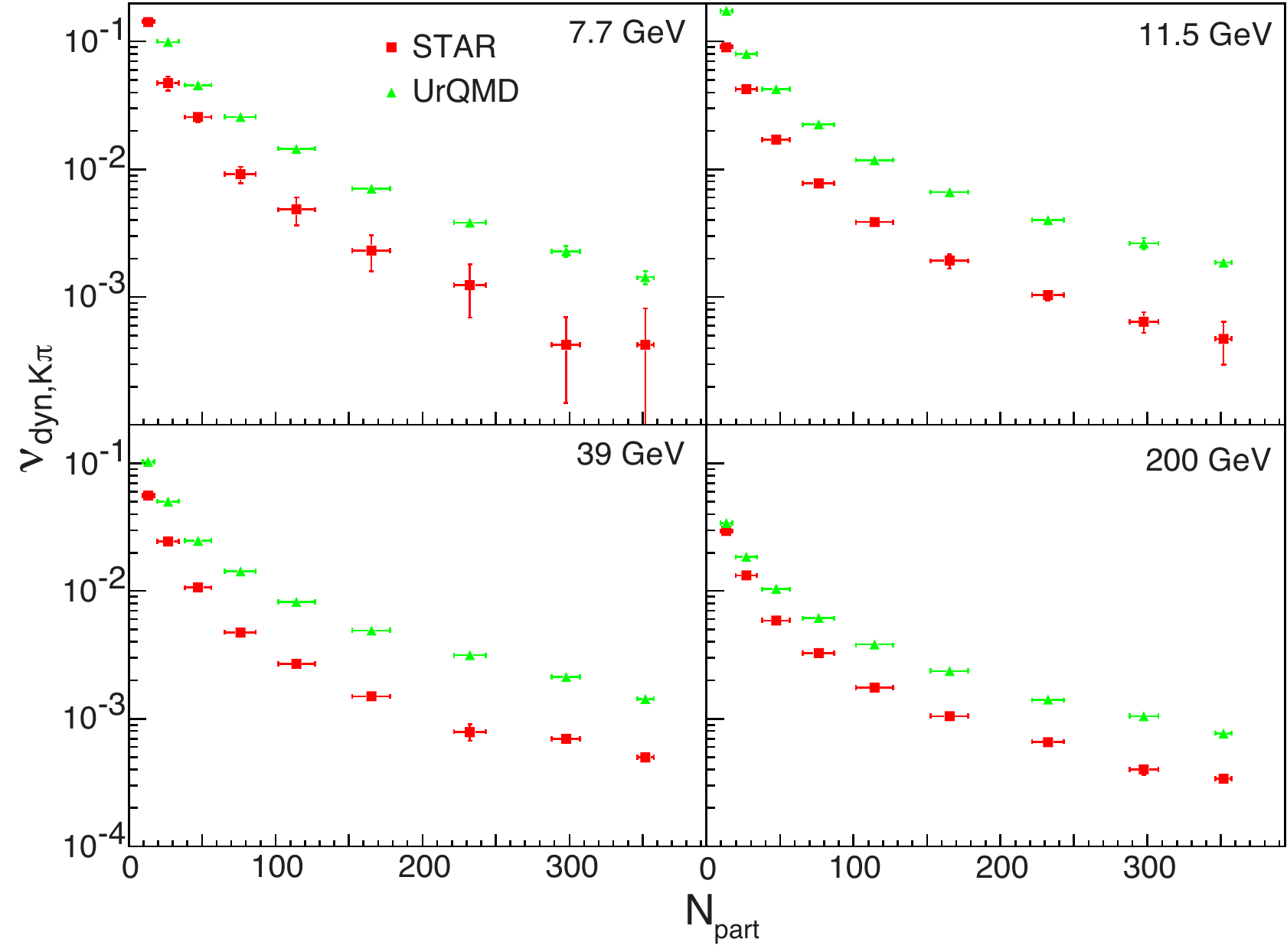}
\caption{\label{fig:fig5_1_01} Centrality dependence of $\nu_{{\rm dyn},  K\pi}$ results from Au+Au collisions at $\sqrt{s_{\rm NN}}$ = 7.7, 11.5, 39 and 200 GeV. The data (red squares) are compared to UrQMD model calculations (green triangles).}
\end{figure}

Figure~\ref{fig:fig5_1_01}  shows $\nu_{{\rm dyn},  K\pi}$ results plotted in terms of the number of participating nucleons from Au+Au collisions at $\sqrt{s_{\rm NN}}$ = 7.7, 11.5, 39 and 200 GeV. $\nu_{{\rm dyn},  K\pi}$  is positive and decreases with increasing number of participating nucleons. A UrQMD model calculation with STAR acceptance cuts is shown in the same figure. The UrQMD results have the same trend as the data but over predict the magnitude for all four energies. 

\begin{figure}
\includegraphics[width=38pc]{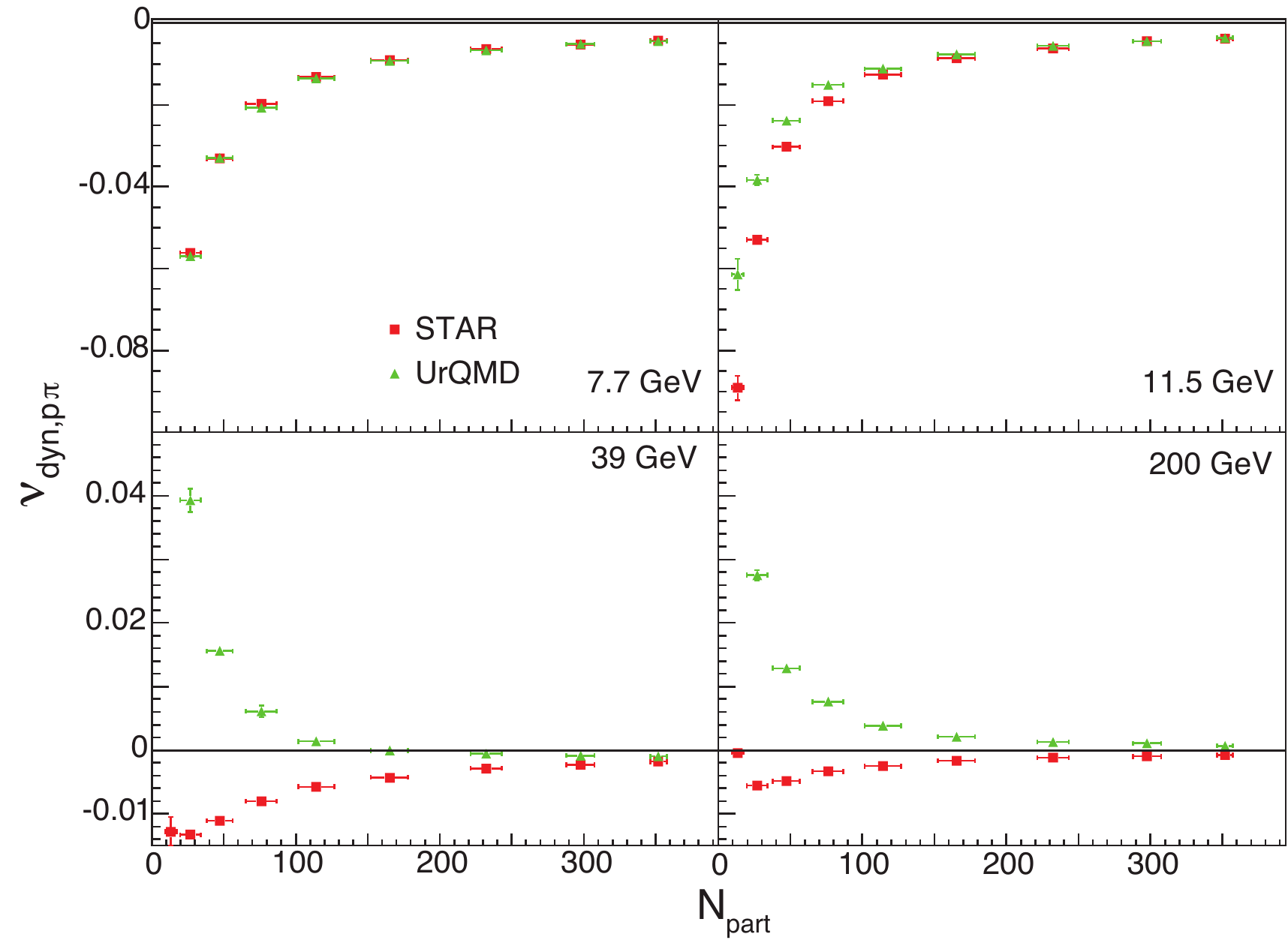}
\caption{\label{fig:fig5_1_02} Centrality dependence of $\nu_{{\rm dyn},  p\pi}$ results from Au+Au collisions at $\sqrt{s_{\rm NN}}$ = 7.7, 11.5, 39 and 200 GeV. Here the $p$ is the sum of protons and anti-protons, and the data (red squares) are compared to UrQMD model calculations (green triangles).}
\end{figure}

Figure~\ref{fig:fig5_1_02} shows the measured $p/\pi$ fluctuations in terms of the number of participating nucleons from Au+Au collisions at $\sqrt{s_{\rm NN}}$ = 7.7, 11.5, 39 and 200 GeV. Here $p$ is the sum of protons and anti-protons. Unlike $\nu_{{\rm dyn},  K\pi}$, the $\nu_{{\rm dyn},  p\pi}$ result is negative and increases with increasing number of participating nucleons. The negative value of  $\nu_{{\rm dyn}}$ means the cross-correlation terms dominate, which could be due to the proton-pion correlation from resonance decay (e.g. $\Delta  \to p + \pi$). The UrQMD model with STAR acceptance cuts agrees well with data at 7.7 GeV, but over predicts the data at higher energies and in peripheral collisions.

\begin{figure}
\includegraphics[width=38pc]{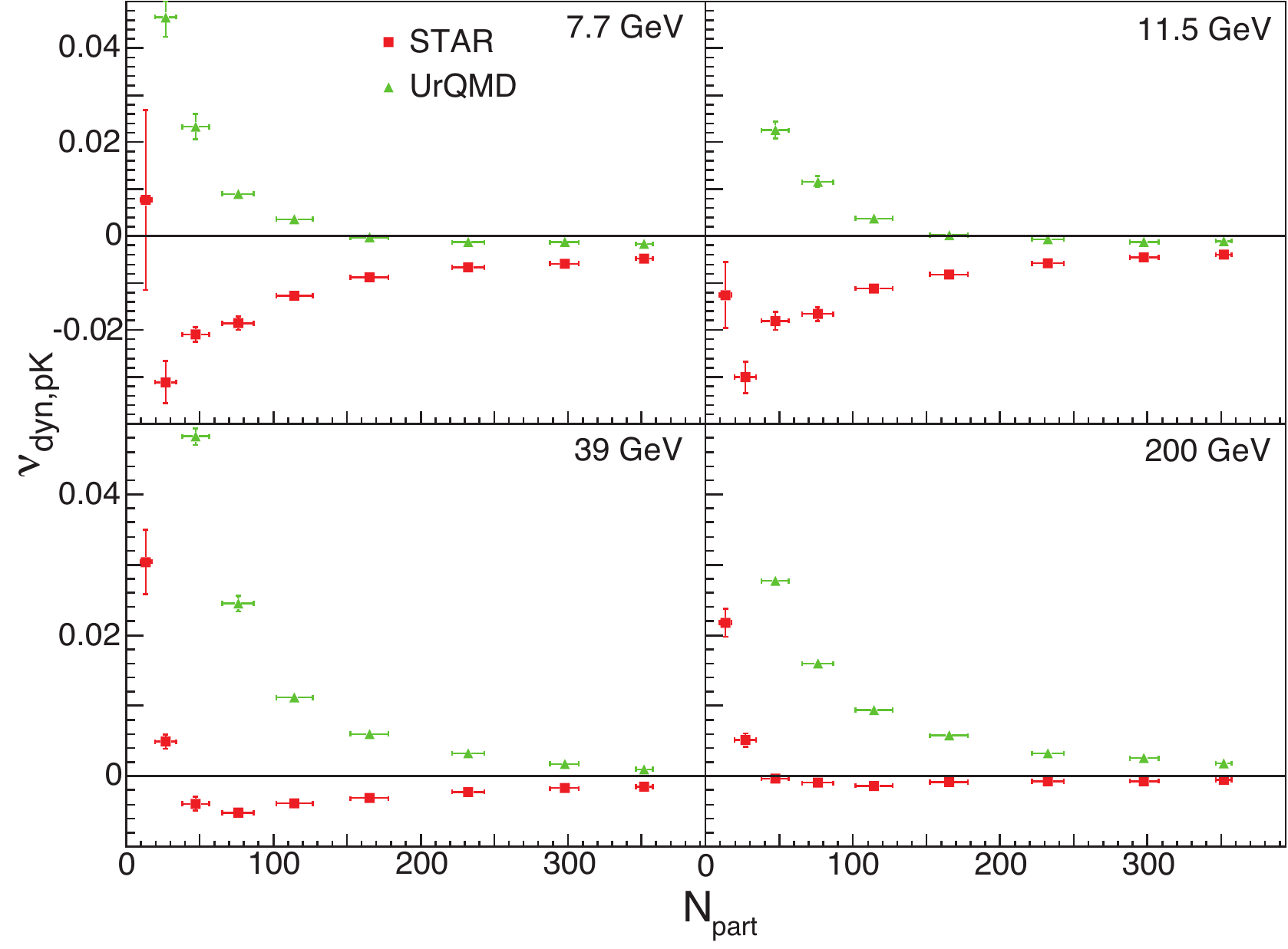}
\caption{\label{fig:fig5_1_03} Centrality dependence of $\nu_{{\rm dyn},  pK}$ results from Au+Au collisions at $\sqrt{s_{\rm NN}}$ = 7.7, 11.5, 39 and 200 GeV. Here the $p$ is the sum of protons and anti-protons, and the data (red squares) are compared to UrQMD (green triangles).}
\end{figure}

Figure~\ref{fig:fig5_1_03}  shows $\nu_{{\rm dyn},  pK}$ results plotted versus the number of participating nucleons from Au+Au collisions at $\sqrt{s_{\rm NN}}$ = 7.7, 11.5, 39 and 200 GeV. For all four energies shown here, the data show a smooth decrease with decreasing number of participating nucleons. For 39 GeV and 200 GeV, the data show smaller centrality dependence and $\nu_{{\rm dyn}, p/K}$ changes from negative to positive values at peripheral collisions, which implies enhanced fluctuations at peripheral collisions. The UrQMD model  with STAR acceptance cuts over predicts the magnitude, especially for peripheral collisions.

\begin{figure}
\centering
\includegraphics[width=32pc]{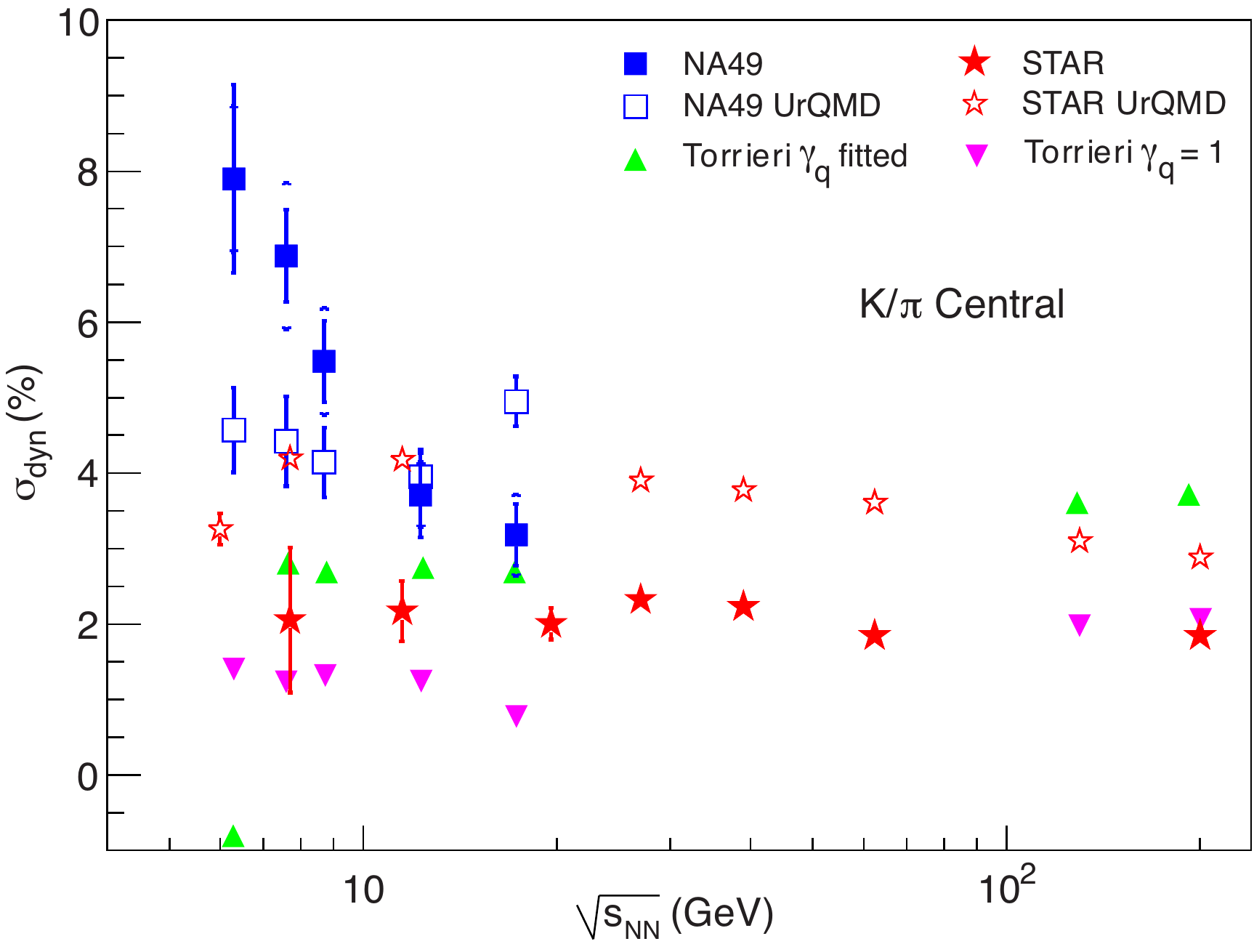}
\caption{\label{fig:fig5_1_04} Energy dependence of $\sigma_{{\rm dyn},  K\pi}$ results. STAR results (red stars) are from Au+Au collisions, 0-5\% centrality, only statistical errors are shown, while NA49 results (blue squares) are from Pb+Pb collisions, 0-3.5\% centrality, both statistical and systematical error are shown.}
\end{figure}

\begin{table}
\begin{center}
  \begin{tabular}{ |c || c | c | c |  }
    \hline
     $\sqrt{s_{\rm NN}}$  (GeV) & $<K>$&  $<\pi>$ & $<p>$ \\ \hline
     200  &38.66 & 468.47 & 29.27 \\ \hline
     62.4 &32.77 & 395.70 & 25.95 \\ \hline
     39     &29.65 & 369.41 & 25.91 \\ \hline
     27     &27.58 & 348.20 & 26.86 \\ \hline
     19.6  &25.15 & 317.24 & 28.83 \\ \hline
     11.5  &19.27 & 243.52 & 33.44 \\ \hline
     7.7     &14.25 & 185.47 & 40.23 \\
    
    \hline

  \end{tabular}
\caption{\label{tab:tab5_3_01} Identified particle numbers used in the $\nu_{\rm dyn}$ calculation, 0-5\% centrality only.}
\end{center}
\end{table}

The energy dependence of particle-ratio fluctuations is also a interesting topic. Previous results from NA49\cite{kpi_NA49} and STAR\cite{kpi_STAR} show that for $K/\pi$ fluctuations, $\sigma_{\rm dyn}$ is constant from RHIC to top SPS energies but rises strongly at lower SPS energies. This increase of fluctuation at low SPS energies could indicate the onset of deconfinement at RHIC top energies. Figure~\ref{fig:fig5_1_04} shows the $\sigma_{\rm dyn}$ results using STAR's new 7.7, 11.5, 39, 62.4, 200 GeV data from Run 10 and 19 and 27 GeV data from Run 11. Note here the STAR results (data and UrQMD) are calculated via  $\nu_{\rm dyn}$ and converted to $\sigma_{\rm dyn}$ using the relation $\sigma _{\rm dyn}  = {\rm sgn}(\nu_{\rm dyn})\sqrt{|\nu _{\rm dyn}|}$.  STAR results (red stars) are approximately independent of collision energy at a level of about 2\%. This disagrees with NA49's results (blue squares), which show a strong increase with decreasing incident energy. Table~\ref{tab:tab5_3_01} shows the identified particle numbers used in this analysis, while the NA49 numbers can be found at Ref.~\cite{pk_NA49}.

The same figure also shows model calculations. The points labeled STAR UrQMD represent a UrQMD calculation with STAR acceptance cuts while the points labeled NA49 UrQMD show a UrQMD calculation with NA49 acceptance cuts applied. These two UrQMD results agree well with each other, which indicates  $\sigma_{{\rm dyn},  K\pi}$ should be independent of experimental acceptance differences between the two experiments. Both UrQMD calculations show little energy dependence and over predict the magnitude of the data. The triangles in Figure~\ref{fig:fig5_1_04}  are statistical hadronization model results from Torrieri \cite{SH_Torrieri}. The magenta triangles stand for the chemical equilibrium model with light quark phase space occupancy $\gamma _q  = 1 $, while the green triangles show the chemical non-equilibrium model in which the value of $\gamma _q $ if varied to reproduce the $K^{+}/\pi^{+}$ yield ratios from RHIC to SPS energies. The equilibrium model agrees well with the data at high incident energies, but slightly under predicts the data at SPS energies.  The non-equilibrium model over predicts the fluctuations at all energies. None of the models presented here fully describe the incident energy dependence of the data.

\begin{figure}
\centering
\includegraphics[width=32pc]{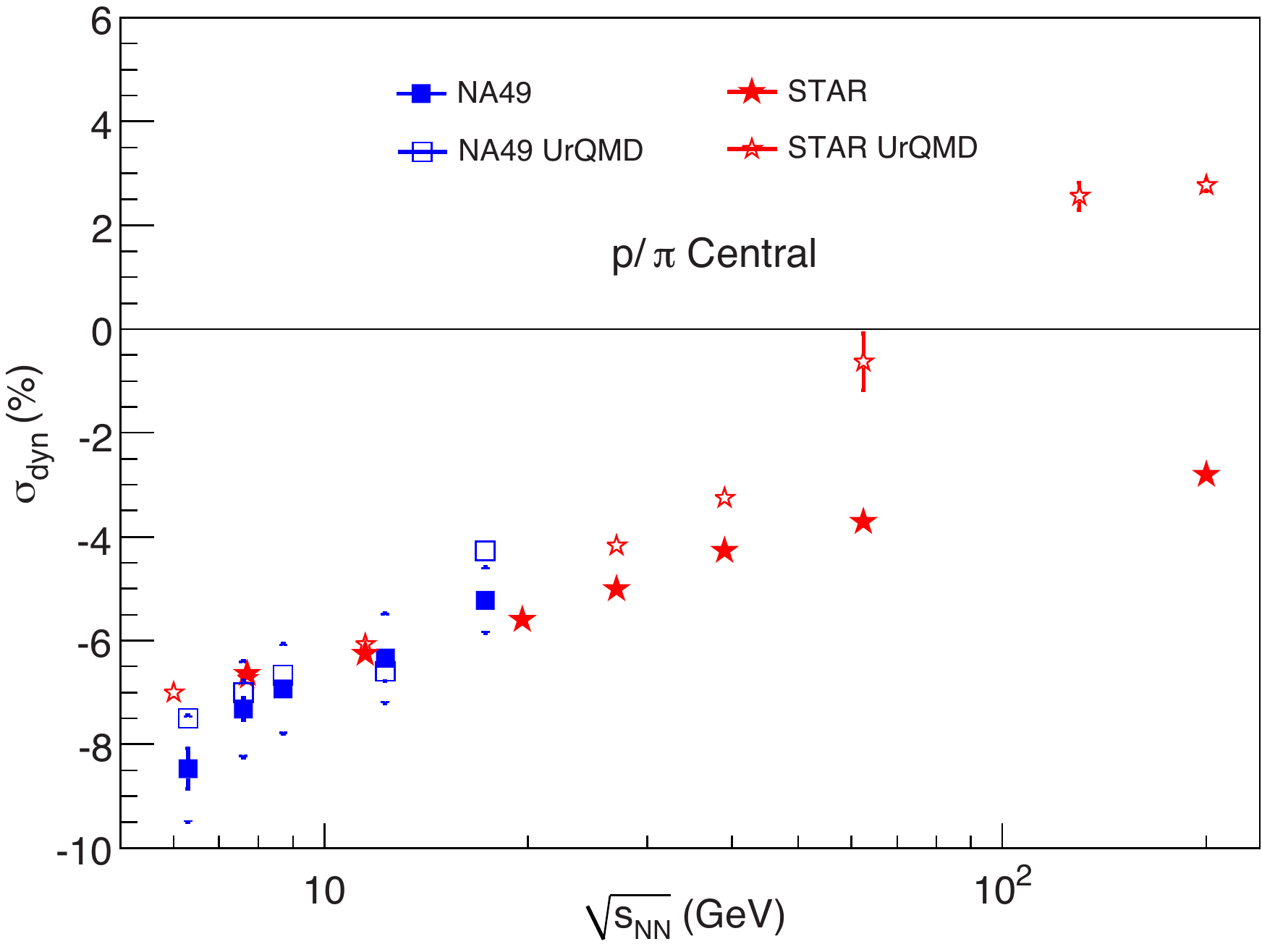}
\caption{\label{fig:fig5_1_05} Energy dependence of $\sigma_{{\rm dyn},  p\pi}$ results. STAR results (red stars) are from Au+Au collisions, 0-5\% centrality, only statistical errors are shown, while NA49 results (blue squares) are from Pb+Pb collisions, 0-3.5\% centrality, both statistical and systematical error are shown.}

\end{figure}

Unlike the results for $K/\pi$ fluctuations, the results for $p/\pi$ fluctuations are affected by resonance correlations (e.g. $\Delta ,\Lambda ,\Sigma$ all decay to $p,\pi$). These correlations increase the cross-correlation terms of $\nu_{\rm dyn}$ and produce a negative $\nu_{\rm dyn}$  value.   Figure~\ref{fig:fig5_1_05} shows the incident energy dependence of $\sigma_{\rm dyn}$, again STAR results (data and UrQMD) are calculated via  $\nu_{\rm dyn}$ and converted to $\sigma_{\rm dyn}$ using the relation $\sigma _{\rm dyn}  = {\rm sgn}(\nu_{\rm dyn})\sqrt{|\nu _{\rm dyn}|}$. Unlike the results for $K/\pi$ fluctuations, the STAR and NA49~\cite{kpi_NA49} results for $p/\pi$ fluctuations show good agreement.  They are both negative and increase with increasing collision energy. The UrQMD model describes the data well at SPS energies, which supports the resonance correlations interpretation because UrQMD is a hadronic transport model. However, UrQMD become positive and over predicts the data at higher energies.

\begin{figure}
\centering
\includegraphics[width=32pc]{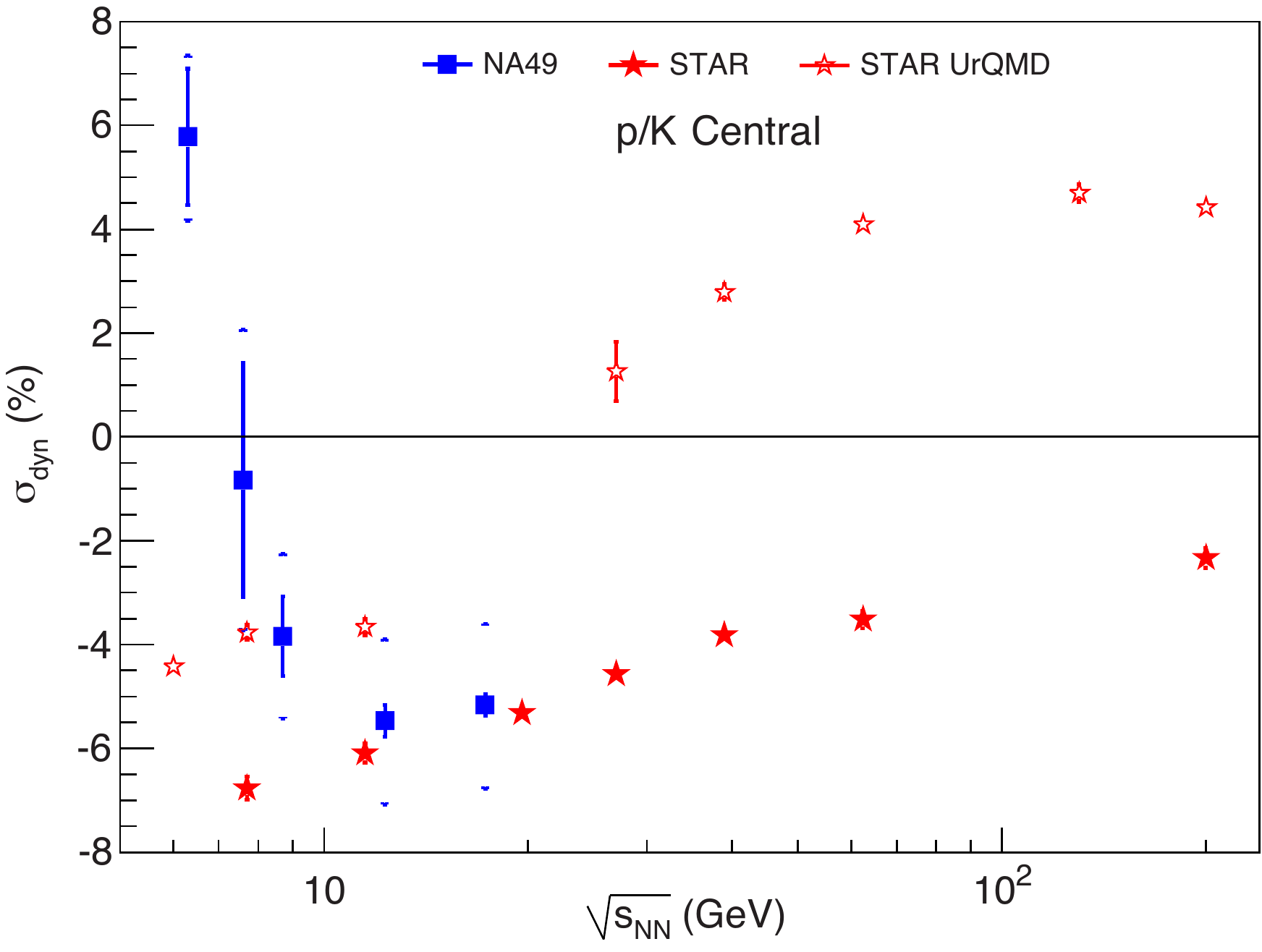}
\caption{\label{fig:fig5_1_06} Energy dependence of $\sigma_{{\rm dyn},  pK}$ results. STAR results (red stars) are from Au+Au collisions, 0-5\% centrality and only statistical errors are shown, while NA49 results (blue squares) are from Pb+Pb collisions, 0-3.5\% centrality, and both statistical and systematical error are shown.}

\end{figure}

$p/K$ fluctuations, which are related to baryon-strangeness correlations, can be used as a tool to study the deconfinement phase transition \cite{pk_correlation}. In a deconfined phase where quarks and gluons are the basic degrees of freedom, strange quarks ($S$ = -1, $B$ = 1/3) are the only carrier of strangeness, which results in a strong correlation between baryon number and strangeness. In a hadron gas, however, strangeness is mainly carried by the kaons which have zero baryon number , and thus greatly reduce the correlation between baryon number and strangeness.  Figure~\ref{fig:fig5_1_06} shows the incident energy dependence of $\sigma_{\rm dyn}$ results, as before the STAR results (data and UrQMD) are calculated via  $\nu_{\rm dyn}$ and converted to $\sigma_{\rm dyn}$ using the relation $\sigma _{\rm dyn}  = {\rm sgn}(\nu_{\rm dyn})\sqrt{|\nu _{\rm dyn}|}$. The NA49 results show a non-trivial increase of fluctuation with decreasing collision energy: from more correlation (negative $\sigma_{\rm dyn}$) to enhanced fluctuation (positive $\sigma_{\rm dyn}$). This suggests a possible change in the baryon number-strangeness correlation at$\sqrt{s_{\rm NN}}  \approx $  8 GeV \cite{pk_NA49}. However, the STAR data in Figure~\ref{fig:fig5_1_06} show a smooth decrease with decreasing collision energy and disagree with NA49 data at 7.7 GeV. Further study is still needed to understand the differences between the two experiments. However, it should be noted that NA49's kaon identification ability might be suspect because a similar trend is observed in $K/\pi$ fluctuations, which also requires kaon identification. 

A UrQMD calculation with the STAR acceptance filter is also shown in the same figure. UrQMD always over predicts fluctuations and becomes positive at high collision energies. 

\section{Separate Sign Fluctuations}

The motivation underlying the study of fluctuations is to identify critical fluctuations.  However, hadronic processes like resonance decay can also influence the particle-ratio fluctuations. To better understand the origin of the observed signal for fluctuations, separate sign fluctuations are also measured for STAR data. Because the $K^{-}$ and $\bar{p}$ yields go to zero at low energies, we will present only $K^{+}$ and $p$ related fluctuations.

\begin{figure}
\centering
\includegraphics[width=38pc]{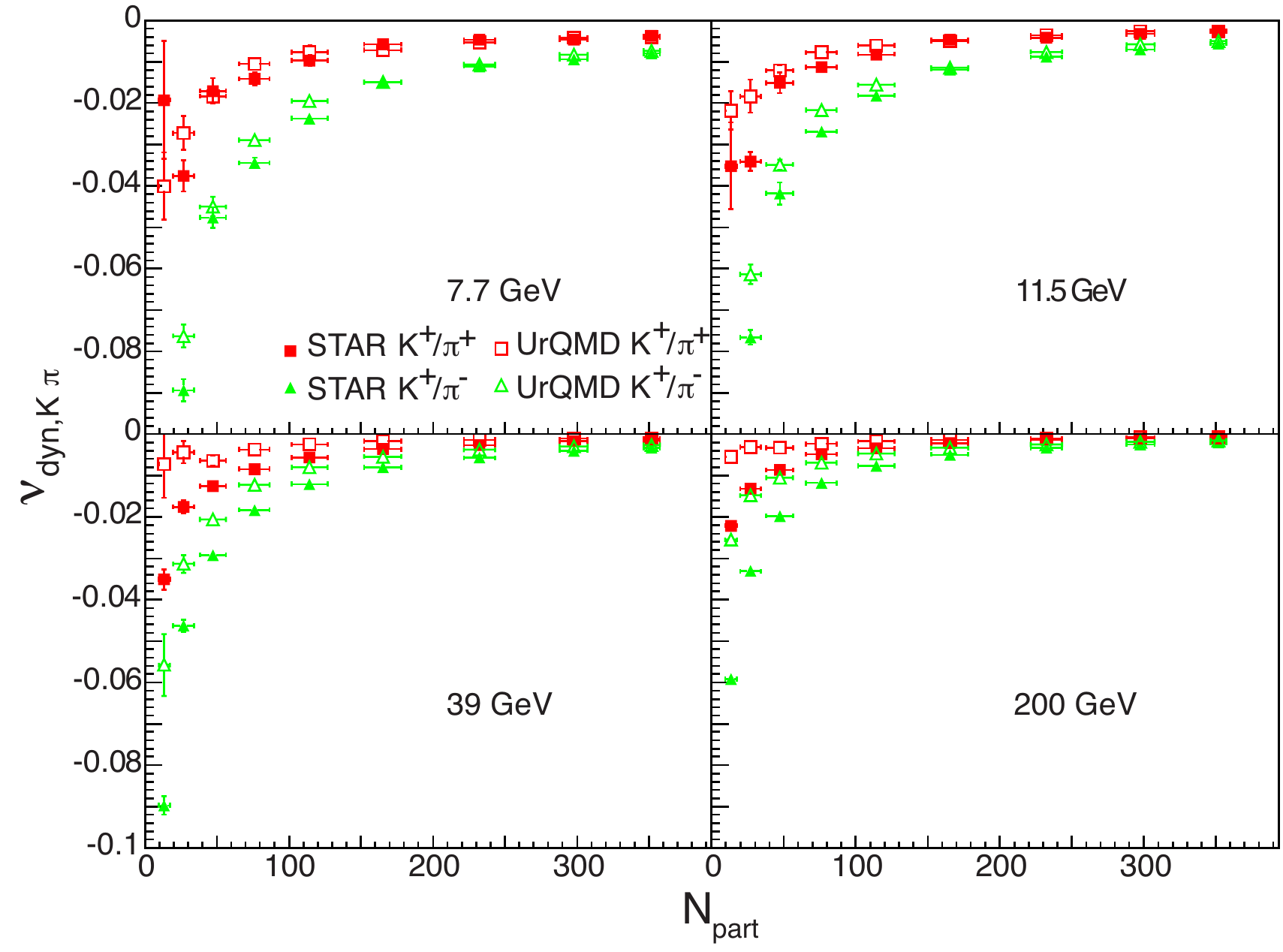}
\caption{\label{fig:fig5_2_01}Centrality dependence of separate sign $\nu_{{\rm dyn},  K^{+}\pi^{+}}$ and $\nu_{{\rm dyn},  K^{+}\pi^{-}}$ results from Au+Au collisions at $\sqrt{s_{\rm NN}}$ = 7.7, 11.5, 39 and 200 GeV. The data (solid symbols) are compared to UrQMD model calculations (hollow symbols).}

\end{figure}

Figure~\ref{fig:fig5_2_01} shows $\nu_{{\rm dyn},  K^{+}\pi^{+}}$ and $\nu_{{\rm dyn},  K^{+}\pi^{-}}$ results plotted in terms of the number of participating nucleons from Au+Au collisions at $\sqrt{s_{\rm NN}}$ = 7.7, 11.5, 39 and 200 GeV. The results for $\nu_{{\rm dyn},  K^{+}\pi^{+}}$  are negative and decrease with decreasing number of participating nucleons. $\nu_{{\rm dyn},  K^{+}\pi^{-}}$  shows a stronger negative value, which could be due to decay process like $ K^* (892) \to K^ +   + \pi ^ +  $. A UrQMD model calculation with STAR acceptance cuts is also shown in the same figure. The UrQMD results agree with data in central collisions, but over predict the signal for peripheral collisions, especially at high collision energies.

\begin{figure}
\centering
\includegraphics[width=38pc]{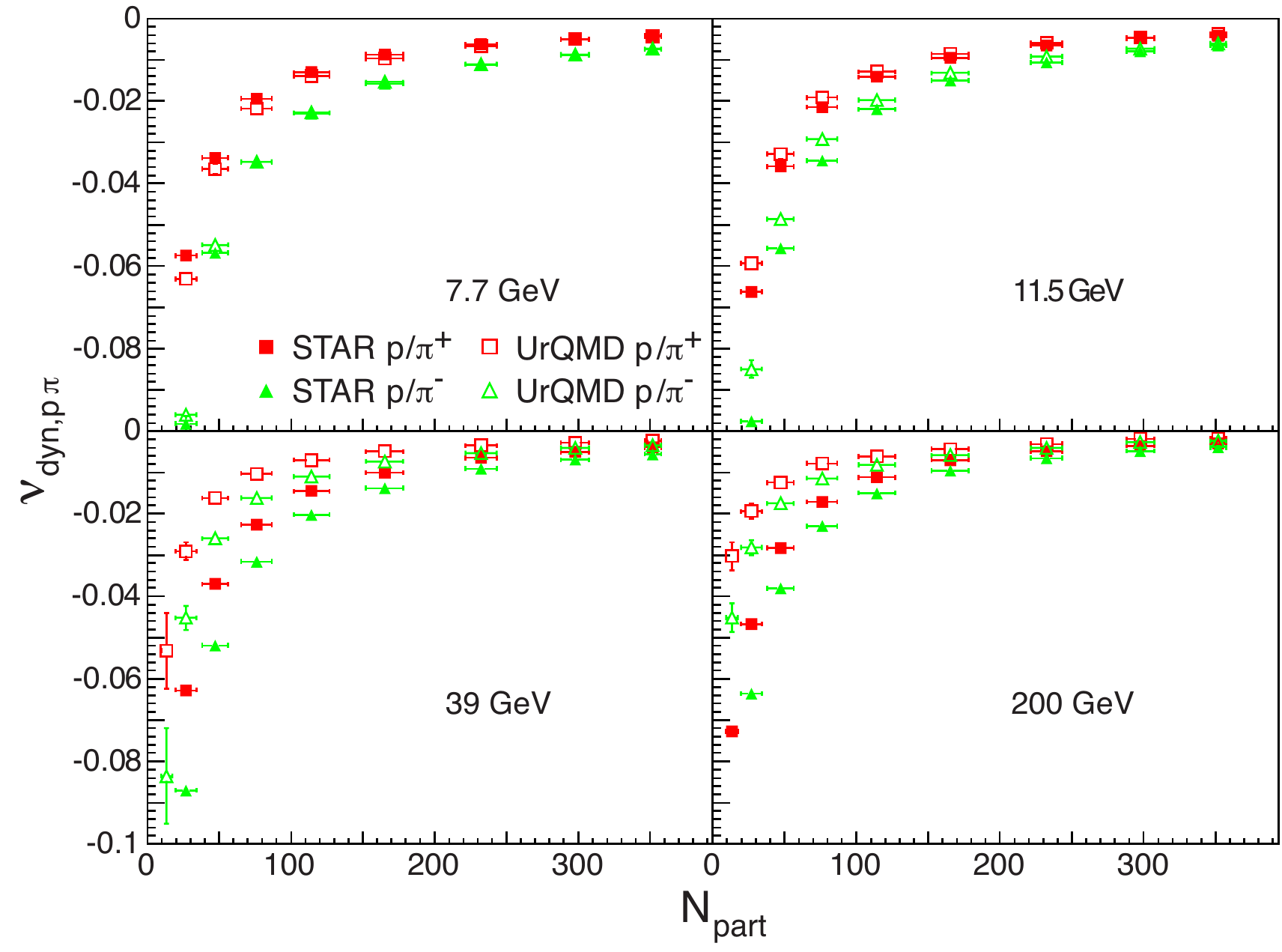}
\caption{\label{fig:fig5_2_02}Centrality dependence of separate sign $\nu_{{\rm dyn},  p\pi}$ results from Au+Au collisions at $\sqrt{s_{\rm NN}}$ = 7.7, 11.5, 39 and 200 GeV. The data (solid symbols) are compared to UrQMD model calculation (hollow symbols).}

\end{figure}

\begin{figure}
\centering
\includegraphics[width=38pc]{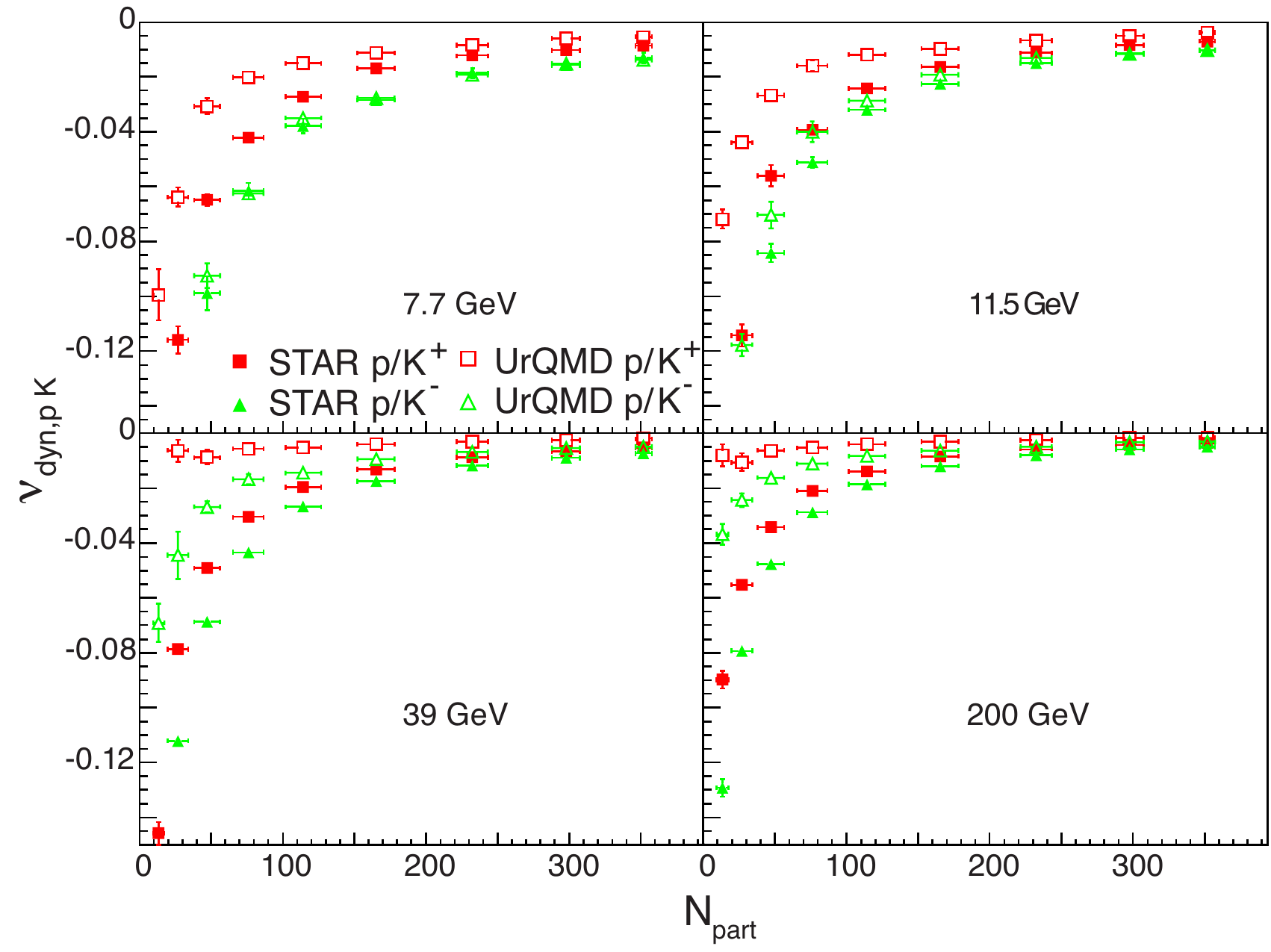}
\caption{\label{fig:fig5_2_03}Centrality dependence of separate sign $\nu_{{\rm dyn},  pK}$ results from Au+Au collisions at $\sqrt{s_{\rm NN}}$ = 7.7, 11.5, 39 and 200 GeV. The data (solid symbols) are compared to UrQMD model calculation (hollow symbols).}

\end{figure}

Similarly, Figure~\ref{fig:fig5_2_02} shows the centrality dependence $\nu_{{\rm dyn},  p\pi^{+}}$ and $\nu_{{\rm dyn},  p\pi^{-}}$ results. $\nu_{{\rm dyn},  p\pi^{+}}$ is negative and decreases with decreasing number of participating nucleons. $\nu_{{\rm dyn},  p\pi^{-}}$  shows a stronger negative value, which could due to $\Lambda$ decay. A UrQMD model calculation with the STAR acceptance cuts is also shown in the same figure. The UrQMD results agree with data at low energies and central collisions, but over predict the signal at peripheral collisions and higher collision energies.  A similar result is also observed for $\nu_{{\rm dyn},  pK^{+}}$ and $\nu_{{\rm dyn},  pK^{-}}$ fluctuations in Figure~\ref{fig:fig5_2_03}.

\begin{figure}
\centering
\includegraphics[width=30pc]{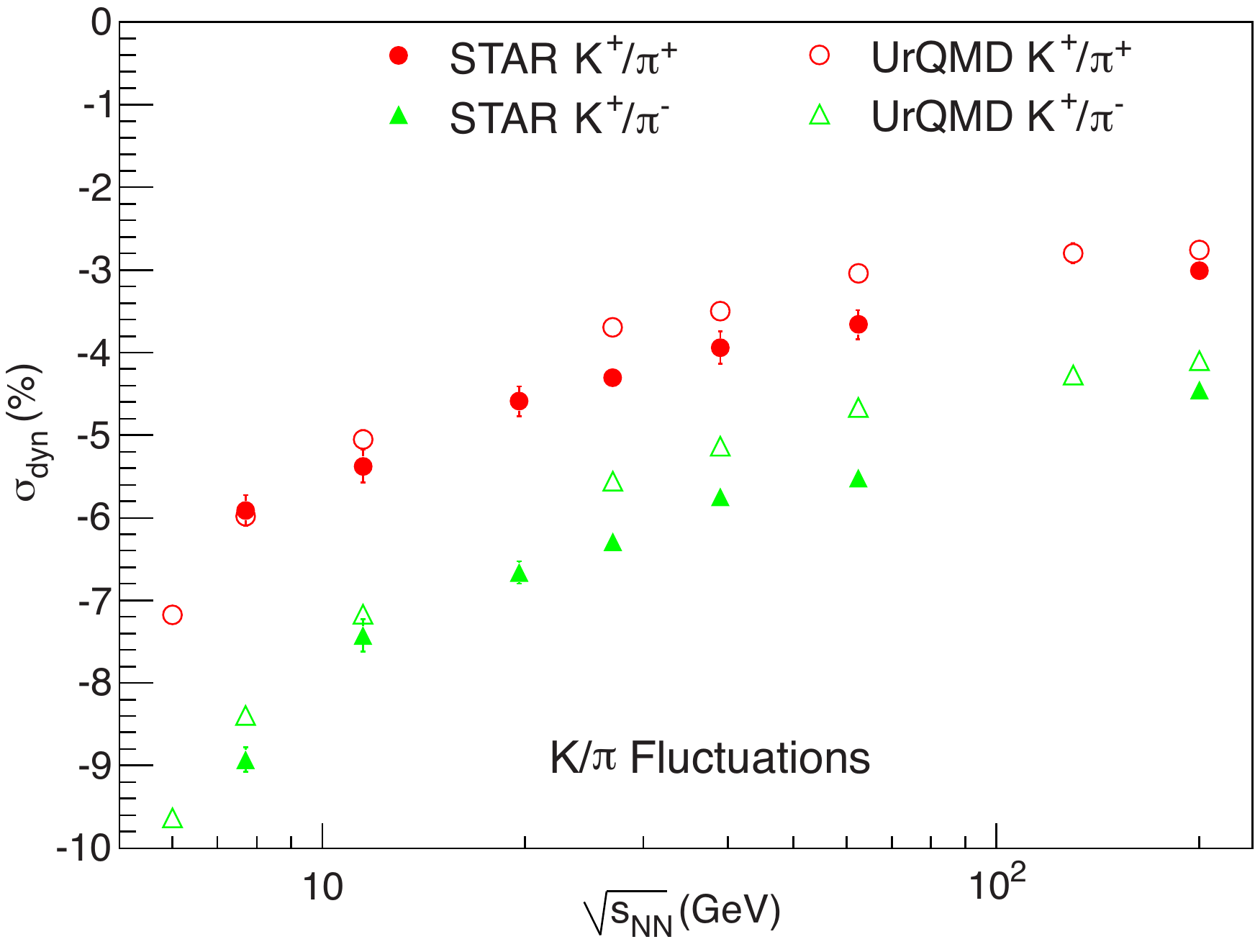}
\caption{\label{fig:fig5_2_04} Energy dependence of separate sign $\sigma_{{\rm dyn},  K\pi}$ results. STAR results (solid symbols) are from Au+Au collisions, 0-5\% centrality. Also shown are UrQMD results (hollow symbols).}

\end{figure}

Because resonance decays are more significant at lower collision energies, it is expected that the correlations should be stronger at low $\sqrt{s_{\rm NN}}$. Figure~\ref{fig:fig5_2_04} shows the incident energy dependence of the $\sigma_{{\rm dyn},  K^{+}\pi^{+}}$ and $\sigma_{{\rm dyn},  K^{+}\pi^{-}}$ results from central Au+Au collisions (0-5\%). Note that the results are calculated via  $\nu_{\rm dyn}$ and converted to $\sigma_{{\rm dyn}}$ using the relation $\sigma _{\rm dyn}  = {\rm sgn}(\nu_{\rm dyn})\sqrt{|\nu _{\rm dyn}|}$. Both  $\sigma_{{\rm dyn},  K^{+}\pi^{+}}$ and $\sigma_{{\rm dyn},  K^{+}\pi^{-}}$ are negative and show a smooth decrease with decreasing incident energy, while the UrQMD results with a STAR acceptance filter slightly over predict the data. Since UrQMD is a hadronic transport model, the good agreement between data and model indicate the hadronic contribution dominates the measured fluctuation signal. As discussed before, decay processes like $ K^* (892) \to K^ +   + \pi ^ -  $ introduce a strong correlation to the $K^+/\pi^-$ fluctuation, while other resonance decays like $ K_1 (1270)^ +   \to K^ +   + \rho ^0  \to K^ +   + \pi ^ +   + \pi ^ -  $ could give negative $K^+/\pi^+$ fluctuations. A study \cite{UrQMD_decay} using UrQMD also confirmed that removal of $K^*$ and $\phi$ decays significantly change the summed sign and separate sign $K/\pi$ fluctuation results.

\begin{figure}
\centering
\includegraphics[width=30pc]{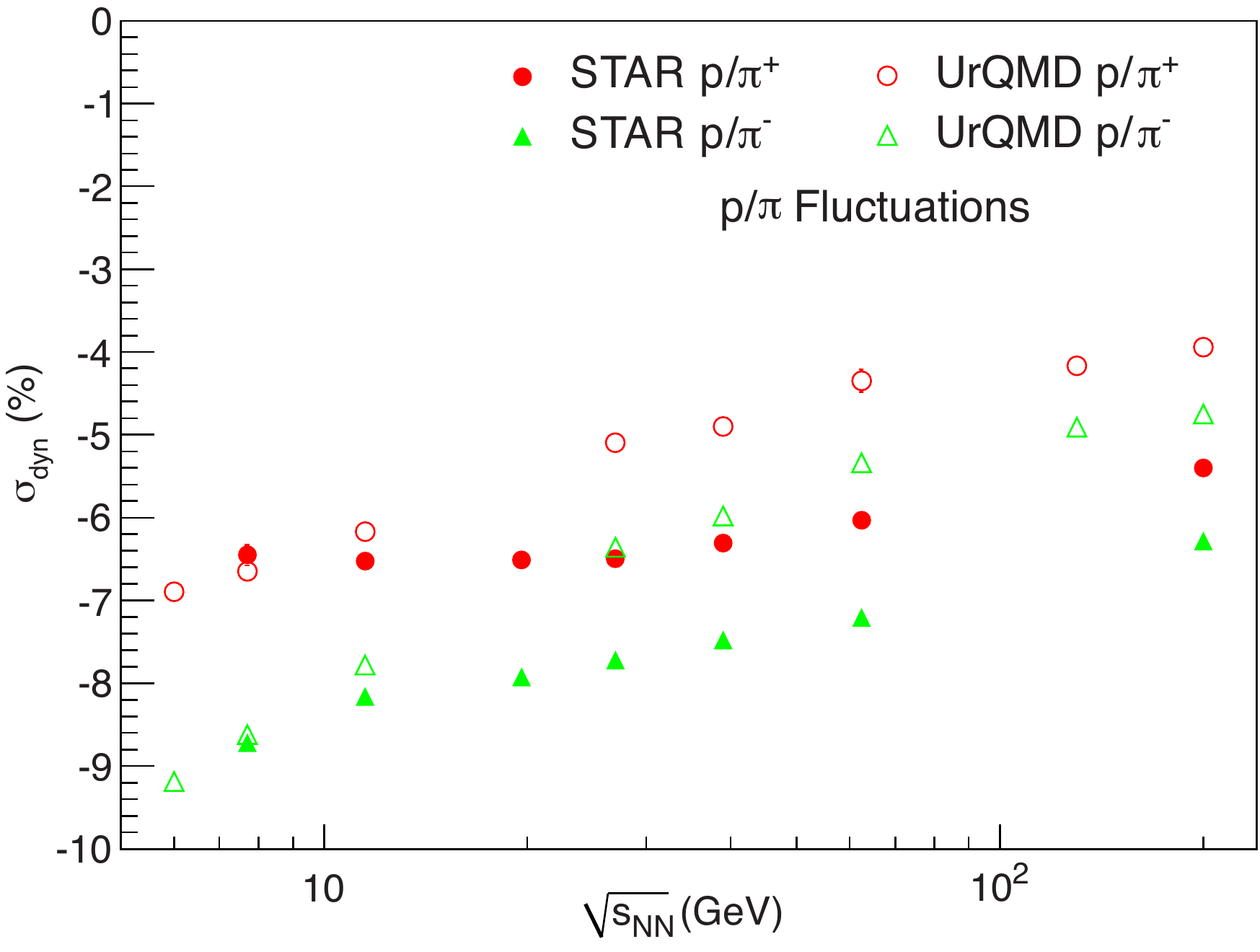}
\caption{\label{fig:fig5_2_05} Energy dependence of separate sign $\sigma_{{\rm dyn},  p\pi}$ results. STAR results (solid symbols) are from Au+Au collisions, 0-5\% centrality, also shown are UrQMD results(hollow symbols).}

\end{figure}

Figure~\ref{fig:fig5_2_05} shows the energy dependence of separate sign $p/\pi$ fluctuations with 0-5\% centrality.  Since the anti-proton yield vanishes at low collision energy, only proton results are shown here. Both $\sigma_{{\rm dyn},  p\pi^{+}}$ and $\sigma_{{\rm dyn},  p\pi^{-}}$ results are negative and decrease with decreasing collision energy. The $\sigma_{{\rm dyn},  p\pi^{-}}$ are more correlated due to $\Delta$ decays: $ \Delta  \to p + \pi ^ -  $, The UrQMD model agrees well with data at low energies but over predicts at high RHIC energies.

\begin{figure}
\centering
\includegraphics[width=30pc]{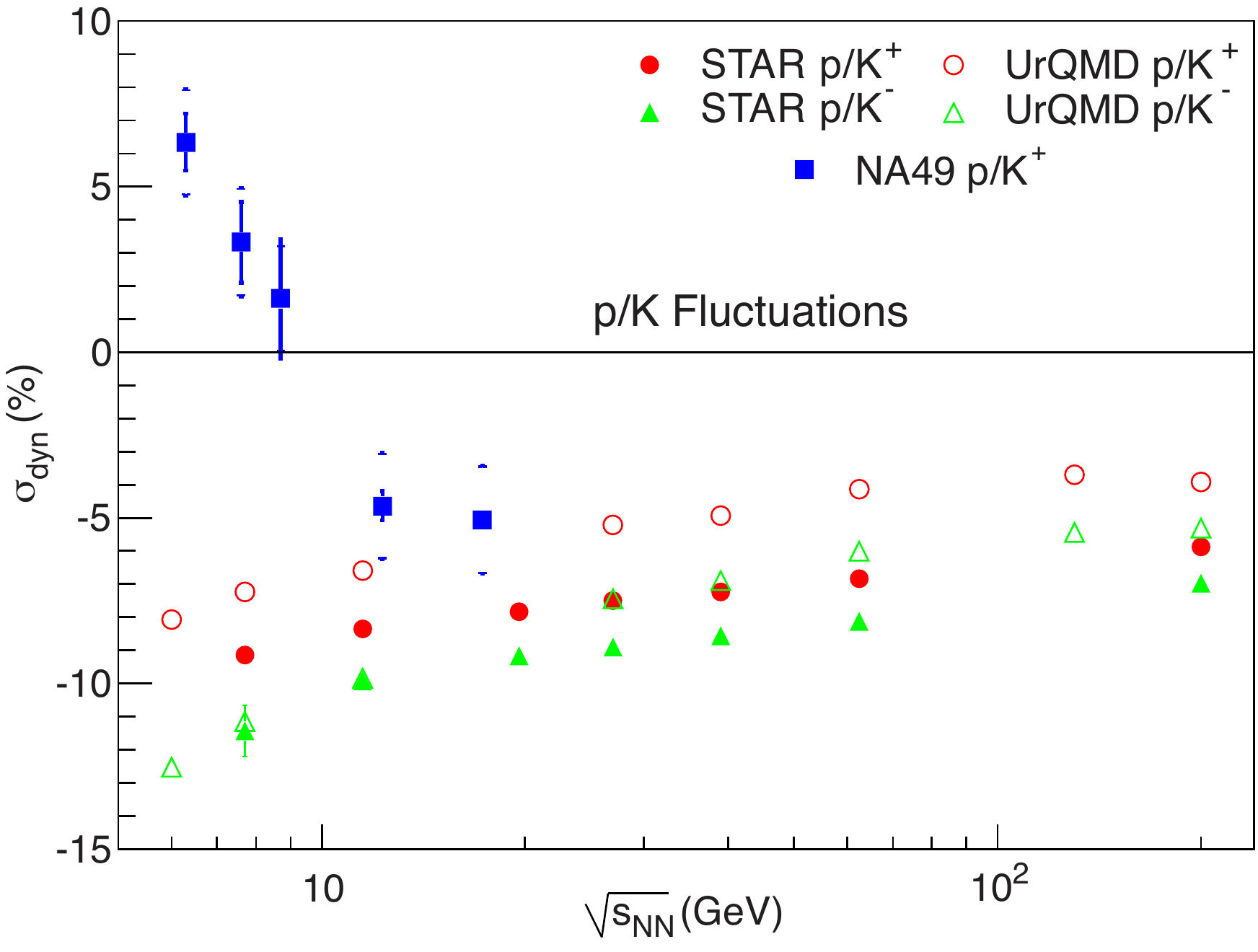}
\caption{\label{fig:fig5_2_06} Energy dependence of separate sign $\sigma_{{\rm dyn},  pK}$ results. STAR results (solid circles) are from Au+Au collisions, 0-5\% centrality. Only statistical error are shown.  NA49 results (blue squares) are from Pb+Pb collisions, 0-3.5\% centrality, with both statistical and systematical error being shown. Also shown are UrQMD results (hollow symbols).}

\end{figure}

Since there is no known resonance decay that feeds into protons and $K^+$, it has been suggested that $p/K^+$ fluctuation could be a good tool to investigate the proton-kaon correlations. Figure~\ref{fig:fig5_2_06} shows the incident energy dependence of separate sign $\sigma_{{\rm dyn}}$ results. Only proton results are shown due to vanishing $\bar{p}$ yield at low energies. STAR results (data and UrQMD) are calculated via  $\nu_{\rm dyn}$ and converted to $\sigma_{{\rm dyn}}$ using the relation $\sigma _{\rm dyn}  = {\rm sgn}(\nu_{\rm dyn})\sqrt{|\nu _{\rm dyn}|}$. Similar to the summed sign $p/K$ fluctuations, the NA49 results show a non-trivial increase of fluctuations with decreasing collision energy from more correlation (negative $\sigma_{{\rm dyn}}$) to enhanced fluctuations (positive $\sigma_{{\rm dyn}}$). This is again suggested as a possible change in the baryon number-strangeness correlation \cite{pk_NA49}. However, STAR results for $\sigma_{{\rm dyn},  p/K^{+}}$ in the same figure show a smooth decrease with decreasing collision energy, which disagrees with NA49 data but generally agrees with the trend of UrQMD. The reason for the negative values of $\sigma_{{\rm dyn},  pK^{+}}$ is still under discussion.

\section{Scaling Properties of Fluctuation Observable}

In previous sections, we discussed the centrality and energy dependence of the particle-ratio fluctuations. It has been suggested that these fluctuation results could depend on the experimental multiplicity, which is the number of actual identified particles used in $\nu_{\rm dyn}$ calculation \cite{kpi_scaling}. Since the fluctuation observable is proposed as a tool to study QCD critical point, it is crucial to understand those trivial effects like multiplicity dependence first. The following  scalings \cite{kpi_scaling} have been suggested to study the energy dependence of $K/\pi$ fluctuations:

200~GeV
Poisson scaling,

\begin{eqnarray}
\sigma _{\rm dyn} (\sqrt s ) = \sigma _{\rm dyn} ({\rm 200~GeV}) \times \frac{{\sqrt {\frac{1}{{ < K > }} + \frac{1}{{ < \pi  > }}} \left| {_{\sqrt s } } \right.}}{{\sqrt {\frac{1}{{ < K > }} + \frac{1}{{ < \pi  > }}} \left| {_{{\rm 200~GeV}} } \right.}}
\end{eqnarray}

Particle number scaling,

\begin{eqnarray}
\sigma _{\rm dyn} (\sqrt s ) = \sigma _{\rm dyn} ({\rm 200~GeV}) \times \frac{{\sqrt { < K >  +  < \pi  > } \left| {_{{\rm 200~GeV}} } \right.}}{{\sqrt { < K >  +  < \pi  > } \left| {_{\sqrt s } } \right.}}
\end{eqnarray}

$N_{K}$ scaling,

\begin{eqnarray}
\sigma _{\rm dyn} (\sqrt s ) = \sigma _{\rm dyn} ({\rm 200~GeV}) \times \frac{{\sqrt { < K > } \left| {_{{\rm 200~GeV}} } \right.}}{{\sqrt { < K > } \left| {_{\sqrt s } } \right.}}
\end{eqnarray}

$N_{\pi}$ scaling,

\begin{eqnarray}
\sigma _{\rm dyn} (\sqrt s ) = \sigma _{\rm dyn} ({\rm 200~GeV}) \times \frac{{\sqrt { < \pi > } \left| {_{{\rm 200~GeV}} } \right.}}{{\sqrt { < \pi > } \left| {_{\sqrt s } } \right.}}
\end{eqnarray}

Geometric scaling, 

\begin{eqnarray} 
\sigma _{\rm dyn} (\sqrt s ) = \sigma _{\rm dyn} ({\rm 200~GeV}) \times \frac{{( < K >  < \pi  > )^{1/4} \left| {_{{\rm 200~GeV}} } \right.}}{{( < K >  < \pi  > )^{1/4} \left| {_{\sqrt s } } \right.}}
\end{eqnarray}

\begin{figure}
\centering
\includegraphics[width=32pc]{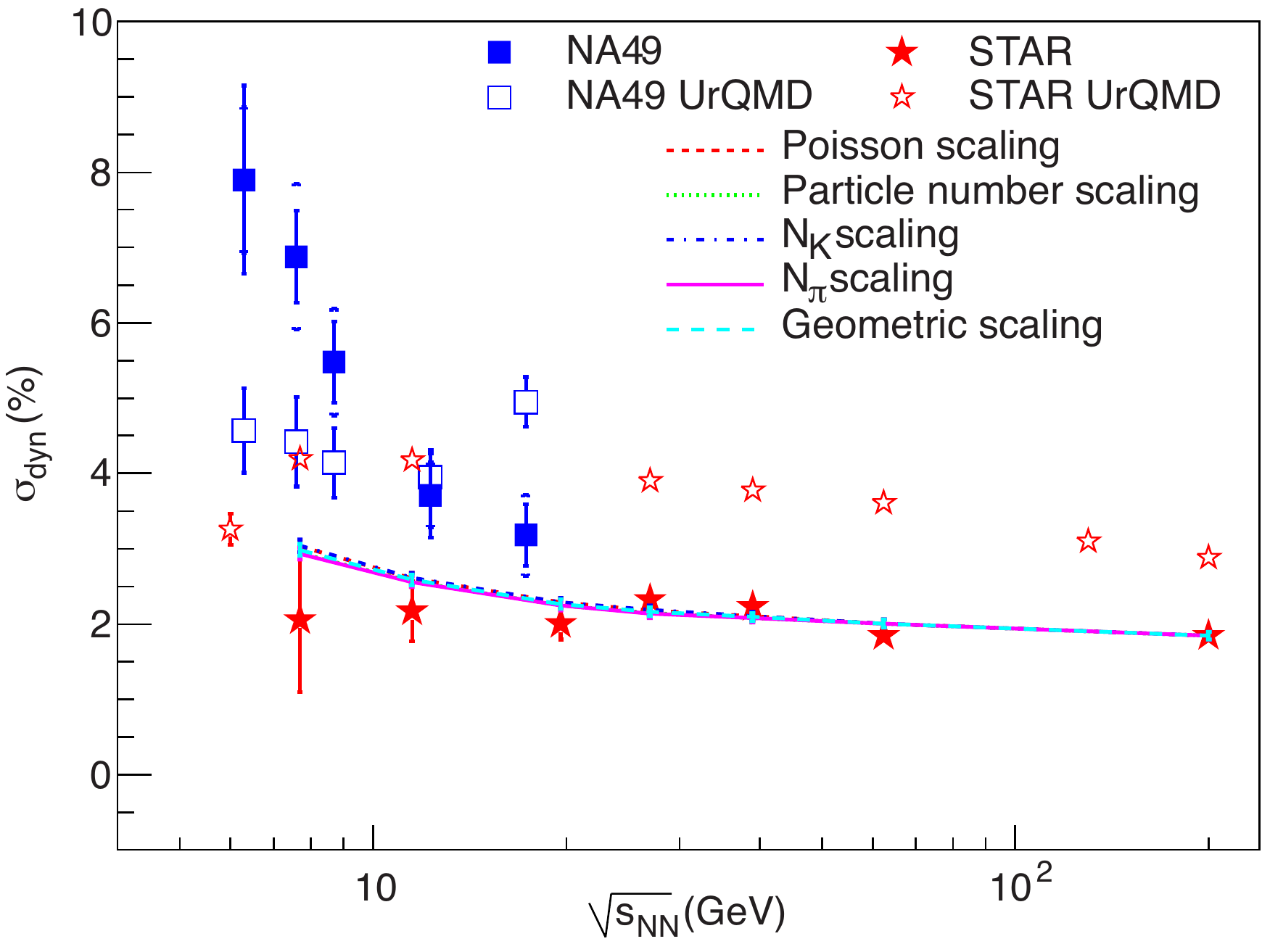}
\caption{\label{fig:fig5_3_01} Energy dependence of $\sigma_{{\rm dyn},  p\pi}$ results. STAR results (red stars) are from Au+Au collisions, 0-5\% centrality, only statistical error are shown, while NA49 results (blue squares) are from Pb+Pb collisions, 0-3.5\% centrality, both statistical and systematical error are shown. The scaling results are shown as lines.}

\end{figure}

Figure~\ref{fig:fig5_3_01} shows the previously discussed energy dependence of the summed sign $\sigma_{{\rm dyn},  K\pi}$, as well as all five scaling methods mentioned above. All methods of scaling start from the 200 GeV $\sigma_{{\rm dyn},  K\pi}$ value and use the number of identified kaons and pions in the STAR acceptance as listed in table ~\ref{tab:tab5_3_01}. We can see that all five scaling methods give a similar result and increase slightly with decreasing  beam energy, while STAR data show little energy dependence. The methods of scaling agree well with data at high RHIC energies but over predict the results at 7.7 GeV and 11.5 GeV. Due to the large statistical error, it is difficult to drawn any conclusion based on the current data. Future study is still necessary to improve the data quality and test if the scaling holds at the lowest STAR energies.

\begin{figure}
\centering
\includegraphics[width=32pc]{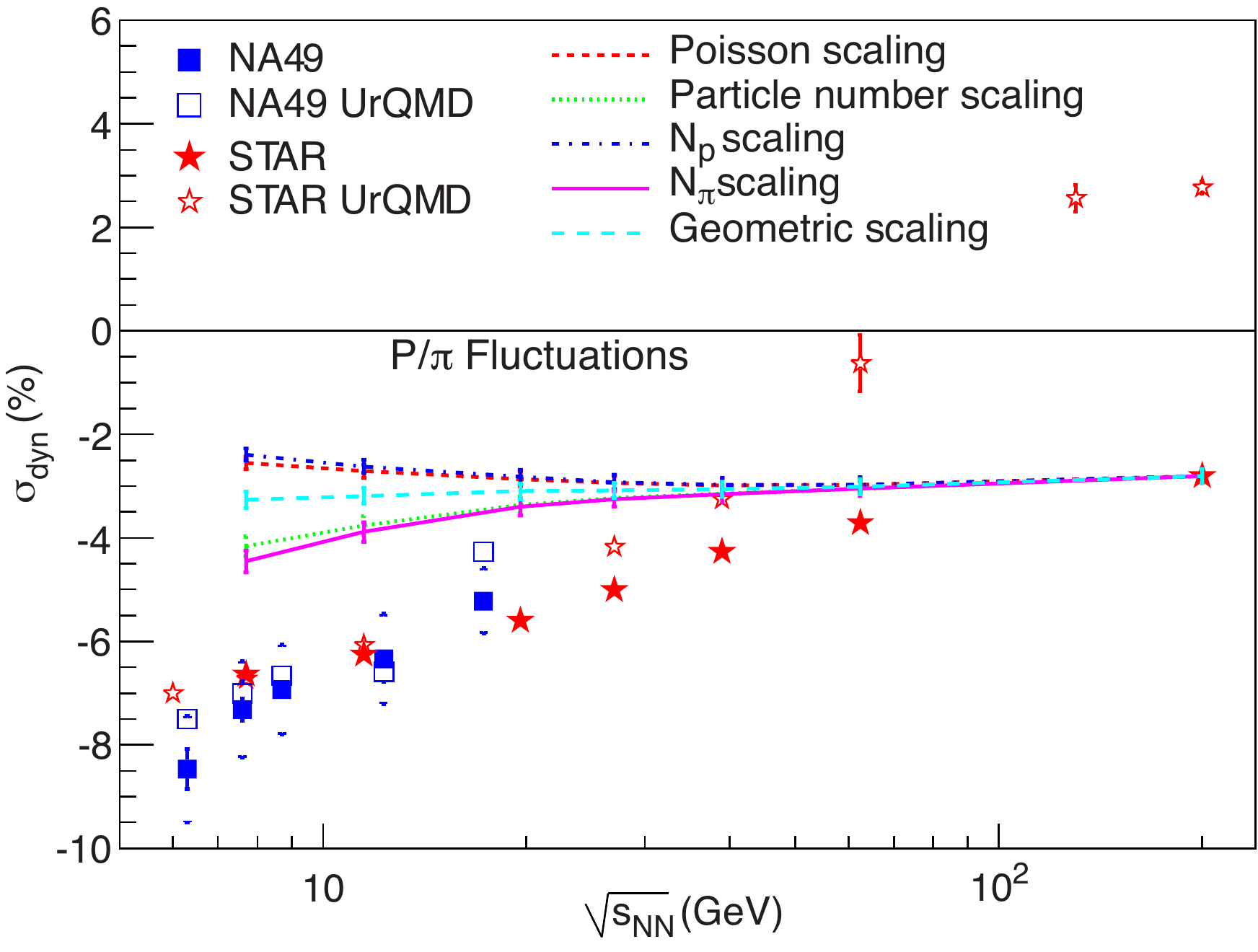}
\caption{\label{fig:fig5_3_02} Energy dependence of $\sigma_{{\rm dyn},  p\pi}$ results. STAR results (red stars) are from Au+Au collisions, 0-5\% centrality. The scaling results are shown as lines.}

\end{figure}

Similarly, we also test the different scaling methods for $p/\pi$ fluctuations.  Figure~\ref{fig:fig5_3_02} shows energy dependence of the $\sigma_{{\rm dyn},  p\pi}$ with all five scaling methods mentioned above. Since $ N_p  \ll N_\pi  $, the Poisson scaling should be close to the $N_{p}$ scaling, and the particle number scaling should be close to the $N_{\pi}$ scaling. This is indeed the case in Figure~\ref{fig:fig5_3_02}, we see both the Poisson scaling and $N_{p}$ scaling  show less negative toward lower energies, which is due to increasing proton yield at lower energies, while the particle number scaling and $N_{\pi}$ scaling show more negative toward lower energies due to smaller pion yield. The geometric scaling shows little beam energy dependence. However, none of the above methods for scaling reproduce the STAR and NA49 data. This disagreement between data and scaling methods implies other sources of correlations such as enhanced resonance production at lower collision energies.

\begin{figure}
\centering
\includegraphics[width=32pc]{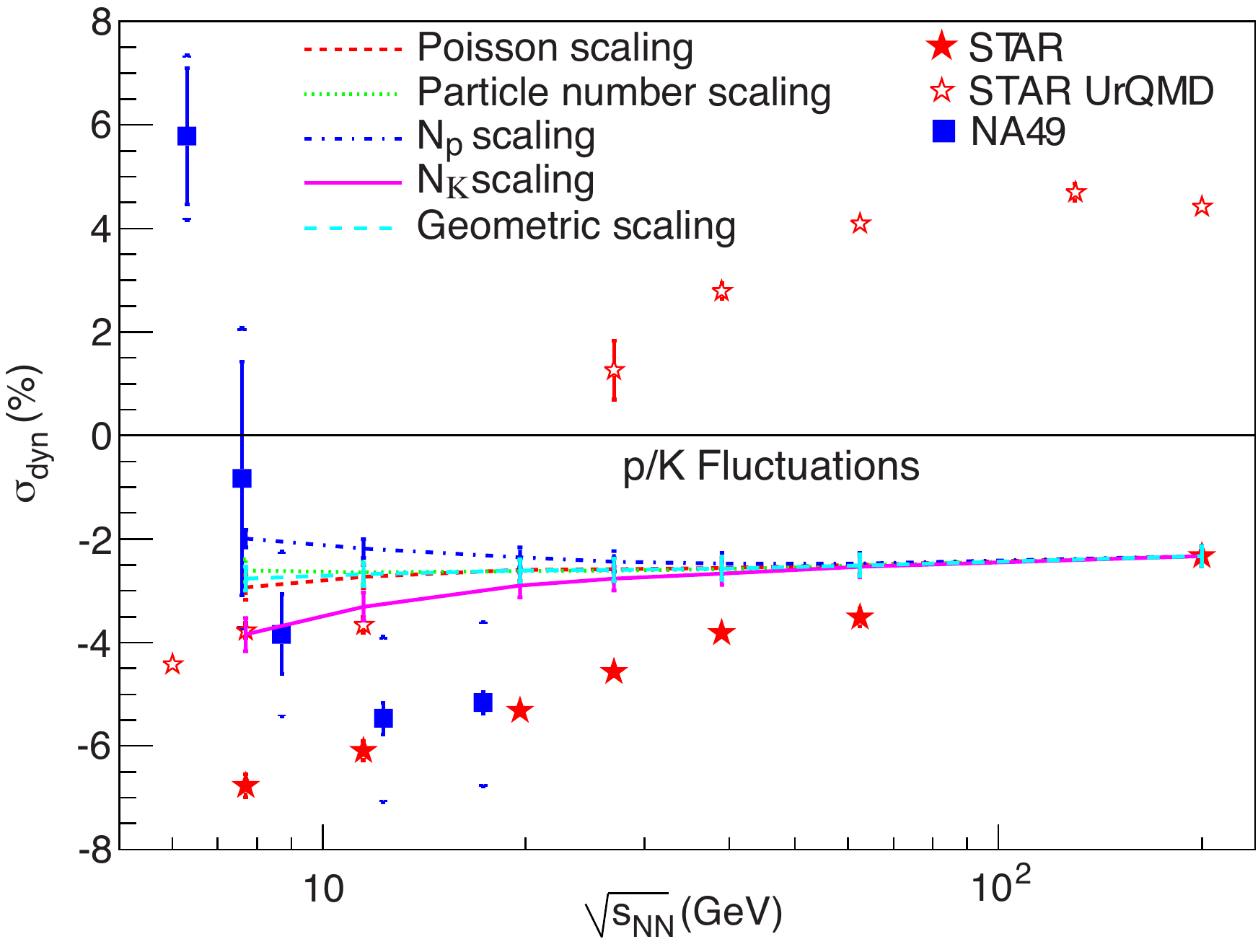}
\caption{\label{fig:fig5_3_03} Energy dependence of $\sigma_{{\rm dyn},  pK}$ results. STAR results (red stars) are from Au+Au collisions, 0-5\% centrality, only statistical error are shown, while NA49 results (blue squares) are from Pb+Pb collisions, 0-3.5\% centrality, both statistical and systematical error are shown.The scaling results are shown as lines.}

\end{figure}

Figure~\ref{fig:fig5_3_03} shows scaling results for $p/K$ fluctuations. Again, none of the scalings could reproduce STAR data, which could due to correlations other than simple multiplicity dependence. 

\section{Compare to $\sigma_{{\rm dyn}}$}

\begin{figure}
\centering
\includegraphics[width=38pc]{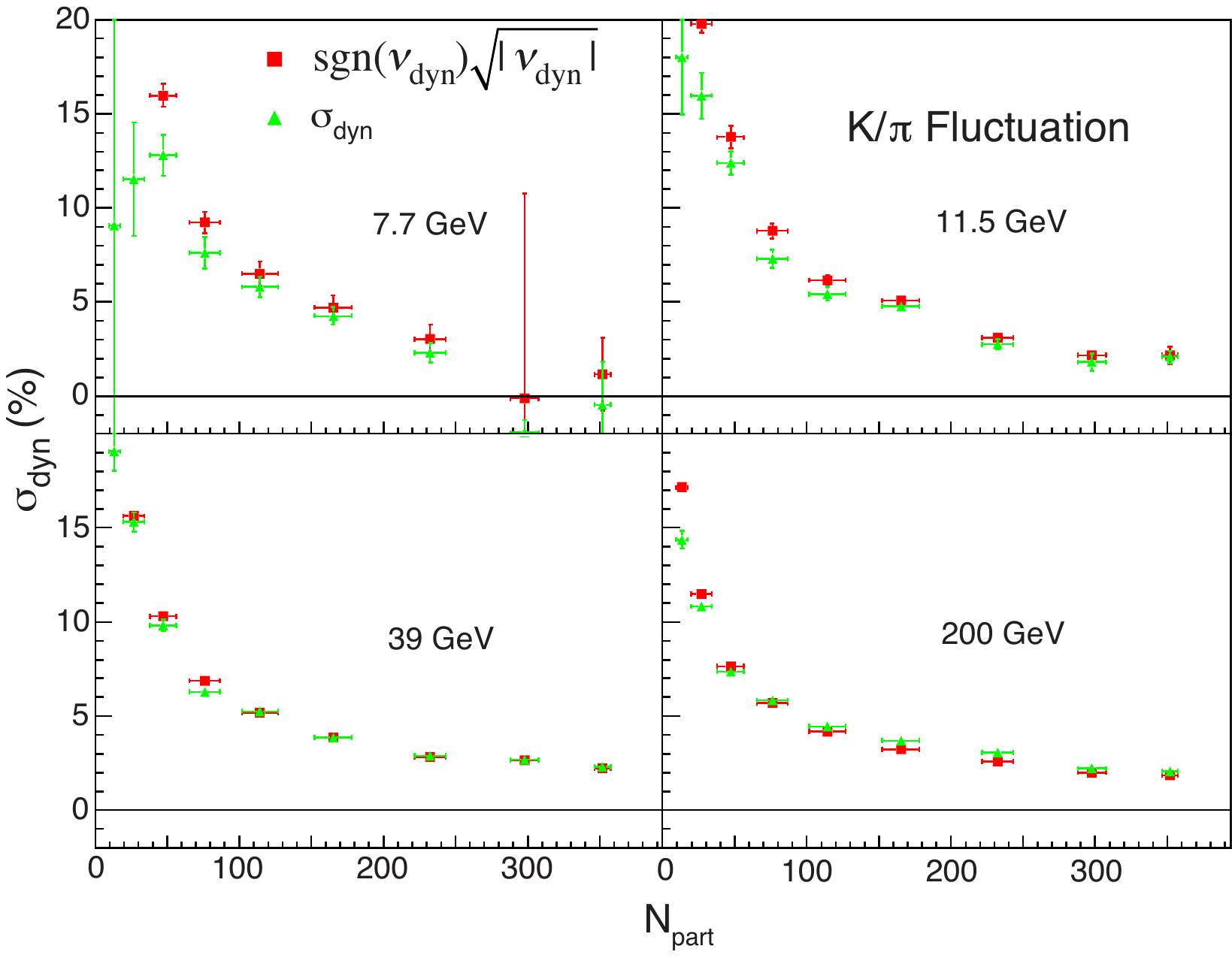}
\caption{\label{fig:fig5_4_01}Centrality dependence of  ${\rm sgn}(\nu_{{\rm dyn}})\sqrt{|\nu_{{\rm dyn,K\pi}}|}$  and $\sigma_{{\rm dyn},  K\pi}$  results from Au+Au collisions at $\sqrt{s_{\rm NN}}$ = 7.7, 11.5, 39 and 200 GeV. }

\end{figure}

As discussed in Section~\ref{fluctuation}, we have $\sigma _{\rm dyn}  = {\rm sgn}(\nu_{\rm dyn})\sqrt{|\nu _{\rm dyn}|}$ for particle-ratio fluctuations if there are a sufficient number of events. To demonstrate this relation, we calculated ${\rm sgn}(\nu_{{\rm dyn}})\sqrt{|\nu_{{\rm dyn}}|}$ and compared it to $\sigma_{{\rm dyn}}$ results calculated using the equation $\sigma_{{\rm dyn}} ={\rm sgn(\sigma_{\rm data}^2 - \sigma_{\rm mixed}^2)}\sqrt{|\sigma_{\rm data}^2 - \sigma_{\rm mixed}^2|}$ . Figure~\ref{fig:fig5_4_01} shows the comparison between $\nu_{{\rm dyn}}$ and $\sigma_{{\rm dyn}}$ for $K/\pi$ fluctuations. The results are plotted versus $N_{\rm part}$ at $\sqrt{s_{\rm NN}}$ = 7.7, 11.5, 39 and 200 GeV. As expected, $\sigma_{{\rm dyn}}$ and ${\rm sgn}(\nu_{{\rm dyn}})\sqrt{|\nu_{{\rm dyn}}|}$ results agree within errors for most cases. The only deviation occurs at the most peripheral collisions when the particle numbers are small and statistical uncertainties are large.

\begin{figure}
\centering
\includegraphics[width=38pc]{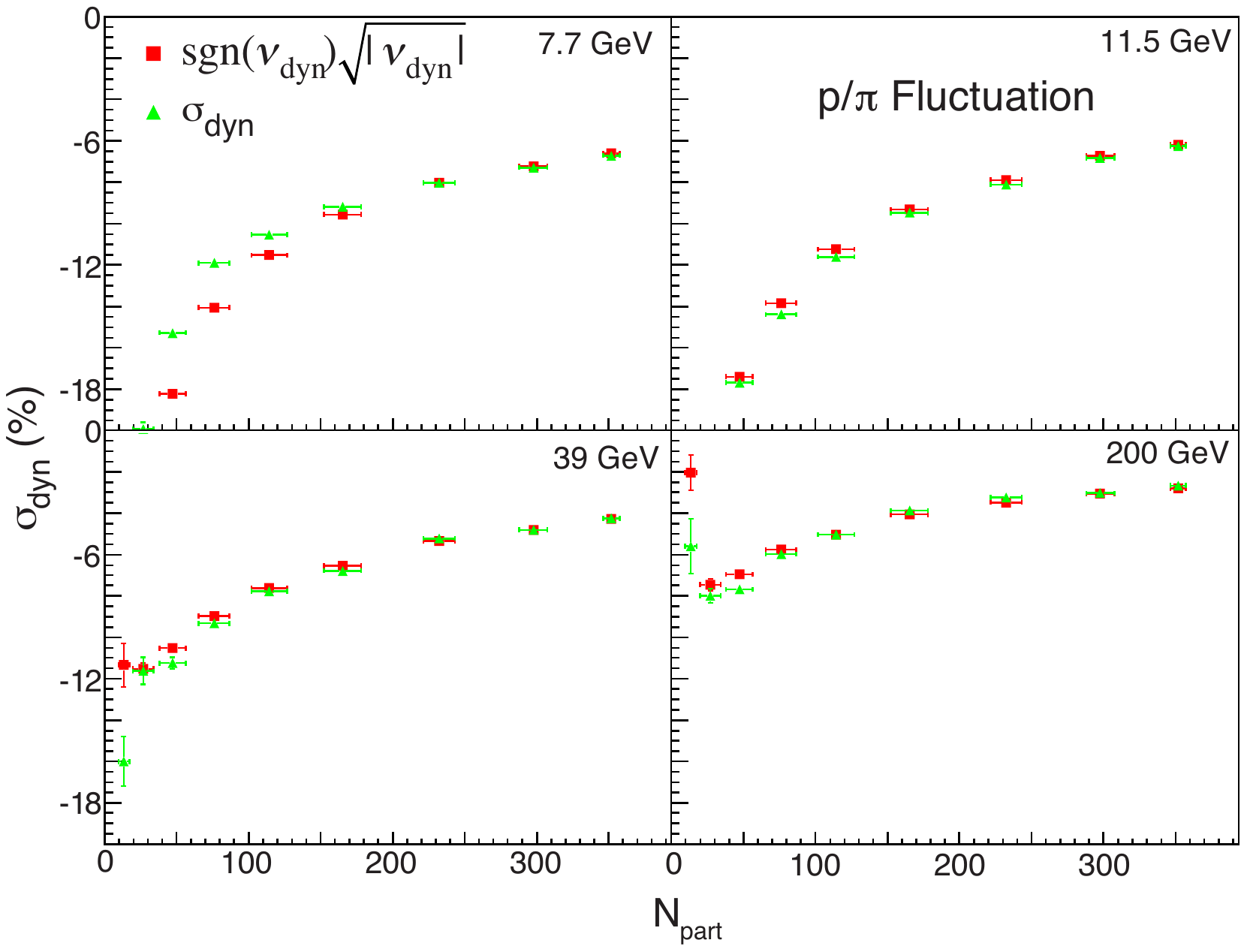}
\caption{\label{fig:fig5_4_02}Centrality dependence of ${\rm sgn}(\nu_{{\rm dyn}})\sqrt{|\nu_{{\rm dyn,p\pi}}|}$   and $\sigma_{{\rm dyn},  p\pi}$ results from Au+Au collisions at $\sqrt{s_{\rm NN}}$ = 7.7, 11.5, 39 and 200 GeV. }

\end{figure}

Figure~\ref{fig:fig5_4_02} shows a similar comparison for $p/\pi$ fluctuations for the centrality dependence of $\nu_{{{\rm dyn}, p\pi}}$  and $\sigma_{{\rm dyn},  p\pi}$ results from Au+Au collisions at $\sqrt{s_{\rm NN}}$ = 7.7, 11.5, 39 and 200 GeV. Again, $\sigma_{{\rm dyn}}$ and ${\rm sgn}(\nu_{{\rm dyn}})\sqrt{|\nu_{{\rm dyn}}|}$ results agree well for most cases.

\begin{figure}
\centering
\includegraphics[width=38pc]{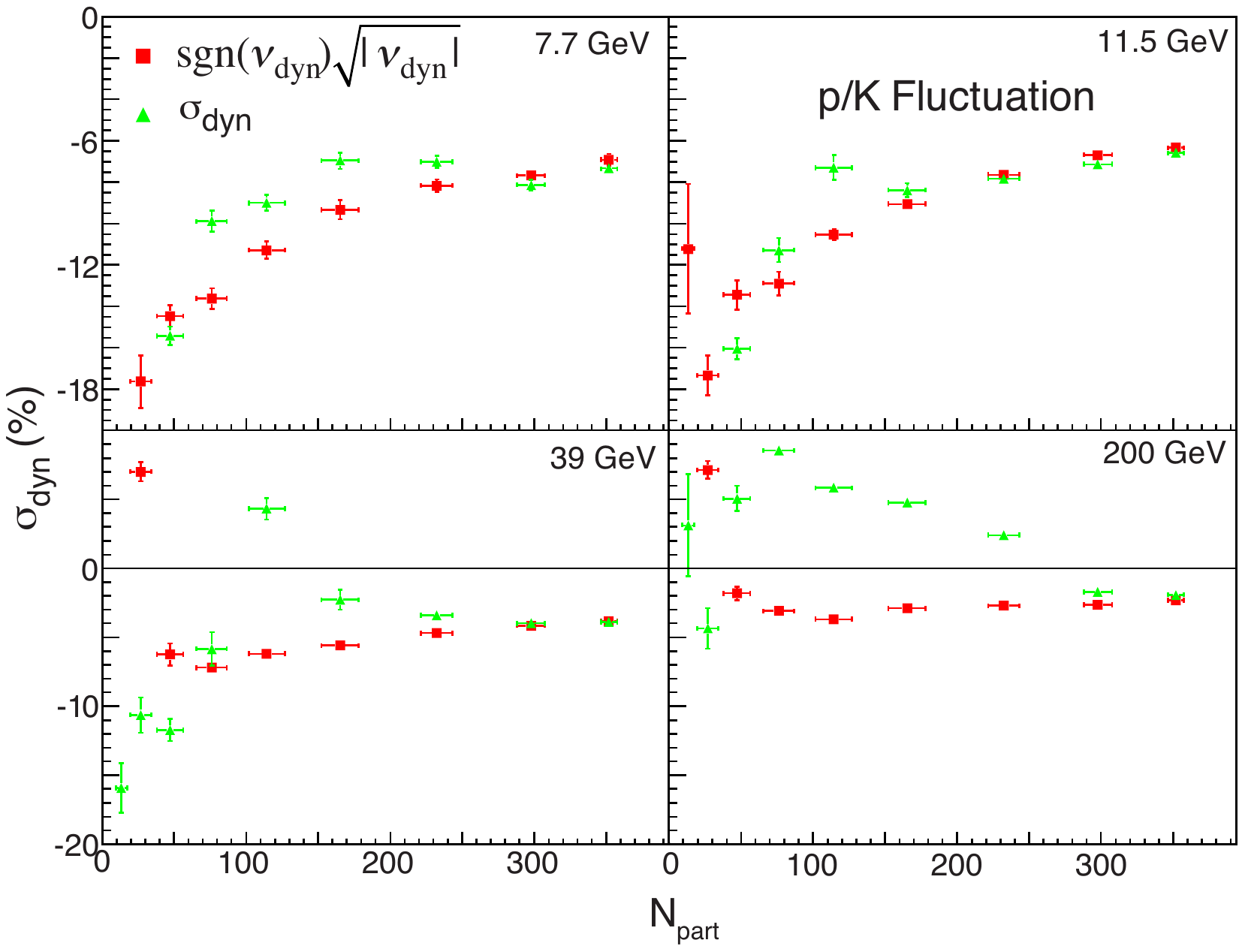}
\caption{\label{fig:fig5_4_03}Centrality dependence of ${\rm sgn}(\nu_{{\rm dyn}})\sqrt{|\nu_{{\rm dyn,pK}}|}$  and $\sigma_{{\rm dyn},  pK}$ from Au+Au collisions at $\sqrt{s_{\rm NN}}$ = 7.7, 11.5, 39 and 200 GeV. }

\end{figure}

Figure~\ref{fig:fig5_4_03} shows the centrality dependence of $\nu_{{\rm dyn, pK}}$  and $\sigma_{{\rm dyn},  pK}$ results from Au+Au collisions at $\sqrt{s_{\rm NN}}$ = 7.7, 11.5, 39 and 200 GeV. Unlike $K/\pi$ and $p/\pi$ fluctuations, which have a relatively large denominator (identified pions), both proton and kaon numbers are relatively small (Table~\ref{tab:tab5_3_01}), especially for peripheral collisions. This requires more events to get stable results for $\sigma_{{\rm dyn}}$. In the figure, we can see that $\sigma_{{\rm dyn}}$ and ${\rm sgn}(\nu_{{\rm dyn}})\sqrt{|\nu_{{\rm dyn}}|}$ results still agree very well in central collisions. However, they start to deviate from each other at mid-centrality. The results for ${\rm sgn}(\nu_{{\rm dyn}})\sqrt{|\nu_{{\rm dyn}}|}$ decrease smoothly from central to peripheral collisions for all four energies shown here, except for the most peripheral points when the multiplicity is low and the background is relatively high. The $\sigma_{{\rm dyn}}$ results show little deviation from ${\rm sgn}(\nu_{{\rm dyn}})\sqrt{|\nu_{{\rm dyn}}|}$ at 7.7 and 11.5 GeV but the difference becomes clear at higher energies. Since $\sigma_{{\rm dyn}}$ requires a mixed events calculation, it is more sensitive to detector issues and the event mixing method.  Therefore ${\rm sgn}(\nu_{{\rm dyn}})\sqrt{|\nu_{{\rm dyn}}|}$ gives a better result in these low multiplicity situations.

\chapter{Balance Function Results}

\section{Balance Functions from Au+Au, $d$+Au, and $p+p$ Collisions at $\sqrt{s_{\rm NN}}$ = 200 GeV}
\label{Run7_balance}

\subsection{Balance Functions in Terms of $\Delta \eta$ and $\Delta y$ }

\begin{figure}
\includegraphics[width=36pc]{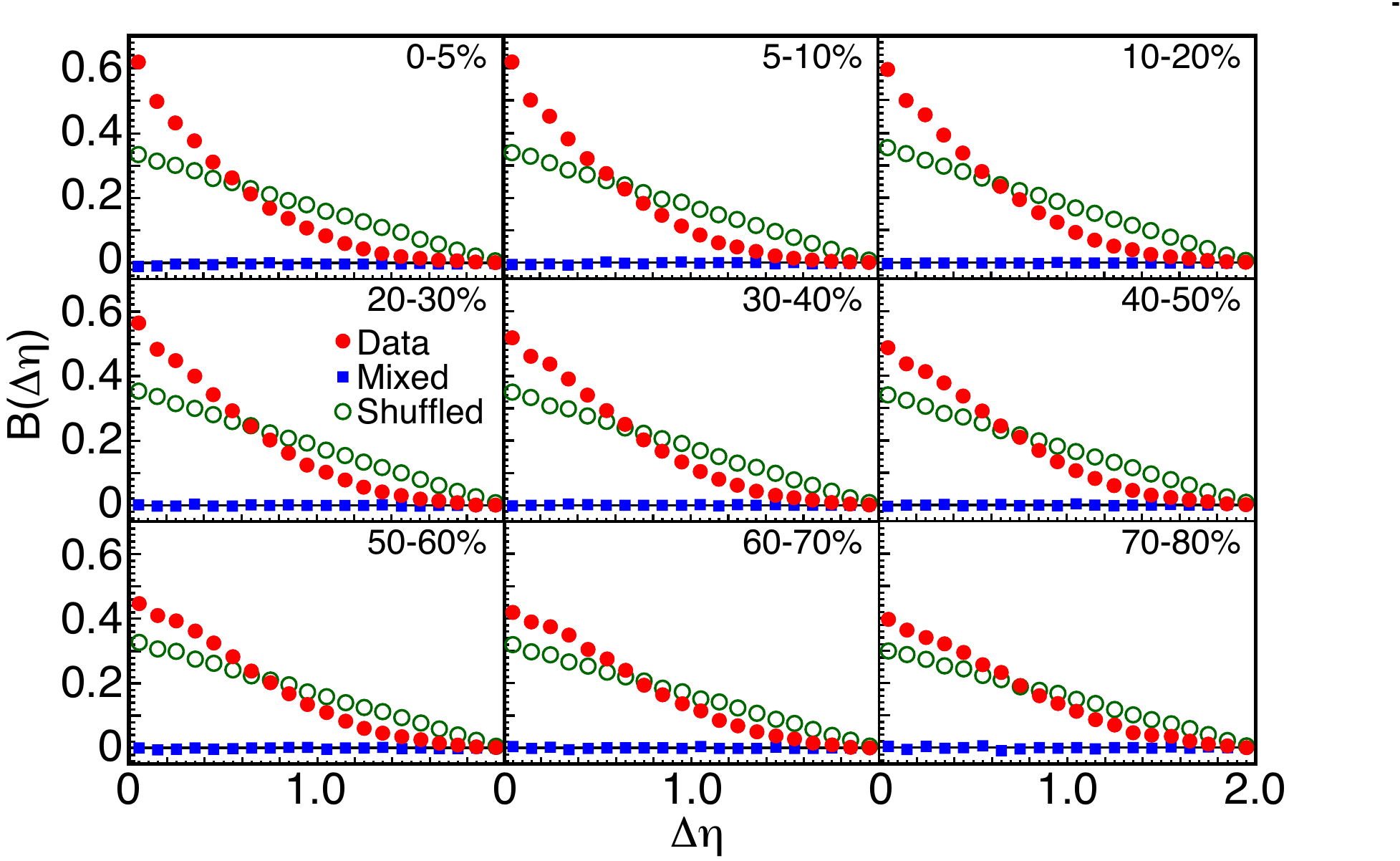}
\caption{\label{fig:fig6_1_01} The balance function in terms of $\Delta \eta$ for all charged particle pairs from Au+Au collisions at$\sqrt{s_{\rm NN}}$ = 200 GeV for nine centrality bins.}

\end{figure}

Figure~\ref{fig:fig6_1_01}  shows the balance function in terms of $\Delta \eta$ for all charged particles from Au+Au collisions at $\sqrt{s_{\rm NN}}$ = 200 GeV for nine centrality bins from most central (0-5\%) to most peripheral (70-80\%). The balance function gets narrower as the collisions become more central. The balance function for mixed events is zero for all centralities and $\Delta \eta$. The balance function for shuffled events is significantly wider than the measured balance functions. Model predictions show that inter-pair correlations (e.g. Hanbury-Brown Twiss (HBT) and final state interactions) should be significant for $\Delta \eta < 0.1 $\cite{balance_distortions}.

\begin{figure}
\centering
\includegraphics[width=34pc]{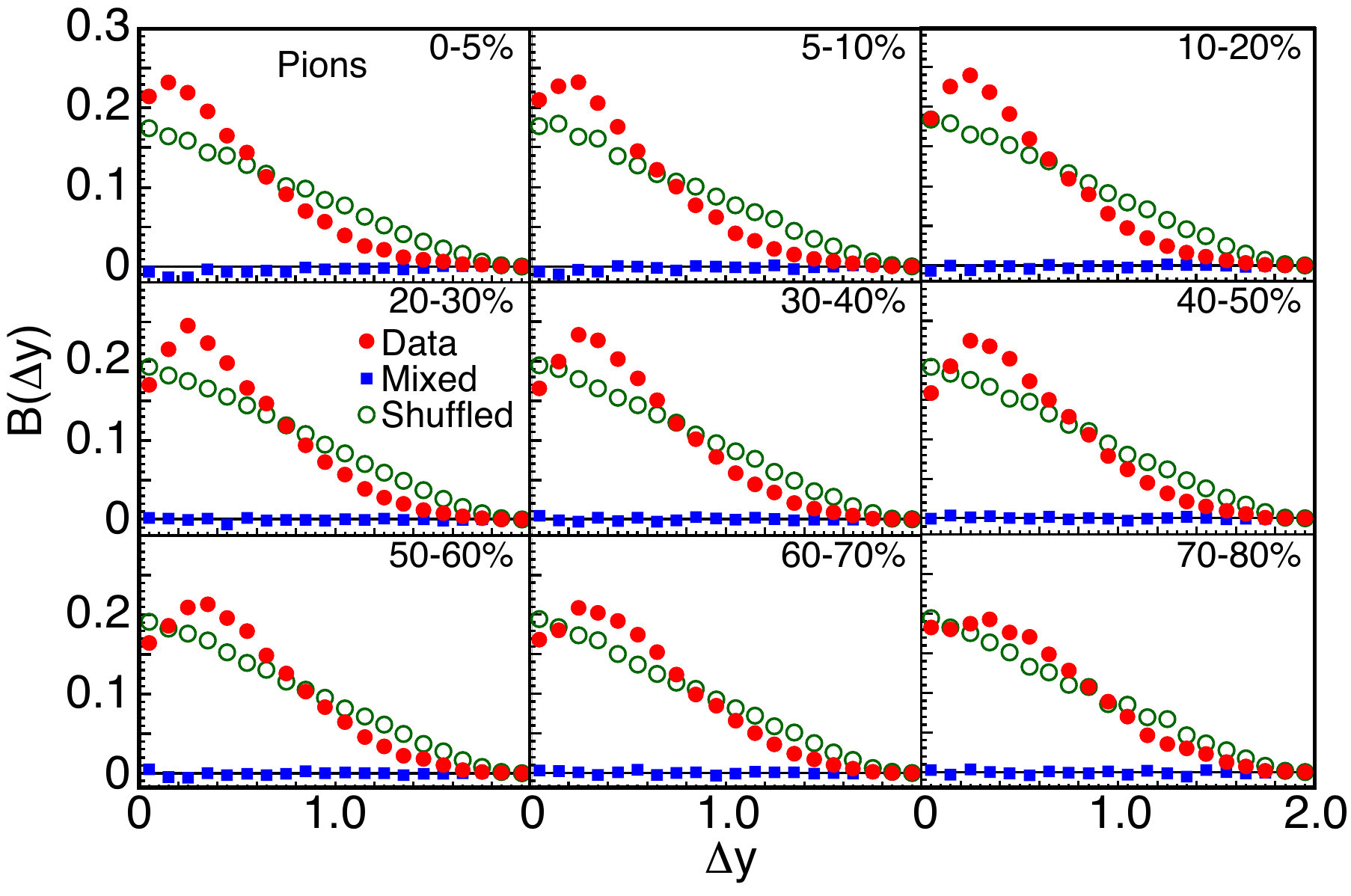}
\caption{\label{fig:fig6_1_02} The balance function in terms of $\Delta y$ for identified charged pion pairs from Au+Au collisions at $\sqrt{s_{\rm NN}}$ = 200 GeV for nine centrality bins.}
\end{figure}

\begin{figure}
\centering
\includegraphics[width=34pc]{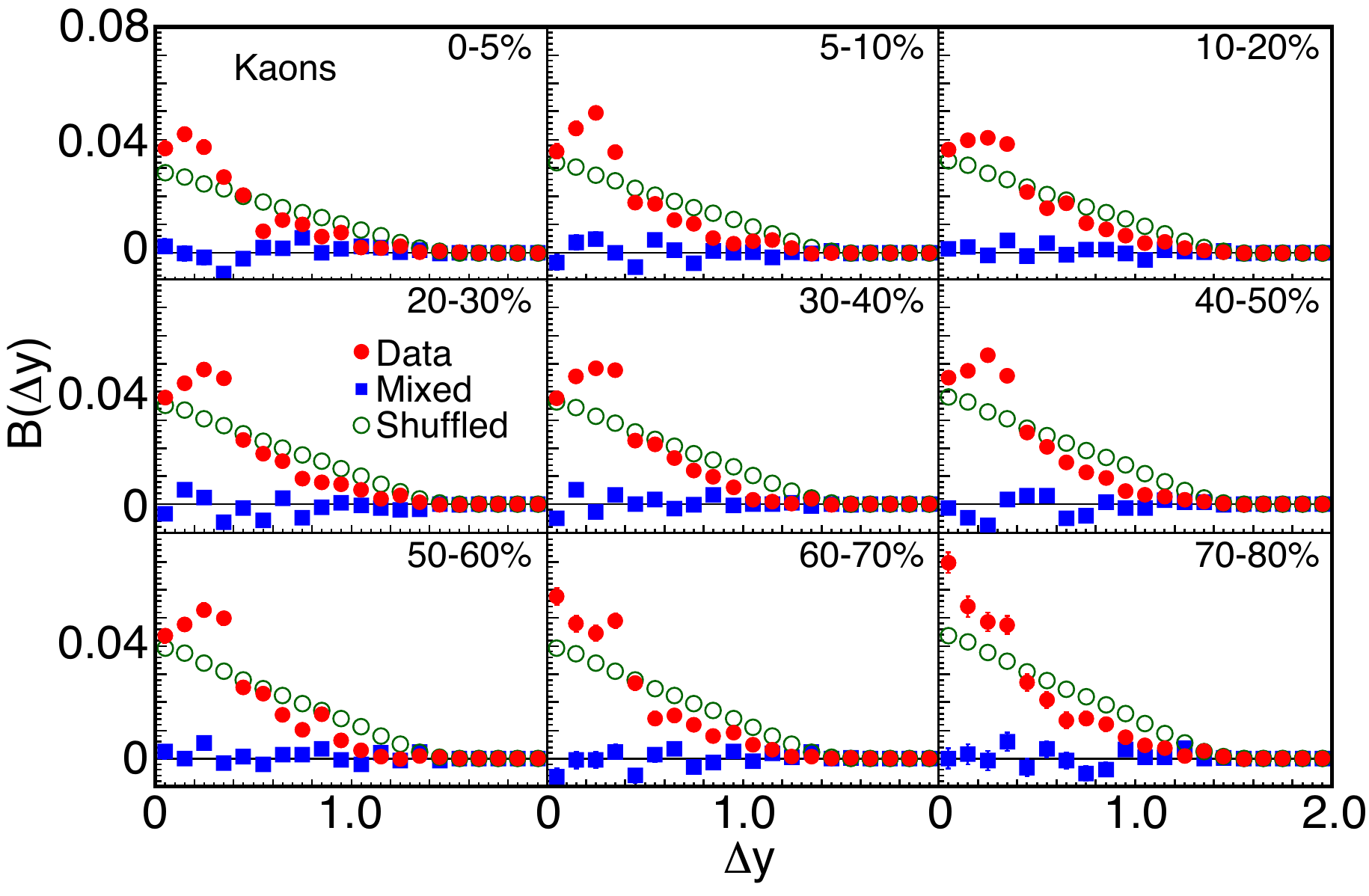}
\caption{\label{fig:fig6_1_03} The balance function in terms of $\Delta y$ for identified charged kaon pairs from Au+Au collisions at $\sqrt{s_{\rm NN}}$ = 200 GeV for nine centrality bins.}
\end{figure}

Figures~\ref{fig:fig6_1_02} and \ref{fig:fig6_1_03} show the balance functions for identified charged pion pairs and kaons pairs, respectively, for Au+Au collisions at $\sqrt{s_{\rm NN}}$ = 200 GeV for nine centrality bins as a function of the relative rapidity, $\Delta y$.  The balance function for identified pion pairs gets narrower in central collisions.  The lower magnitude of the balance function for pion pairs and kaon pairs compared with the balance function for all charged particles is due to the fact that the efficiency of observing an identified pion or kaon is lower than for unidentified charged particles. The balance function calculated from mixed events is zero for all centralities and $\Delta y$ for both pions and kaons.  The balance functions calculated using shuffled events are substantially wider than the measured balance functions.  The discontinuity in $B(\Delta y)$ for kaons around $\Delta y$ = 0.4 visible at all centralities is due to $\phi$ decay, which was verified using HIJING calculations.
Model predictions show that inter-pair correlations should be significant for $\Delta y < 0.2$ \cite{balance_distortions}.  These effects scale with the multiplicity and thus are more apparent in central collisions.

\begin{figure}
\centering
\includegraphics[width=18pc]{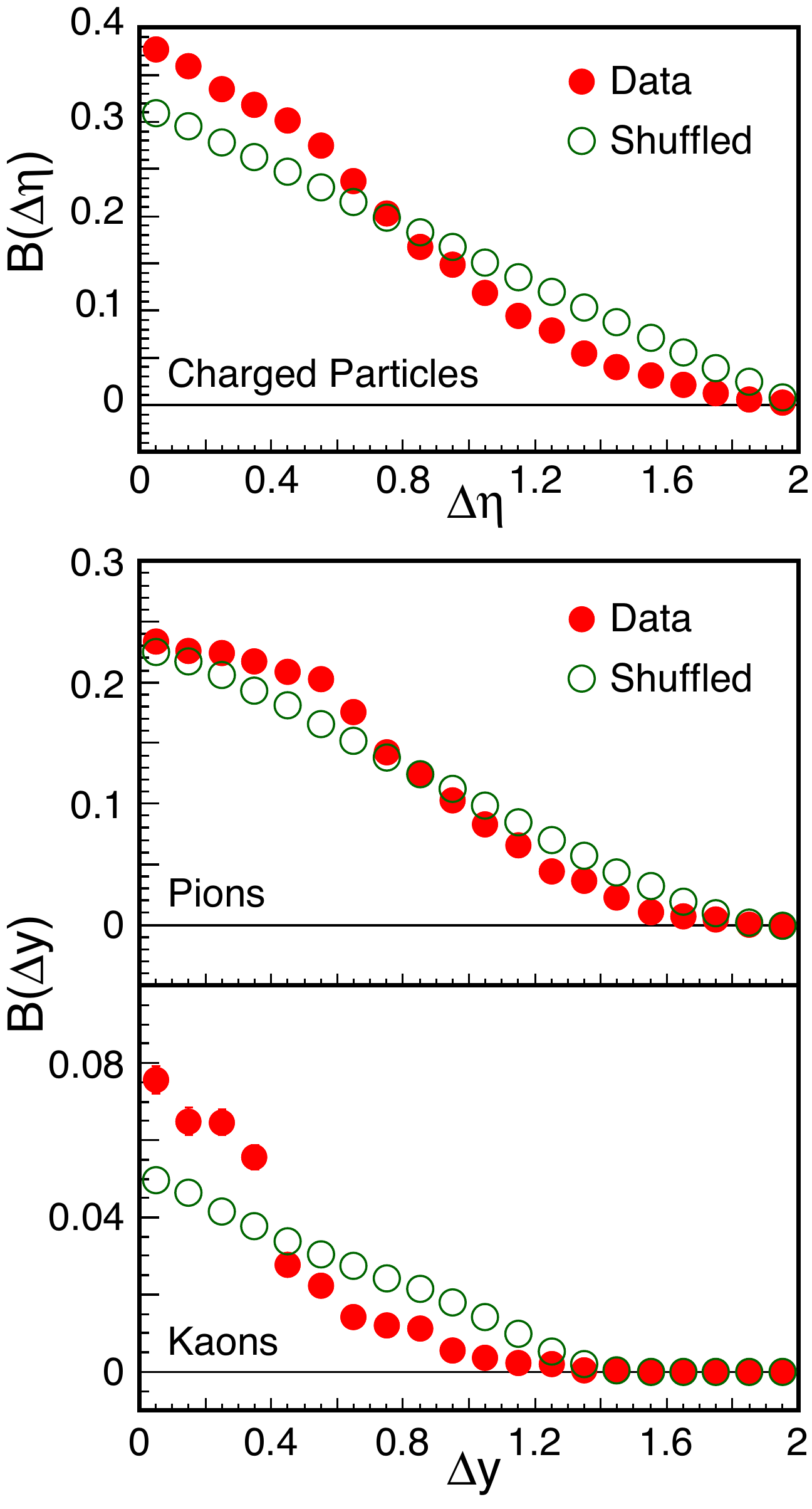}
\caption{\label{fig:fig6_1_04}The balance function for $p+p$ collisions at $\sqrt{s}$ = 200 GeV.  The top panel shows the balance function for all charged particles in terms of $\Delta \eta$.  The bottom panel gives the balance function for charged pion pairs and charged kaon pairs in terms of $\Delta y$.}
\end{figure}

\begin{figure}
\centering
\includegraphics[width=18pc]{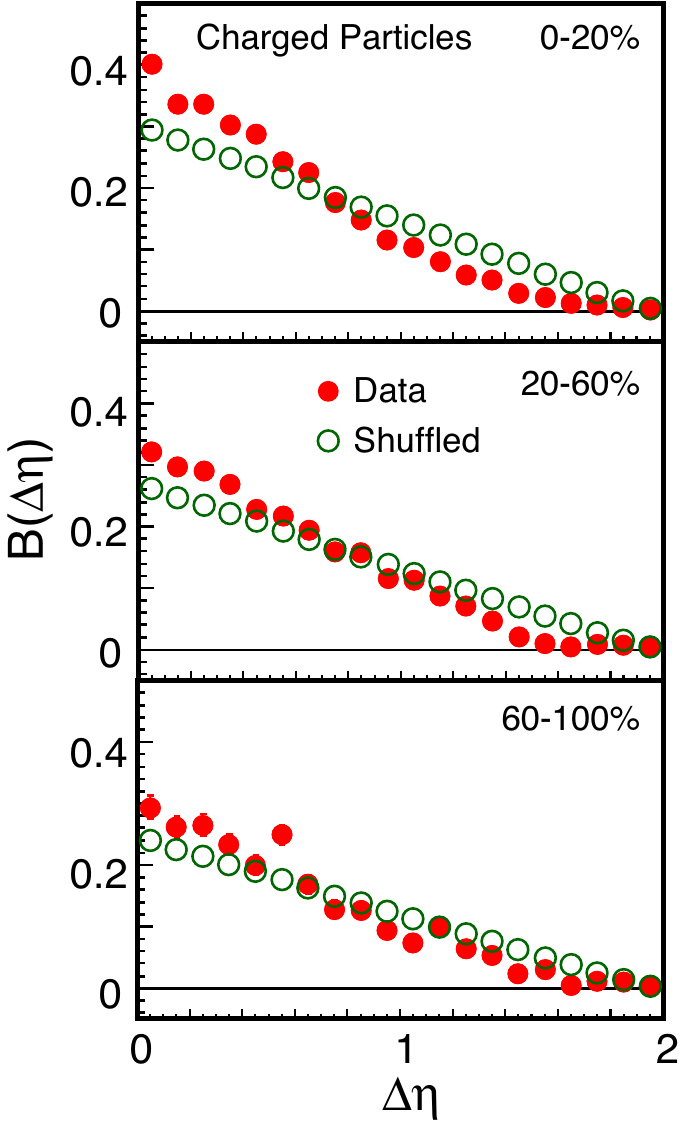}
\caption{\label{fig:fig6_1_05} The balance function in terms of $\Delta\eta$ for all charged particles from $d$+Au collisions at $\sqrt{s_{\rm NN}}$ = 200 GeV for three centrality bins.}
\end{figure}

To investigate the system-size dependence of the balance function and to provide a nucleon-nucleon reference for the balance functions extracted from Au+Au collisions, we measured the balance functions for $p+p$ and $d$+Au collisions at $\sqrt{s_{\rm NN}}$ = 200 GeV.  Figure~\ref{fig:fig6_1_04} shows the balance functions for all charged particles for $p+p$ collisions at  $\sqrt{s}$ = 200 GeV.  The balance functions for $p+p$ collisions are integrated over all observed event multiplicities to allow comparison with centrality-selected $d$+Au and Au+Au results.  Note that the width of the balance function in terms of $\Delta \eta$ for $p+p$ collisions is independent of the multiplicity of tracks in the event. The top panel of Figure~\ref{fig:fig6_1_04} shows the balance function for all charged particles in terms of $\Delta \eta$. In the bottom panel of Figure~\ref{fig:fig6_1_04}, the balance functions are shown for identified charged pion pairs and identified charged kaon pairs in terms of $\Delta y$ from $p+p$ collisions at $\sqrt{s}$ = 200 GeV.   The balance function for mixed events is zero for all $\Delta \eta$ and all $\Delta y$. The observed shapes of the balance functions for the identified charged pions and kaons are similar to those observed in peripheral (70 - 80\%) Au+Au collisions.  The fact that the balance function for kaon pairs has a lower magnitude than the balance function for pion pairs reflects the lower efficiency for identifying charged kaons versus identifying charged pions in STAR.

Figure~\ref{fig:fig6_1_05} shows the balance functions in terms of $\Delta \eta$ for all charged particles from $d$+Au collisions at $\sqrt{s_{\rm NN}}$ = 200 GeV for three centrality bins, 0-20\%, 20-60\%, and 60-100\%.

\subsection{Balance Functions in Terms of $q_{\rm inv}$}
\label{balance_qinv}
The balance function in terms of $\Delta \eta$ and $\Delta y$ is observed to narrow in central collisions and model calculations have been used to interpret this narrowing in terms of delayed hadronization \cite{balance_theory,balance_distortions_jeon,balance_distortions,balance_blastwave}.  However, in a thermal model, the width of the balance function in terms of $\Delta \eta$ and $\Delta y$ can be influenced by radial flow.  In the absence of detector efficiency and acceptance considerations, the width of the balance function in terms of the Lorentz invariant momentum difference between the two particles, $q_{\rm inv}$, is determined solely by the breakup temperature if the balancing charges are emitted from the same point in coordinate space.   However, when detector acceptance is taken into account, some  dependence on collective flow is introduced~\cite{balance_distortions}. Thus, analyzing the balance function in terms of $q_{\rm inv}$ avoids some of the complications associated with collective flow, and the balance function calculated with a breakup temperature should be the narrowest possible balance function if the particles are assumed to be emitted from the same position in coordinate space. In addition, contributions to the balance function from the decay of particles are more identifiable.  For example, the decay of $K_S^0$ produces a sharp peak in $B(q_{\rm inv})$ for charged pions, while the contribution to $B(\Delta y)$ for charged pions from the decay of  $K_S^0$ is spread out over several bins in $\Delta y$.

To study balance functions in terms of $q_{\rm inv}$, we use identified charged pions and identified charged kaons.  For pion pairs, we observe a peak from the decay $K^{0}_{S} \rightarrow \pi^{+} + \pi^{-}$.  For kaon pairs, we observe a peak from the decay $\phi \rightarrow K^{+} + K^{-}$.  These peaks are superimposed on the balance function of correlated charge/anti-charge pairs not resulting from the decay of a particle.

\begin{figure}
\centering
\includegraphics[width=34pc]{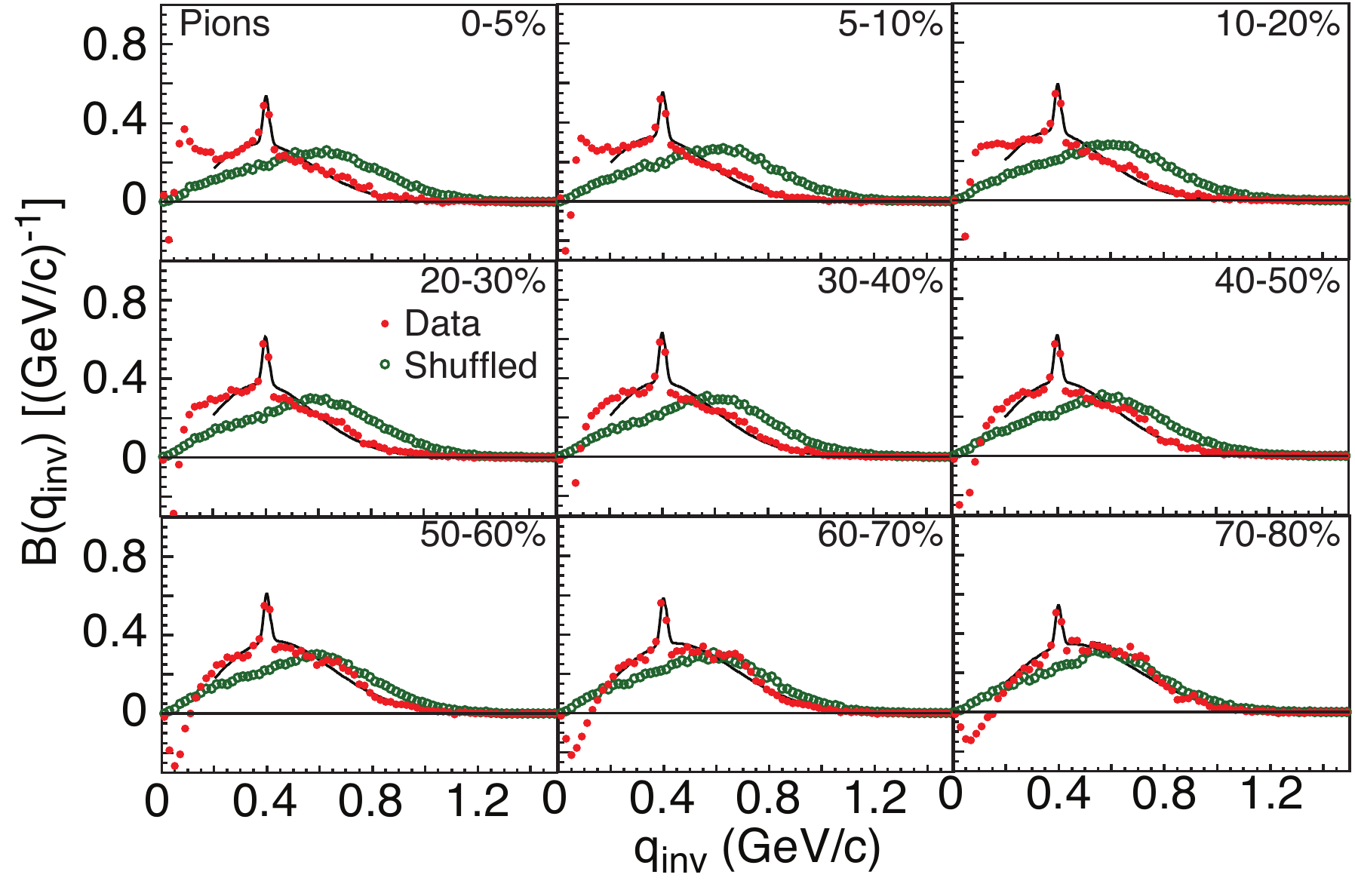}
\caption{\label{fig:fig6_1_06}The balance function in terms of $q_{\rm inv}$ for charged pion pairs from Au+Au collisions at $\sqrt{s_{\rm NN}}$ = 200 GeV in nine centrality bins.  Curves correspond to a thermal distribution (Equation~\ref{thermal}) plus $K_{S}^{0}$ decay.}
\end{figure}

\begin{figure}
\centering
\includegraphics[width=34pc]{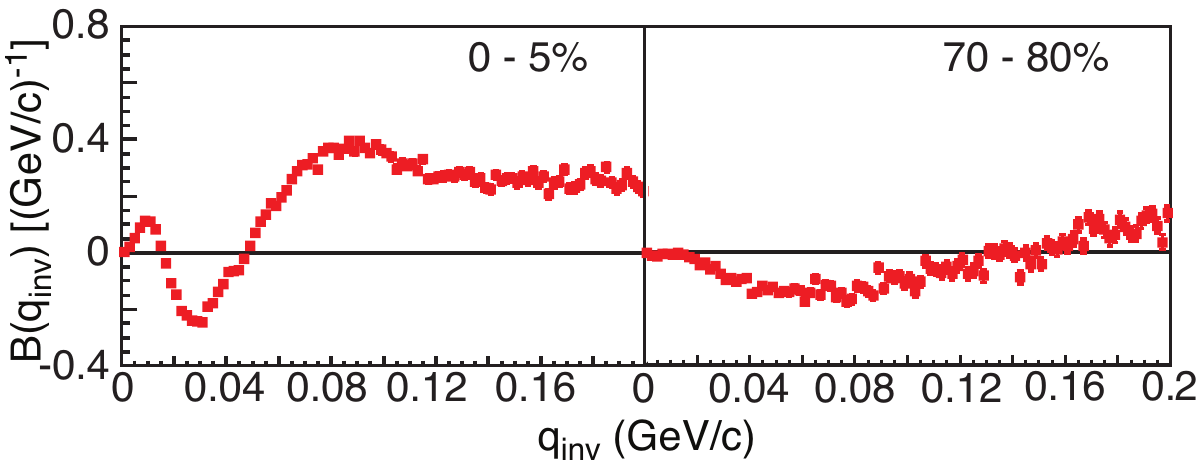}
\caption{\label{fig:fig6_1_07} The balance function in terms of
$q_{\rm inv}$ for charged pion pairs in two centrality bins over the range $0 < q_{\rm inv} < 0.2$ GeV/$c$.}
\end{figure}

Figure~\ref{fig:fig6_1_06} shows the balance function for identified charged pions in terms of $q_{\rm inv}$ for Au+Au collisions at $\sqrt{s_{\rm NN}}$ = 200 GeV for nine centrality bins. These balance functions have been corrected by subtracting the balance functions calculated using mixed events.  These mixed events are not zero for all $q_{\rm inv}$ because of differences in the tracking at TPC sector boundaries for opposite charges.  The balance functions calculated for mixed events integrate to zero as one would expect and the subtraction of the mixed events from the measured balance functions does not affect the integral of the resulting balance functions.   At each centrality, a peak is observed corresponding to charged pion pairs resulting from $K_{S}^{0} \to \pi ^{+} + \pi ^{-}$.  The solid curves represent a fit consisting of the sum of two terms.  The first term consists of a non-relativistic thermal distribution of the form 
\begin{eqnarray}
\label{thermal}
B(q_{\rm inv})=aq_{\rm inv}^2e^{-q_{\rm inv}^2/(2\sigma^{2})}
\end{eqnarray}
where $a$ is a constant, the pre-factor $q_{\rm inv}^2$ accounts for the phase-space effect, and $\sigma$ is a width parameter.  The second term of the fit is a Gaussian distribution in $q_{\rm inv}$ describing the $K_S^0$ decay.  Note that no peak from the decay of the $\rho^{0}$ is visible in central collisions around $q_{\rm inv}$ = 0.718 GeV/$c$ where one would expect to observe the $\rho^{0}$.  This non-observation of the $\rho^{0}$ is in contrast to HIJING, which predicts a large $\rho^{0}$ peak.  The $\rho^{0}$ peak is visible in the most peripheral collisions, which is consistent with our previous study of $\rho^{0}$ production at higher $p_{t}$ \cite{star_rho}.  The authors of Ref.~\cite{balance_blastwave} attribute the apparent disappearance of the $\rho^{0}$ in central collisions to the cooling of the system as it expands, which lowers the production rate of $\rho^{0}$ compared with pions.  The measured balance functions for pions are distinctly different from the balance functions calculated using shuffled events.  In particular, the sharp peak from the $K_S^0$ decay is not present in the balance functions calculated using shuffled events. 

HBT/Coulomb  effects are visible for $q_{\rm inv} < 0.2$ GeV/$c$ in Figure~\ref{fig:fig6_1_06}. Figure~\ref{fig:fig6_1_07} shows the balance function over the range of $0 < q_{\rm inv} < 0.2$ GeV/$c$ for the most central bin (0 - 5\%) and the most peripheral bin (70 -80\%). The Coulomb force pulls opposite charges closer together and pushes same charges apart, leading to an enhancement of opposite-sign and a suppression of same-sign pairs at small $q_{\rm inv}$.  This effect leads to a rise in the balance function at small $q_{\rm inv}$, which is larger in central collisions, where the long-range Coulomb force affects more particles \cite{balance_blastwave}.  In peripheral collisions, because the Coulomb interaction is less important and the HBT correction is larger because of the smaller source size, the Coulomb enhancement disappears and the balance function becomes negative at small $q_{\rm inv}$ \cite{balance_blastwave}.  

\begin{figure}
\centering
\includegraphics[width=34pc]{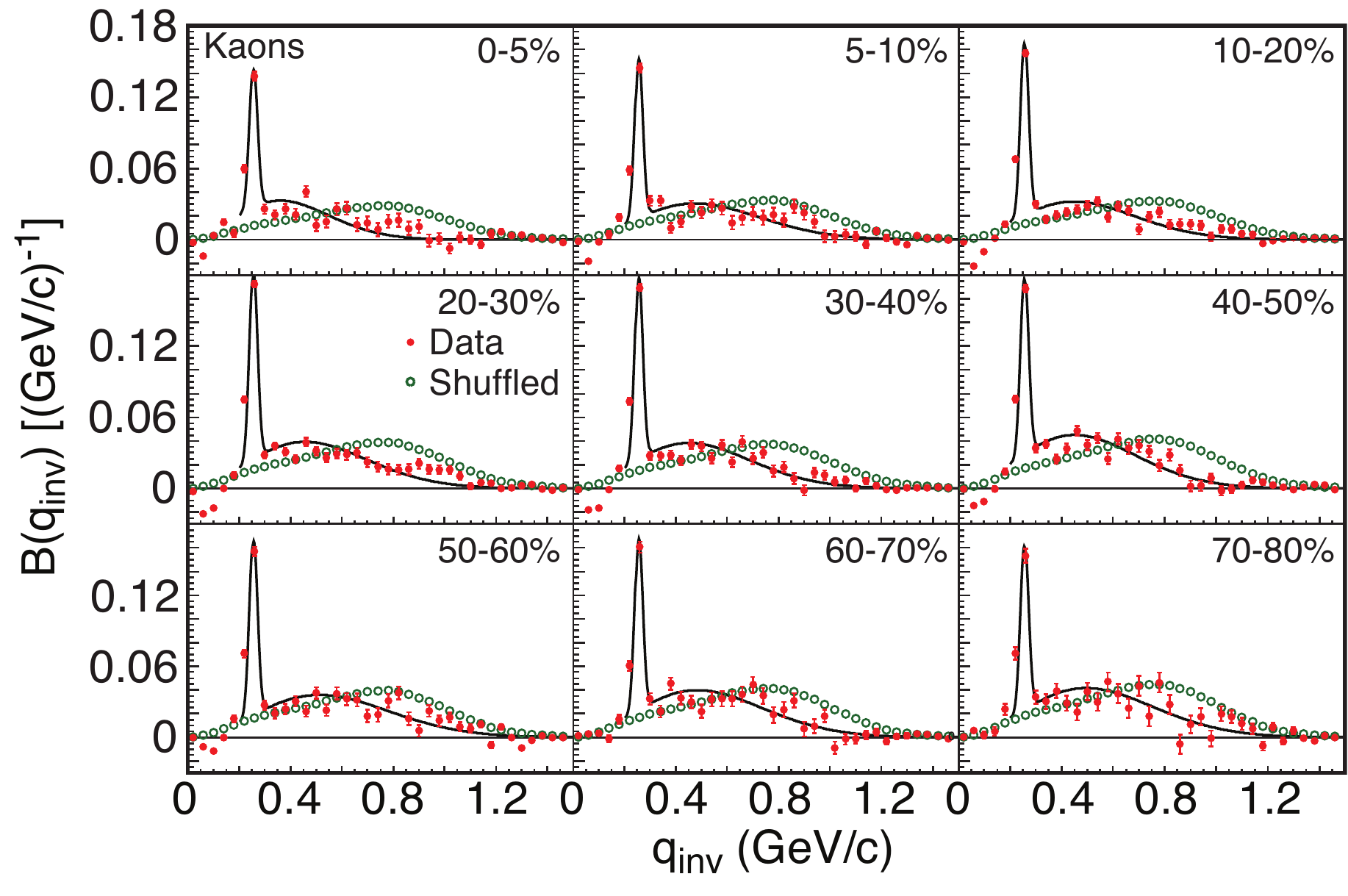}
\caption{\label{fig:fig6_1_08} The balance function
in terms of $q_{\rm inv}$ for charged kaon pairs from Au+Au collisions at
$\sqrt{s_{\rm NN}}$ = 200 GeV in nine centrality bins.  Curves correspond to a thermal (Equation~\ref{thermal})
distribution plus $\phi$ decay.}
\end{figure}

Figure~\ref{fig:fig6_1_08} shows the balance function for identified charged kaons in terms of $q_{\rm inv}$ for Au+Au collisions at $\sqrt{s_{\rm NN}}$ = 200 GeV in nine centrality bins.  These balance functions were corrected by subtracting mixed events as was done for the charged pion results.  At each centrality, a peak is observed corresponding to charged kaon pairs resulting from $\phi \to K ^{+} + K ^{-}$.  The solid curves represent fits consisting of a non-relativistic thermal distribution (Equation~\ref{thermal}) plus a Gaussian distribution in $q_{\rm inv}$ for the $\phi$ decay.  HBT/Coulomb effects at low $q_{\rm inv}$ for kaon pairs are not as strong as those observed for pion pairs.  The measured balance functions are distinct from the balance functions calculated from shuffled events.

Several differences between $B(q_{\rm inv})$ for charged pions and charged kaons are evident.  The observed HBT/Coulomb effects at low $q_{\rm inv}$ are much stronger for pions than for kaons.  The HBT/Coulomb  effects for pions change dramatically with centrality while the HBT  effects for kaons are small and change little with centrality.   The overall normalization for kaons is lower than the overall normalization for pions, reflecting the lower efficiency for detecting identified kaons.  The contribution to $B(q_{\rm inv})$ for pions from $K_S^0$ decay  is approximately 7\%, independent of centrality. The contribution to  $B(q_{\rm inv})$ for kaons from $\phi$ decay is approximately 50\%,  independent of centrality.

\begin{figure}
\centering
\includegraphics[width=28pc]{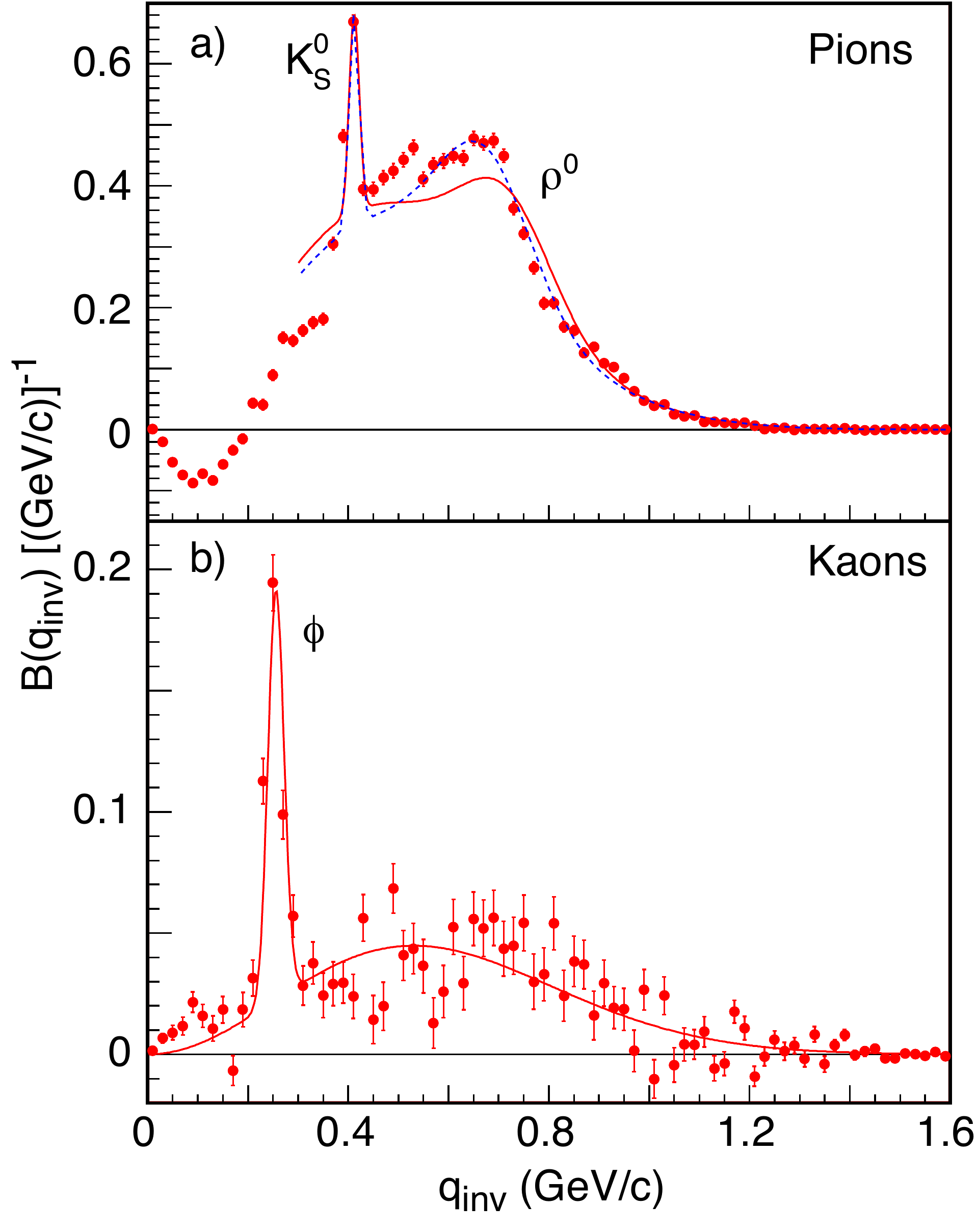}
\caption{\label{fig:fig6_1_09} The balance function in terms of $q_{\rm inv}$ for charged pion pairs [part a)] and charged kaon pairs [part b)] from $p+p$ collisions at $\sqrt{s}$ = 200 GeV integrated over all multiplicities. Solid curves correspond to a thermal distribution (Equation~\ref{thermal}) plus $K_{S}^{0}$ and $\rho^{0}$ decay for pions and $\phi$ decay for kaons.  The dashed curve for pions represents a fit to a thermal distribution (Equation~\ref{thermal}) plus $K_{S}^{0}$ decay and $\rho^{0}$ decay, with the $\rho^{0}$ mass shifted down by 0.04 GeV/$c^2$.}
\end{figure}

Figure~\ref{fig:fig6_1_09} shows the balance functions in terms of $q_{\rm inv}$ for $p+p$ collisions at $\sqrt{s}$ = 200 GeV.   Figure~\ref{fig:fig6_1_09}a shows the balance function for charged pion pairs and Figure~\ref{fig:fig6_1_09}b shows the balance function for charged kaon pairs.  The solid curves are thermal fits (Equation~\ref{thermal}) plus a peak for $K_{S}^{0}$  and $\rho^{0}$ decay in the case of charged pions, and for $\phi$ decay  in the case of charged kaons.  The thermal fit does not reproduce the charged pion results, while it works well for the charged kaon data. The mass of the $\rho^{0}$ used in the fit shown for pion pairs was assumed to be 0.77 GeV/$c^2$.  A better fit can be obtained if the mass of the $\rho^{0}$ is lowered by 0.04 GeV/$c^2$, as was observed previously in studies of $\rho^{0}$ production in $p+p$ collisions at $\sqrt{s}$ = 200 GeV \cite{star_rho}.  This fit is shown as a dashed curve in the upper panel of Figure~\ref{fig:fig6_1_09}.  Note that the $\rho^{0}$ peak visible in $B(q_{\rm inv})$ for pions from $p+p$ collisions is not observed in $B(q_{\rm inv})$ for pions from central Au+Au collisions, but is observed for pions from peripheral Au+Au collisions, as shown in Figure~\ref{fig:fig6_1_06}.

\subsection{Balance Function in Terms of Components of $q_{\rm inv}$}

Here we present results for the three components of $q_{\rm inv}$. These components are $q_{\rm long}$, the component along the beam direction; $q_{\rm out}$, the component in the direction of the transverse momentum of the observed pair; and $q_{\rm side}$, the component perpendicular to $q_{\rm long}$ and $q_{\rm out}$.

Analysis of the balance function for these three components can address the question of what causes the balance function to narrow in central Au+Au collisions.  In a thermal model where the balancing particles are emitted from the same position in coordinate space, the widths would be identical for the three components.  On the other hand, charge separation associated with string dynamics should result in balance functions that are wider in $q_{\rm long}$ than in $q_{\rm side}$ or $q_{\rm out}$ \cite{balance_distortions,balance_blastwave}.   Also because the velocity gradient is much higher in the longitudinal direction, diffusion should broaden the balance function more in $q_{\rm long}$ \cite{balance_blastwave}.

\begin{figure}
\centering
\includegraphics[width=34pc]{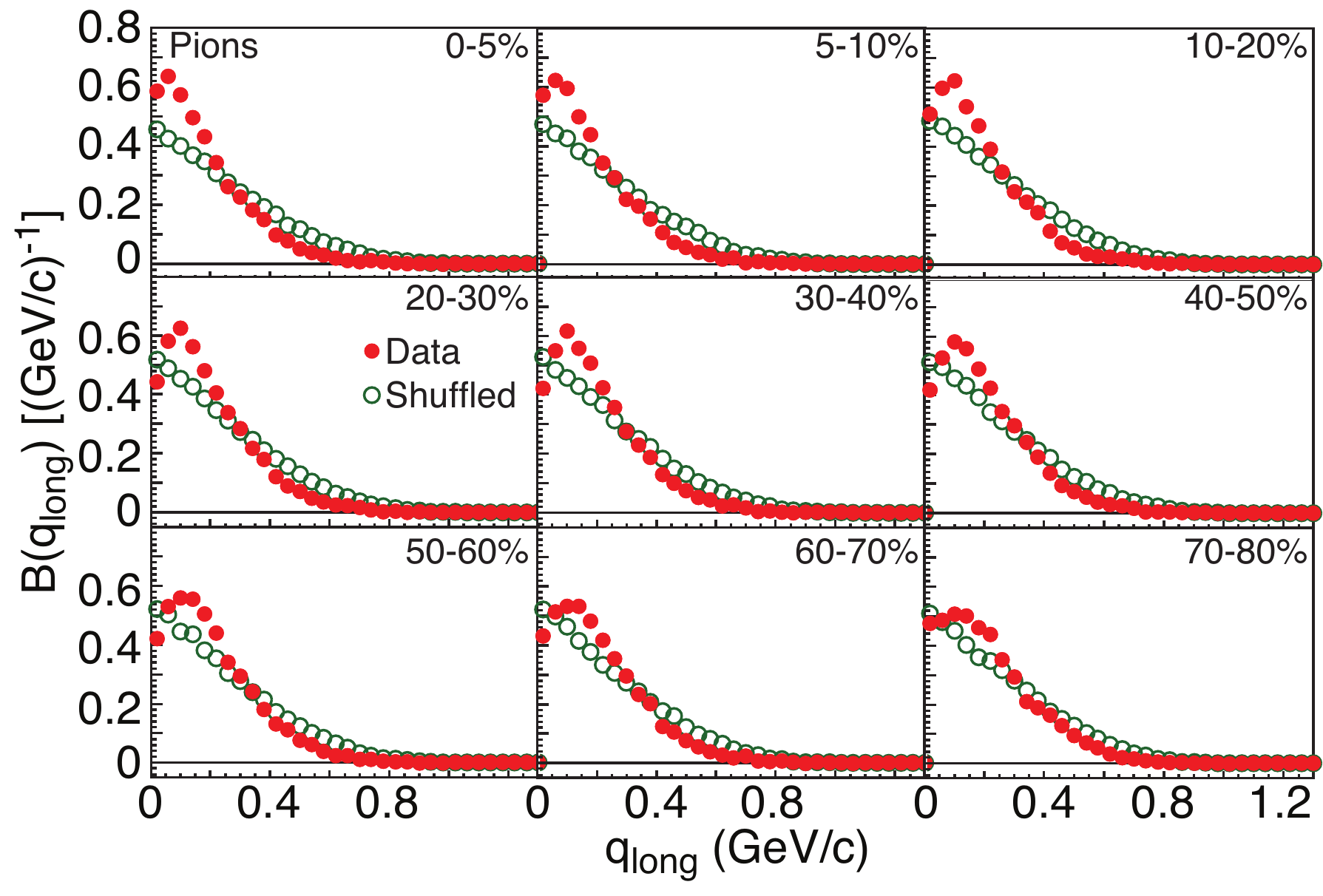}
\caption{\label{fig:fig6_1_10} The balance function in terms of $q_{\rm long}$ for charged pion pairs from Au+Au collisions at $\sqrt{s_{\rm NN}}$ = 200 GeV in nine centrality bins.}
\end{figure}

\begin{figure}
\centering
\includegraphics[width=34pc]{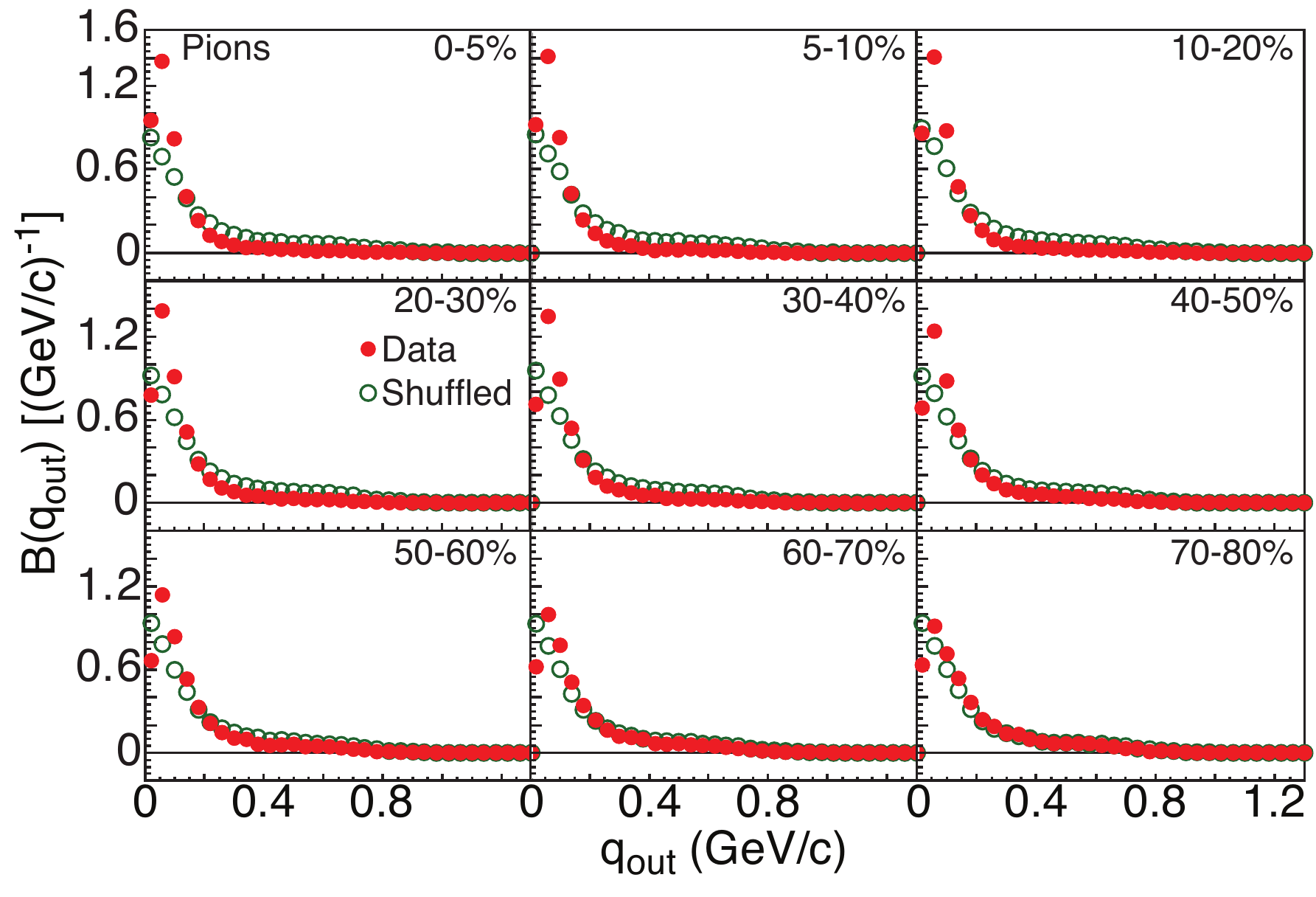}
\caption{\label{fig:fig6_1_11} The balance function in terms of $q_{\rm out}$ for charged pion pairs from Au+Au collisions at $\sqrt{s_{\rm NN}}$ = 200 GeV in nine centrality bins.}
\end{figure}

\begin{figure}
\centering
\includegraphics[width=34pc]{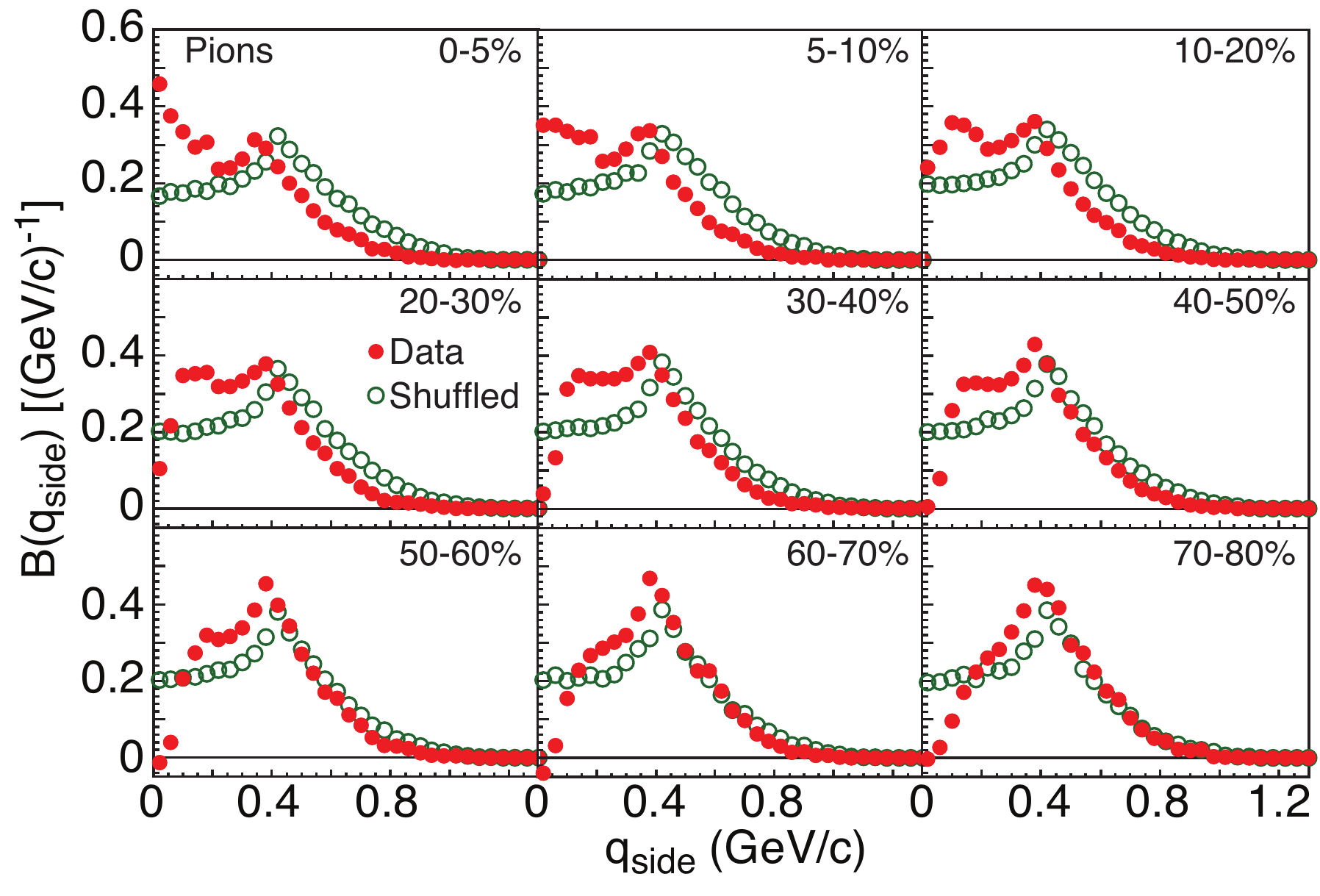}
\caption{\label{fig:fig6_1_12}The balance function in terms of $q_{\rm side}$ for charged pion pairs from Au+Au collisions at $\sqrt{s_{\rm NN}}$ = 200 GeV for nine centrality bins.}
\end{figure}

Figs.~\ref{fig:fig6_1_10}, \ref{fig:fig6_1_11}, and \ref{fig:fig6_1_12} show the balance functions for charged pion pairs from Au+Au collisions at $\sqrt{s_{\rm NN}}$ = 200 GeV in terms of $q_{\rm long}$,   $q_{\rm out}$, and  $q_{\rm side}$ respectively. The balance functions calculated using mixed events are subtracted from the measured balance functions. The balance functions for all three components are narrower in central collisions than in peripheral collisions.

The balance functions in terms of $q_{\rm side}$ do not look like those measured using $q_{\rm long}$ or $q_{\rm out}$ because the lower momentum cut-off of STAR strongly affects
$B(q_{\rm side})$ for $q_{\rm side} < 0.38$ GeV/$c$, which underscores the importance of performing comparisons with models that have been put through detailed efficiency and acceptance filters.

\subsection{Balance Functions in Terms of  $\Delta \phi$}
\label{BDeltaPhi}

The balance function in terms of $\Delta \phi$ may yield information related to transverse flow at freeze-out \cite{Bozek} and may be sensitive to jet production. One might expect that jet-like phenomena would involve the emission of correlated charge/anti-charge pairs at small relative azimuthal angles. We present balance functions for all charged particles with $0.2 < p_{t} < 2.0$ GeV/$c$ from Au+Au collisions at $\sqrt{s_{\rm NN}}$ = 200 GeV as a function of the relative azimuthal angle, $\Delta \phi$.  In addition, we present $B(\Delta\phi)$ for all charged particles with $1.0 < p_{t} < 10.0$ GeV/$c$ to enhance any possible jet-like contributions to the balance function.

\begin{figure}
\centering
\includegraphics[width=34pc]{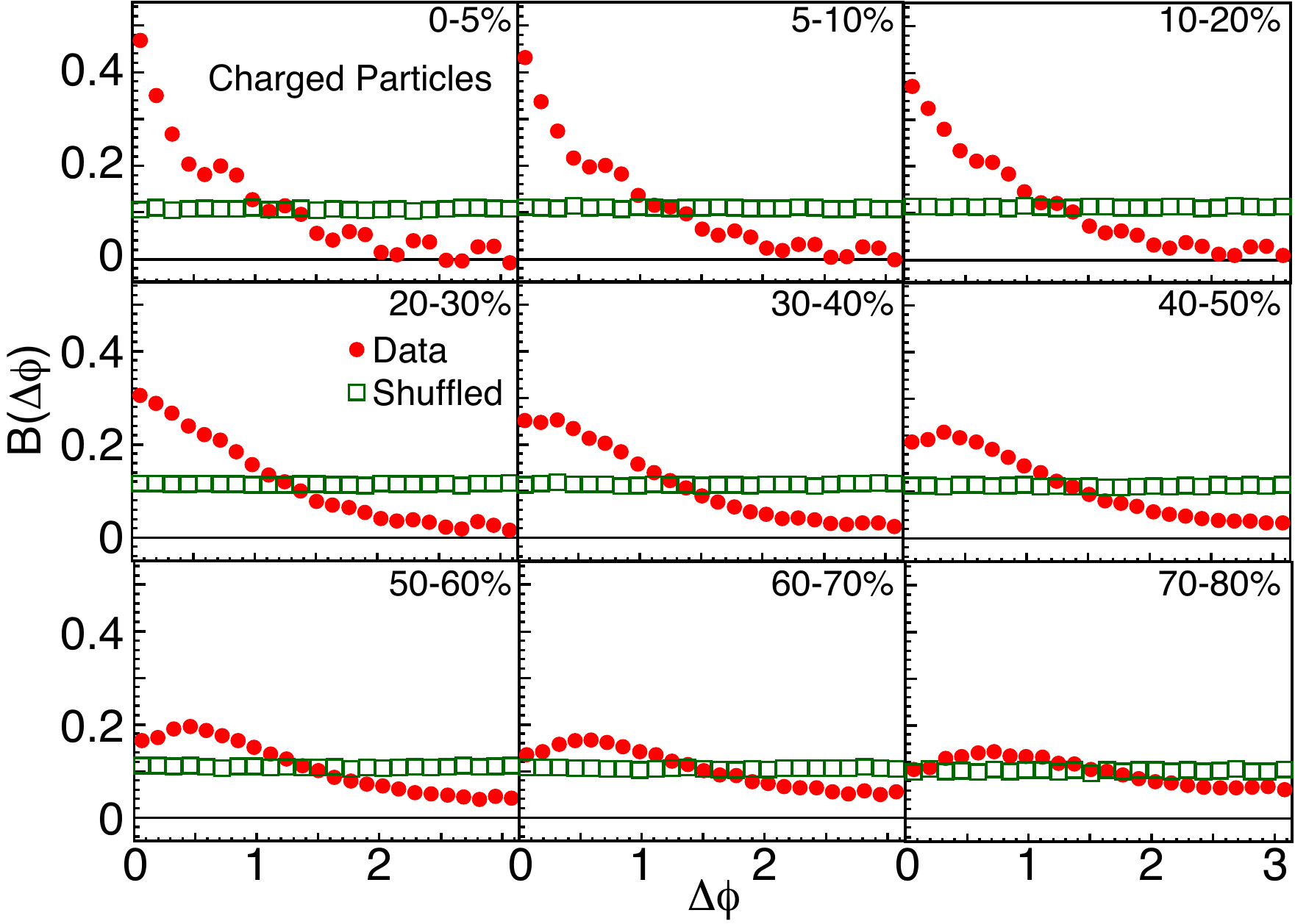}
\caption{\label{fig:fig6_1_13} The balance function in terms of $\Delta \phi$ for all charged particles with $0.2 < p_{t} < 2.0$ GeV/$c$ from Au+Au collisions at $\sqrt{s_{\rm NN}}$ = 200 GeV in nine centrality bins. The closed circles represent the real data minus the mixed events.}
\end{figure}

Fig.~\ref{fig:fig6_1_13} shows the balance functions as a function of $\Delta \phi$ for all charged particles with $0.2 < p_{t} < 2.0$ GeV/$c$ in nine centrality bins.  The balance functions for mixed events were subtracted.  Note that some structure in $\Delta \phi$ related to the  sector boundaries of the STAR TPC is still visible after the subtraction of the mixed events. We observe a peaking at $\Delta \phi$ = 0 in central collisions, while in peripheral collisions, the balance functions are almost flat.  Fig.~\ref{fig:fig6_1_13} also shows the balance functions calculated using shuffled events.  The balance functions from shuffled events are constant with $\Delta\phi$ and show no centrality dependence.

\begin{figure}
\centering
\includegraphics[width=34pc]{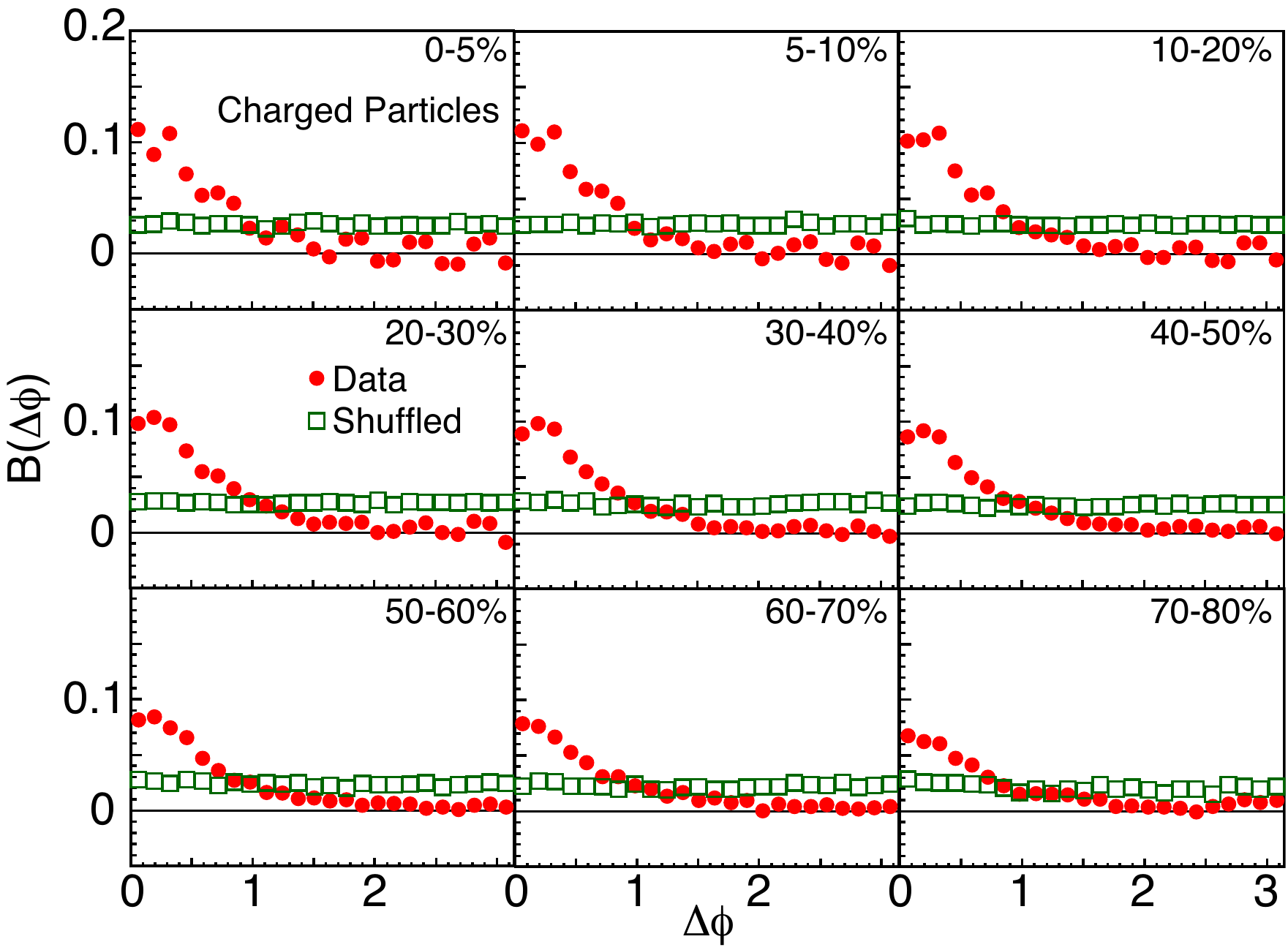}
\caption{\label{fig:fig6_1_14} The balance function in terms of $\Delta \phi$ for all charged particles with $1.0 < p_{t} < 10.0$ GeV/$c$ from Au+Au collisions at $\sqrt{s_{\rm NN}}$ = 200 GeV in nine centrality bins. The closed circles represent the real data minus the mixed events.}
\end{figure}

To augment this result, Fig.~\ref{fig:fig6_1_14} presents balance functions  in which we use only particles with $1.0 < p_{t} < 10.0$ GeV/$c$.  For this case, we see that the measured balance functions vary little with centrality.  Again the balance functions calculated with shuffled events are constant with $\Delta\phi$ and show no centrality dependence. HIJING calculations for $B(\Delta\phi)$ for all charged particles with $0.2 < p_{t} < 2.0$ GeV/$c$ exhibit little dependence on $\Delta\phi$, while HIJING calculations for particles with $1.0 < p_{t} < 10.0$ GeV/$c$ are peaked at $\Delta\phi$ = 0, suggesting that the balance functions for this higher $p_{t}$ range show jet-like characteristics.

The dramatically tight correlations in $\Delta \phi$ in central collisions of Au+Au shown in Fig.~\ref{fig:fig6_1_13} are qualitatively consistent with the radial flow of a perfect liquid.  In a liquid with very short mean free path, the balancing particles would remain in close proximity throughout the reaction.  A large mean free path, which would necessitate a large viscosity, would damp the correlations in $\Delta \phi$ \cite{Teaney}.  This trend is also consistent with a picture where charges are not created until after the flow has been established.

\subsection{Comparison with Models}

\begin{figure}
\centering
\includegraphics[width=24pc]{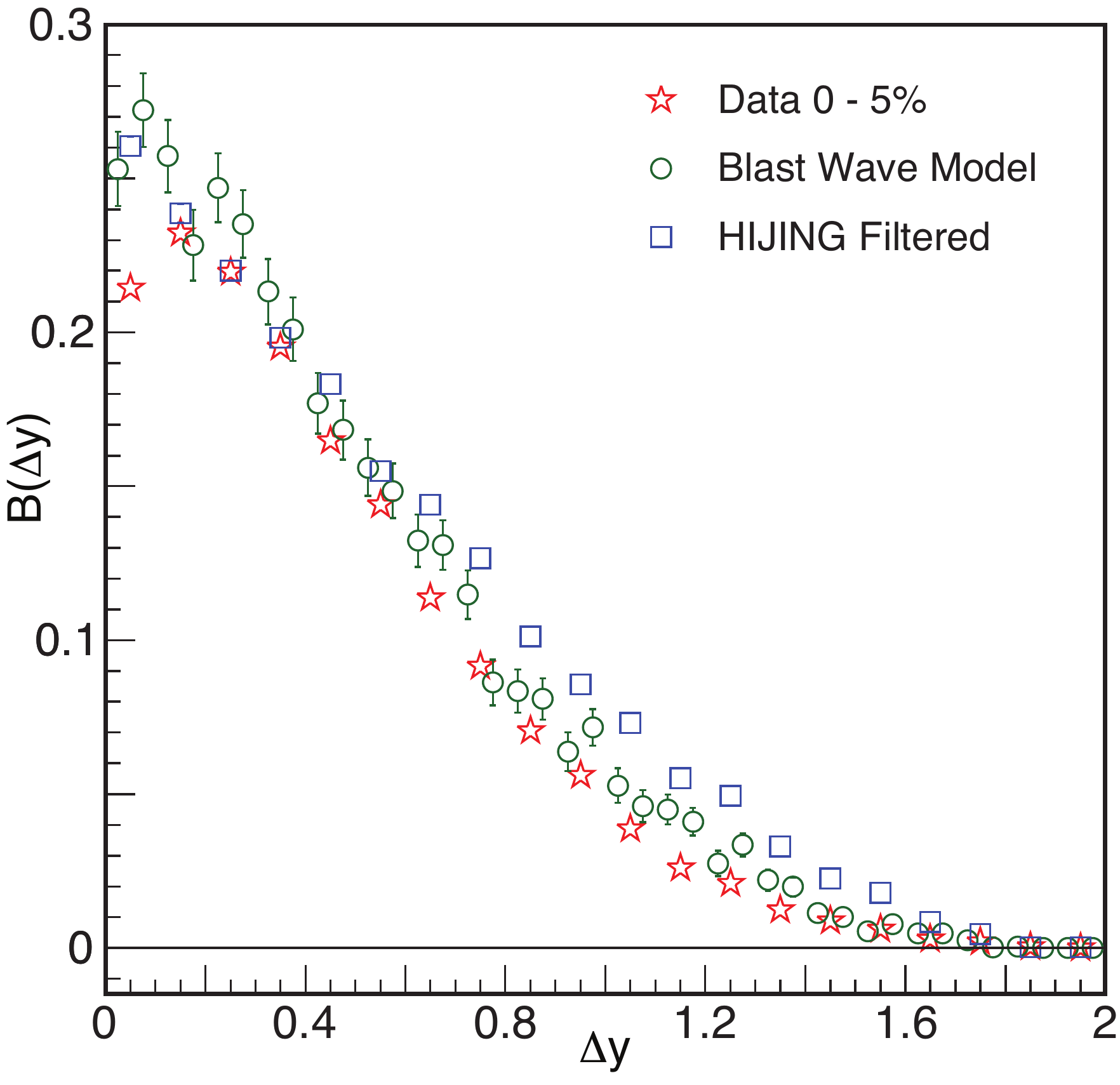}
\caption{\label{fig:fig6_1_15} The balance function in terms of $\Delta y$ for charged pions from central collisions of Au+Au at $\sqrt{s_{\rm NN}}$ = 200 GeV compared with predictions from the blast-wave model from Ref.~\cite{balance_blastwave} and filtered HIJING calculations taking into account acceptance and efficiency.}
\end{figure}

Fig.~\ref{fig:fig6_1_15} compares the measured balance function $B(\Delta y)$ for charged pion pairs from central collisions of Au+Au at $\sqrt{s_{\rm NN}}$ = 200 GeV to the predictions of the blast-wave model \cite{balance_blastwave} and to filtered HIJING calculations taking into account acceptance and efficiency.  The blast-wave model includes radial flow, emission of charge/anti-charge pairs of particles close together in space and time, resonances, HBT and Coulomb effects, strong force effects, inter-domain interactions, and a STAR experimental filter.  The blast-wave calculations shown in Fig.~\ref{fig:fig6_1_15} include the acceptance cuts in the current paper.  The resulting absolute predictions of the blast-wave model agree well with the measured balance function.  In contrast, the balance function predicted by HIJING is significantly wider than the measured balance function.  The widths of the balance functions predicted by the blast-wave and HIJING are compared with the experimental values in Fig.~\ref{fig:fig6_1_18}.

The width of the balance function predicted by the blast-wave model is close to the width observed in central collisions.  
The blast-wave model assumes that the charge/anti-charge pairs of particles are created close together in space and at the same time, and contains no scattering or longitudinal expansion that would widen the balance function in terms of $\Delta y$.  Thus, the agreement of the predicted width from the blast-wave model and the data are consistent with the idea of delayed hadronization in that delayed hadronization in central collisions would minimize the contribution of diffusion effects to the width of the balance function.

\begin{figure}
\centering
\includegraphics[width=24pc]{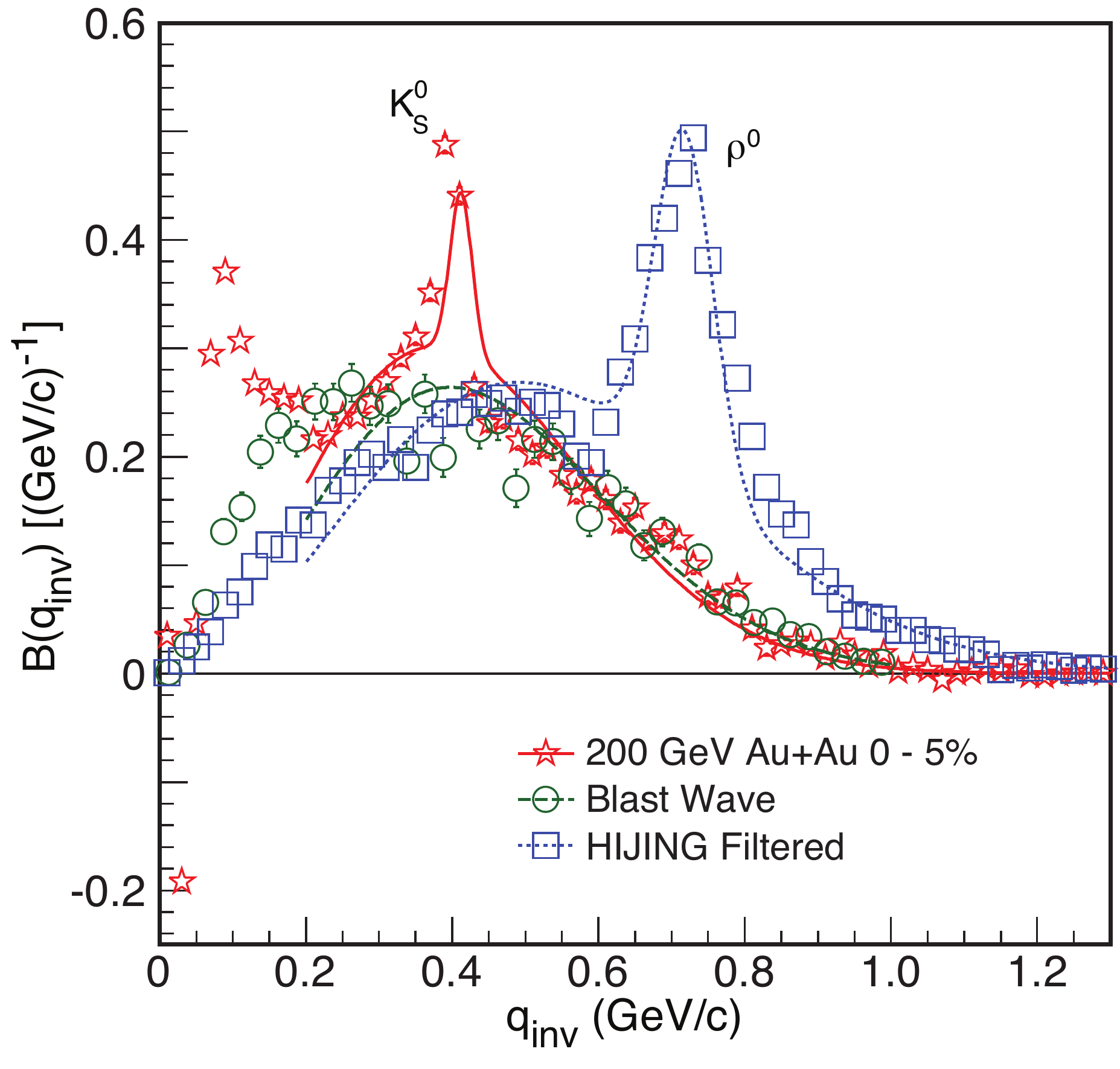}
\caption{\label{fig:fig6_1_16} The balance function in terms of $q_{\rm inv}$ for charged pions from central collisions of Au+Au at $\sqrt{s_{\rm NN}}$ = 200 GeV compared with predictions from the blast-wave model from Ref.~\cite{balance_blastwave} and predictions from filtered HIJING calculations including acceptance and efficiency. For the blast-wave calculations, HBT is not included and the decays of the $K_S^0$ and $\rho^{0}$ are not shown. }
\end{figure}

The balance function in terms of $q_{\rm inv}$ provides the most direct way to study the dependence of the balance function on temperature.  Fig.~\ref{fig:fig6_1_16} compares the balance function in terms of $q_{\rm inv}$ for charged pion pairs from central collisions of Au+Au at $\sqrt{s_{\rm NN}}$ = 200 GeV to the predictions of the blast-wave model and to filtered HIJING calculations.  For the blast-wave model calculations, HBT is not included and the decays of the $K^{0}$ and $\rho^{0}$ are not shown.  The solid curve for the data represents a fit comprised of a thermal distribution (Equation~\ref{thermal}) plus $K^{0}$ decay.  The dashed curve for the blast-wave model calculations represents a thermal fit (Equation~\ref{thermal}).  The dotted curve for the HIJING calculations represents a thermal distribution (Equation~\ref{thermal}) plus $\rho^{0}$ decay.  All the fits are carried out over a range in $q_{\rm inv}$ that is not affected by HBT/Coulomb effects.  The width extracted from the thermal fit to the blast-wave model calculations is compared with the width extracted from experimental data in Fig.~\ref{fig:fig6_1_19}.  The blast-wave model reproduces the observed width in central collisions.  The HIJING calculations show a strong $\rho^{0}$ peak that is not present in the data.

Future analyses should be able to disentangle the effects of cooling and diffusion in driving the narrowing of the balance function.  Diffusive effects should largely manifest themselves in the $q_{\rm long}$ variable because the initial velocity is in the longitudinal direction and some creation mechanisms, such as strings, preferentially separate the pairs in the longitudinal direction.

\subsection{Balance Function Widths}
\label{Widths}

The balance functions presented in the previous section provide insight into the correlation of charge/anti-charge pairs in collisions at RHIC.  This approach complements the approach of studying these phenomena using charge-dependent correlation functions in two dimensions,
$(\Delta \eta,\Delta \phi)$ \cite{star_deta_dphi_cf_200, star_deta_dphi_cf}.
The balance function can be related to these correlation functions and to other two-particle observables.   $B(\Delta y)$ can be interpreted as the distribution of relative rapidities of correlated charge/anti-charge pairs.  The width of  $B(\Delta y)$ then can be used to determine whether correlated charge/anti-charge pairs of particles are emitted close together or far apart in rapidity.  The width of the balance function $B(q_{\rm inv})$ can be used to study thermal distributions because this balance function can be related to the temperature, and is largely unaffected by any radial expansion.

\begin{figure}
\centering
\includegraphics[width=24pc]{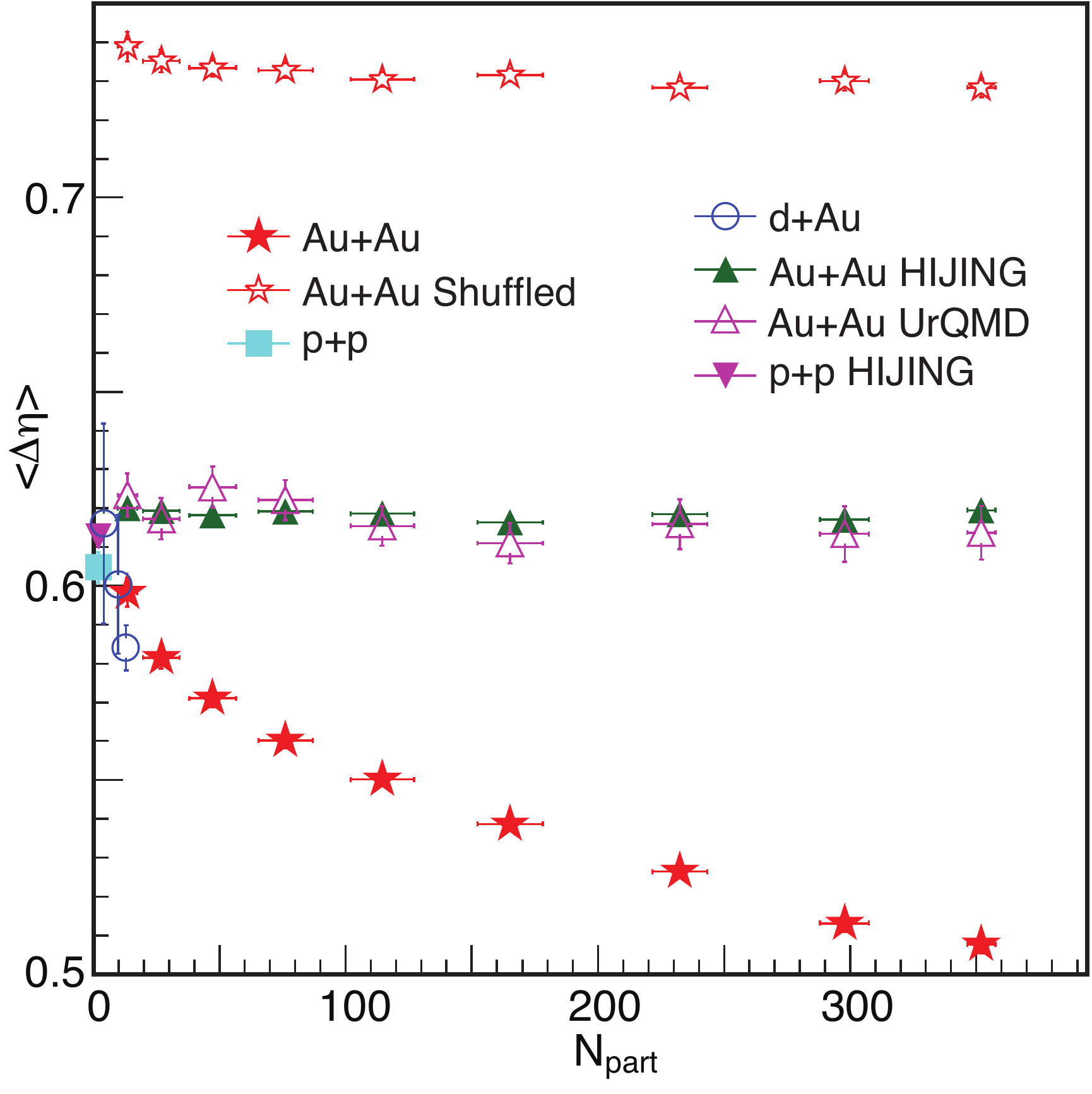}
\caption{\label{fig:fig6_1_17} The balance function width $\langle \Delta \eta \rangle$ for all charged particles from Au+Au collisions at $\sqrt{s_{\rm NN}}$ = 200 GeV compared with the widths of balance functions calculated using shuffled events.  Also shown are the balance function widths for $p+p$ and $d$+Au collisions at $\sqrt{s_{\rm NN}}$ = 200 GeV.  Filtered HIJING calculations are also shown for the widths of the balance function from $p+p$ and Au+Au collisions. Filtered UrQMD calculations are shown for the widths of the balance function from Au+Au collisions.}
\end{figure}

To quantify the evolution of the balance functions $B(\Delta y)$ and $B(\Delta\eta)$ with centrality, we extract the width, $\langle \Delta y \rangle$ and $\langle \Delta \eta \rangle$, using a weighted average.

\begin{eqnarray}
\label{WA}
\left\langle {\Delta \eta } \right\rangle  = \frac{{\sum\limits_{i = i_{{\rm{lower}}} }^{i_{{\rm{upper}}} } {B\left( {\Delta \eta _i } \right)\Delta \eta _i } }}{{\sum\limits_{i = i_{{\rm{lower}}} }^{i_{{\rm{upper}}} } {B\left( {\Delta \eta _i } \right)} }}
\end{eqnarray}

For $B(\Delta \eta)$, the weighted average is calculated for $ 0.1 \le \Delta \eta \le 2.0$ and for $B(\Delta y)$, the weighted average is calculated for $ 0.2 \le \Delta y \le 2.0$.

Fig.~\ref{fig:fig6_1_17} shows the balance function widths for all charged particles from Au+Au, $d$+Au, and $p+p$ collisions at $\sqrt{s_{\rm NN}}$ = 200 GeV plotted in terms of the number of participating nucleons, $N_{\rm part}$.  In addition, we present the widths of the balance functions from Au+Au collisions for shuffled events.  The widths of the shuffled events are considerably larger than those from the measured data and represent the largest width we can measure using the STAR acceptance for the system under consideration. 

The balance function widths scale smoothly from $p+p$ through the three centrality bins for $d$+Au and down to the nine Au+Au collision centrality data points. This figure also shows filtered HIJING calculations for $p+p$ and Au+Au calculations for HIJING and UrQMD. The HIJING calculations for $p+p$ reproduce the measured width. The Au+Au HIJING and UrQMD calculations, however, show little centrality dependence and are comparable to those calculated from the HIJING $p+p$ simulations. This is despite the fact that HIJING does not predict any appreciable radial flow while UrQMD predicts radial flow in Au+Au collisions but less than that observed experimentally. This radial flow should produce a narrower balance function in central collisions where radial flow is the largest, while hadronic scattering should lead to a wider balance function.  The fact that the measured widths from Au+Au collisions narrow in central collisions is consistent with trends predicted by models incorporating late hadronization \cite{balance_theory,balance_blastwave}.

\begin{figure}
\centering
\includegraphics[width=24pc]{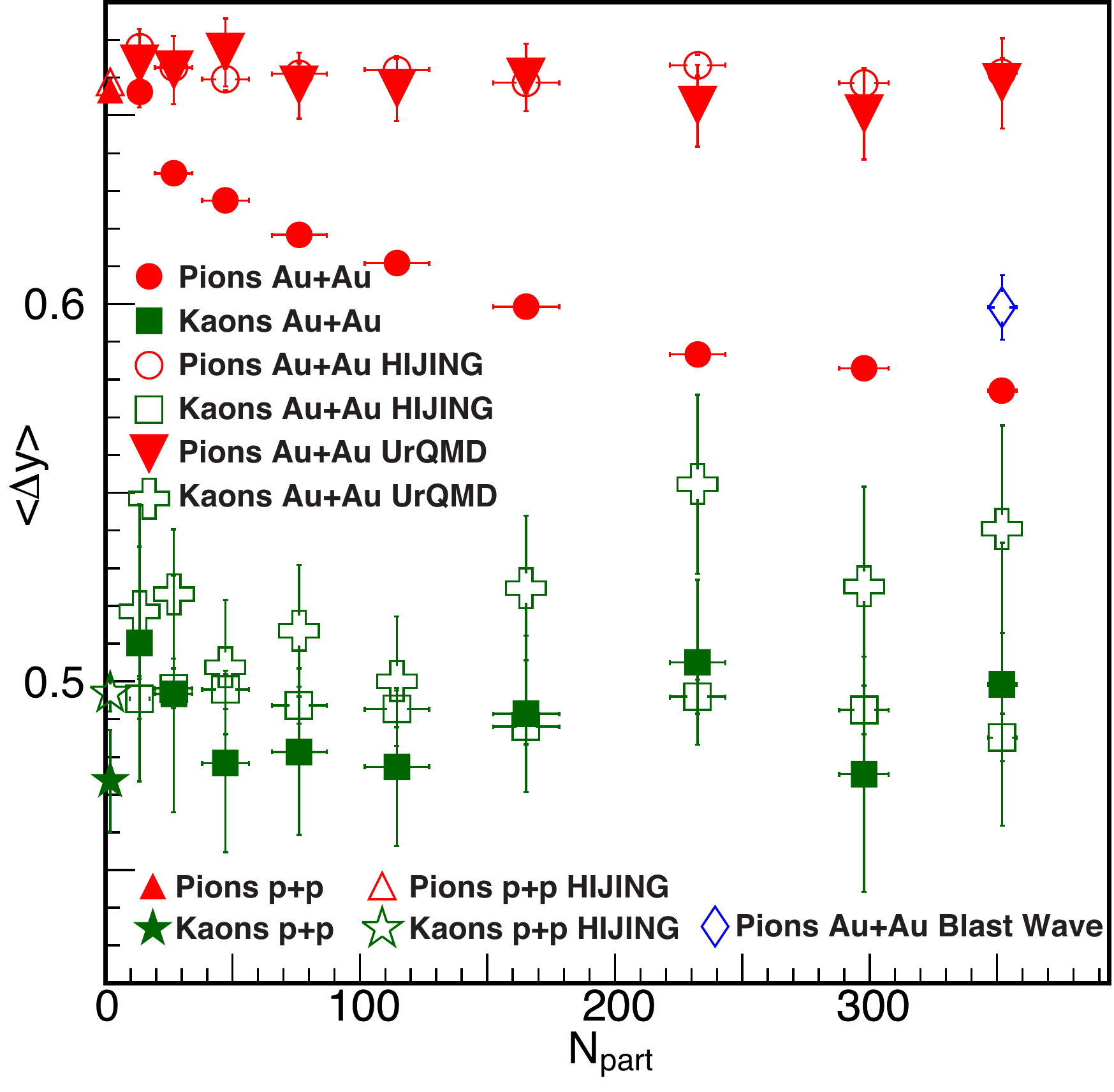}
\caption{\label{fig:fig6_1_18} The balance function widths for identified charged pions and charged kaons from Au+Au collisions at $\sqrt{s_{\rm NN}}$ = 200 GeV and $p+p$ collisions at $\sqrt{s}$ = 200 GeV.  Filtered HIJING calculations are shown for the same systems.  Filtered UrQMD calculations are shown for Au+Au. Also shown is the width of the balance function for pions predicted by the blast-wave model of Ref. \cite{balance_blastwave}.}
\end{figure}

\begin{figure}
\centering
\includegraphics[width=24pc]{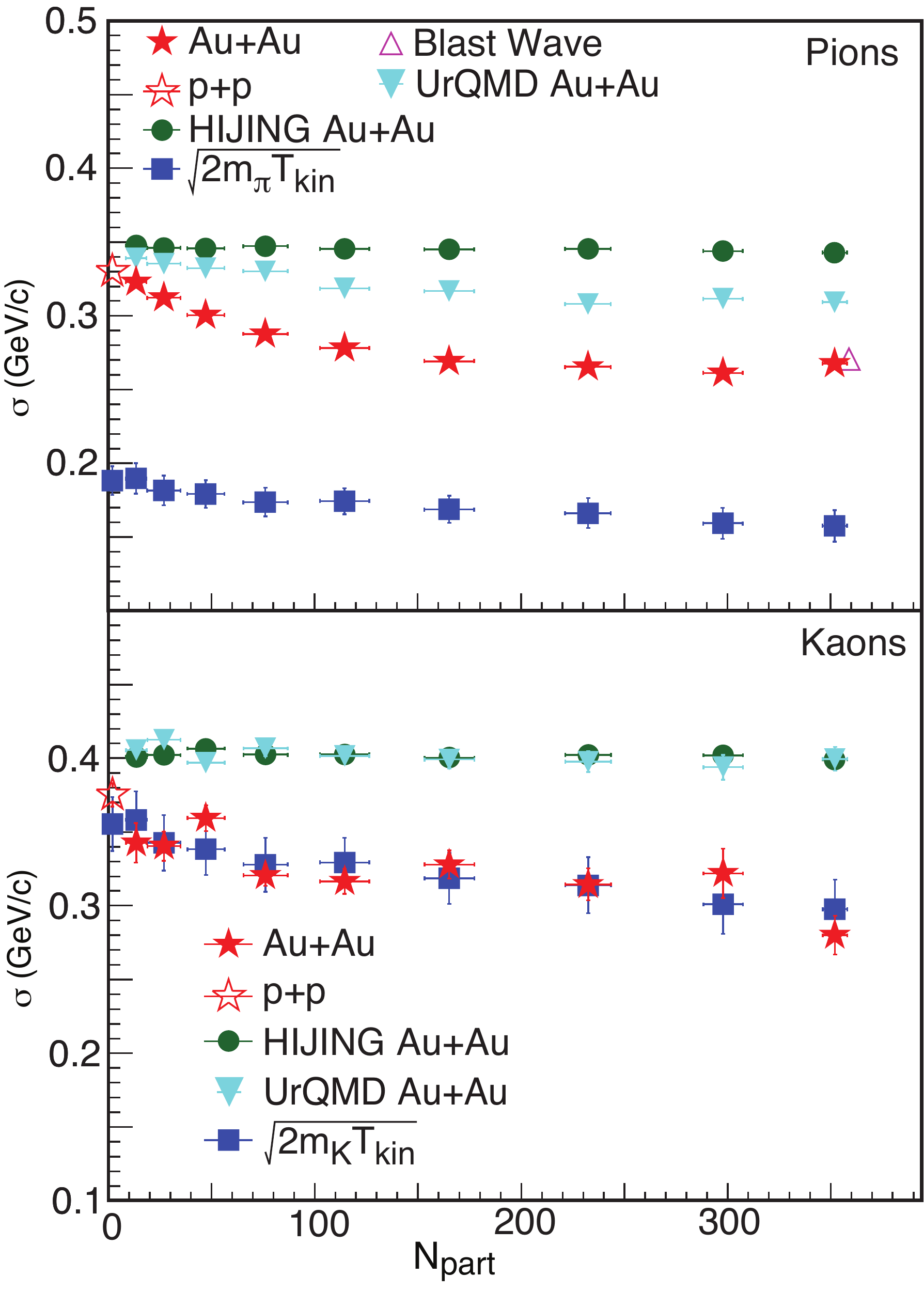}
\caption{\label{fig:fig6_1_19} The balance function width $\sigma$ extracted from
$B(q_{\rm inv})$ for identified charged pions and kaons from Au+Au collisions at $\sqrt{s_{\rm NN}}$ = 200 GeV 
and $p+p$ collisions at $\sqrt{s}$ = 200 GeV using a thermal fit (Equation~\ref{thermal}) where $\sigma$ is the width.  Filtered HIJING and UrQMD calculations are shown for pions and kaons from Au+Au collisions at $\sqrt{s_{\rm NN}}$ = 200 GeV.  Values are shown for $\sqrt{2mT_{\rm kin}}$ from Au+Au collisions, where $m$ is the mass of a pion or a kaon, and $T_{\rm kin}$ is calculated from identified particle spectra \cite{STAR_identified_spectra}.  The width predicted by the blast-wave model of Ref. \cite{balance_blastwave} is also shown for pions.}
\end{figure}

Fig.~\ref{fig:fig6_1_18} presents the widths of the balance function, $B(\Delta y)$, for identified charged pions and identified charged kaons from $p+p$ collisions at $\sqrt{s}$ = 200 GeV and Au+Au collisions at $\sqrt{s_{\rm NN}}$ = 200 GeV.  Also shown are filtered HIJING and UrQMD calculations.  For charged pions, the measured balance function widths for Au+Au collisions get smaller in central collisions, while the filtered HIJING and UrQMD calculations for Au+Au again show no centrality dependence.  The HIJING calculations for $p+p$ collisions reproduce the observed widths.

In contrast, the widths of the measured balance function for charged kaons from Au+Au collisions show little centrality dependence.  The extracted widths for charged kaons are consistent with the predictions from filtered HIJING calculations and are consistent with the $p+p$ results.  The widths for charged kaons predicted by UrQMD are somewhat larger than the data.  The agreement with HIJING and the lack of centrality dependence may indicate that kaons are produced mainly at the beginning of the collision rather than during a later hadronization stage \cite{balance_theory}.  The larger widths predicted by UrQMD for kaons may reflect the hadronic scattering incorporated in UrQMD, although the statistical errors are large for both the data and the model predictions.
 
Fig.~\ref{fig:fig6_1_19} shows the widths extracted from $B(q_{\rm inv})$ for identified charged pions and kaons from Au+Au collisions at $\sqrt{s_{\rm NN}}$ = 200 GeV and $p+p$ collisions at  $\sqrt{s}$ = 200 GeV using a thermal distribution (Equation~\ref{thermal}) where $\sigma$ is the width.  The widths for the pions are somewhat smaller than the widths for the kaons, although the kaon widths have a large statistical error.  This width is related to the temperature of the system when the pions and kaons are formed.  Filtered HIJING calculations show no centrality dependence and predict a difference between the widths for pions and kaons.  The widths predicted by UrQMD for pions are smaller than those predicted by HIJING but are still larger than the measured widths.  In addition, the widths predicted by UrQMD for pions seem to show a centrality dependence, although it is not as strong as that for the data.  The widths predicted by UrQMD for kaons show no centrality dependence and agree with HIJING.

For a thermal system in the non-relativistic limit ($m \gg T$), the balance function has the functional form  given in Equation~\ref{thermal} where $\sigma=\sqrt{2mT}$. For kinetic freeze-out temperatures $T \sim 0.1$ GeV \cite{STAR_identified_spectra}, kaons are non-relativistic, and this functional form was seen to describe the balance function in Fig.~\ref{fig:fig6_1_08}.  Indeed, as seen in the right panel of Fig.~\ref{fig:fig6_1_19}, the evolution in the width of the balance function may be understood in terms of the evolution of the freeze-out temperature as a function of centrality  \cite{STAR_identified_spectra}.

In the ultra-relativistic case ($m \ll T$), the balance function from a thermal system is exponential rather than Gaussian, $B(q_{\rm inv}) \sim q_{\rm inv}^2e^{-q_{\rm inv}/T}$.
The proper functional form for pions, being neither non-relativistic nor ultra-relativistic, is more complicated.  Indeed, we found that neither the Gaussian form nor the exponential form fully describe the pion balance
function in Fig.~\ref{fig:fig6_1_06}.  Thus, to get a feeling for whether the evolution in freeze-out temperature can explain the narrowing of the balance function for pions, we turn to numerical calculations. Calculations in Ref. \cite{balance_distortions} show a 27\% reduction in the Gaussian width of  $B(q_{\rm inv})$ as the temperature is varied from 120 to 90 MeV, the temperatures inferred from fits to peripheral and central collisions, respectively \cite{STAR_identified_spectra}.  As seen in Fig.~\ref{fig:fig6_1_19}, the measured width for peripheral (central) collisions is 0.33 GeV/$c$ (0.27 GeV/$c$), a 18\% reduction. Thus, the centrality evolution in freeze-out temperature may help explain much of the narrowing of the
balance function in terms of $q_{\rm inv}$ for pions as well as for kaons.  However, firm conclusions require more complete calculations including all detector effects.

\begin{figure}
\centering
\includegraphics[width=24pc]{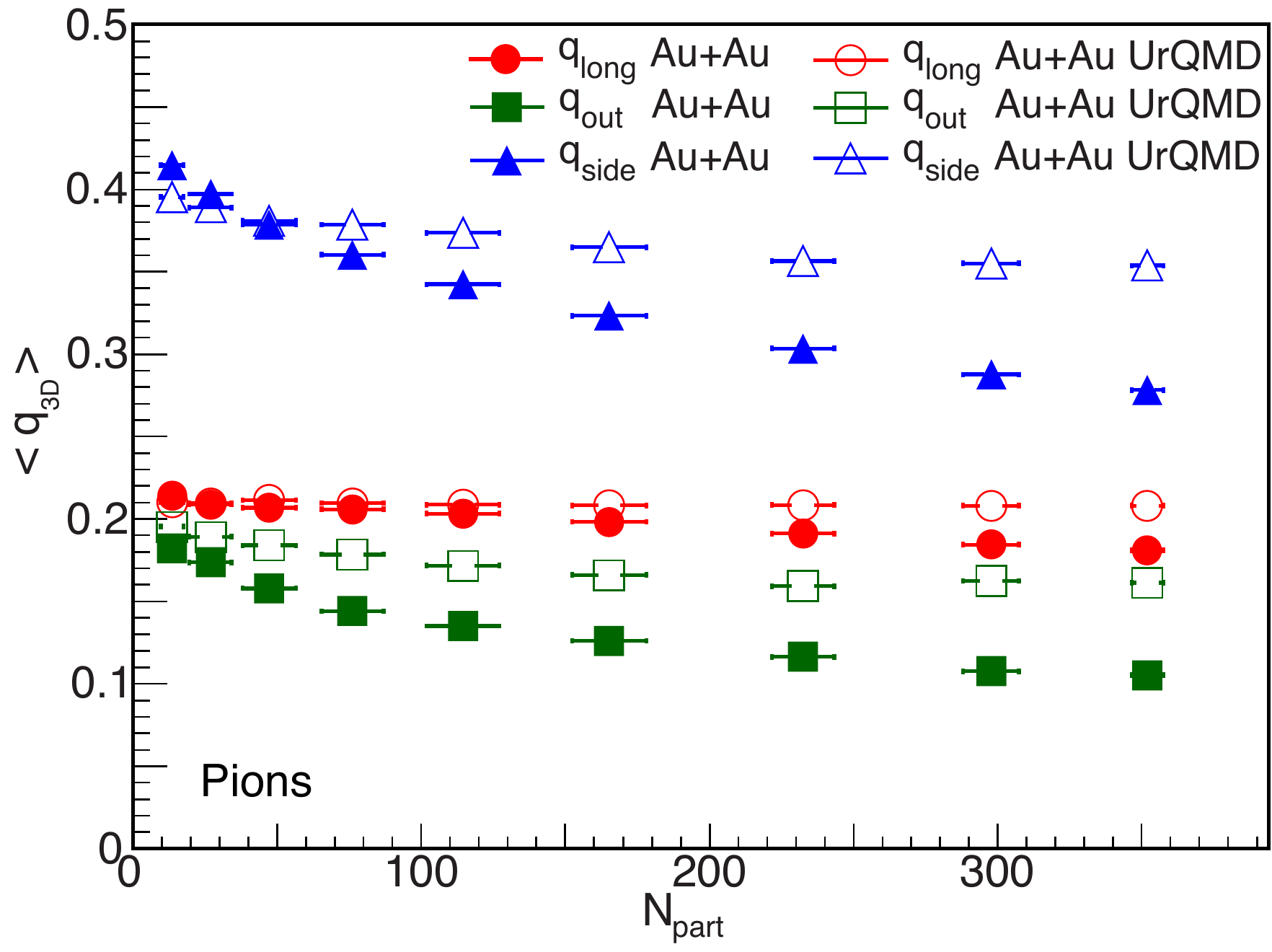}
\caption{\label{fig:fig6_1_20}The widths for the balance functions for pions in terms of $q_{\rm long}$, $q_{\rm out}$, and $q_{\rm side}$ compared with UrQMD calculations.}
\end{figure}

Fig.~\ref{fig:fig6_1_20} shows the widths of the balance functions in terms of $q_{\rm long}$, $q_{\rm out}$, and $q_{\rm side}$ for charged pion pairs in Au+Au collisions at $\sqrt{s_{\rm NN}}$ = 200 GeV compared with the results of filtered UrQMD calculations.  These widths were extracted by taking the weighted average over the $q_{\rm long}$, $q_{\rm out}$, and $q_{\rm side}$ range from 0.0 to 1.3 GeV/$c$.  The width $\langle q_{\rm side} \rangle$ is larger than $\langle q_{\rm long} \rangle$ and $\langle q_{\rm out} \rangle$ because the lower $p_{t}$ threshold of STAR affects it more strongly.  In the most peripheral collisions, the widths $\langle q_{\rm long} \rangle$ and $\langle q_{\rm out} \rangle$ are comparable to each other.  As the collisions become more central, both $\langle q_{\rm long} \rangle$ and $\langle q_{\rm out} \rangle$ decrease.  The change in $\langle q_{\rm long} \rangle$ is less than the change of $\langle q_{\rm out} \rangle$ with increasing centrality.  Thus it seems that the two transverse widths, $\langle q_{\rm out} \rangle$, and $\langle q_{\rm side} \rangle$, decrease in central collisions more strongly than the longitudinal width, $\langle q_{\rm long} \rangle$.  This may imply that string dynamics and diffusion due to longitudinal expansion may keep $\langle q_{\rm long} \rangle$ from decreasing as much in more central collisions \cite{balance_blastwave}.  The decrease in the transverse widths is consistent with the decrease in $T_{\rm kin}$ as the collisions become more central. In the most peripheral collisions, the widths predicted by UrQMD are consistent with the data.  As the collisions become more central, the predicted widths decrease slightly, but not as much as observed in the data.  This is consistent with results using the balance function in terms of $q_{\rm inv}$.  Additional theoretical input is required to draw more conclusions from the analysis of the balance function in terms of the components of $q_{\rm inv}$.

\begin{figure}
\centering
\includegraphics[width=24pc]{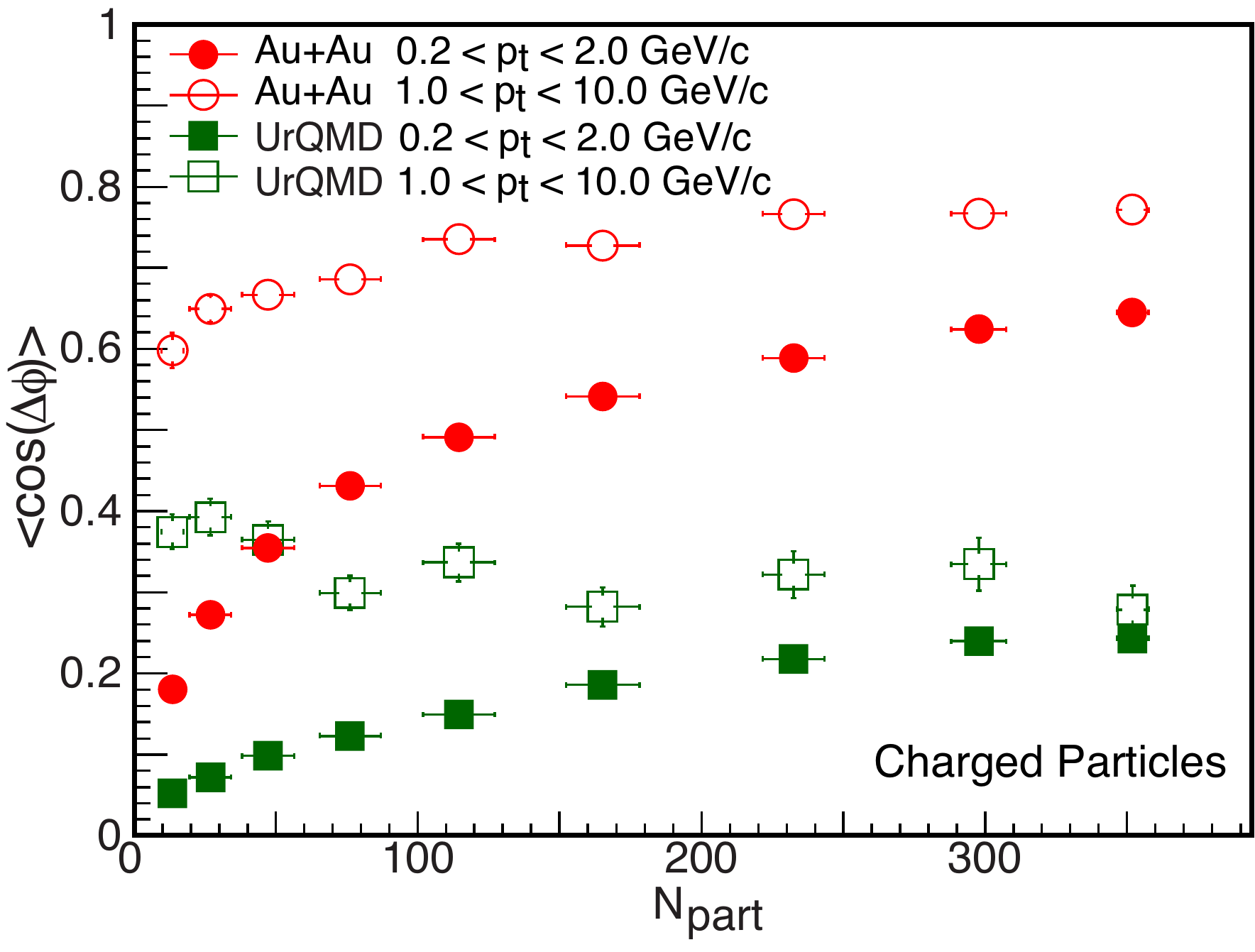}
\caption{\label{fig:fig6_1_21} The weighted average cosine of the relative azimuthal angle, $\langle \cos{(\Delta \phi)} \rangle$,  extracted from $B(\Delta \phi)$ for all charged particles with $0.2 < p_{t} < 2.0$ from Au+Au collisions at $\sqrt{s_{\rm NN}}$ = 200 GeV and from all charged particles with $1.0 < p_{t} < 10.0$ GeV/$c$, compared with predictions using filtered UrQMD calculations.}
\end{figure}

Fig.~\ref{fig:fig6_1_21} shows the weighted average cosine of the relative azimuthal angle, $\langle \cos{(\Delta \phi)} \rangle$, extracted from the balance
functions $B(\Delta \phi)$ for all charged particles from Au+Au
collisions at $\sqrt{s_{\rm NN}}$ = 200 GeV with $0.2 < p_{t} < 2.0$ GeV/$c$ and $1.0 < p_{t} < 10.0$ GeV/$c$. The values for $\langle \cos{(\Delta \phi)} \rangle$  are extracted over the range $0 \le \Delta \phi \le \pi$.  For the lower $p_{t}$ particles, the balance function narrows dramatically in central collisions (large positive values of $\langle \cos{(\Delta \phi)} \rangle$).  The narrow balance functions observed in central collisions may be a signature of the flow of a perfect liquid, as discussed above.  For the higher $p_{t}$ particles, $\langle \cos{(\Delta \phi)} \rangle$ in Au+Au collisions shows less centrality dependence.

Fig.~\ref{fig:fig6_1_21} also shows UrQMD calculations for  $\langle \cos{(\Delta \phi)} \rangle$.  The predictions for the $0.2 < p_{t} < 2.0$ GeV/$c$ data set are much lower than the measured values, which is consistent with the observation that UrQMD underpredicts radial flow.  The predictions for $\langle \cos{(\Delta \phi)} \rangle$ for the $1.0 < p_{t} < 10.0$ GeV/$c$ data set show no centrality dependence and are also much lower than the measured values.

\section{Beam Energy Dependent Balance Functions}

\begin{figure}
\centering
\includegraphics[width=34pc]{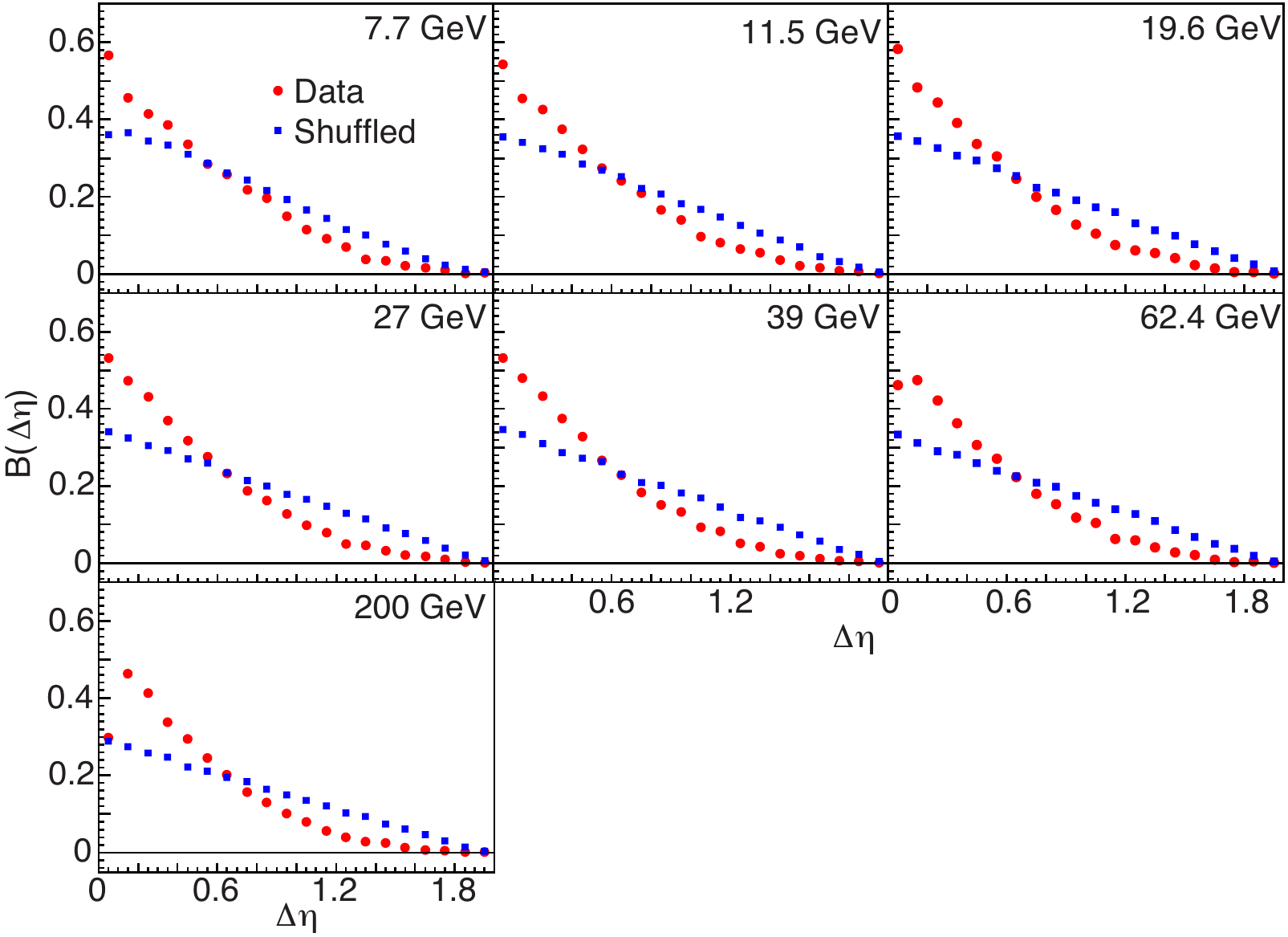}
\caption{\label{fig:fig6_2_01} The balance function in terms of $\Delta \eta$ for all charged particles. Central events (0-5\%) are shown here with $\sqrt{s_{\rm NN}}$ from 7.7 to 200 GeV.}
\end{figure}

Figure~\ref{fig:fig6_2_01} shows the balance function in terms of $\Delta \eta$ for all charged particles. The most central events (0-5\%) are shown for seven incident energies. The data (red circles) in the figure are the balance function results from real data corrected by subtracting the balance function calculated using mixed events. We can see that, for all the energies shown here, the balance functions from data are narrower than the ones from shuffled events (blue squares). 

\begin{figure}
\centering
\includegraphics[width=38pc]{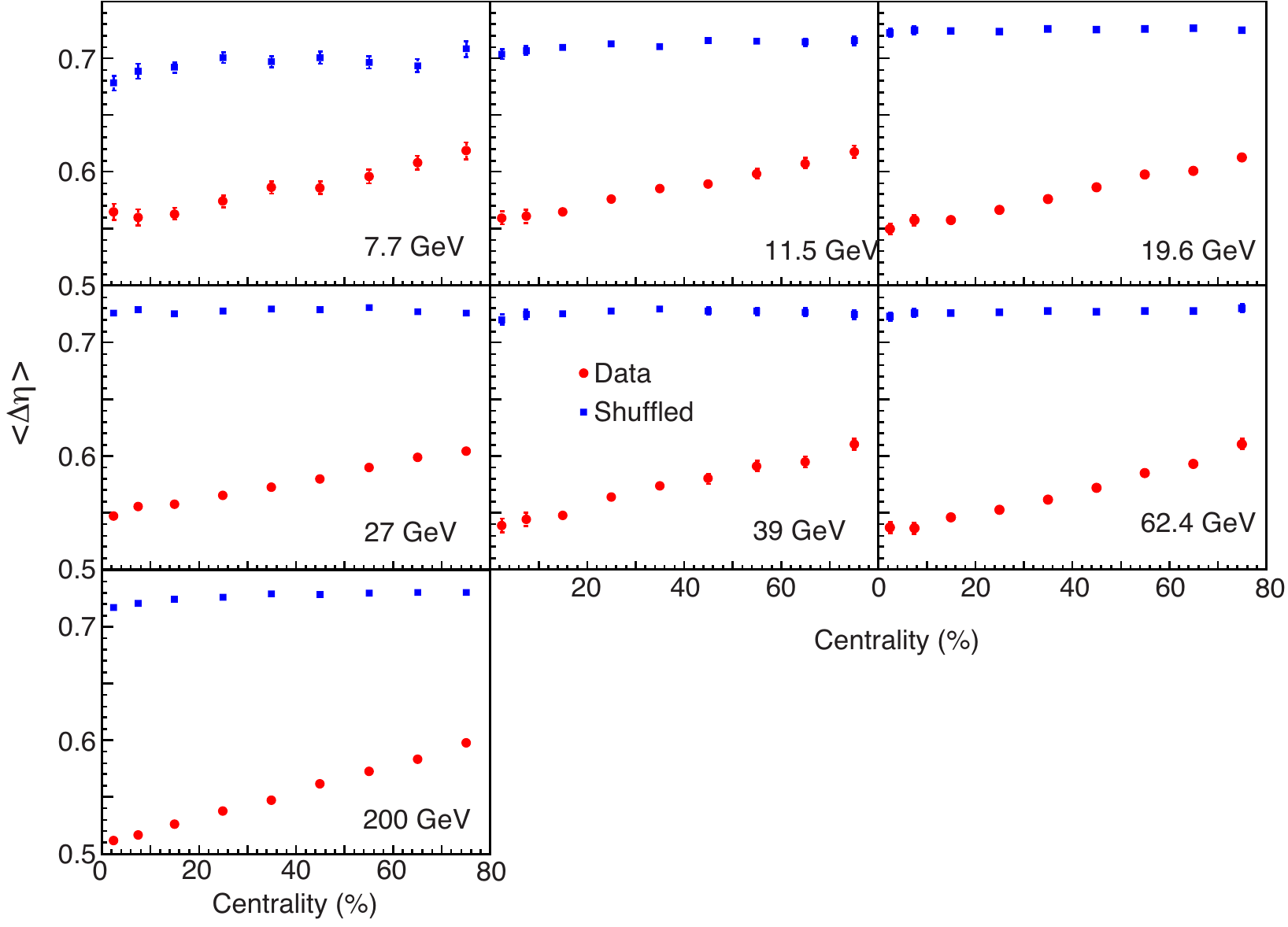}
\caption{\label{fig:fig6_2_01_2}Centrality dependence of the balance function width $\left\langle {\Delta \eta } \right\rangle$ for seven collision energies compared with shuffled events.  }
\end{figure}

To quantify the balance function widths for all the energies, Figure~\ref{fig:fig6_2_01_2} shows the centrality dependence of the weighted average (Equation \ref{WA}) of the balance function for seven collision energies.  The weighted average is calculated for $0.1 < \Delta \eta  < 2.0 $ to reduce the contributions from inter-pair correlations (HBT and Columb)\cite{balance_distortions}. For all the energies in the figure, shuffled events show no centrality dependence, while the data show narrower balance functions in central collisions. This narrowing of the balance function at central collisions may imply delayed hadronization.~\cite{parity_soeren}

\begin{figure}
\centering
\includegraphics[width=32pc]{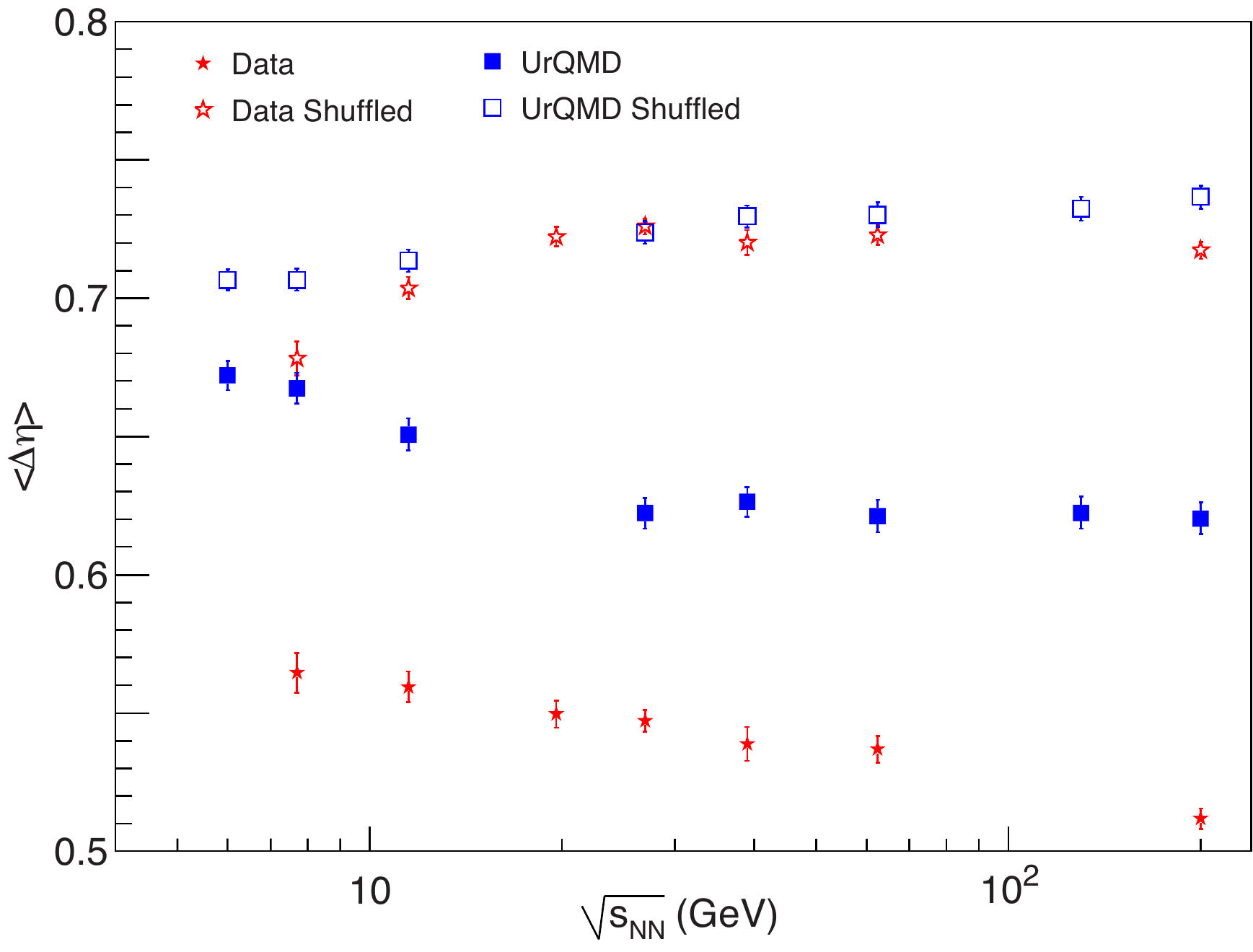}
\caption{\label{fig:fig6_2_02}Energy dependence of the balance function width $\left\langle {\Delta \eta } \right\rangle$ for central Au+Au collisions (0-5\%) compared with shuffled events. Both data and UrQMD calculations are shown here. }
\end{figure}

Figure~\ref{fig:fig6_2_02} shows the energy dependence of the balance function width for central Au+Au collisions. The data show a smooth decrease of $\left\langle {\Delta \eta } \right\rangle$ with increasing energy. UrQMD calculations predict a similar trend but over predict the observed results. Since the balance function is sensitive to the hadronization time and relative diffusion after hadronization, this decrease in balance function width could be a signal of the onset of deconfinment or radial flow. The UrQMD model is a hadronic model that does not have a deconfined phase transition and has little flow. This early hadronization time combined with strong interaction between final particles leads to a wider balance function. In the same figure, the shuffled events from both data and UrQMD show a wider balance function that slightly increases with increasing energy. Since the shuffled events represent the widest balance function within STAR's acceptance, the change of the balance function calculated using shuffled events is due to the slight changes in STAR's $\eta$ acceptance with energy. 

Due to a calibration issue, one sector out of twenty-four sectors in the TPC has been turned off for Run 10 200 GeV data.  For the Run 10 7.7 GeV data, the wide $z$ vertex cut ($|z| < 70$ cm) introduce a non-uniform $\eta$ distribution. These detector acceptance effects cause a difference in the shuffled events between data and UrQMD simulations. 

To reduce acceptance effects and make a better comparison of the balance function width between different incident energies and different experiments, NA49 has proposed a normalized parameter $W$~\cite{NA49_PRC_2007}, which is defined as:

\begin{eqnarray}
\label{eq:bfmoments}
W = \frac{{100 \cdot ( < \Delta \eta  > _{\rm shuffled}  -  < \Delta \eta  > _{\rm data} )}}{{ < \Delta \eta  > _{\rm shuffled} }}
\end{eqnarray}

\begin{figure}
\centering
\includegraphics[width=30pc]{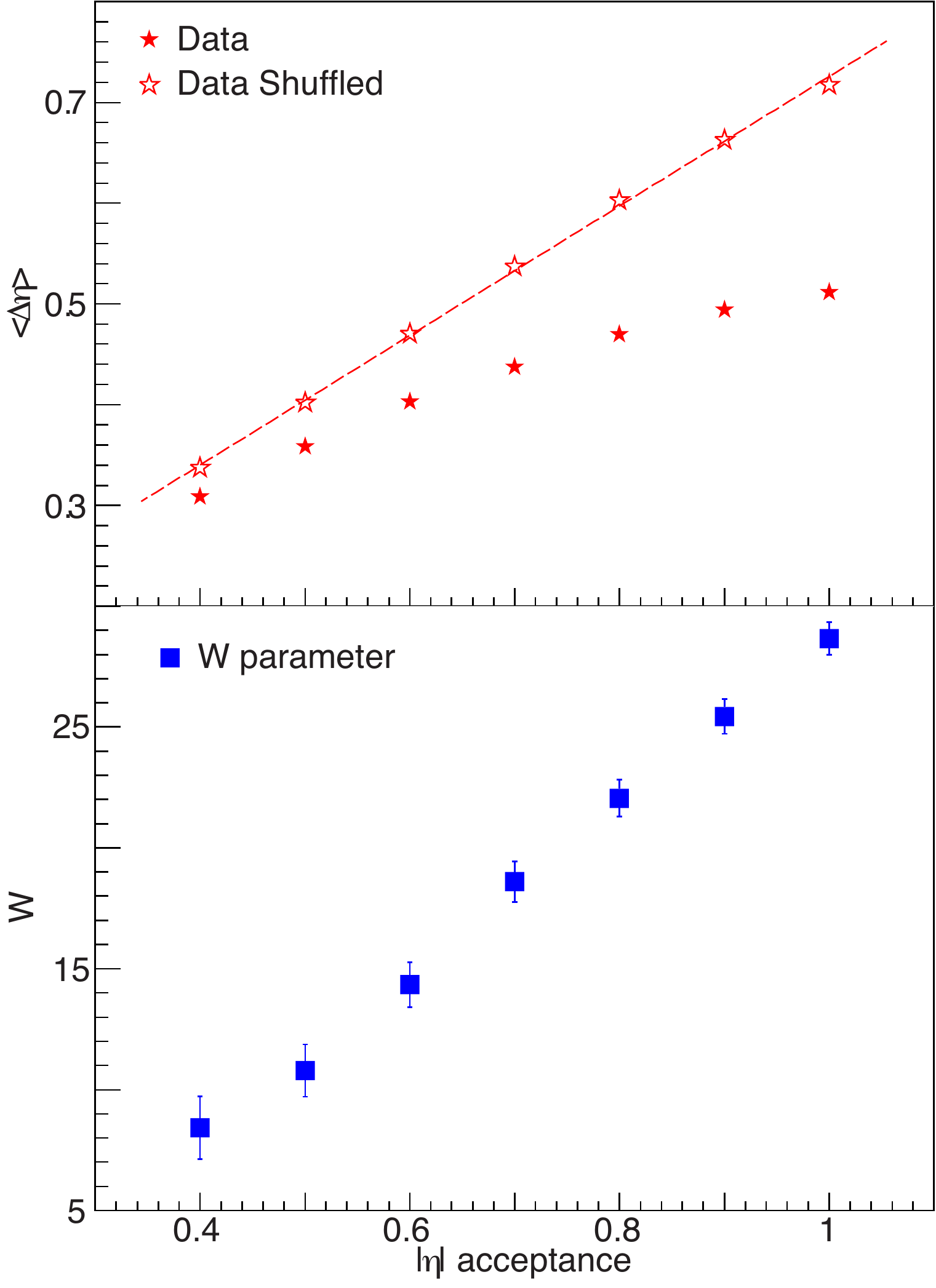}
\caption{\label{fig:fig6_2_03} Upper panel: acceptance dependence of the balance function width $\left\langle {\Delta \eta } \right\rangle$ for both data and shuffled events at $\sqrt{s_{\rm NN}}$ = 200 GeV. Lower panel: Acceptance dependence of the normalized $W$ parameter at $\sqrt{s_{\rm NN}}$ = 200 GeV.}
\end{figure}

The $W$ parameter represents the deviation of the data from shuffled events.  Thus a narrower measured balance function means a stronger deviation from shuffled events, which will give a larger $W$ value. Although the $W$ parameter was proposed to be insensitive to acceptance effects, we find a strong dependence of $W$ on the acceptance. Figure~\ref{fig:fig6_2_03} shows the acceptance dependence of the $W$ parameter at $\sqrt{s_{\rm NN}}$ = 200 GeV. To study the effect, we reduced STAR's pseudorapidity acceptance from $|\eta| < 1.0$ to $|\eta| < 0.4$. The upper panel shows the acceptance dependence $\left\langle {\Delta \eta } \right\rangle$ of the data and shuffled events. The shuffled events have a relatively linear dependence of $\left\langle {\Delta \eta } \right\rangle$ on $|\eta|$. The fit for shuffled events has a $\chi^2$ of 16.9 for 5 degrees of freedom, while the $\left\langle {\Delta \eta } \right\rangle$ from data shows clear deviation from this linear correlation. The lower panel of  Figure~\ref{fig:fig6_2_03}  shows the $W$ parameter calculated used the $\left\langle {\Delta \eta } \right\rangle$ from data and shuffled events at each acceptance. Clearly $W$ decreases with decreasing acceptance.

\begin{figure}
\centering
\includegraphics[width=30pc]{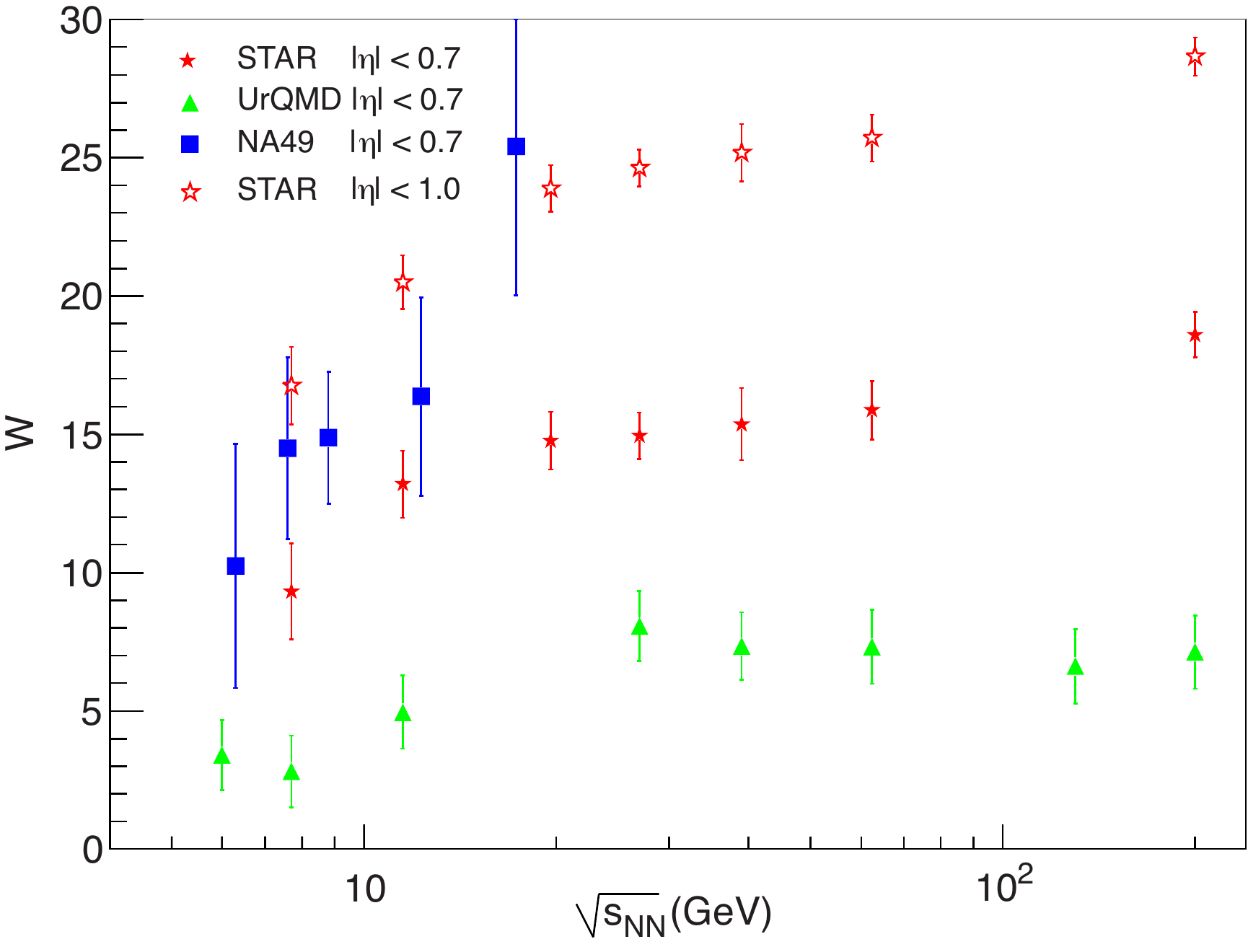}
\caption{\label{fig:fig6_2_04}Energy dependence of the normalized parameter $W$ for most central Au+Au collisions (0-5\%) from SPS energy to RHIC energy. Also shown are UrQMD results.}
\end{figure}

To study the energy dependence of the $W$ parameter, Figure~\ref{fig:fig6_2_04} shows the $W$ parameter with a reduced acceptance. Since NA49 has a pesudorapidity acceptance of approximately 1.4 units~\cite{NA49_PRC_2007}, we reduce the acceptance of STAR data and UrQMD calculations to $|\eta| < 0.7$. The STAR data show a smooth increase with increasing energy, which is anticipated from Figure~\ref{fig:fig6_2_02}. The NA49 results agree with STAR data, except for the NA49 highest energy. The UrQMD calculations predict the increasing trend but under predict the magnitude. Since a large $W$ means a narrower balance function, the results from the normalized $W$ parameter are consistent with weighted average results from~\ref{fig:fig6_2_02}. To illustrate the magnitude of the acceptance effect, STAR data with full $|\eta| < 1.0$ acceptance are also shown in same figure (open stars). Similar to Figure~\ref{fig:fig6_2_03}, a larger $|\eta|$ acceptance leads to a larger $W$ value.  Overall, within the same acceptance, the $W$ parameter shows  a smooth increase from lowest SPS energy to top RHIC energy.

\section{Reaction-Plane-Dependent Balance Functions}

\begin{figure}
\centering
\includegraphics[width=40pc]{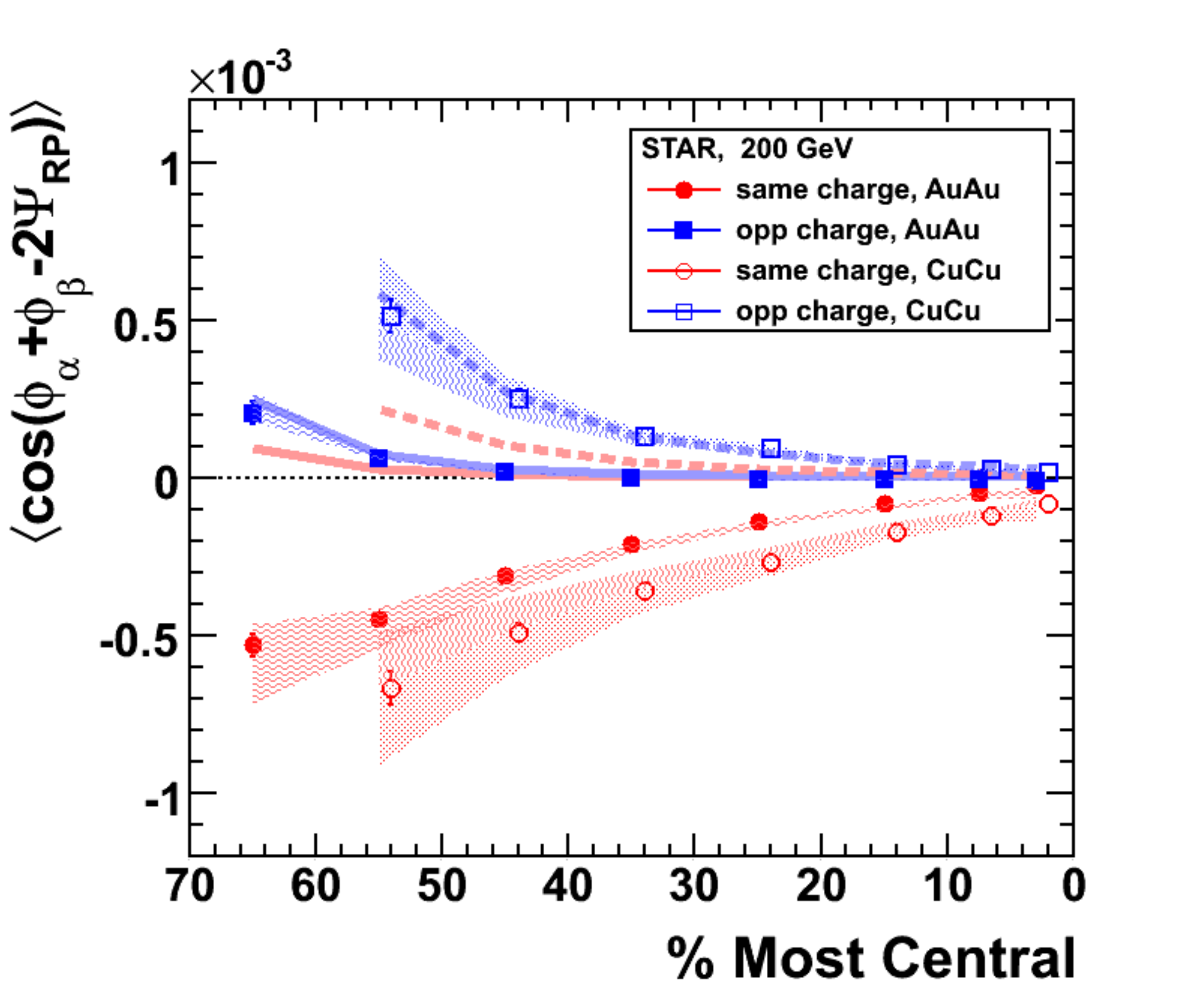}
\caption{\label{fig:fig6_3_00}Centrality dependence of the three point correlator in Au+Au and Cu+Cu collisions at $\sqrt{s_{\rm NN}}$ = 200 GeV. Shaded bands represent uncertainty from the measurement of $v_2$. The figure is from Ref.~\cite{parity_PRL}}

\end{figure}

Recently, it has been proposed that the hot and dense matter created in heavy ion collisions may form metastable domains where parity is locally violated.  This possible local parity violation \cite{LPV} coupled with the strong magnetic field produced by passing charged nuclei in such a collision could cause a charge separation across the reaction plane in non-central collisions called the chiral magnetic effect (CME) \cite{CME_1,CME_2,CME_3}. One observable proposed to measure the CME is the three point correlator  \cite{3_point_correlator}

\begin{eqnarray}
\gamma_{\alpha, \beta}=\left\langle\cos (\phi _\alpha  + \phi _\beta  - 2\psi _{RP} )\right\rangle.
\end{eqnarray}

Current theoretical understanding suggest that the Chiral Magnetic Effect will cause a charge separation across the reaction plane, thus a negative $\gamma$ for same-sign correlations and positive $\gamma$ for opposite-sign correlations. It is also expected that the opposite-sign correlations might be suppressed since it need to go through the hot and dense medium. Indeed, figure~\ref{fig:fig6_3_00} shows centrality dependence of the three point correlator in Au+Au and Cu+Cu collisions at $\sqrt{s_{\rm NN}}$ = 200 GeV~\cite{parity_PRL}. For the Au+Au collisions, same charge correlations are clearly positive and opposite charge correlations are negative as expected. The magnitude of opposite charge correlations are smaller compare to same-sign correlation, which agrees with the possible suppression of back-to-back charge correlations.

On the other hand, the balance function, which measures the correlation between the opposite-sign charge pairs, is sensitive to the mechanisms of charge formation and the subsequent relative diffusion of the balancing charges \cite{balance_PRL}. The reaction-plane-dependent balance function can be written as

\begin{eqnarray}
B(\phi ,\Delta \phi ) = \frac{1}{2}\{ \frac{{\Delta _{ +  - } (\phi ,\Delta \phi ) - \Delta _{ +  + } (\phi ,\Delta \phi )}}{{N_ +  (\phi )}} + \frac{{\Delta _{ -  + } (\phi ,\Delta \phi ) - \Delta _{ -  - } (\phi ,\Delta \phi )}}{{N_ -  (\phi )}}\} .
\end{eqnarray}

\noindent
Here  $N_{+(-)}(\phi)$ is the total number of positive(negative) particles that have an azimuthal angle $\phi$ with respect to the event plane and $\Delta_{+-}(\phi,\Delta \phi)$ represents the total number of pairs summed over all events where the first (positive) particle has an azimuthal angle $\phi$ with respect to event plane and the second (negative) particle has a relative azimuthal angle $\Delta\phi$ with respect to the first particle. Similarly we can express $\Delta_{++}(\phi,\Delta \phi)$, $\Delta_{-+}(\phi,\Delta \phi)$ and $\Delta_{--}(\phi,\Delta \phi)$.

The data used in this analysis are from Au+Au collisions at $\sqrt{s_{\rm NN}}$ = 200 , 62.4, 39, 11.5, and 7.7 GeV measured using the STAR detector.  A transverse momentum cut of $0.2 < p_{t} < 2.0$ GeV/$c$ was applied as well as a pseudorapidity cut of $|\eta| < 1.0$. The second order event plane from TPC is used here and electrons are suppressed using the specific energy loss inside the TPC.

\begin{figure}
\centering
\includegraphics[width=30pc]{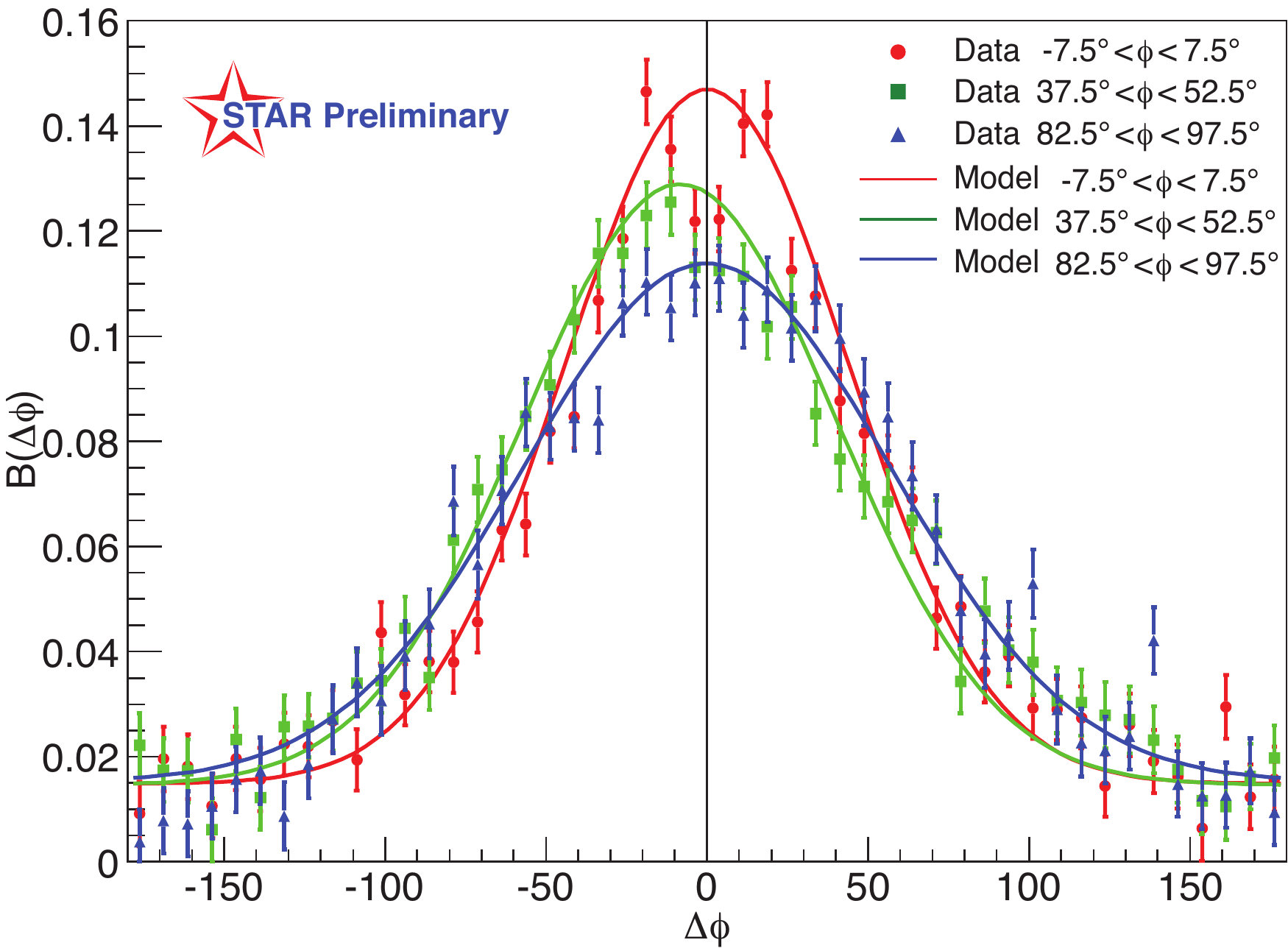}
\caption{\label{fig:fig6_3_01}The balance function for $\phi = 0^{\circ}$ (in-plane), $\phi = 45^{\circ}$, and $\phi = 90^{\circ}$ (out-of-plane) particles pairs (not corrected for event plane resolution). The 40-50\% centrality bin is shown.}

\end{figure}

Figure~\ref{fig:fig6_3_01} shows $\phi = 0^{\circ}$ (in-plane), $\phi = 45^{\circ}$, and $\phi = 90^{\circ}$ (out-of-plane) balance function for 40-50\% centrality only. The in-plane balance function is narrower than the out-of-plane balance function, which is caused by the strong collective flow in-plane. The $\phi = 45^{\circ}$ balance function is asymmetric and peaked at negative $\Delta \phi$ because charge pairs are more correlated on the in-plane side due to strong elliptic flow.  Also shown are the blast-wave model calculations from Ref. \cite{parity_soeren}.

\begin{figure}
\centering
\includegraphics[width=36pc]{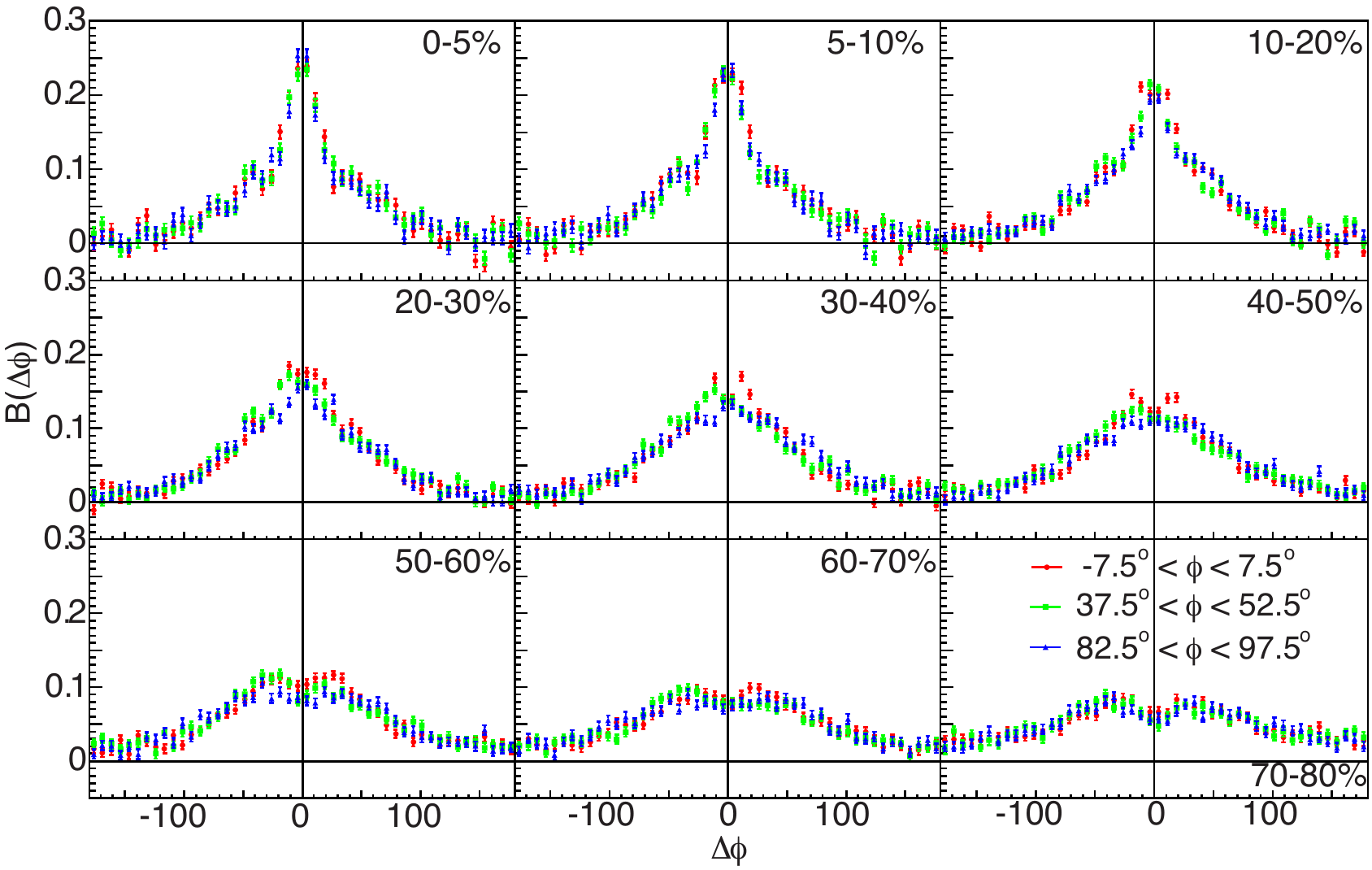}
\caption{\label{fig:fig6_3_01_2}The balance function for $\phi = 0^{\circ}$ (in-plane), $\phi = 45^{\circ}$, and $\phi = 90^{\circ}$ (out-of-plane) particles pairs (not corrected for event plane resolution). Nine centrality bins are shown.}

\end{figure}

A similar trend is also observed for other centralities. Figure~\ref{fig:fig6_3_01_2} shows the same analysis for nine centralities from most central (0-5\%) to most peripheral (70\%-80\%) collisions. All three cases show peak structures in central collisions, which is due to strong collective flow. The asymmetry of the $\phi = 45^{\circ}$ balance function is most significant at mid-central collisions, where the elliptic flow is strongest.

To quantify the effect of collective flow on the balance function, we present the weighted average cosine, $c_b(\phi)$, and weighted average sine, $s_b(\phi)$, extracted from the balance functions.

\begin{eqnarray}
\nonumber
c_b (\phi ) = \frac{1}{{z_b (\phi )}}\int {d\Delta \phi } B(\phi ,\Delta \phi )\cos (\Delta \phi ),\\
\nonumber
s_b (\phi ) = \frac{1}{{z_b (\phi )}}\int {d\Delta \phi } B(\phi ,\Delta \phi )\sin (\Delta \phi ), \\
z_b (\phi ) = \int {d\Delta \phi B(\phi ,\Delta \phi )}.
\end{eqnarray}

\begin{figure}
\centering
\includegraphics[width=30pc]{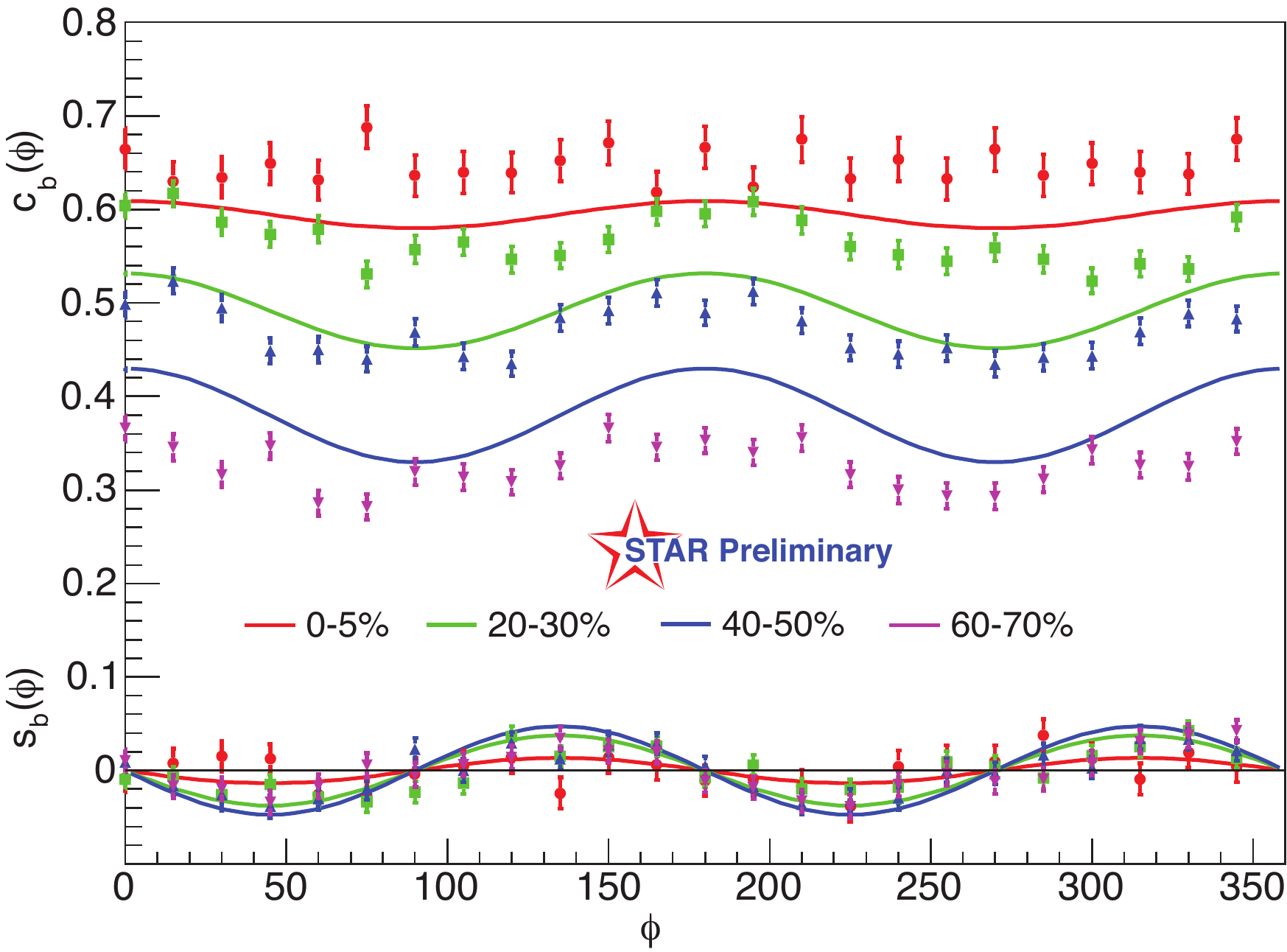}
\caption{\label{fig:fig6_3_02}The weighted average cosine and sine of balance function. Four centralities are shown here. The points are from the experimental data(not corrected for event plane resolution), while solid lines are from the blast-wave model of Ref. \cite{parity_soeren}.}

\end{figure}

$c_b(\phi)$ represents the width of balance function. If charges are created at the same point and do not diffuse due to strong collective flow, $c_b(\phi)$ would be close to unity.  $s_b(\phi)$ is an odd function of $\Delta\phi$, so it quantifies the asymmetry of the balance function. Figure~\ref{fig:fig6_3_02} shows $c_b(\phi)$ and $s_b(\phi)$ for Au+Au  collisions at  $\sqrt{s_{\rm NN}}$ = 200 GeV.   $c_b(\phi)$ is closer to unity in the 0-5\% centrality bin, which is due to a stronger collective flow in central collisions, while in mid-peripheral and peripheral collisions,  $c_b(\phi)$ shows a difference between the in-plane and out-of-plane balance functions, which is caused by stronger elliptic flow in the event plane.  $s_b(\phi)$ reaches a maximum at $\phi {\rm{  =  }}135^{\circ},315^{\circ}$ and a minimum at $\phi {\rm{  =  }}45^{\circ},225^{\circ}$, which demonstrates that charged pairs are more likely to be emitted in-plane.

Figure~\ref{fig:fig6_3_02} also shows a comparison with the blast-wave model of Ref. \cite{parity_soeren}. The blast-wave model includes a breakup temperature $T_{\rm kin}$, the maximum collective velocities in the in-plane and out-of-plane directions, the spatial anisotropy of the elliptic shape by fitting STAR published $v_{2}$ and spectra \cite{STAR_v2}.  This model also assumes local charge conservation and initial separation of balancing charges at freeze-out by fitting experimental results \cite{balance_PRC}. The difference between data and the blast-wave model predictions could be due to the finite event plane resolution for the data.

The difference between the same-sign and opposite-sign three point correlator $\gamma_{\alpha \beta}$ can be expressed as \cite{parity_soeren} 
\begin{eqnarray}
\gamma _p  = \frac{1}{2}(2\gamma _{ +  - }  - \gamma _{ +  + }  - \gamma _{ -  - } ) = \frac{2}{M}[v_2  < c_b (\phi ) >  + v_{2c}  - v_{2s} ],
\end{eqnarray}
\vspace{1pc}

where
\[
\begin{array}{l}
 v_{2c}  = < c_b (\phi )\cos (2\phi ) >  - v_2  < c_b (\phi ) >,  \\ 
 v_{2s}  =  < s_b (\phi )\sin (2\phi ) >,  \\ 
 \end{array}
\]

\begin{figure}
\centering
\includegraphics[width=32pc]{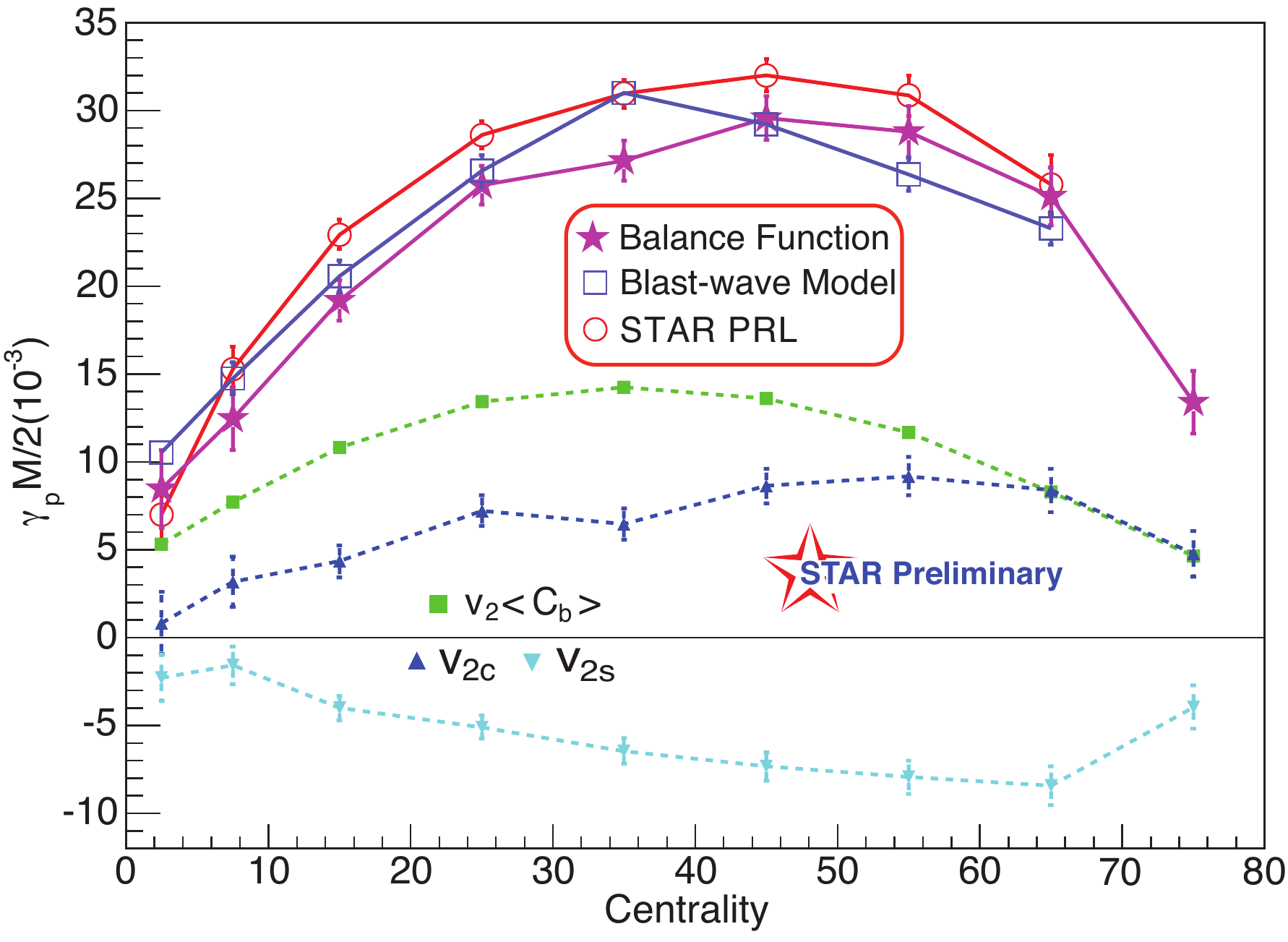}
\caption{\label{fig:fig6_3_03}Parity observable $\gamma_P$ scaled by experimental multiplicity. }

\end{figure}

and the bracket represents
$$ 
< f(\phi ) >  = \frac{1}{M}\int {d\phi \frac{{dM}}{{d\phi }}z_b (\phi )f(\phi )} . 
$$

In this equation, $v_2 \left\langle {c_b (\phi )} \right\rangle $ will be positive if there are more charge pairs in-plane than out-of-plane.  $v_{2c}$ will be positive if the charge pairs are more correlated in-plane than out-of-plane, while $v_{2s}$ will be negative if the charge pairs are more correlated on the in-plane side.

Figure~\ref{fig:fig6_3_03} shows the parity observable calculated from balance functions as well as its three components. All data points are corrected for the event-plane resolution here. To compare with previous results, we also plot the $\gamma_P$  from STAR published data (figure~\ref{fig:fig6_3_00})~\cite{parity_PRL} scaled by the measured uncorrected multiplicity in the same plot. Mathematically, the balance function result should equal the one from $\gamma_P$ and they do agree well. Thus a thermal blast-wave model \cite{parity_soeren} incorporating local charge conservation and flow reproduces most of the signal.

\begin{figure}
\centering
\includegraphics[width=38pc]{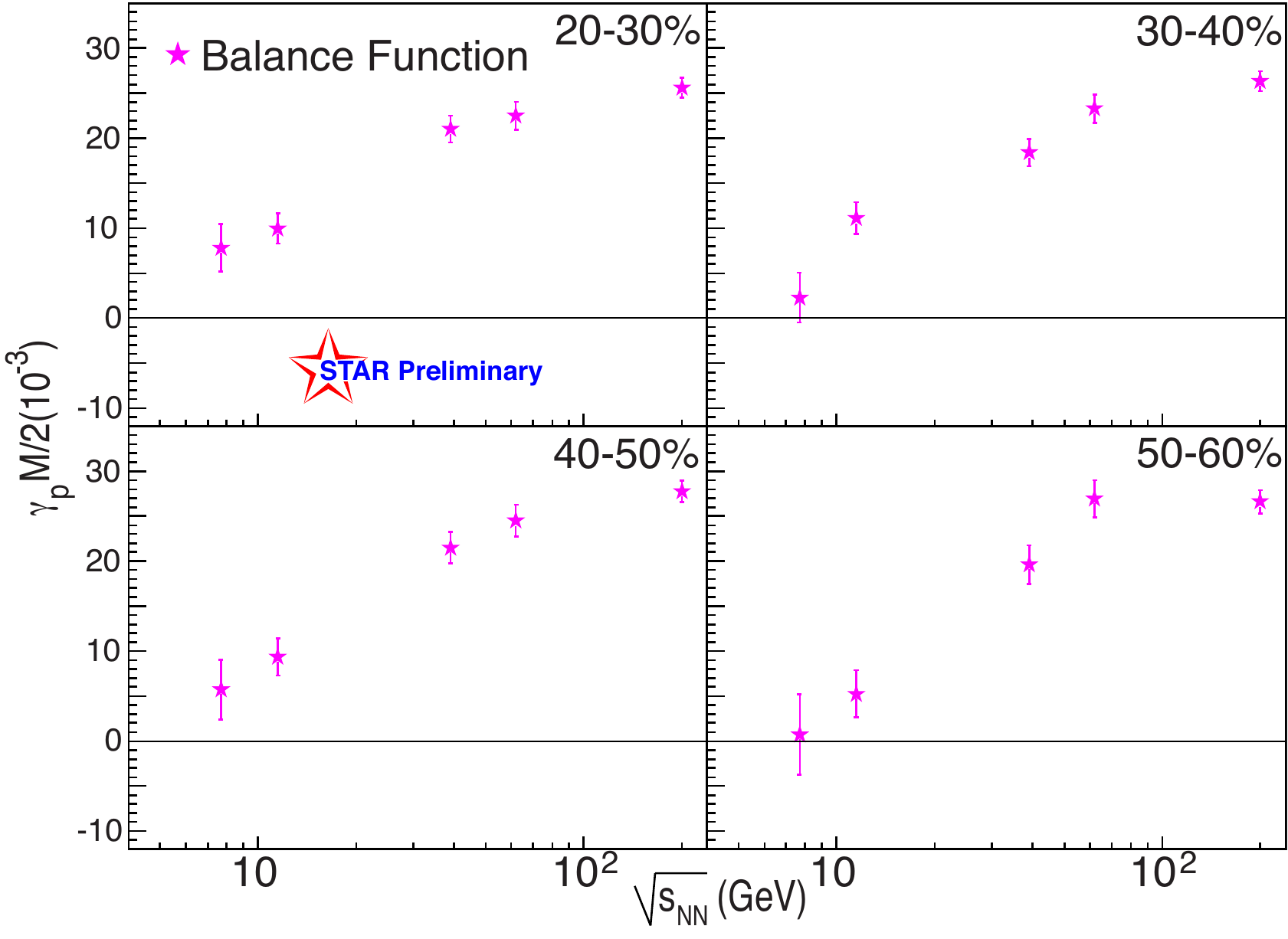}
\caption{\label{fig:fig6_3_04}The beam energy dependence of balance function.  Four centralities are shown here. }

\end{figure}

Another topic of interest is the beam energy dependence of the CME.  Recent calculations show it only exists in the deconfined, chirally symmetric phase \cite{CME_3} and decreases with increasing beam energy \cite{CME_BES}. In addition, the CME effect should disappear at energies below $\sqrt{s_{\rm NN}}$ = 10 GeV.  Figure~\ref{fig:fig6_3_04} shows the same parity observable calculated from balance functions at $\sqrt{s_{\rm NN}}$ = 200 , 62.4, 39, 11.5, and 7.7 GeV. We can see that, for all four centralities shown here, the data show a smooth decrease with decreasing collision energy. This smooth decrease differs from the CME calculation, which predicts an increasing signal with decreasing beam energy and sharply disappearance of signal near the top energy of SPS\cite{CME_BES}.  However, these results are consistent with the fact that $v_{2}$ decreases smoothly with decreasing beam energy in the same energy range.  

\section{Balance Functions for Identified Kaons and Protons with RHIC Run 10 Data}

In Section~\ref{Run7_balance}, we discussed the balance function for identified kaons using STAR's Run 7 data. Here we continue the discussion of balance functions for identified kaons and protons with RHIC Run 10 data. There are two major benefits of this new 200 GeV data set. First, the TOF detector, which was not available during Run 7, was fully installed and operational during Run 10. The usage of the TOF detector greatly enhanced STAR's particle identification ability. The identified kaons' momentum range was extended from $0.2 < p_{t} < 0.6$ GeV/$c$ (TPC only PID) to $0.2 < p_{t}$, $p < 1.6$ GeV/$c$ (TPC+TOF PID), while the identified protons' momentum range was extended from $0.4 < p_{t} < 1.0$ GeV/$c$ (TPC only PID) to $0.4 < p_{t}$, $p < 3.0$ GeV/$c$ (TPC+TOF PID). Second, due to new faster readout electronics, more events were recorded during Run 10. In Section~\ref{Run7_balance}, about 28 million events were analyzed, while here we report our results using 240 million Au+Au minimum bias data recorded during the RHIC Run 10 period.  

\subsection{Balance Functions in Terms of $\Delta y$}

\begin{figure}
\centering
\includegraphics[width=36pc]{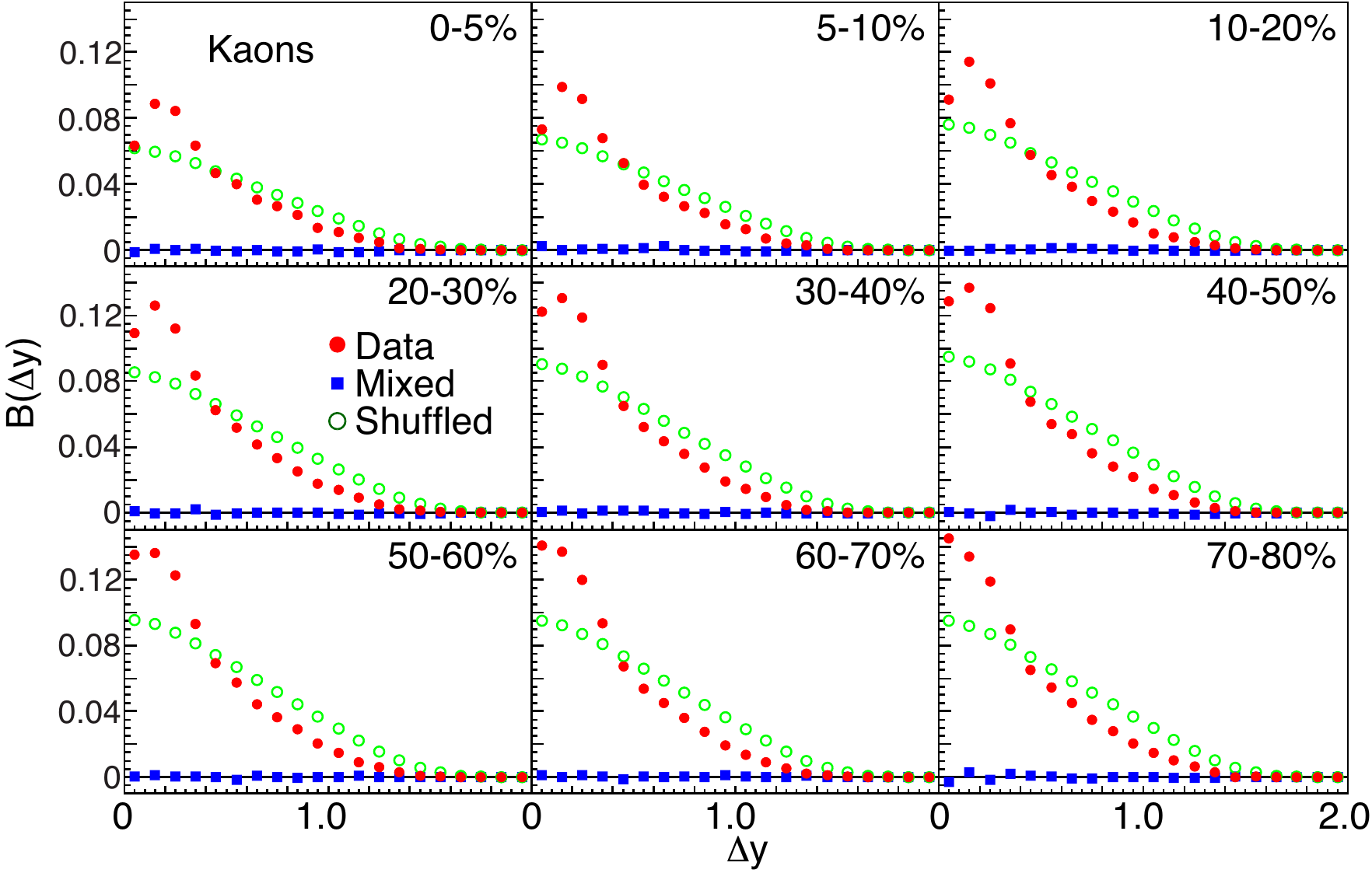}
\caption{\label{fig:fig6_4_01}The balance function in terms of $\Delta y$ for identified charged kaon pairs from STAR Run 10 Au+Au collisions at $\sqrt{s_{\rm NN}}$ = 200 GeV for nine centrality bins.}

\end{figure}

Figure~\ref{fig:fig6_4_01} shows the balance functions for identified charged kaons pairs, for Au+Au collisions at $\sqrt{s_{\rm NN}}$ =  200 GeV for nine centrality bins as a function of the relative rapidity. The balance function calculated from mixed events is zero for all centralities. The balance functions calculated using shuffled events are substantially wider than the measured balance functions. Again, a discontinuity around $\Delta y = 0.4$ is visible in data at all centralities due to $\phi$ decay. Model predictions show that inter-pair correlations should be significant for $\Delta y < 0.2$ \cite{balance_distortions}.

Note that the magnitudes of the balance functions in Figure~\ref{fig:fig6_4_01} are much larger than the magnitudes in Figure~\ref{fig:fig6_1_03}, which is due to the fact that the kaon acceptance is much larger in Run 10 than in Run 7 because of the TOF detector. 

\begin{figure}
\centering
\includegraphics[width=38pc]{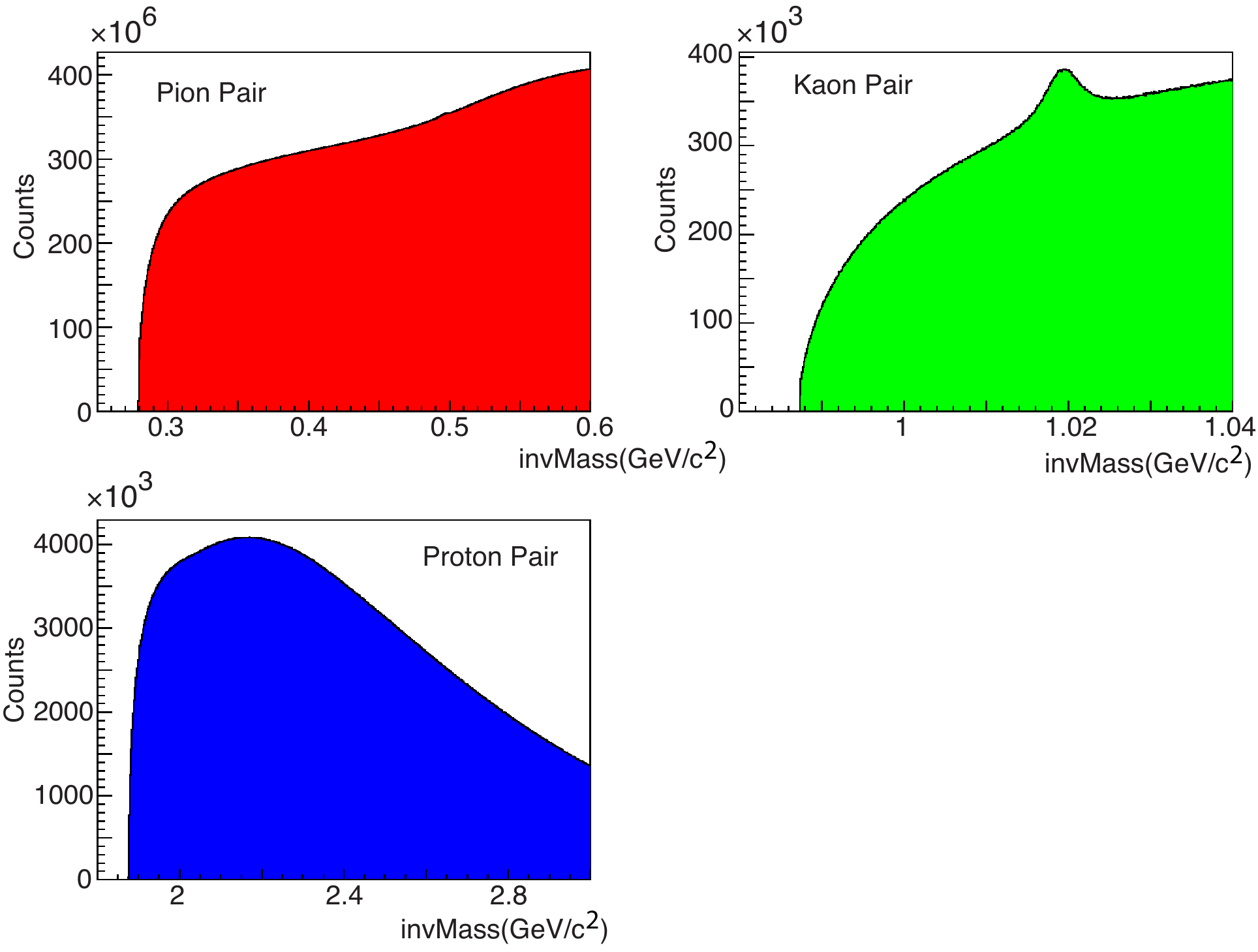}
\caption{\label{fig:fig6_4_02}The invariant mass distribution of identified pion, kaon and proton pairs from STAR Run 10 Au+Au collisions at $\sqrt{s_{\rm NN}}$ = 200 GeV.}

\end{figure}

Since the balance function is designed to measure the correlation between opposite-sign charge pairs, it is sensitive to correlations such as decays. As discussed in Section~\ref{Run7_balance}, a peak from the decay $K^{0}_{s} \rightarrow \pi^{+} + \pi^{-}$ is observed from pion balance function in terms of $q_{\rm inv}$.  For kaon pairs,  a peak from the decay $\phi \rightarrow K^{+} + K^{-}$ is also observed. To better understand the decay effect, Figure shows~\ref{fig:fig6_4_02} shows the invariant mass distributions of identified pion, kaon, and proton pairs from Run 10 Au+Au collisions. A mass peak around 0.5 GeV/$c^2$ is observed in the $\pi^+\pi^-$ invariant mass distribution, which corresponds to $K^{0}_{s}$ decay. For $K^+K^-$, a more visible peak from $\phi$ decay is also seen.  For protons, no obvious mass peak exists in the range plotted.

In order to reduce the $\phi$ decay contribution to the balance function, a suppression technique is applied to the balance function calculations. For each event, we calculate the invariant mass of every $K^+K^-$ pair, remove pairs that have an invariant mass between 1.01 and 1.03 GeV/$c^2$, then calculate the balance function with the remaining pairs. This method removes most of the $\phi$ decay daughters. However, this method also removes random pairs that fall into the invariant mass range by chance. To compensate for this, we also applied the same reduction to mixed events and use a mixed event subtraction to correct for it.

\begin{figure}
\centering
\includegraphics[width=38pc]{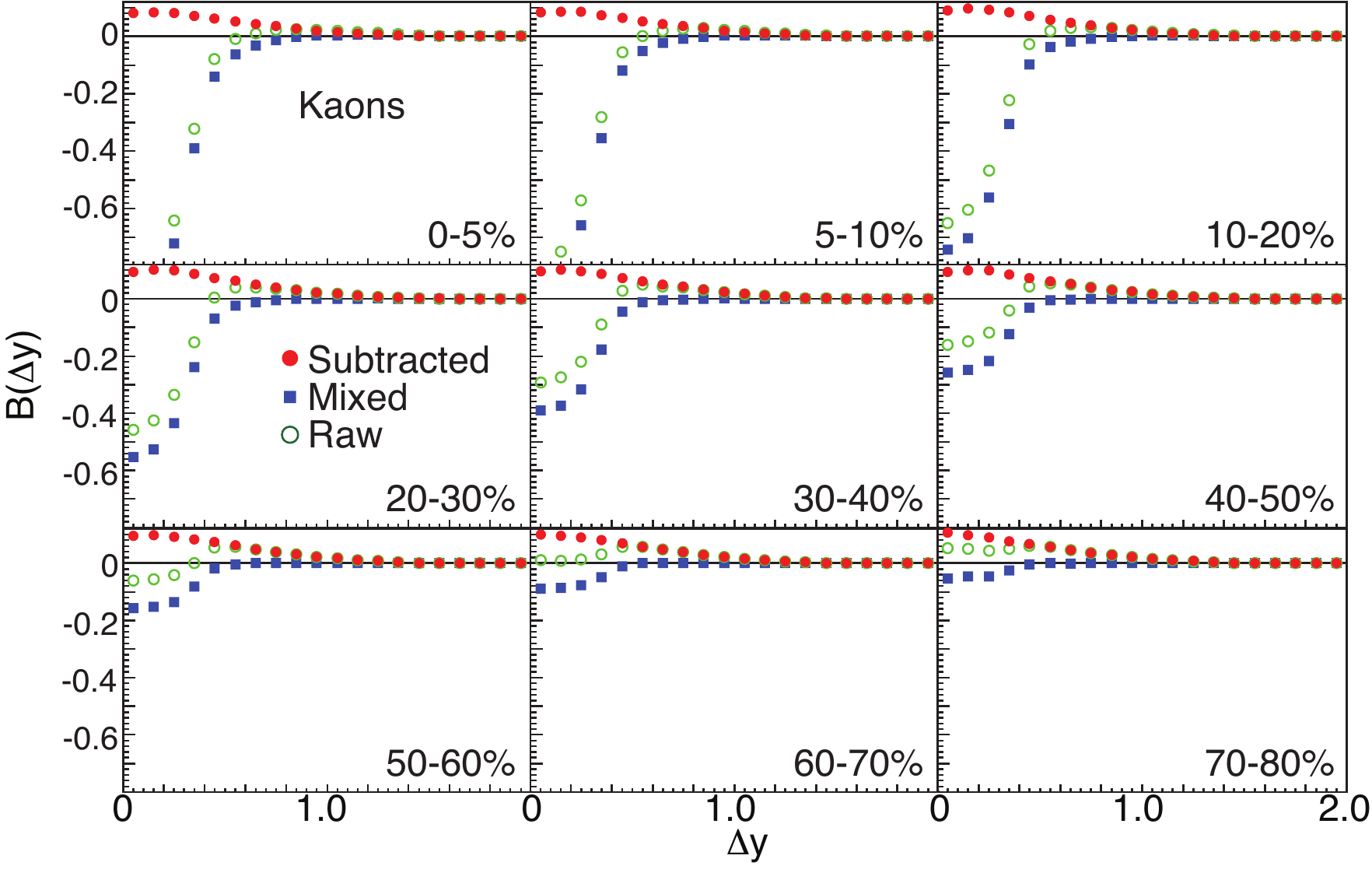}
\caption{\label{fig:fig6_4_03}The balance function in terms of $\Delta y$ for identified charged kaon pairs after removal of the $\phi$. Data are from STAR Run 10 Au+Au collisions at $\sqrt{s_{\rm NN}}$ = 200 GeV for nine centrality bins.}

\end{figure}

Figure~\ref{fig:fig6_4_03} shows the balance function in terms of $\Delta y$ for identified charged kaon pairs after removal of the $\phi$ decays. The green hollow circles represent $B(\Delta y)$ from raw data (before mixed event subtraction) while the blue squares show results from mixed events. It is clear that the removal of $\phi$ decay daughters, which removed about 21\% of kaon pairs in real data and 20\% of kaon pairs in mixed events, has significantly decreased the balance function results at small $\Delta y$. This effect is larger in central collisions where the event multiplicity is high and the probability of having $K^+K^-$ pairs coincidentally fall into the $\phi$ mass range is high. However, the final results after mixed event subtraction (red circles in Figure~\ref{fig:fig6_4_03}) are smooth and independent of these effects.

\begin{figure}
\centering
\includegraphics[width=38pc]{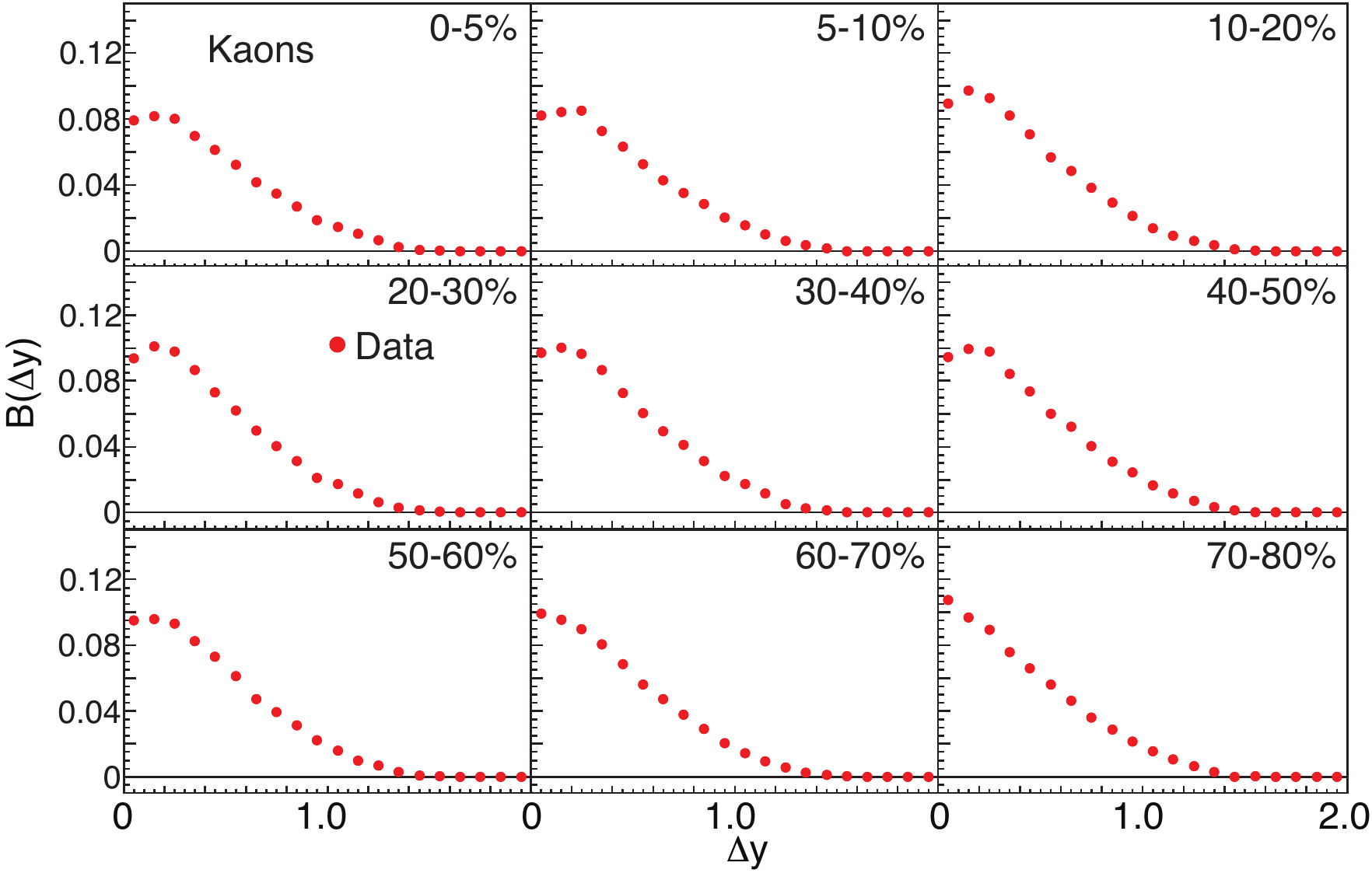}
\caption{\label{fig:fig6_4_04}The balance function in terms of $\Delta y$ for identified charged kaon pairs after removal of $\phi$. Data are from STAR Run 10 Au+Au collisions at $\sqrt{s_{\rm NN}}$ = 200 GeV for nine centrality bins. }

\end{figure}

Figure~\ref{fig:fig6_4_04} shows the same balance function results after mixed events subtraction but with a different scale. We can see that, different from Figure~\ref{fig:fig6_4_01} and Figure~\ref{fig:fig6_1_03}, the discontinuity around $\Delta y = 0.4$ which means that the $\phi$ decay has been eliminated at all centralities. HBT effects are still significant for $\Delta y < 0.2$.

\begin{figure}
\centering
\includegraphics[width=34pc]{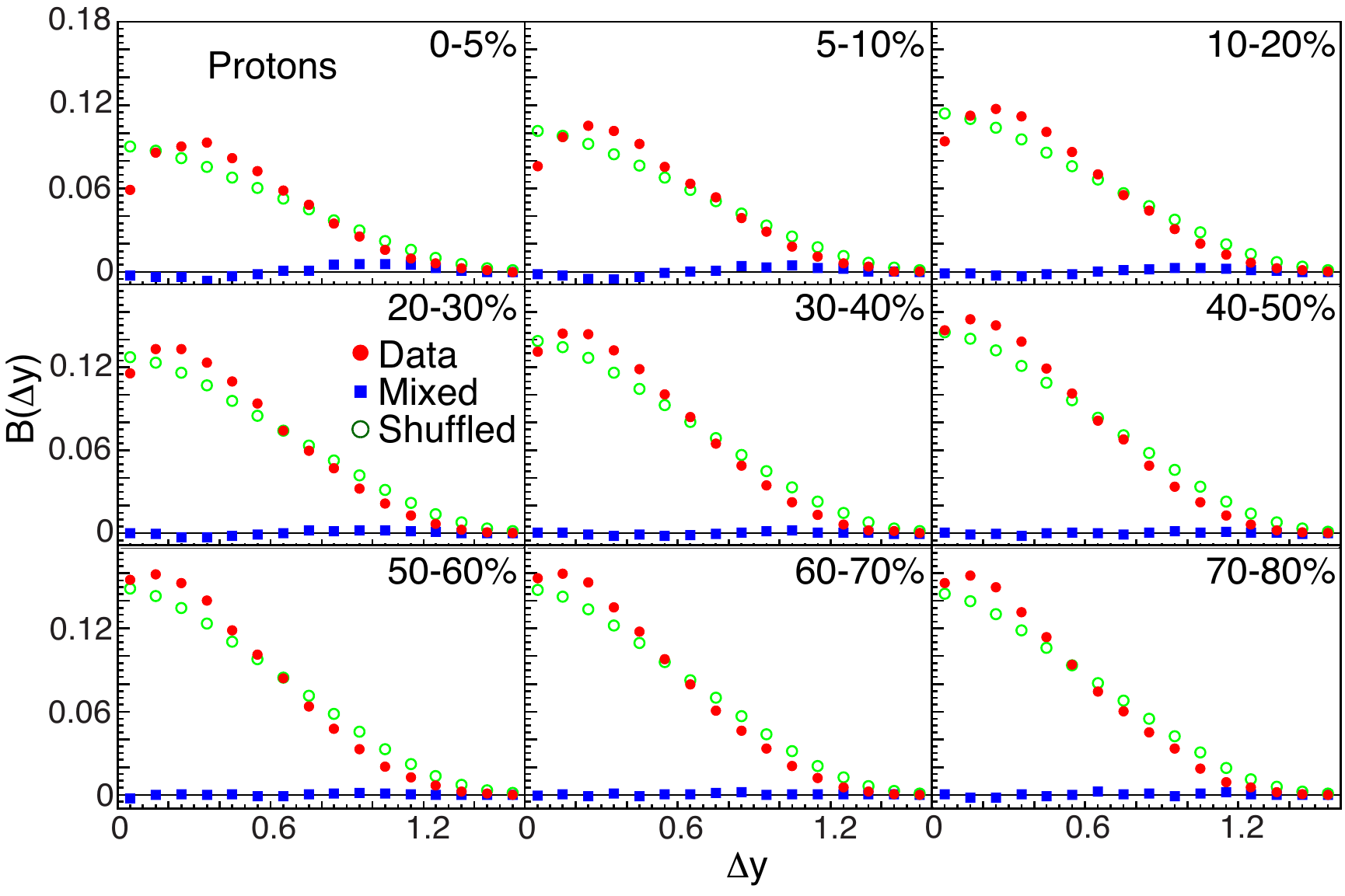}
\caption{\label{fig:fig6_4_05}The balance function in terms of $\Delta y$ for identified charged proton pairs from STAR Run 10 Au+Au collisions at $\sqrt{s_{ NN}}$ = 200 GeV for nine centrality bins. }

\end{figure}

Figure~\ref{fig:fig6_4_05} shows the balance functions for identified charged proton pairs for Au+Au collisions at $\sqrt{s_{\rm NN}}$ =  200 GeV for nine centrality bins as a function of the relative rapidity. The balance function calculated from mixed events is close to but not equal to zero for all centralities, which could due to the fact that more protons than anti-protons are produced at this energy. The balance functions calculated using shuffled events are substantially wider than the measured balance functions. A dip from inter-pair correlations is also observed at $\Delta y < 0.2$.

\subsection{Balance Functions in Terms of $q_{\rm inv}$}

\begin{figure}
\centering
\includegraphics[width=38pc]{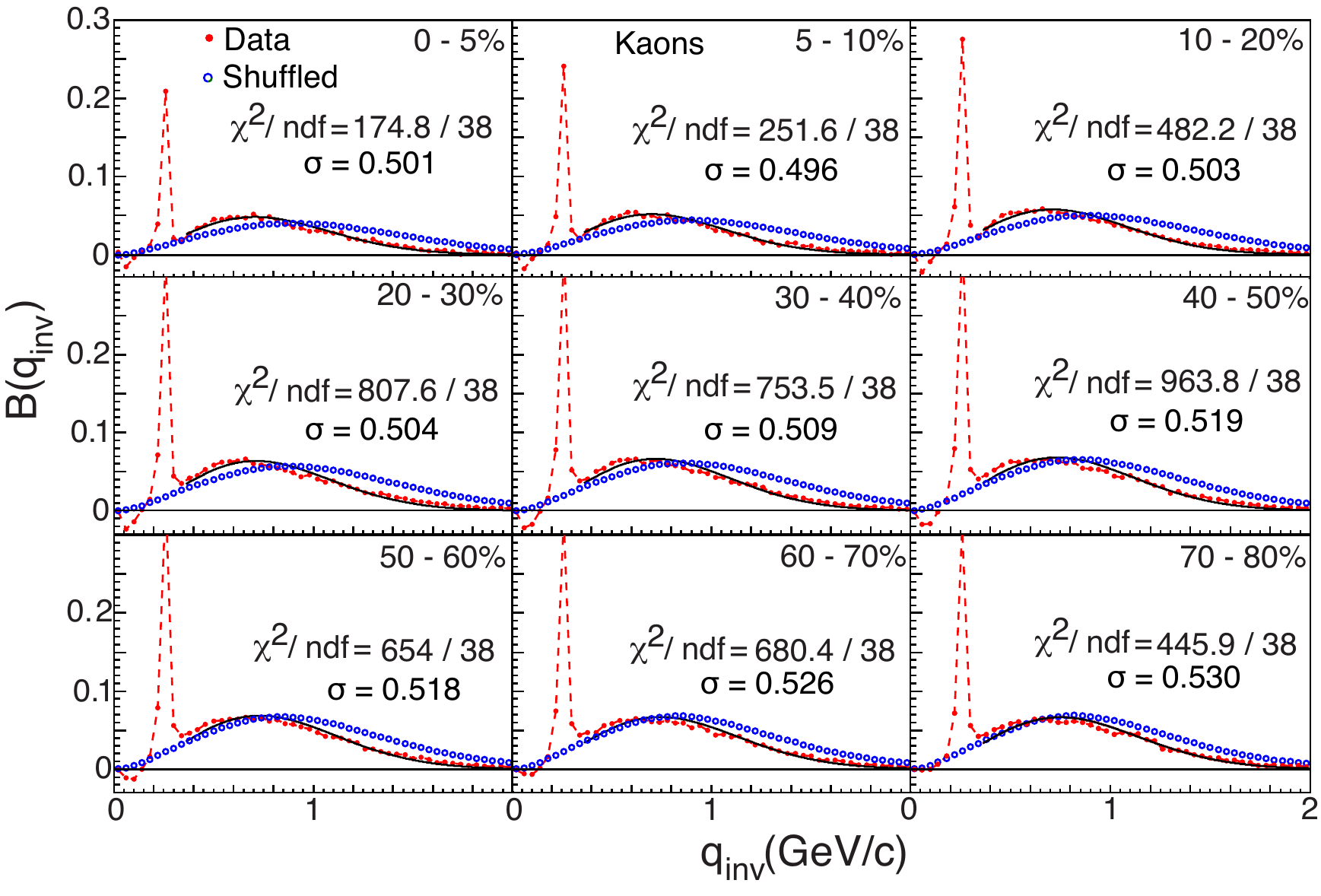}
\caption{\label{fig:fig6_4_06}The balance function in terms of $q_{\rm inv}$ for charged kaon pairs from Au+Au collisions at $\sqrt{s_{\rm NN}}$ = 200 GeV in nine centrality bins. Solid lines correspond to a thermal fit (Equation~\ref{thermal}).  Dashed lines are drawn to guide the eye.}

\end{figure}

In Section~\ref{balance_qinv}, we discussed the balance function in terms of the Lorentz invariant momentum difference ($q_{\rm inv}$) between the two particles. Unlike the balance function in terms of $\Delta y$, the balance function in terms of $q_{\rm inv}$ is much less affected by the radial flow effect and the contributions from decay feed down should be easier to identify and correct. Figure~\ref{fig:fig6_4_06} shows the balance function for identified  charged kaons in terms of $q_{\rm inv}$ for Au+Au collisions with Run 10 200 GeV data. Nine centrality bins are shown here, and all data points are corrected by mixed event subtraction. Compared to Figure~\ref{fig:fig6_1_08}, the increased event number and $p_t$ acceptance (TPC+TOF PID) significantly decreased the statistical uncertainty. The overall normalization in Run 10 is also higher than Run 7 due to large $p_t$ acceptance.  A peak corresponding to the decay of  $\phi  \to K^ +   + K^ -  $ is still observed in each centrality bin. The HBT/Coulomb effects at low $q_{\rm inv}$ is also observable, although it is relatively small compared to the results from pion pairs. The balance functions from shuffled events are clearly different than the balance functions for the data for all nine centrality bins.

To extract the width, which is possibly related to the system temperature, we fit the data using a non-relativistic thermal distribution from Equation~\ref{thermal}. The fitting range is chosen to be $ 0.36 < q_{\rm inv}  < 2.0$ (GeV/$c$) in order to remove HBT/Coulomb effects and $\phi$ decay feed down. The fits are shown in Figure~\ref{fig:fig6_4_06}. Although the fits looks reasonable, the $\chi ^2 /{\rm ndf} $ are relatively large for most cases due to small error bars in data. The width extracted from fits show a slightly narrower distribution at central collisions.

\begin{figure}
\centering
\includegraphics[width=38pc]{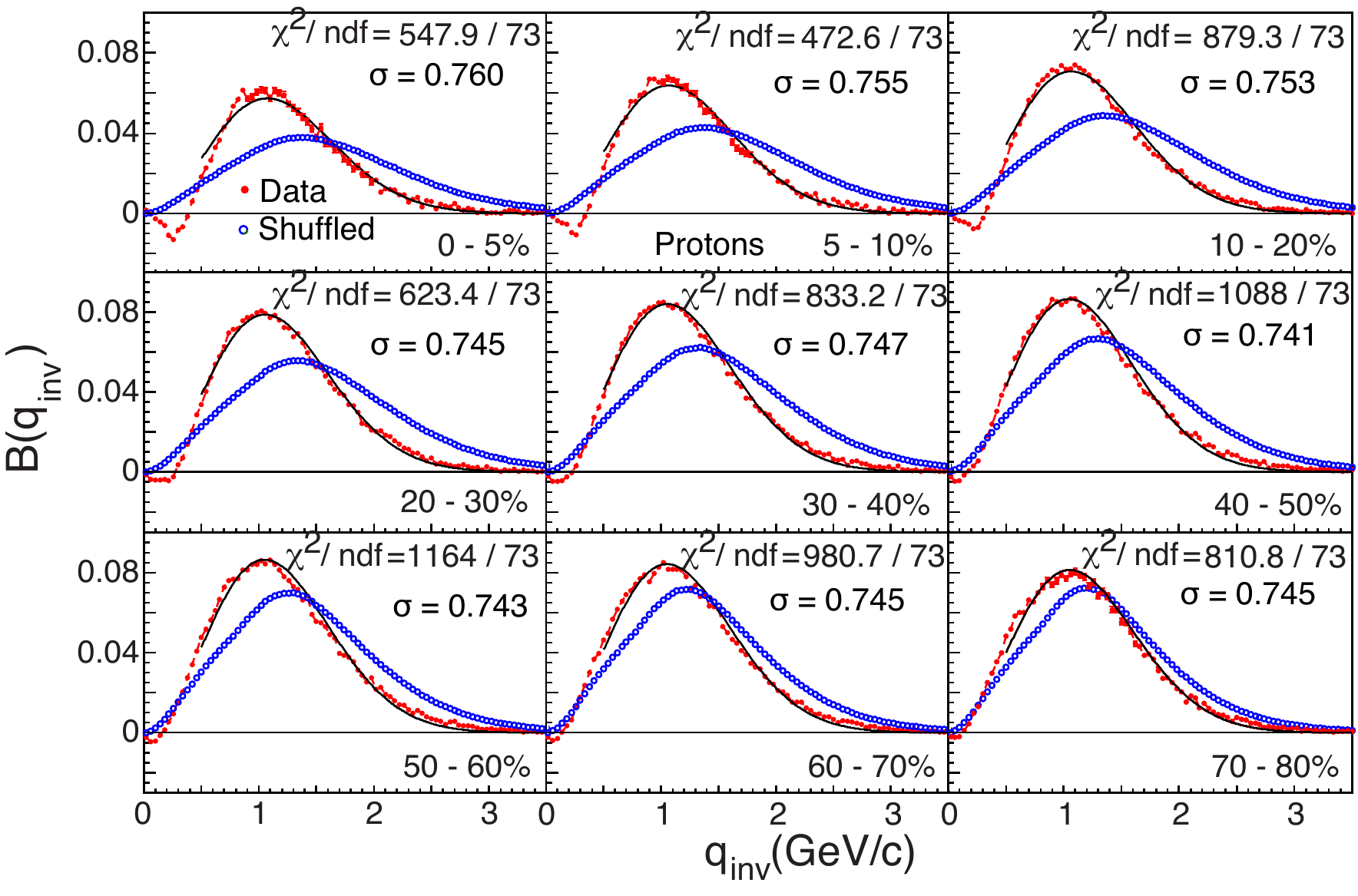}
\caption{\label{fig:fig6_4_07}The balance function in terms of $q_{\rm inv}$ for charged proton pairs from Au+Au collisions at $\sqrt{s_{\rm NN}}$ = 200 GeV in nine centrality bins. Solid lines correspond to a thermal fit (Equation~\ref{thermal}).)}

\end{figure}

Figure~\ref{fig:fig6_4_07} shows the balance functions for identified charged proton pairs in terms of $q_{\rm inv}$. Unlike pions or kaons, there are no major resonances/unstable particles decay into proton anti-proton pairs, resulting in no clear mass peak in $B(q_{\rm inv})$ distribution. The HBT/Coulomb effects is still seen at low $q_{\rm inv}$. Again, we use the non-relativistic thermal distribution from Equation~\ref{thermal} to fit data within $ 0.5 < q_{\rm inv}  < 3.5$ (GeV/$c$). The $\chi ^2 /{\rm ndf} $ is still large for all centralities but better than the fits for kaons. This might due to the fact that proton mass is larger than kaon mass, which would make this non-relativistic approximation more realistic ($m \gg T $, $T \sim 0.1$ GeV for kinetic freeze-out). The $\sigma$s from the fits indicate a wider balance function in central collisions. The shuffled events again differ significantly from the real data for all centralities. 

\subsection{Balance Function Widths}

In previous sections, we have discussed the balance function for identified kaons and protons in terms of $\Delta y$ and $q_{\rm inv}$. To quantify the centrality dependence and compare with models, here we extract the width using a weighted average from Equation~\ref{WA}. Since we have discussed that the thermal fit for $q_{\rm inv}$ has a relatively large $\chi ^2 /{\rm ndf} $, we also calculated the weighted average for $q_{\rm inv}$ to show model independent results.

\begin{figure}
\centering
\includegraphics[width=32pc]{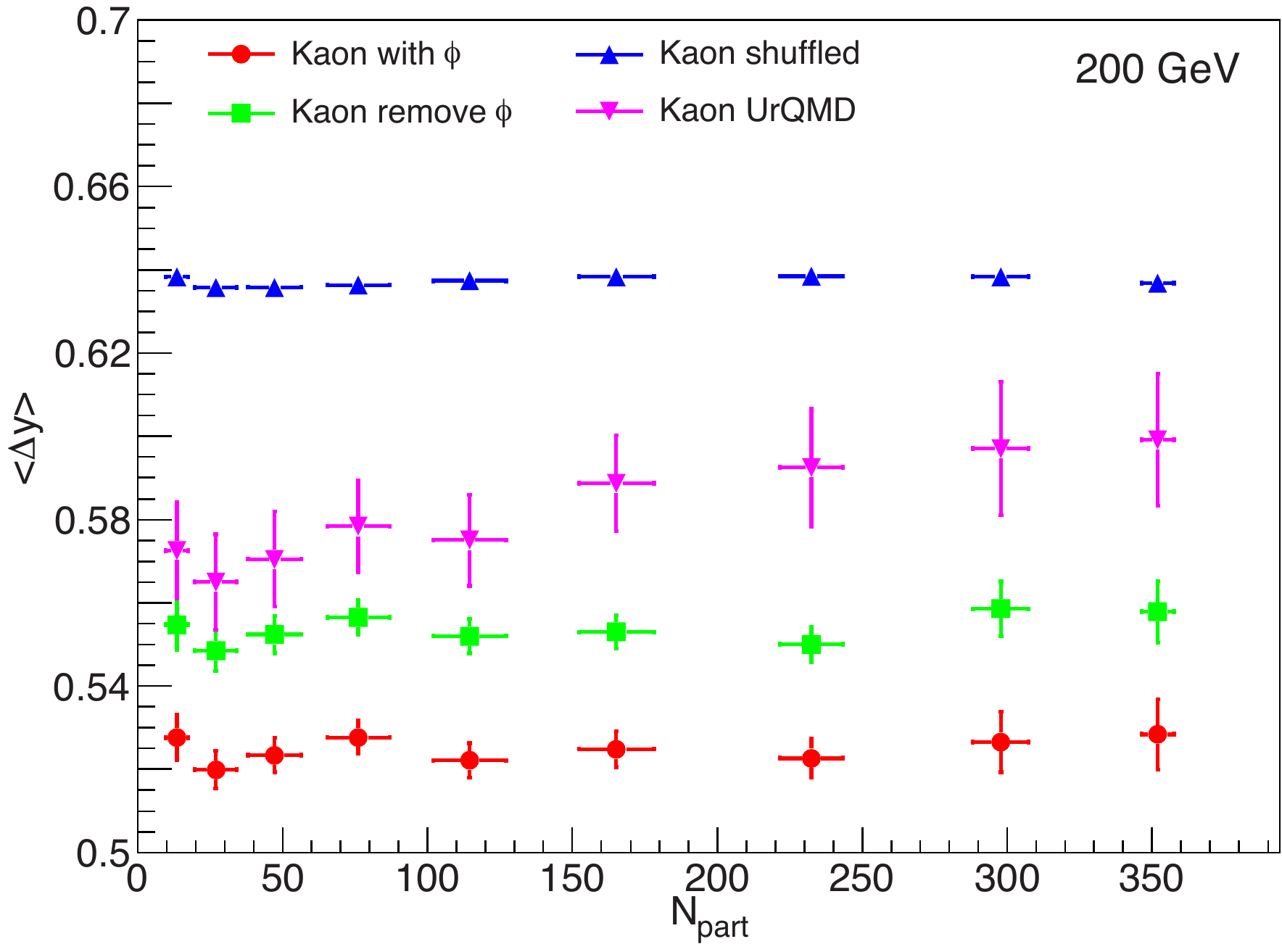}
\caption{\label{fig:fig6_4_08}Centrality dependence of the balance function widths for identified charged kaons from Au+Au collisions at $\sqrt{s_{\rm NN}}$ = 200 GeV.}

\end{figure}

Figure~\ref{fig:fig6_4_08} shows the balance function widths for identified kaons $\Delta y$ from Run 10 Au+Au collision at $\sqrt{s_{ \rm NN}}$ = 200 GeV. The results are plotted in terms of the number of participating nucleons, $N_{\rm part}$. To reduce the HBT/Coulomb effects, we calculated the width for $0.2 \le \Delta y \le 2.0$. The width from measured data without $\phi$ suppression (red circles) shows no centrality dependence. The suppression of $\phi$ decays daughters, which removed extra correlations at small $\Delta y$, increases the balance function width for all centralities (green squares). However, both cases show no centrality dependence from measured data. A UrQMD calculation incorporating the experimental acceptance is also shown in the same figure. The UrQMD widths are larger than data and increase with increasing centrality, which could be due to the hadronic scattering incorporated in UrQMD. The shuffled events widths are larger than both data and models and show no centrality dependence.

\begin{figure}
\centering
\includegraphics[width=32pc]{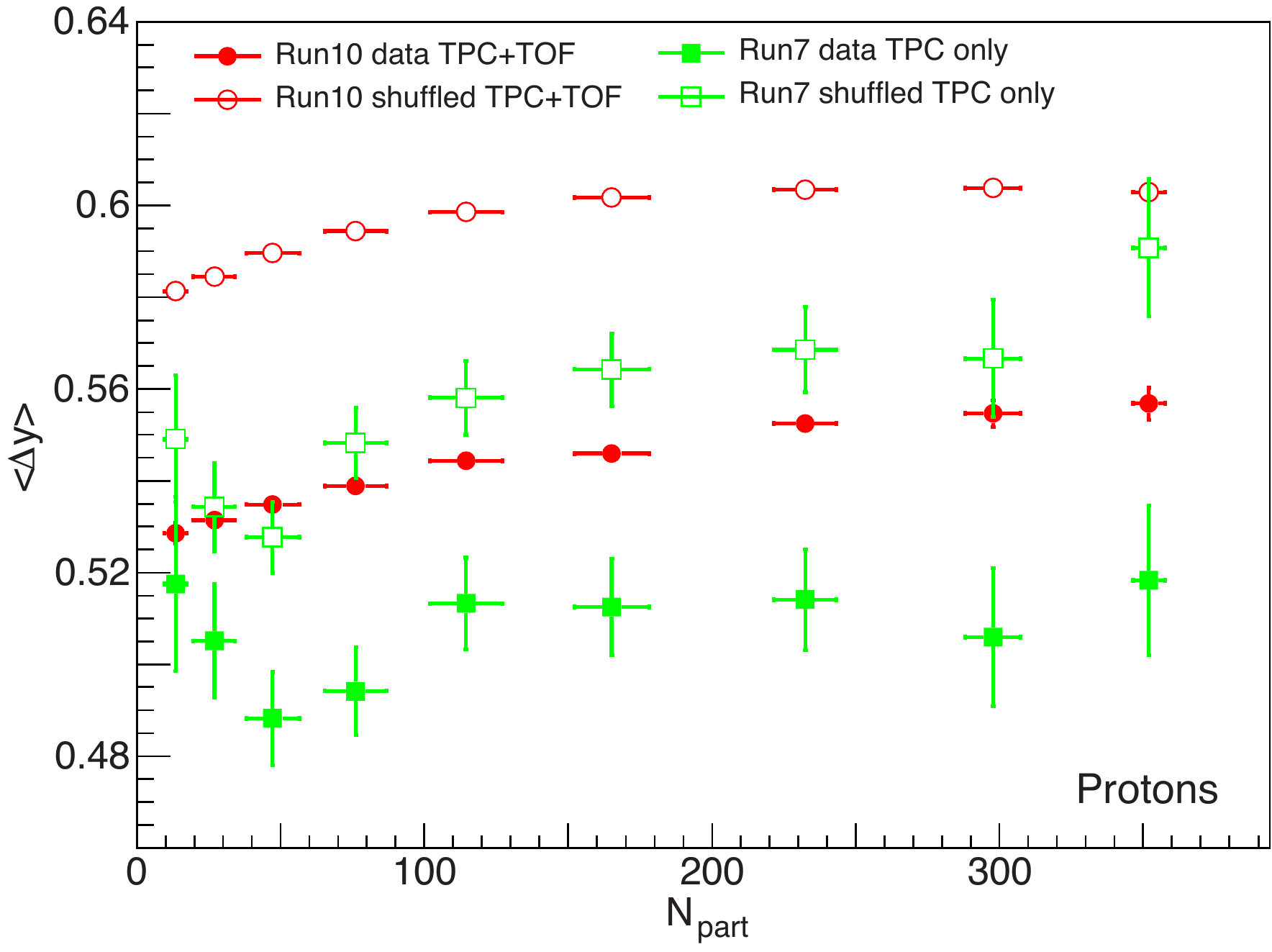}
\caption{\label{fig:fig6_4_09}Centrality dependence of the balance function widths for identified protons from Au+Au collisions at $\sqrt{s_{\rm NN}}$ = 200 GeV compared with the widths of balance functions calculated using shuffled events. }

\end{figure}

Figure~\ref{fig:fig6_4_09} shows the balance function widths for identified protons in terms of $\Delta y$ at $\sqrt{s_{\rm NN}}$ = 200 GeV from both Run 7 and Run 10 data. Since the TOF detector was not fully installed before year 2009, the Run 7 data use the TPC as the only particle identification detector and has a momentum acceptance of  $ 0.4 < p_t  < 1.0 $ GeV/$c$, while the Run 10 data used both the TPC and TOF as  particle identification detectors and has an extended momentum acceptance of $ p_t  > 0.4,p < 3.0 $ GeV/$c$. Both Run 7 data and shuffled events are lower than Run 10 results due to the different acceptance. The statistical error in Run 7 is also large due to smaller acceptance and fewer events, which makes it hard to determine the centrality dependence. However, the Run 10 data show that the balance function width for identified kaons increases in central collisions. The shuffled events are much wider but show a similar trend. The increase of the shuffled events widths is due to the changing proton momentum spectra as a function of centrality. 

\begin{figure}
\centering
\includegraphics[width=32pc]{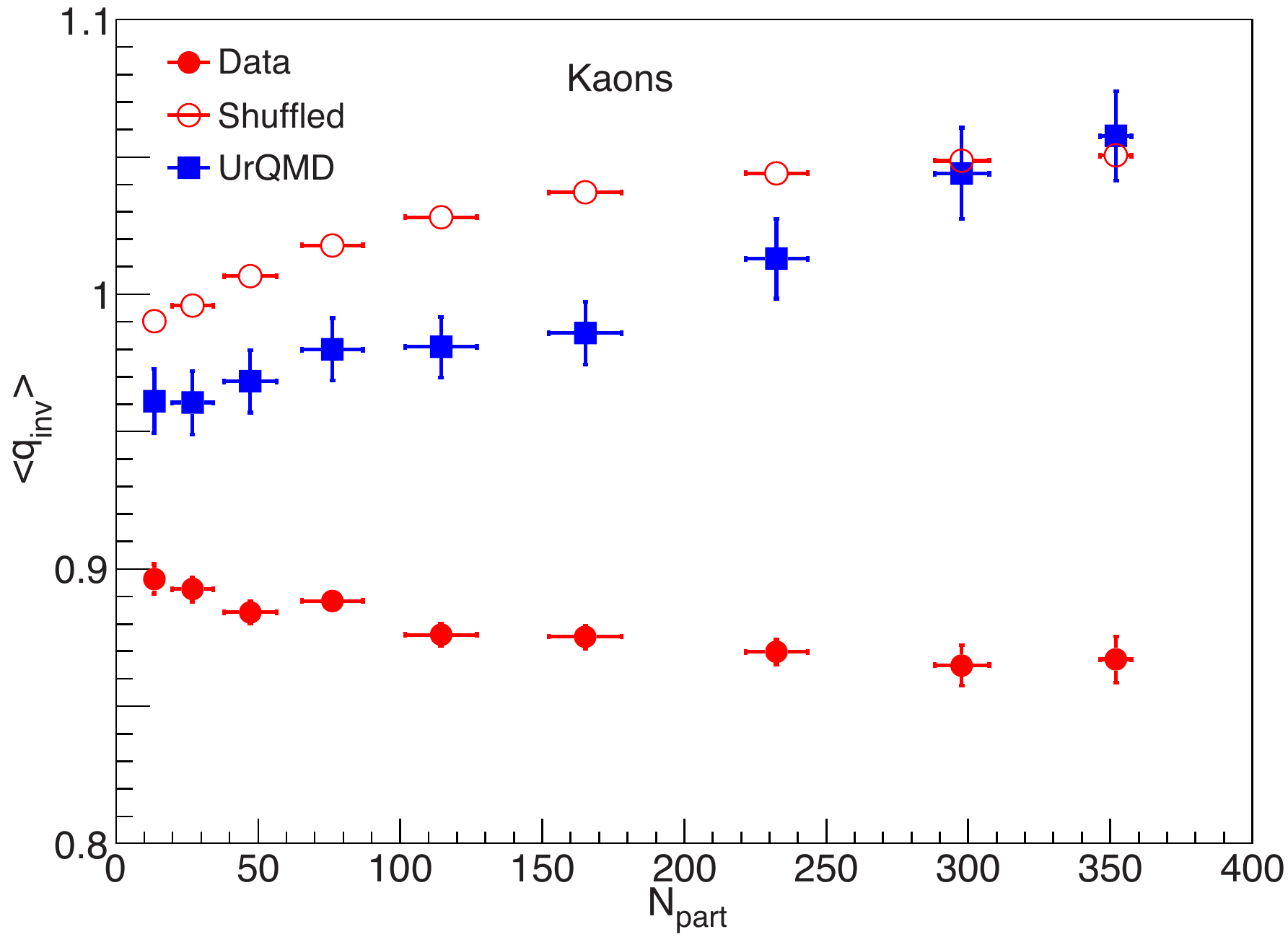}
\caption{\label{fig:fig6_4_10}Centrality dependence of the balance function widths of $q_{\rm inv}$ for identified charged kaons from Au+Au collisions at $\sqrt{s_{\rm NN}}$ = 200 GeV. }

\end{figure}

Figure~\ref{fig:fig6_4_10} shows the weighted average from Equation~\ref{WA} for identified charged kaons from Au+Au collisions at $\sqrt{s_{\rm NN}}$ = 200 GeV.  This determination of the width of the balance function does not assume a functional form.  A range of $0.36 < q_{\rm inv}  < 2.0$ GeV/$c$ is chosen to remove HBT/Coulomb effects and $\phi$ decay effects. Similar to the extracted widths from the thermal distribution, the weighted average shows a slight decrease at central collisions. while both UrQMD and shuffled events are much wider and increase in central collisions.

\chapter{Conclusions and Outlook}

In this paper, we present dynamical $K/\pi$, $p/\pi$, and $K/p$ ratio fluctuations from Au+Au collisions at $\sqrt{s_{\rm NN}}$ = 7.7 to 200 GeV. Overall, no non-monotonic behavior is observed in the energy dependence of fluctuations. The $K/\pi$ fluctuations results show little energy dependence, while the $p/\pi$, and $K/p$ fluctuations results are negative and decrease with decreasing collision energy.  The energy dependence of $p/\pi$, and $K/p$ can be understand in terms of stronger resonance production at lower energies. However, the disagreement  between STAR and NA49 data for $K/\pi$ and $K/p$ is still under discussion and no conclusion can be made yet. NA49 is a fixed target experiment and its acceptance changes dramatically as a function of the incident energy.  The fact that model comparisons between experiments using the respective acceptance of the detectors agree indicate that simple acceptance effects along can't explain the differences between two experiments. Other effects such as NA49's particle identification method might introduce correlations and could be the source of disagreement. 

We also discussed that simple multiplicity scaling failed to reproduce the energy dependence of $p/\pi$ and $K/p$ fluctuations. For $K/\pi$ fluctuation the question of whether the scaling holds is not conclusive due to the large statistical uncertainties at low energies. The failure of multiplicity scaling is expected because it assumes that the only change with beam energy is the system volume, which is not true because of effects like resonance production that become more dominant at lower energies.

To study correlations introduced by hadronic process like resonance decay, separate sign fluctuations are also measured. We observed that for all three cases, same-sign fluctuations are negative and become more negative with decreasing energy. The origin of negative same-sign fluctuations for $K/\pi$ and $K/p$ fluctuations is still under investigation since there is no particle that directly decays to same-sign pairs. Processes such as associated production could play an important role in $K/\pi$ and $K/p$ fluctuations.

We have measured the balance function for $p+p$, $d$+Au, and Au+Au collisions at $\sqrt{s_{\rm NN}}$ = 200 GeV for all charged particles, identified charged pions, and identified charged kaons.  We observe that the balance functions in terms of $\Delta \eta$ for all charged particles and in terms of $\Delta y$ and $q_{\rm inv}$ for charged pions narrow in central Au+Au collisions.  This centrality dependence is consistent with trends predicted by models incorporating delayed hadronization. The balance functions
$B(\Delta \eta)$ and $B(\Delta y)$ can be affected by radial flow while the balance function $B(q_{\rm inv})$ is largely unaffected by the implied reference frame transformation.  We observe that the system size dependence of the width of the balance function for charged particles scales with $N_{\rm part}$ as was observed at $\sqrt{s_{\rm NN}}$ = 17.3 GeV.  In contrast, HIJING and UrQMD model calculations for the width of the balance function in terms of $\Delta y$ or $\Delta \eta$ show no dependence on system size or centrality.

For charged kaons we observe that the width of the balance function $B(\Delta y)$ shows little dependence on centrality for Au+Au collisions at $\sqrt{s_{\rm NN}}$ = 200 GeV.  This lack of dependence on centrality may indicate that strangeness is created early in the collision rather than in a later hadronization stage.  The fact that the width of the balance function for charged kaons in terms of $\Delta y$ is independent of centrality was verified by removing $\phi$ decays using high statistics data.

For both pions and kaons, the width of the balance function in terms of $q_{\rm inv}$ decreases with increasing centrality.  This narrowing may be affected by the evolution of the kinetic freeze-out temperature with centrality.  This explanation is strengthened by the observation that the widths of the balance functions for pions in terms of the two transverse components of $q_{\rm inv}$, $q_{\rm out}$ and $q_{\rm side}$, decrease in central collisions.  However, more quantitative conclusions require more complete theoretical studies.

A comparison with a blast-wave model \cite{balance_blastwave} suggests that the balance function $B(\Delta y)$ for pion pairs in central Au+Au collisions at $\sqrt{s_{\rm NN}}$ = 200 GeV is as narrow as one could expect, as the model assumes that the balancing charges are perfectly correlated in coordinate space at breakup.  This correlation might be explained either by having the charges created late in the reaction, thus denying them the opportunity to separate in coordinate space, or having them created early, but maintaining their close proximity through very limited diffusion. Whereas the first explanation is motivated by a picture of delayed hadronization, the idea of limited diffusion is consistent with the matter having a very small viscosity, which also requires a small mean free path. Furthermore, both these explanations account for the observation that the balance function narrows with centrality, since the breakup temperature, which determines the width, falls with increasing centrality.  The additional information provided here concerning the decomposition of the balance function into $q_{\rm out}$, $q_{\rm side}$, and $q_{\rm long}$ may provide the basis for a more stringent test of competing theoretical pictures.

For the beam energy dependence, the balance function in terms of $\Delta \eta$ narrows in central collisions for all the energies from 7.7 to 200 GeV. The balance function width $\left\langle {\Delta \eta } \right\rangle$ shows a smooth decrease with increasing collision energy. This narrowing of the balance function and along with the difference between data and the UrQMD model could be a signal of onset of deconfinement or radial flow. A normalized width parameter is discussed.  We find that although designed to be independent of acceptance, the $W$ parameter shows a strong acceptance dependence. By applying the same acceptance cut, we find a good agreement between STAR and NA49. 

The reaction-plane-dependent balance function analysis gives the same difference between the like-sign and unlike-sign charge dependent azimuthal correlations as the three point correlator results published by STAR. A thermal blast-wave model incorporating local charge conservation and flow reproduces most of the difference between like- and unlike-sign charge-dependent azimuthal correlations. The good agreement between model and data indicates that, unlike previous interpretation as possible local parity violation, most of the observed signal is due to local charge conservation and flow. The reaction-plane-dependent balance function results show a smooth decrease with decreasing collision energy, which contradicts chiral magnet effect predictions but is consistent with the fact that event anisotropy $v_2$ decreases with decreasing energy.

\end{doublespace}




\begin{thebibliography}{???}

\bibitem{QCD_book}
G. Dissertori, I. Knowles, and M. Schmelling. {\it Quantum Chromodynamics - High Energy Experiments and Theory}. Oxford Universidy Press, 2003.

\bibitem{QCD_gauge}
Manfred B�hm, Ansgar Denner, and Hans Joos. {\it Gauge Theories of the strong and electroweak interaction}. B. G. Teubner, 2001.

\bibitem{Confinement}
Barger, R. Phillips,
{\it Collider Physics}. Addison�Wesley. ISBN 0201149451


\bibitem{Asymptotic_freedom}
D. Gross and F. Wilczek, 
Phys. Rev. Lett. {\bf 30}, 1343 (1973);

\bibitem{alpha_paper}
S. Bethke,
Prog. Part. Nucl. Phys. {\bf 58}, 351 (2007), hep-ex/0606035.

\bibitem{QCD_eos_paper}
Szabolcs Borsanyi  {\it et al}.
JHEP 1011, 077 (2010)

\bibitem{cross_over}
F. R. Brown  {\it et al}.
Phys. Rev. Lett. {\bf 65}, 2491 (1990)

\bibitem{collision_geometry}
Thomas Sch�fer
10.1103/Physics.2.88


H. Politzer,
Phys. Rev. Lett. {\bf 30}, 1346 (1973)


\bibitem{Color_super}
Mark G. Alford, Andreas Schmitt, Krishna Rajagopal and Thomas Sch�fer
Rev. Mod. Phys. {\bf 80}, 1455 (2008)

\bibitem{1st_order}
Shinji Ejiri, 
Phys. Rev. D {\bf 78} 074507 (2008)

\bibitem{critical_theory}
M. A. Stephanov
PoS LAT2006:024,2006

\bibitem{BES_write_up}
M.M.Aggarwal {\it et al.} [STAR Collaboration],
arXiv:1007.2613v1

\bibitem{NSAC}
2007 NSAC Long Range Plan, the figure is taken from 

http://rhig.physics.yale.edu/\textasciitilde ullrich/lrpfigs/


\bibitem{sss_thesis}
Shusu Shi,
Event anisotropy v2 at STAR, PhD Thesis.

\bibitem{glauber}
Glauber R J 1959 in Lectures in Theoretical Physics, 
edited by W E Brittin and L G Dunham (Interscience,N.Y), Vol. 1 315

\bibitem{glauber_review}
Michael L. Miller, Klaus Reygers, Stephen J. Sanders and Peter Steinberg,
Annu. Rev. Nucl. Part. Sci., {\bf 57}, 205-43 (2007).


\bibitem{sergei_flow_old}
S. Voloshin and Y. Zhang, 
Z. Phys. C 70, 665 ~1996

\bibitem{LQCD_critical}
 Phys. Rev. D {\bf 79}, 074505 (2009)

\bibitem{LQCD_pro}
PoS (CPOD07) 026

\bibitem{p_k_core} 
Phys. Rev. Lett. {\bf 95}, 182301 (2005)

\bibitem{sigma_dyn}
Phys. Rev. Lett. {\bf 86}, 1965 (2001)

\bibitem{nu_dyn}
C. Pruneau, S. Gavin, and S. Voloshin, Phys. Rev. C {\bf66}, 044904 (2002).


\bibitem{charge_fluct}
S.~Jeon and V.~Koch,
Phys.\ Rev.\ Lett.\  {\bf 85}, 2076 (2000).
  
\bibitem{balance_PRL}
S.A. Bass, P. Danielewicz, and S. Pratt,
Phys. Rev. Lett. {\bf 85}, 2689 (2000).

\bibitem{balance_PRC}
B. I. Abelev {\it et al}. [STAR Collaboration],
Phys. Rev. C {\bf 82}, 024905 (2010).

\bibitem{star_balance_130}
J. Adams {\it et al.} [STAR Collaboration],
Phys. Rev. Lett. {\bf 90}, 172301 (2003).  

\bibitem{hydro_model}
P. Huovinen and P.V. Ruuskanen,
Annu. Rev. Nucl. Part. Sci. 2006. 56:163�206

\bibitem{UrQMD_1}
S. A. Bass, {\it et al.} 
Prog. Part. Nucl. Phys. 41 (1998) 225-370


\bibitem{UrQMD_2}
M. Bleicher,{\it et al.} 
J. Phys. G: Nucl. Part. Phys. 25 (1999) 1859-1896

\bibitem{UrQMD_web}
http://urqmd.org/

\bibitem{Pythia}
Torbjorn Sjostrand, Stephen Mrenna, and Peter Skands. 
PYTHIA 6.4 Physics and Manual. JHEP, 05:026, 2006.

\bibitem{HIJING_2}
Miklos Gyulassy and Xin-Nian Wang, 
Comput. Phys. Commun. {\bf 83}, 307 (1994).

\bibitem{HIJING}
Xin-Nian Wang and Miklos Gyulassy, 
Phys.Rev.D {\bf 44}, 3501 (1991).

\bibitem{RHIC}
Nucl. Instrum. Meth. A  {\bf499} (2003) 235�244


\bibitem{e_cooling}
T. Satogata {\it et al.}, PoSCPOD07:051,2007

\bibitem{run_summary}
table from http://www.agsrhichome.bnl.gov/RHIC/Runs/

\bibitem{RHIC_5years}
Achim Franz,
Nucl. Instrum. Meth. A  {\bf 566} (2006) 54�61

\bibitem{STAR_nim}
K.H. Ackermann et al. Nucl. Instrum. Meth. A  {\bf499} (2003) 624�632

\bibitem{SVT_nim}
R. Bellwied et al. Nucl. Instrum. Meth. A  {\bf499}  (2003) 640-651

\bibitem{SSD_nim}
L. Arnold et al. Nucl. Instrum. Meth. A  {\bf499}  (2003) 652-658

\bibitem{BEMC_nim}
M. Beddo et al. Nucl. Instrum. Meth. A  {\bf499}  (2003) 725-739

\bibitem{FTPC_nim}
K.H. Ackermann et al. Nucl. Instrum. Meth. A  {\bf499}  (2003) 713�719

\bibitem{EEMC_nim}
C.E. Allgower et al. Nucl. Instrum. Meth. A  {\bf499}  (2003) 740-750

\bibitem{TPC_nim}
K.H. Ackermann et al. Nucl. Instrum. Meth. A  {\bf499}  (2003) 659�678

\bibitem{TPC_readout_nim}
M. Anderson et al. Nucl. Instrum. Meth. A  {\bf499}  (2003) 679�691

\bibitem{TPC_alex}
Figure produced by  Alexander Schmah

\bibitem{TOF_Kohei}
A Large Area Time of Flight Detector for the STAR Experiment at RHIC
Kohei Kajimoto, PhD Thesis.

\bibitem{TOF_nim}
W.J. Llope, Nucl. Instr. and Meth. A (2010), doi:10.1016/j.nima.2010.07.086

\bibitem{TOF_eff_xing}
Xing Dong, private communication

\bibitem{flow_sergei}
Phys. Rev. C {\bf 58},1671 (1998).



\bibitem{kpi_NA49}
C. Alt et al., 
Phys. Rev. C {\bf 79}, 044910 (2009).


\bibitem{kpi_STAR}
J. Adams {\it et al.} [STAR Collaboration],
Phys. Rev. Lett. {\bf 103}, 092301 (2009)


\bibitem{SH_Torrieri}
G. Torrieri, 
Int. J. Mod. Phys. E {\bf 16}, 1783 (2007).

\bibitem{pk_correlation}
V. Koch, A. Majumder, and J. Randrup
Phys. Rev. Lett. {\bf 95}, 182301 (2005)

\bibitem{pk_NA49}
 T. Anticic et al. [NA49 collaboration],
 Phys. Rev. C {\bf 83}, 061902(R) (2011) 
 
 \bibitem{UrQMD_decay}
 Dmytro Kresan and Volker Friese
 PoS(CFRNC2006)017
 
  \bibitem{kpi_scaling}
V. Koch and T. Schuster,
Phys. Rev. C {\bf 81}, 034910 (2010)
 
\bibitem{balance_theory}
S. A. Bass, P. Danielewicz, and S. Pratt  ,
Phys. Rev. Lett. {\bf 85}, 2689 (2000).

\bibitem{balance_distortions_jeon}
S. Jeon and S. Pratt,
Phys. Rev. C {\bf 65}, 044902 (2002).

\bibitem{balance_distortions}
S. Pratt and S. Cheng,
Phys. Rev. C {\bf 68}, 014907 (2003).

\bibitem{balance_blastwave}
S. Cheng, S. Petriconi, S. Pratt, M. Skoby,
C. Gale, S. Jeon, V. Topor Pop, and Q. Zhang,
Phys. Rev. C {\bf 69}, 054906 (2004).

\bibitem{star_rho}
J. Adams {\it et al.} [STAR Collaboration],
Phys. Rev. Lett. {\bf 92}, 092301 (2004).

\bibitem{Bozek}
P. Bo\.{z}ek,
Phys. Lett. B {\bf 609}, 247 (2005).

\bibitem{Teaney}
D. Teaney,
Phys. Rev. C {\bf 68}, 034913 (2003).

\bibitem{STAR_identified_spectra}
J. Adams {\it et al.} [STAR Collaboration],
Phys. Rev. Lett. {\bf 92}, 112301 (2004).

\bibitem{star_deta_dphi_cf_200}
J. Adams {\it et al.} [STAR Collaboration],
J. Phys. G {\bf 32}, L37 (2006).

\bibitem{star_deta_dphi_cf}
J. Adams {\it et al.} [STAR Collaboration],
Phys. Lett. B {\bf 634}, 347 (2006).


\bibitem{NA49_PRC_2007}
C. Alt {\it et al.} [NA49 Collaboration],
Phys. Rev. C {\bf 76}, 024914 (2007).

\bibitem{LPV}
D. Kharzeev, R. D. Pisarski and M. H. G. Tytgat, 
Phys. Rev. Lett. {\bf 81}, 512 (1998).

\bibitem{CME_1}
D. Kharzeev, 
Phys. Lett. B {\bf 633} 260 (2006).

\bibitem{CME_2}
D. Kharzeev, L. McLerran, and H. Warringa, 
Nucl. Phys. A {\bf 803} 227 (2008).

 \bibitem{CME_3}
K. Fukushima, D. Kharzeev, and H. Warringa,
Phys. Rev. D {\bf 78} 074033 (2008).

\bibitem{3_point_correlator}
S. A. Voloshin,
Phys. Rev. C {\bf 70} 057901 (2004).


\bibitem{parity_soeren}
S. Schlichting and S. Pratt,
Phys. Rev. C {\bf 83}, 014913 (2011).

\bibitem{STAR_v2}
B. I. Abelev {\it et al}. [STAR Collaboration],
Phys. Rev. C {\bf 72},14904 (2005).

\bibitem{parity_PRL}
B. I. Abelev {\it et al}. [STAR Collaboration],
Phys. Rev. Lett. {\bf 103}, 251601 (2009).

\bibitem{CME_BES}
V. Toneev and V. Voronyuk, 
EPJ Web of Conferences {\bf 13}, 02005 (2011).

 \end{thebibliography}
\end{document}